\documentclass{aa}  

\usepackage{graphicx}
\usepackage[utf8]{inputenc}
\usepackage[varg]{txfonts}
\usepackage[breaklinks, colorlinks, citecolor=blue, linkcolor=blue]{hyperref}
\usepackage[usenames, dvipsnames]{color}
\usepackage{amsmath}
\usepackage{graphicx}
\usepackage[english]{babel}
\usepackage{siunitx}
\usepackage{float}
\usepackage{url}
\usepackage{longtable}
\usepackage{fancyhdr}
\usepackage{footmisc}
\usepackage{natbib}
\usepackage[normalem]{ulem}
\usepackage{caption}
\usepackage{multirow}
\usepackage{color}
\usepackage{array} 
\usepackage{ulem}
\usepackage[flushleft]{threeparttable}

\usepackage{txfonts}
\usepackage[table]{xcolor}
\usepackage{orcidlink}
\makeatletter
\def\instrefs#1{{\def\scsep{\def\scsep{,}}\@for\w:=#1\do{\scsep\ref{inst:\w}}}}
\renewcommand{\inst}[1]{\unskip$^{\instrefs{#1}}$}

\renewcommand*\aa@pageof{, page \thepage{} of \pageref*{LastPage}} % https://tex.stackexchange.com/questions/345764/journal-class-shows-package-hyperref-warning-suppressing-link-with-empty-targe

\makeatother

\begin{document} 

\title{The GAPS programme at TNG: TBD}
\subtitle{Studies of atmospheric FeII winds in ultra-hot Jupiters KELT-9b and KELT-20b using the HARPS-N spectrograph}

   %\subtitle{I. Overviewing the $\kappa$-mechanism}
    \titlerunning {FeII in the atmospheres of ultra-hot Jupiters KELT-9b and KELT-20b}
     \authorrunning{M. Stangret et al.}
    
   \author{M.~Stangret\inst{inaf_padova}\orcidlink{0000-0002-1812-8024} 
   \and
   L.~Fossati\inst{graz} %comments in the overleaf
   \and
   M.~C.~D’Arpa\inst{inaf_palermo,uni_palermo} % comments send in pdf in the 1st round , 
   \and
   F.~Borsa\inst{inaf_brera} %comments send, comments almost all implemented
   \and
   V.~Nascimbeni\inst{inaf_padova} 
   \and
   L.~Malavolta\inst{dep_padova,inaf_padova} %to be send, but agreed 
   \and
   D.~Sicilia\inst{inaf_catania} %comments to be send in 2nd version
   \and
   L.~Pino\inst{inaf_arcetri} %comments send
   \and
   F.~Biassoni\inst{inaf_brera} 
   \and
   A.~S.~Bonomo\inst{inaf_torino} %comments send
   \and
   M.~Brogi\inst{uni_torino,inaf_torino}\orcidlink{0000-0002-7704-0153} % comments send
    \and
    R.~Claudi\inst{inaf_padova}
    \and
    M.~Damasso\inst{inaf_torino} %additional author
    \and
    C.~Di Maio\inst{inaf_palermo} %additional author
 %additional author
    \and
    P.~Giacobbe\inst{inaf_torino} %additional author
    \and
    G.~Guilluy\inst{inaf_torino} %comments send, comments almost all implemented
    \and
    A.~Harutyunyan\inst{Galileo}
    \and
    A.~F.~Lanza\inst{inaf_catania} %comments received, grammar comments, to come back to this pdf    \and
     A.~F.~Martínez Fiorenzano\inst{Galileo}\orcidlink{0000-0002-4272-4272}
    \and  
    L.~Mancini\inst{uni_roma,max_planck,inaf_torino}\orcidlink{0000-0002-9428-8732} %comments received (small - mail, not pdf)
    \and
    D.~Nardiello\inst{dep_padova,inaf_padova}\orcidlink{0000-0003-1149-3659}  % comments send
    \and
    G.~Scandariato\inst{inaf_catania} % comments send
    \and
    A.~Sozzetti\inst{inaf_torino} % comments send
    \and
    T.~Zingales\inst{dep_padova,inaf_padova}\orcidlink{0000-0001-6880-5356}  % comments send, comments almost all implemented
          }

\institute{
\label{inst:inaf_padova}INAF – Osservatorio Astronomico di Padova, Vicolo dell'Osservatorio 5, 35122, Padova, Italy 
\\ \email{monika.beata.stangret@gmail.com}%1
\and
\label{inst:graz}Space Research Institute, Austrian Academy of Sciences, Schmiedlstrasse 6, 8042 Graz, Austria
\and
\label{inst:inaf_palermo}INAF – Osservatorio Astronomico di Palermo, Piazza del Parlamento, 1, I-90134 Palermo, Italy
\and
\label{inst:uni_palermo}University of Palermo, Department of Physics and Chemistry “Emilio Segrè, Via Archirafi 36, Palermo, Italy
\and
\label{inst:inaf_brera}INAF – Osservatorio Astronomico di Brera, Via E. Bianchi 46, 23807 Merate, Italy
\and
\label{inst:dep_padova}Dipartimento di Fisica e Astronomia "Galileo Galilei" – Università degli Studi di Padova, Vicolo dell’Osservatorio 3, I-35122 Padova, Italy
\and
\label{inst:inaf_catania}INAF – Osservatorio Astrofisico di Catania, Via S. Sofia 78, 95123 Catania, Italy
\and
\label{inst:inaf_arcetri}INAF – Osservatorio Astrofisico di Arcetri, Largo Enrico Fermi 5, 50125 Firenze, Italy
\and
\label{inst:inaf_torino}INAF – Osservatorio Astrofisico di Torino, Via Osservatorio 20, 10025, Pino Torinese, Italy
\and
\label{inst:uni_torino}Dipartimento di Fisica, Università degli Studi di Torino, via Pietro Giuria 1, I-10125, Torino, Italy 
\and
\label{inst:Galileo} Fundación Galileo Galilei - INAF, Rambla José Ana Fernandez Pérez 7, E-38712 Breña Baja, Tenerife, Spain
\and
\label{inst:uni_roma}Department of Physics, University of Rome ``Tor Vergata``, Via
della Ricerca Scientifica 1, I-00133, Rome, Italy
\and
\label{inst:max_planck}Max Planck Institute for Astronomy, K\"{o}nigstuhl 17,D-69117, Heidelberg, Germany
}

%   \institute{
%   \label{inst:inaf_padova}INAF – Osservatorio Astronomico di Padova, Vicolo dell'Osservatorio 5, 35122, Padova, Italy\\
              %\email{monika.stangret@inaf.it}}

%\date{Received XXX; accepted XXX}

% \abstract{}{}{}{}{} 
% 5 {} token are mandatory
 
  \abstract{Ultra-hot Jupiters (UHJs) are gas giant planets orbiting close to their host star, with equilibrium temperatures exceeding 2000~K, and among the most studied planets in terms of their atmospheric composition. Thanks to a new generation of ultra-stable high-resolution spectrographs, it is possible to detect the signal from the individual lines of the species in the exoplanetary atmospheres.
  
   We employed two techniques in this study. First, we used transmission spectroscopy, which involved examining the spectra around single lines of \ion{Fe}{ii}. Then we carried out a set of cross-correlation studies for two UHJs: KELT-9b and KELT-20b. Both planets orbit fast-rotating stars, which resulted in the detection of the strong Rossiter-McLaughlin (RM) effect and center-to-limb variations in the transmission spectrum. These effects had to be corrected to ensure a precise analysis. 
  
  Using the transmission spectroscopy method, we detected 21 single lines of \ion{Fe}{ii} in the atmosphere of KELT-9b. All of the detected lines are blue-shifted, suggesting strong day-to-night side atmospheric winds. The cross-correlation method leads to the detection of the blue-shifted signal with a signal-to-noise ratio (S/N) of 13.46. Our results are in agreement with models based on  non-local thermodynamical equilibrium (NLTE) effects, with a mean micro-turbulence of $ \nu_{\rm mic}$ = 2.73 $\pm$ 1.5 km\,s$^{-1}$ and macro-turbulence of $\nu_{\rm mac}$ = 8.22 $\pm$ 3.85 km\,s$^{-1}$.  

  In the atmosphere of KELT-20b, we detected 17 single lines of \ion{Fe}{ii}. Considering different measurements of the systemic velocity of the system, we conclude that the existence of winds in the atmosphere of KELT-20b cannot be determined conclusively. The detected signal with the cross-correlation method presents a S/N of 11.51. The results are consistent with NLTE effects, including means of $\nu_{\rm mic}$ = 3.04 $\pm$ 0.35 km\,s$^{-1}$ and $\nu_{\rm mac}$ = 6.76 $\pm$ 1.17 km\,s$^{-1}$.  
  }
   
   


   \maketitle
%
%-------------------------------------------------------------------

\section{Introduction}\label{Section:introduction}

%High-resolution spectroscopic observations allow us to differentiate the signal coming from the atmosphere of the exoplanet, its host star, and the Earth's atmosphere. One of the most suitable targets to study the chemical composition of their atmospheres using both transmission and emission spectroscopy are hot Jupiters and ultra-hot Jupiters (UHJs). Thanks to their short orbital period and hot extended atmospheres, in which the equilibrium temperature ($T_{eq}$) exceeds 2000 K for UHJs \citep{Parmentier_2018}, these planets are an ideal laboratory for those studies. Due to its tidally locked nature which leads to extreme temperature differences between the day and night side of the atmosphere, the different chemical composition on both sides is expected and in addition, the strong day-to-night side winds are observed, reported in several UHJs.

High-resolution spectroscopic observations allow us to differentiate the signals coming from an exoplanet's atmosphere, its host star, and the Earth's atmosphere. Hot Jupiters and ultra-hot Jupiters (UHJs) are among the best targets for studying the chemical composition of their atmospheres through the application of both transmission and emission spectroscopy.

Thanks to their short orbital periods and hot extended atmospheres, where the equilibrium temperature ($T_{eq}$) exceeds 2000 K \citep{Parmentier_2018}, UHJs are ideal laboratories for atmospheric studies. Due to their tidally locked nature, which leads to extreme temperature differences between the day and night sides, a different chemical composition on both sides is expected and, in addition, strong day-night side winds are observed, as reported for several UHJs. In theoretical simulations, \citet{Helling_2019} demonstrated that clouds cannot form due to the extremely high temperatures observed on the day side of the atmospheres. Furthermore, atoms are found in their ionized state, while  molecules, due to their total or partial dissociation, typically have a low abundance \citep{Arcangeli_2018}.

In recent years, studies of UHJ atmospheres have been performed using high-resolution transmission and emission spectroscopy, revealing the presence of a diverse range of species, including neutral and ionized atoms, such as \ion{Ba}{ii}, \ion{Ca}{i}, \ion{Ca}{ii}, \ion{Cr}{i}, \ion{Cr}{ii}, \ion{Fe}{i}, \ion{Fe}{ii}, \ion{H}{i}, \ion{Mg}{i}, \ion{Mg}{ii}, \ion{Na}{i}, \ion{Ni}{i}, \ion{Ni}{ii}, \ion{Rb}{i}, \ion{Si}{i} \ion{Sm}{i}, \ion{Ti}{i}, \ion{Ti}{ii}, \ion{V}{i}, and \ion{Y}{ii}; for example: MASCARA-2b/KELT-20b (\citealt{Casasayas_M2_2018,M2_Casasayas_2019}, \citealt{Langeveld_M2_NaD1D2}, \citealt{Yan_K9_CaII}, \citealt{Cauley_K9} \citealt{Wyttenbach_K9_H}, \citealt{Borsa_K9_O}, \citealt{Pai_K9_FeII}, \citealt{Langeveld_M2_NaD1D2}, \citealt{Sanchez_K9_pashen}, \citealt{Fossati_2023_M2}),  WASP-12b \citep{w12_Jensen}, WASP-33b (\citealt{w33_Nugroho_1,w33_Nugroho_2}, \citealt{w33_Yan}), WASP-121b (\citealt{W121_Cabot}, \citealt{Borsa_w121_RM}, \citealt{W121_Young}, \citealt{w121_Maguire}), WASP-189b (\citealt{W189_Yan_1,W189_Yan_2},
 \citealt{Stangret_6plan}, \citealt{W189_Prinoth_1,W189_Prinoth_2}), and WASP-76b (\citealt{Seidel_2019_w76}, \citealt{Ehrenreich_2020_w76}).

Furthermore, studies of UHJs suggest the presence of a thermal inversion in the middle atmosphere. This phenomenon has been observed in several planets including WASP-33b (\citealt{w33_Nugroho_2}, \citealt{Cont_M2}), WASP-103b \citep{Kreidberg_w103}, KELT-9b \citep{Pino_K9_2022}, and WASP-189b \citep{W189_Yan_1}. These observational results are also supported by forward atmospheric models \citep[e.g.,][]{Lothringer2018,Fossati_NLTE_K9,Fossati_2023_M2},  which suggest that the thermal inversion is due to metal line heating, particularly for planets orbiting stars hotter than $\approx$8300\,K that do not present the high-energy (X-ray and extreme ultraviolet) emission responsible for hydrogen photoionization. With respect to the particular cases of KELT-9b and KELT-20b, which are the focus of this work, \citet{Fossati_NLTE_K9,Fossati_2023_M2} showed that non-local thermodynamical equilibrium (NLTE) effects magnify the inversion through metal-line absorption of stellar UV radiation. Specifically, NLTE leads to significant Fe{\sc ii} overpopulation and Mg{\sc ii} underpopulation, which respectively drive atmospheric heating and cooling, increasing the magnitude of the thermal inversion. These results are supported by the fact that transmission spectra computed accounting for NLTE effects have been so far the only forward models capable of fitting the observed hydrogen Balmer line profiles of these two UHJs \citep{Fossati_NLTE_K9,Fossati_2023_M2}.

In this study, we present an analysis of the chemical composition of two the UHJs KELT-9b and KELT-20b, focusing on the transmission spectroscopic studies of \ion{Fe}{ii} with high-resolution spectroscopic data. The paper is organized as follows. In Sect. \ref{sec:planets} we present a detailed literature review of the studied planets, while in Sect. \ref{Section:Observations} we describe the observations used in this work. In Sect. \ref{Section:TS}, the transmission spectroscopy analysis is presented, with the results obtained for KELT-9b and KELT-20b shown in Sects. \ref{Section:K9-detected_lines} and \ref{Section:M2-detected_lines}, respectively. In Sect. \ref{SEC:TS_models} we focus on the comparison of the results with atmospheric models. The analysis of the transmission spectrum using the cross-correlation method is described in Sect. \ref{SEC:CC}.

\section{Studied planets}\label{sec:planets}

KELT-9b is a UHJ ($M_p=2.88$ M$_J$, $R_p=1.936$ R$_J$) with $T_{eq}=3921$~K. The planet orbits a bright ($V=7.56$~mag) A0-type star in 1.48 days (\citealt{Gaudi_K9_2017}, \citealt{Borsa_K9}). 
The considered stellar and planetary parameters can be found in the Table \ref{tab:params_K9_K20}. The orbital and physical parameters make KELT-9b an ideal candidate for atmospheric studies using transmission and emission spectroscopy. The first detection of the atmospheric signature of KELT-9b was presented by \citet{Yan_Henning_2018}, revealing an extended atmosphere through the H$\alpha$ line. Transmission spectroscopic studies focusing on single lines additionally showed the presence of \ion{Ca}{ii}, \ion{Fe}{ii} ($\lambda$4923.9 $\AA$, $\lambda$5018.4, $\AA$, $\lambda$5169 $\AA$, $\lambda$5197.5 $\AA$, $\lambda$5234.6, $\AA$ $\lambda$5276 $\AA$, $\lambda$5316.6 $\AA$, $\lambda$5363.9 $\AA$, and $\lambda$6456.3 $\AA$), H$\beta$, \ion{Na}{i} (D1 and D2), \ion{O}{i} ($\lambda$7774. 0 $\AA$ triplet), and Pashen $\beta$ (\citealt{Yan_K9_CaII}, \citealt{Cauley_K9}, \citealt{Turner2020_CaII_K9}, \citealt{Wyttenbach_K9_H}, \citealt{Borsa_K9_O}, \citealt{Pai_K9_FeII}, \citealt{Langeveld_M2_NaD1D2}, \citealt{Sanchez_K9_pashen}). Emission and transmission spectroscopy studies using the cross-correlation method have reported the detection of \ion{Ca}{i}, \ion{Ca}{ii}, \ion{Cr}{ii}, \ion{Fe}{i}, \ion{Fe}{ii}, \ion{Mg}{i}, \ion{Mg}{ii}, \ion{Na}{i}, \ion{Ni}{i}, \ion{Sc}{ii}, \ion{Si}{i}, \ion{Sr}{ii}, \ion{Tb}{ii}, \ion{Ti}{ii}, and \ion{Y}{ii} (\citealt{Hoeijmakers_K9_1,Hoeijmakers_K9_2}, \citealt{Kasper_K9_Fe_CC}, \citealt{Pino_K9_2022}, \citealt{Lowson_K9}, \citealt{Ridden-Harper_K9}, \citealt{Borsato_K9_2,Borsato_K9_1}). In addition, using space-based observations, \citet{Changeat_K9} and \citet{Jacobs_K9} presented the detection of TiO, VO, FeH, and H$^-$. 

\begin{table}[h!]
\small\small
\centering
\caption{Physical and orbital parameters of the KELT-9 and KELT-20 systems. }
\begin{tabular}{lc c}
\hline \hline
\\[-1em]
 Parameter  & KELT-9 & KELT-20   \\ \hline
 \\[-1em]
  \multicolumn{3}{c}{\dotfill\it Stellar parameters \dotfill}\\\noalign{\smallskip} 

 %\multicolumn{3}{c}{\it Stellar parameters}\\\noalign{\smallskip}
   \\[-1em]
\quad  $T_{\rm eff}$ [K] & 9600 $\pm$ 400 & 8980$^{+90}_{-130}$ \\
  \\[-1em]
\quad  $\log g$ [cgs]& 4.1 $\pm$ 0.3 & 4.31 $\pm$ 0.02\\
  \\[-1em]
\quad  [Fe/H] & 0.14 $\pm$ 0.30 & -0.02 $\pm$ 0.07  \\
  \\[-1em]
\quad $M_{\star}$ [$\rm{M_{\odot}}$]& 2.32$\pm$ 0.16 & 1.89$^{+0.06}_{-0.05}$ \\
  \\[-1em]
\quad $R_{\star}$ [$\rm{R_{\odot}}$]& 2.418$\pm$ 0.058 &1.60 $\pm$ 0.06   \\
  \\[-1em]
 %\quad  $v\sin i_{\star}$$^a$ [km\,s$^{-1}$]& $0.90^{+0.09}_{-0.12}$ \\
  \quad  $v\sin i_{\star}$ [km\,s$^{-1}$]& 111.4 $\pm$ 1.3 \tablefoottext{a} &114 $\pm$ 3 \\
    \quad  $K_{\star}$ [m\,s$^{-1}$]&  293 $\pm$ 32 & <311.3  \\
  \\[-1em]
  \multicolumn{3}{c}{\dotfill\it Planet parameters \dotfill}\\\noalign{\smallskip}
% \multicolumn{3}{c}{\it Planet parameters}\\\noalign{\smallskip}
   \\[-1em]
 \quad $M_{\rm p}$ [$\rm{M_{J}}$]& 2.88 $\pm$ 0.35 &<3.510  \\
  \\[-1em]
 \quad $R_{\rm p}$ [$\rm{R_{J}}$]&   1.936 $\pm$ 0.047 & 1.83 $\pm$ 0.07\\
 \\[-1em]
  \quad $T_{\rm eq}$ [$\rm K$]& 3921$^{+182}_{-174}$ & 2260 $\pm$ 50 \\
 \\[-1em]
\quad $K_{\rm p}$ [km\,s$^{-1}$]& 236.4 $\pm$ 9.0 \tablefoottext{e} & 157.4 $\pm$ 17.1\tablefoottext{f}\\
  \\[-1em]
  \multicolumn{3}{c}{\dotfill\it Transit parameters \dotfill}\\\noalign{\smallskip}
% \multicolumn{3}{c}{\it Transit parameters}\\\noalign{\smallskip}
   \\[-1em]
 \quad $T_{\rm 0}$ [BJD$_{\rm TDB}$] & 
 2,459,006.3289  \tablefoottext{b} &2,457,503.120120\tablefoottext{g} \\
   & $\pm$ 0.0001 &$\pm$ 0.00018 \\
  \\[-1em]
\\[-1em]
 \quad $P$ [d] & 1.4811188 \tablefoottext{b}  & 3.47410196\tablefoottext{g} \\
 &  $\pm$ 0.0000003 &  $\pm$ 0.00000106 \\
  \\[-1em]
 \quad $T_{14}$ [hr] & 0.15949 $\pm$ 0.00011\tablefoottext{c} & 0.14882$^{+0.00092}_{-0.00090}$  \\
 \\[-1em]
  \multicolumn{3}{c}{\dotfill\it System parameters \dotfill}\\\noalign{\smallskip}
% \multicolumn{3}{c}{\it System parameters}\\\noalign{\smallskip}
   \\[-1em]
  \quad $a$ [au]& 0.03368 $\pm$ 0.00078& 0.0542$^{+0.0014}_{-0.0021}$  \\
 \\[-1em]
 \quad $i_{\rm p}$ [deg]& 86.79 $\pm$ 0.25\tablefoottext{d}  & 86.15$^{+0.28}_{-0.27}$   \\
 \\[-1em]
 %\quad $e$& - & \\
%\\[-1em]
%\quad $K_{\star}$ [m\,s$^{-1}$]& $179.8 \pm 9.0$  & \citet{kelt-14_2016_Rodrigez} \\
%  \\[-1em]
% \quad $\lambda$$^a$ [deg]& $-1.71^{+2.72}_{-2.31}$   \\
   
  \quad $\lambda$ [deg]& -85.78 $\pm$ 0.23 \tablefoottext{b} &   3.9   $\pm$   1.1 \tablefoottext{h}    \\
   \\[-1em]
  \quad e [deg]& 0 (fixed) &   0 (fixed)    \\
   \\[-1em]
%\quad $v_{\rm sys}$ [km\,s$^{-1}$]&   \\
%\\[-1em]
\end{tabular}
\tablefoot{The physical and orbital parameters for KELT-9 were adopted from \citet{Borsa_K9}, except for \tablefoottext{a} {adopted from \citet{Gaudi_K9_2017},} \tablefoottext{b} {\citet{Pino_K9_2022},} \tablefoottext{c}{\citet{Pai_K9_FeII},} \tablefoottext{d}{\citet{Lambda_K9},} and \tablefoottext{e}{retrieved in this work from single line analysis.} The physical and orbital parameters for KELT-20 were adopted from \citet{Talens_M2}, except for \tablefoottext{f}{retrieved in this work}, \tablefoottext{g}{adopted from \citet{M2_Casasayas_2019},} and \tablefoottext{h}{adopted from \citet{lambda_M2}.}}
\label{tab:params_K9_K20}
\end{table}

The first work focusing on the analysis of single \ion{Fe}{ii} lines, presented by \citet{Cauley_K9}, described the detection of nine lines with velocity shifts ranging from $-$1.02 to 3.0 km s$^{-1}$. The most recent studies presented by \citet{Pai_K9_FeII} focus on wind measurements in the atmosphere of KELT-9b, concentrating on single lines of \ion{Fe}{ii}, showing significant day-to-night winds from $-$4 to $-$10 km s$^{-1}$. The substantial differences in the detected wind velocities come from the different choices for the mid-transit time.%, which differs from the first measurements by up to 10 min. 

\begin{figure}[h!]
\centering
\includegraphics[width=0.95\hsize]{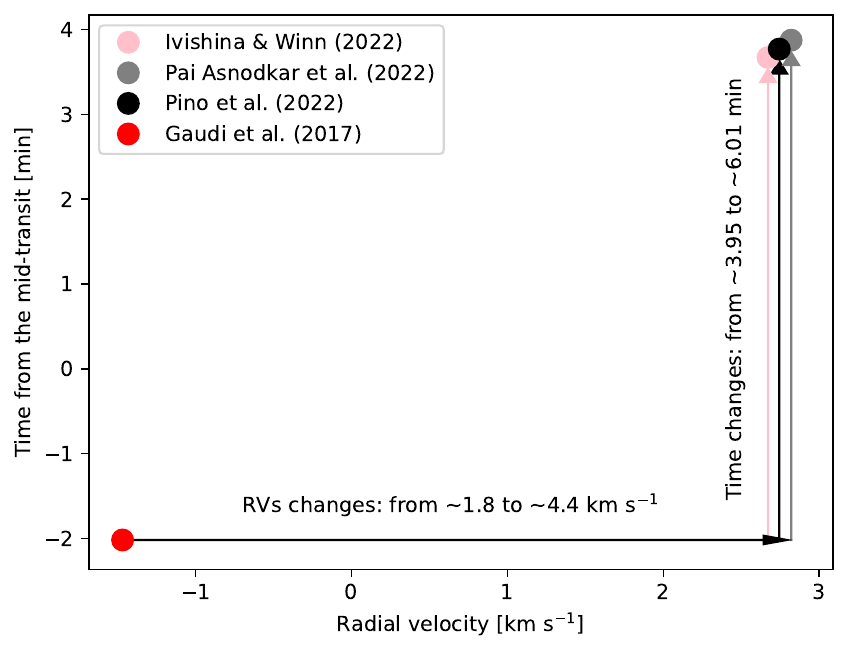}
\caption{Illustrative example displaying the dependence of time from the mid-transit and the radial velocity of the planet change on the T0 and orbital period (for only one exposure), as calculated by \citealt{Gaudi_K9_2017}, \citealt{Pai_K9_FeII}, \citealt{Pino_K9_2022}, and \citealt{Ivshina_Winn_2022_K9}. The values radial velocities of the planet change between 1.8 to 4.4 km~s$^{-1}$ and the time difference from the mid-transit changes from 3.95 to 6.01 min depending on the exposure used in this work, in comparison with the values assuming T0 and period from \citealt{Gaudi_K9_2017}.}
\label{Fig:T0}
\end{figure}

In Fig. \ref{Fig:T0}, we present an example for one exposure (close to the mid-transit) of the measurement of the time from the mid-transit and the radial velocity of one of the exposures assuming four different literature parameters of T0 and orbital period from  \citealt{Gaudi_K9_2017}, \citealt{Pai_K9_FeII}, \citealt{Pino_K9_2022}, and \citealt{Ivshina_Winn_2022_K9}.  The difference between the exposures used in this work varies from 1.8 to 4.4 km~s$^{-1}$ and 3.95 to 6.01 min from the first published measurements by \citealt{Gaudi_K9_2017}.  

KELT-20b \citep{Lund_M2}, also known as MASCARA-2b \citep{Talens_M2}, is an UHJ ($M_p<3.51$ M$_J$, $R_p=1.83$ R$_J$) with $T_{eq}=2260$~K, orbiting a fast-rotating A-type star in 3.47 days. The stellar and planetary parameters can be found in Table \ref{tab:params_K9_K20}. The chemical composition of the atmosphere of KELT-20b has been studied by several groups with low- and high-resolution spectrographs using both transmission and emission spectroscopy. \citet{Casasayas_M2_2018,M2_Casasayas_2019}, \citet{Langeveld_M2_NaD1D2}, and \citet{Fossati_2023_M2} presented single line detection of \ion{Ca}{ii} ($\lambda$8498 $\AA$, $\lambda$8542 $\AA$, and $\lambda$8662 $\AA$), \ion{Fe}{ii} ($\lambda$5018 $\AA$, $\lambda$5169 $\AA$, and $\lambda$5316 $\AA$), \ion{Mg}{i} ($\lambda$5173 $\AA$), \ion{Na}{i} (D1 and D2), H$\alpha$, H$\beta$, and H$\gamma$. In addition, using the cross-correlation method the existence of \ion{Ca}{ii}, \ion{Cr}{ii}, \ion{Fe}{i}, \ion{Fe}{ii}, \ion{Na}{i}, \ion{Ni}{i}, \ion{Ni}{ii}, \ion{Mg}{i}, and \ion{Si}{i} have been reported (\citealt{Stangret_M2}, \citealt{Nugroho_M2}, \citealt{Hoeijmakers_M2}, \citealt{Cont_M2}, \citealt{Yan_M2}, \citealt{Bello-Arufe_M2}, \citealt{Borsa_M2}, \citealt{Johnson_M2}, \citealt{Kasper_M2}, \citealt{Petz_M2}).  Evidence of H$_2$O and CO in the emission spectrum was presented by \citet{Fu_M2}, using observations from TESS, HST WFC3/G141, and the \textit{Spitzer} 4.5 $\mu$m channel.

The first single-line analysis of \ion{Fe}{ii} lines was presented by \citet{M2_Casasayas_2019}, revealing the discovery of three \ion{Fe}{ii} lines ($\lambda$5018 $\AA$, $\lambda$5169 $\AA$, and $\lambda$5316 $\AA$) using high-resolution observations from the HARPS-N spectrograph, reporting day-side winds of $\sim$ $-$2.8 km s$^{-1}$. \citet{Pai_K9_FeII} presented the detection of three additional \ion{Fe}{ii} lines ($\lambda$4923.9 $\AA$,  $\lambda$5276 $\AA$, and $\lambda$5363.9 $\AA$). Consistent with the previous works, day-to-night side winds with a blue shift of $\sim$ $-$2.55 km s$^{-1}$ were detected. 

\begin{figure}[h!]
\centering
\includegraphics[width=0.97\hsize]{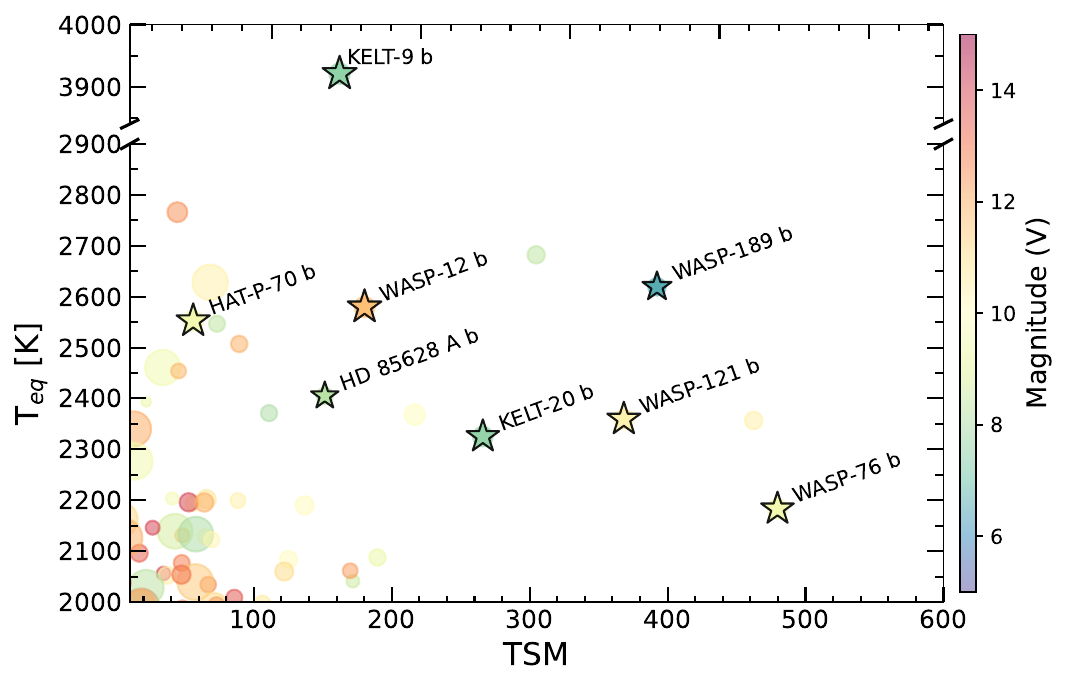}
\caption{Equilibrium temperature as the function of the transmission spectroscopy metrics for all known ultra-hot Jupiters ($T_{eq}$ > 2000 K). Planets with the \ion{Fe}{ii} detected in their atmospheres are marked with the star symbol (the list of planets was taken from the \texttt{ExoAtmospheres} database).}
\label{Fig:TSM}
\end{figure}

KELT-9b and KELT-20b are two of the eight UHJs (nine exoplanets in total), where \ion{Fe}{ii} has been detected in their atmospheres. In Fig. \ref{Fig:TSM}, we plot the transmission spectroscopy metrics for all known UHJs ($T_{eq}$ > 2000 K) as a function of their $T_{eq}$ values and the brightness of the host star.

In this work, we focus on the detection of \ion{Fe}{ii} in the atmosphere of two UHJs KELT-9b and KELT-20b using both single-line analysis and cross-correlation method with high-resolution observations in the visible part of the spectrum.
%where the strongest lines of this ion are expected.

\section{Observations}\label{Section:Observations}

Our work focuses on high-resolution observations of the transits of two exoplanets, KELT-9b and KELT-20b, with the HARPS-N spectrograph (\citealt{harpsn_2012, HARPSN2014}) mounted on the 3.6~m Telescopio Nazionale \textit{Galileo} (TNG) at the Observatorio del Roque de los Muchachos (ORM) in La Palma, Spain. HARPS-N is a high-dispersion spectrograph with a resolving power of $\Re \sim$ 115,000 and continuous wavelength coverage from 3800 to 6900~$\AA$. 

\begin{figure}[h!]
\centering
\includegraphics[width=0.95\hsize]{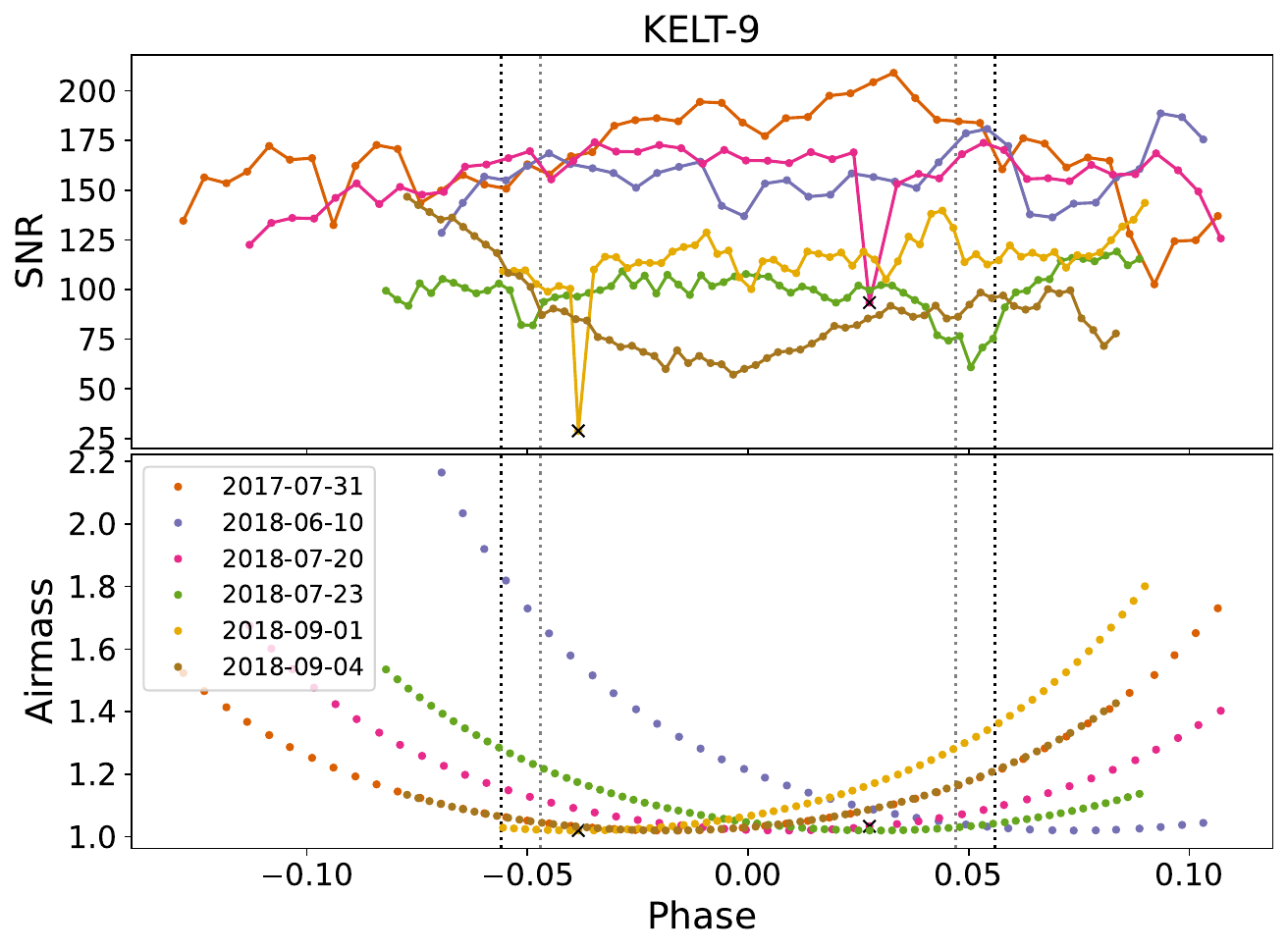} 
\includegraphics[width=0.95\hsize]{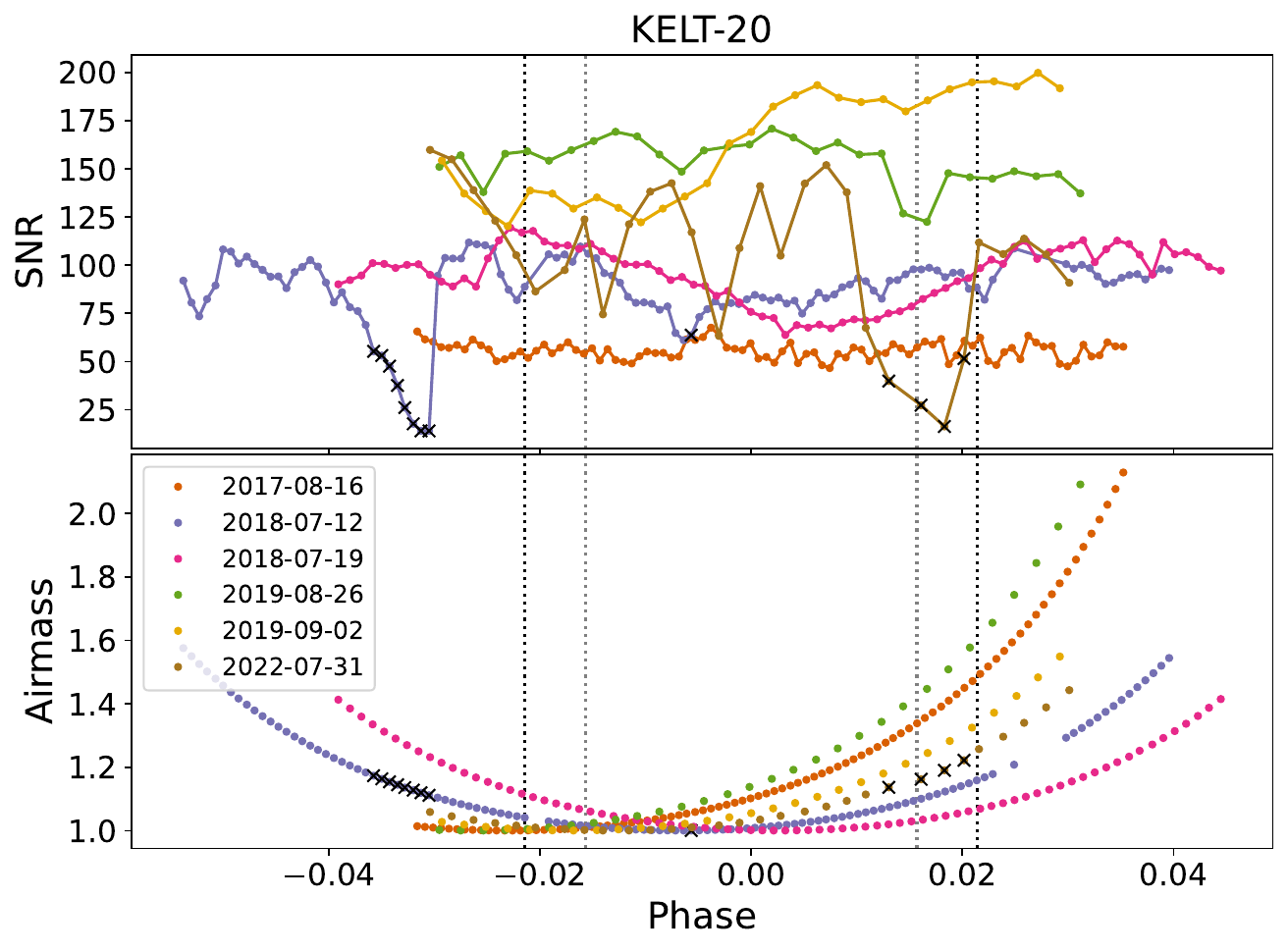}
\caption{Changes in S/N and airmass during the observations of KELT-9b (upper panel) and KELT-20b (lower panel). The black vertical dotted lines indicate T1 and T4, and the gray vertical dotted lines indicate T2 and T3. The black crosses indicate the low S/N spectra that were excluded from the analysis.}
\label{Fig:SNR_K9_M2}
\end{figure}

We have analyzed six nights of primary transit observations of KELT-9b, two of which were public, and the remaining four were taken as part of the GAPS (Global Architecture of Planetary System, \citealt{GAPS_2013}) program. The observing log can be found in Table \ref{tab:obsK9K20}; in brief, all observations cover the orbital phase before, during, and after the transit. The exposure time varies between 300 and 600~s, and the signal-to-noise ratio (S/N), on the spectral order of 53, around the \ion{Na}{i} doublet varies between 57.2 and 209. During data analysis, we decided to remove one spectrum from the night of 2018-07-20 and one from the night of 2018-09-01, because of a significantly lower S/N compared to the other spectra taken during the same night. The S/N and airmass changes during the observations are presented in Fig. \ref{Fig:SNR_K9_M2} (top panel), where black crosses indicate the low S/N spectra that were excluded from the analysis.

\begin{table*}[h!]
\centering
\caption{Observing log of the KELT-9b and KELT-20b transit observations. The table does not include spectra that were excluded from the analysis.}
\begin{tabular}{cccccccc}
\hline\hline
Night & Date of &      & Start & End\tablefoottext{a}  &$T_\mathrm{exp}$ & $N_\mathrm{obs}$ & S/N\tablefoottext{b} \\
         & observation &  &[UT] & [UT]  & [s] &         &    \\ \hline
\\[-1em]
\multicolumn{8}{c}{ KELT-9 }\\\noalign{\smallskip}
\\[-1.5em]
\hline
1 &  2017-07-31 & Public &20:59 & 05:19 &   600  & 49 &  102.6 -- 209.0 \\
\\[-1em]
2 & 2018-06-10  & GAPS &23:08 & 05:17 & 600  & 36  & 128.5 -- 188.6  \\ 
\\[-1em]
3 & 2018-07-20  & Public &21:20  & 05:09 & 600 & 45   &  122.5 -- 174.1 \\ 
\\[-1em]
4 & 2018-07-23  & GAPS &21:34 & 03:38 & 300 & 68  & 60.9 -- 119.1  \\ 
\\[-1em]
5 & 2018-09-01  & GAPS &22:17 & 03:27 & 300 & 57 & 98.9 -- 143.6  \\ 
\\[-1em]
6 & 2018-09-04  & GAPS &20:36 & 02:19  & 300 & 64 &57.2 -- 146.7 \\ 
\\[-1em]
\hline
\hline
\\[-1em]
\multicolumn{8}{c}{ KELT-20 }\\\noalign{\smallskip}
\\[-1.5em]
\hline
1 & 2017-08-16  & Public&22:21 & 03:56  & 200 & 90 & 46.6 -- 67.6 \\
\\[-1em]
2 & 2018-07-12  &Public& 21:27 & 05:15 & 200 & 107 & 61.1 -- 111.8  \\ 
\\[-1em]
3 & 2018-07-19 & Public& 21:25 &  04:23  & 300& 78  & 63.8 -- 119.5  \\ 
\\[-1em]
4 & 2019-08-26 & GAPS&22:05 & 03:09  & 600& 30  & 122.5 -- 170.9  \\ 
\\[-1em]
5 & 2019-09-02 &GAPS &20:52 & 01:45 & 600& 29 & 120.3 -- 199.9  \\ 
\\[-1em]
6 & 2022-07-31 & GAPS&22:34 & 03:36 & 600 & 26  & 63.3 -- 159.9   \\ 
\\[-1em]
\hline
\end{tabular}\\
\tablefoot{\tablefoottext{a}{Time of the beginning of the last exposure.} \tablefoottext{b}{Minimum and maximum S/N for each night, calculated in the order containing the \ion{Na}{i}\,D lines (order 53).}}
\label{tab:obsK9K20}
\end{table*}

In the case of KELT-20b, we analyzed six transit observations of HARPS-N, three of which were taken as part of the GAPS collaboration and three were public. The observing log can be found in Table \ref{tab:obsK9K20}. The exposure time varies from 200 to 600~s and the S/N around the \ion{Na}{i} doublet varies from 46.6 to 199.9. During the analysis, we removed the spectra from the nights 2018-07-12 and 2022-07-31 with S/N around \ion{Na}{i} lower than 60. This removal was not applied to the night of 2022-07-31 due to the lower signal for all data. The S/N and airmass changes during the observations are presented in Fig. \ref{Fig:SNR_K9_M2} (bottom panel).

\section{Transmission spectroscopy}\label{Section:TS}

In the first step of data analysis, we employed the Spectral Lines Of Planets with python
%\LEt{ First set the full name, then acronym in parentheses, please.***}
(\texttt{SLOPpy}\footnote{\url{https://github.com/LucaMalavolta/SLOPpy}}, \citealt{Sloppy_2022}) pipeline to correct for sky emission, differential refraction, and telluric lines. The telluric correction was performed with \texttt{Molecfit} \citep{Molecfit_2015_1,Molecfit_2015_2}, an ESO software tool that fits synthetic transmission spectra to the data in order to correct for atmospheric absorption features. As this step is part of the \texttt{SLOPpy} analysis flow,  we refer to \citet{Sloppy_2022} for more details. Furthermore, we tested our analysis using spectra non-corrected by tellurics for the following lines: \ion{Fe}{II} $\lambda$4489 $\AA$, $\lambda$4555 $\AA$, $\lambda$5018 $\AA$, and $\lambda$5316 $\AA$ (see an example of transmission spectrum with and without correction for KELT-9b in Fig. Figure \ref{Fig:molec_tell}). In general, due to the fact that the lines in our work lie far away from strong telluric lines, the differences between corrected and uncorrected spectra are negligible.  

In the next step, we shifted the spectra to the stellar rest frame using the barycentric correction, as well as the stellar motions around the center of mass using the literature measurements of the semi-amplitude of the radial velocities, $K_*$. The data were not corrected for the systemic velocity ($v_{\rm sys}$), due to the different values reported in the literature for both KELT-9b and KELT-20b. A detailed discussion can be found in Sects. \ref{Section:K9-detected_lines} and \ref{Section:M2-detected_lines}. The $v_{\rm sys}$ value needs to be calculated with good precision to correct for the center-to-limb variation (CLV) and the Rossiter-McLaughlin (RM) effect and, thus, to accurately calculate the wind in the atmosphere of the exoplanet, which is one of the goals of this work. The calculation of $v_{\rm sys}$ is presented and explained in the next sections of this work. The systemic velocity is constant during the observations, which consequently allows us to apply this correction at every step in the analysis.

 %Because the broad-band continuum of the HARPS spectrograph varies slightly over the course of the observations, we performed a colour-correction of the spectral orders by normalising the mean flux of each spectral order over wavelength to the mean flux of each spectrum over time.

In the next step, due to computational limitations, the s1d spectrum was divided into 46 segments and in the following analysis, each of the steps was applied to each segment separately. We normalized the spectrum for each of the segments by the average spectrum over time, following the approach presented by \citet{Hoeijmakers_2020_normalization}. To this end, for each of the segments, we calculate the mean value of the averaged spectrum, next the averaged spectrum is divided by this value. Then, each of the spectra was divided by the corresponding value from the previous calculations so that all spectra remained at a similar flux level. Next, by looking at the changes of each pixel over time, the outliers that varied more than 3$\sigma$ from the mean value were linearly interpolated to the nearest pixels. The stellar signal was corrected by dividing all spectra by the master-out spectrum, which was calculated as the average spectrum of all out-of-transit spectra. The remaining residuals were used for the analysis of the single lines of \ion{Fe}{ii}, as well as for the cross-correlation method presented in Sect. \ref{SEC:CC}.

Here, we examine the \ion{Fe}{ii} single lines in the transmission spectrum. For each line, we created a residual map which consists of the residuals for each of the orbital phases and the wavelength from $\pm 5$~$\AA$ from the laboratory wavelength of the line (obtained from the National Institute of Standards and Technology via \texttt{petitRADTRANS}, \citealt{PetitRadTrans}). The analysis was conducted separately for each night and subsequently combined, assuming equal weight for each night. In the case of KELT-20b, due to the low S/N of the first and fifth nights, an additional analysis was performed using only the second, third, and fourth nights. However, the results were found to be consistent with those obtained using all nights; thus we present the results using all the nights.

By applying a Markov chain Monte Carlo (MCMC, \citealt{MCMC_emcee}) algorithm we fitted the planetary signal and the CLV and RM effect simultaneously. In the case of the planetary signal, we assume a Gaussian shape of the signal, which moves with the radial velocity of the planet in its orbit. The velocity of the planet on its orbit $v_{\rm pl}$ assuming circular orbit was calculated as:

\begin{equation}
    v_{\rm pl}= K_{\rm p} \sin{2 \pi \phi (t)} + v_{\rm sys+wind},
\end{equation}

where $K_{\rm p}$ is the semi-amplitude of the exoplanet radial velocity, $\phi$ is the orbital phase, and  $v_{\rm sys+wind}$ is the velocity of the signal, which consists of the systemic velocity and the velocity of the wind in the atmosphere ($v_{\rm wind}$), remembering that  $v_{\rm sys}$ has not yet been corrected. In addition, using the \texttt{LDTk} tool \citep{LDTK_1,LDTK_2}, we calculated the limb-darkening coefficient for each star. Then, using \texttt{PyTransit} \citep{PyTransit} assuming the parameters from Table \ref{tab:params_K9_K20}, we calculated the transit models. The fitted signal of the planet was then scaled using the transit model.

Both KELT-9b and KELT-20b orbit fast-rotating stars; therefore, a strong RM effect is expected, especially influencing the shape of the \ion{Fe}{ii} lines, which are also present in the stellar spectra. To correct for this effect together with the CLV, we created models of the CLV and RM effect following \citet{Yan_CLV} and \citet{Yan_Henning_2018}, and later used in several works of atmospheric studies such as \citet{M2_Casasayas_2019}, \citet{Chen_RM}, \citet{Borsa_w121_RM}, and \citet{Stangret_6plan}. The synthetic spectrum for our planets was computed with \texttt{Spectroscopy Made Easy} (\texttt{SME}, \citealt{SME}) using the VALD3 line list \citep{vald3} and Kurucz ATLAS9 stellar model atmospheres \citep{ATLAS9}. Assuming local thermodynamical equilibrium (LTE) and solar abundance, we calculated models for 21 different limb-darkening angles. Next, we calculated the stellar spectra at the different orbital phases of the planet, keeping in mind that the planet covers different parts of the stellar spectra. In addition, the models were created for several radii of the planet, from 0.7 to 1.5, with a step of 0.1, taking into consideration that the observed radius of the planet depends on the wavelength at which the observations were performed. Finally, all of the spectra were divided by the out-of-transit spectra where we did not expect the influence of the planet on the spectrum, leaving us with the CLV and RM effect at each of the orbital phases and different radii of the planet. The models were then linearly interpolated to the different orbital phases depending on the considered night and planet, and also  broadened to the resolution of the HARPS-N spectrograph using \texttt{instrBroadGaussFast}\footnote{\url{https://pyastronomy.readthedocs.io/en/latest/pyaslDoc/aslDoc/broad.html}} from \texttt{PyAstronomy} \citep{pyastronomy}. Finally, the models were interpolated with the third-degree polynomial to the different radii of the planet.  

In the MCMC analysis, the CLV and RM effect models were used to fit the radius of the planet depending on the wavelength ($R_\lambda$), as well as $v_{\rm sys}$, where the models were calculated assuming $v_{\rm sys}=0$ km\,s$^{-1}$. In the final MCMC, we used 20 walkers and 5000 steps to fit the amplitude, $h$, $K_{\rm p}$, $\sigma$, and $v_{\rm sys+wind}$ for the planetary signal and $R_\lambda$ and $v_{\rm sys}$ using the CLV and RM models. Each of these parameters had a uniform prior. All of the corner plots of the analysis are presented in the appendix. 

At the high temperatures found in the atmospheres of UHJs, \ion{Fe}{ii} presents strong absorption features. In this work, to identify the possible signals in the transmission spectrum,  we created a synthetic model of \ion{Fe}{ii} for a uniform atmospheric temperature of 4000 K with \texttt{PetitRadTrans}. Next, we created a ``fast'' transmission spectrum (TS) assuming theoretical $K_{\rm p}$, literature $v_{\rm sys}$, and $R_{\lambda}=R_{\rm p}$. Then, we shortened the list of possible detections by visual inspection. For each of the lines, nights, and combinations of nights for each planet, we performed the MCMC analysis described in Sect. \ref{Section:TS}.

%The \ion{Fe}{ii} shows strong absorption lines for extremely high temperatures, such as those found in the atmospheres of ultra-hot Jupiters. In our work in order to identify the possible signal in the transmission spectrum,  we created the synthetic model of \ion{Fe}{II} for the temperature of 4000 K with \textit{PetitRadTrans}. Next, we created a 'fast' transmission spectrum (TS) assuming theoretical $K_p$, literature $v_{sys}$, and $R_{\lambda}=R_p$. By visual inspection, we shortened the list of possible detections. For each of the lines, nights, and combinations of the nights for each planet, we performed the MCMC analysis described in Section \ref{Section:TS}.

Tables \ref{tab:detection_separated_nights_K9} and \ref{tab:detection_separated_nights_M2} present a brief overview of the detected and non-detected lines for each night, as well as for the nights combined for KELT-9b and KELT-20b, respectively. The lines for which all the parameters converged in the MCMC analysis are represented in green. Those for which only one parameter did not converge in the MCMC analysis are marked in yellow, while non-detections (i.e., more than two parameters were not converged in the MCMC analysis) are marked in red. In the case of single-night analysis, the results are presented for visualization purposes only.  

\begin{table}[h!]
\centering
\caption{Detected and non-detected \ion{Fe}{ii} lines for KELT-9b for each of the nights. Detected lines are colored green, non-detected lines are colored red, and possible detections are colored yellow.}
\small
\begin{tabular}{|c|c|c|c|c|c|c|c|}
\hline
Line&  All & N1&N2&N3&N4&N5&N6\\[0.1em]
\hline
$\lambda$4173&\cellcolor{green!25}& \cellcolor{yellow!25}&\cellcolor{red!25}&\cellcolor{green!25}&\cellcolor{red!25}&\cellcolor{green!25}& \cellcolor{yellow!25}     \\[0.1em]
\hline
$\lambda$4233&\cellcolor{green!25} &\cellcolor{yellow!25}& \cellcolor{green!25}&\cellcolor{yellow!25} &\cellcolor{red!25}&\cellcolor{yellow!25}&\cellcolor{yellow!25}\\[0.1em]

%\hline
%$\lambda$4303&\cellcolor{green!25} &\cellcolor{yellow!25} & \cellcolor{red!25}&\cellcolor{red!25}&\cellcolor{red!25}&\cellcolor{green!25}& \cellcolor{red!25}\\[0.1em]

\hline
$\lambda$4351&\cellcolor{green!25}& \cellcolor{green!25}& \cellcolor{red!25}&\cellcolor{red!25}&\cellcolor{yellow!25}&\cellcolor{green!25}&\cellcolor{green!25}\\[0.1em]

\hline
$\lambda$4385&\cellcolor{green!25}&\cellcolor{red!25} & \cellcolor{green!25}&\cellcolor{red!25}&\cellcolor{red!25}&\cellcolor{red!25}&\cellcolor{red!25}\\[0.1em]

\hline
$\lambda$4489&\cellcolor{green!25}& \cellcolor{red!25}& \cellcolor{red!25} &\cellcolor{red!25}&\cellcolor{red!25}&\cellcolor{green!25}&\cellcolor{red!25}\\[0.1em]

\hline
$\lambda$4508&\cellcolor{green!25}& \cellcolor{red!25}& \cellcolor{green!25} &\cellcolor{red!25}&\cellcolor{red!25}&\cellcolor{yellow!25}&\cellcolor{red!25}\\[0.1em]

\hline
$\lambda$4515&\cellcolor{green!25}&\cellcolor{green!25} &\cellcolor{green!25} &\cellcolor{red!25}&\cellcolor{red!25}&\cellcolor{yellow!25}&\cellcolor{red!25}\\[0.1em]

\hline
$\lambda$4520&\cellcolor{green!25}&\cellcolor{red!25}&\cellcolor{red!25} &\cellcolor{red!25}&\cellcolor{red!25}&\cellcolor{yellow!25}&\cellcolor{red!25}\\[0.1em]

\hline
$\lambda$4522&\cellcolor{green!25}&\cellcolor{yellow!25} & \cellcolor{yellow!25}&\cellcolor{red!25}&\cellcolor{yellow!25}&\cellcolor{yellow!25}&\cellcolor{green!25}\\[0.1em]

\hline
$\lambda$4555&\cellcolor{green!25}&\cellcolor{red!25} & \cellcolor{red!25}&\cellcolor{green!25}&\cellcolor{red!25}&\cellcolor{red!25}&\cellcolor{green!25}\\[0.1em]

\hline
$\lambda$4583&\cellcolor{green!25}&\cellcolor{green!25} &\cellcolor{green!25} &\cellcolor{green!25}&\cellcolor{green!25}&\cellcolor{green!25}&\cellcolor{green!25}\\[0.1em]

\hline
$\lambda$4620&\cellcolor{green!25}& \cellcolor{red!25}&\cellcolor{green!25}&\cellcolor{green!25}&\cellcolor{green!25}&\cellcolor{red!25}&\cellcolor{green!25}\\[0.1em]

\hline
$\lambda$4629&\cellcolor{green!25}&\cellcolor{green!25} &\cellcolor{green!25} &\cellcolor{yellow!25}&\cellcolor{yellow!25}&\cellcolor{green!25}&\cellcolor{green!25}\\[0.1em]

\hline
$\lambda$4923&\cellcolor{green!25}&\cellcolor{green!25} & \cellcolor{green!25}&\cellcolor{yellow!25}&\cellcolor{green!25}&\cellcolor{green!25}&\cellcolor{green!25}\\[0.1em]

\hline
$\lambda$5018&\cellcolor{green!25}&\cellcolor{green!25} & \cellcolor{green!25}&\cellcolor{green!25}&\cellcolor{green!25}&\cellcolor{green!25}&\cellcolor{green!25}\\[0.1em]

\hline
$\lambda$5169&\cellcolor{green!25}&\cellcolor{green!25} & \cellcolor{green!25}&\cellcolor{green!25}&\cellcolor{green!25}&\cellcolor{green!25}&\cellcolor{green!25}\\[0.1em]

\hline
$\lambda$5197&\cellcolor{green!25}&\cellcolor{red!25} &\cellcolor{green!25} &\cellcolor{green!25}&\cellcolor{red!25}&\cellcolor{red!25}&\cellcolor{red!25}\\[0.1em]

\hline
$\lambda$5234&\cellcolor{green!25}& \cellcolor{yellow!25}&\cellcolor{green!25} &\cellcolor{yellow!25}&\cellcolor{green!25}&\cellcolor{green!25}&\cellcolor{red!25}\\[0.1em]

\hline
$\lambda$5276&\cellcolor{green!25}& \cellcolor{green!25}&\cellcolor{green!25} &\cellcolor{yellow!25}&\cellcolor{green!25}&\cellcolor{green!25}&\cellcolor{red!25}\\[0.1em]
\hline
$\lambda$5316&\cellcolor{green!25}&\cellcolor{green!25}&\cellcolor{green!25}&\cellcolor{green!25}&\cellcolor{green!25}&\cellcolor{green!25}&\cellcolor{green!25} \\[0.1em]
\hline
$\lambda$5362&\cellcolor{green!25}&\cellcolor{green!25}&\cellcolor{green!25}&\cellcolor{red!25}&\cellcolor{yellow!25}&\cellcolor{yellow!25}&\cellcolor{yellow!25} \\[0.1em]

\hline
\end{tabular}\\
\label{tab:detection_separated_nights_K9}
\end{table}

\begin{table}[h]
\centering
\caption{Same as Table \ref{tab:detection_separated_nights_K9}, but for KELT-20b.}
\small
\begin{tabular}{|c|c|c|c|c|c|c|c|}
\hline
Line&  All & N1&N2&N3&N4&N5&N6\\[0.1em]
\hline
$\lambda$4173&\cellcolor{red!25}&\cellcolor{red!25} & \cellcolor{red!25}&\cellcolor{red!25} &\cellcolor{red!25}& \cellcolor{red!25}&\cellcolor{red!25} \\[0.1em]
\hline

$\lambda$4233&\cellcolor{green!25} & \cellcolor{red!25}&\cellcolor{red!25} & \cellcolor{red!25}&\cellcolor{green!25}&\cellcolor{green!25}&\cellcolor{red!25}\\[0.1em]

\hline
$\lambda$4351&\cellcolor{green!25}& \cellcolor{red!25}&\cellcolor{red!25}& \cellcolor{yellow!25}&\cellcolor{red!25}&\cellcolor{red!25}&\cellcolor{red!25}\\[0.1em]

\hline
$\lambda$4385&\cellcolor{green!25}& \cellcolor{red!25}&\cellcolor{red!25}&\cellcolor{red!25}&\cellcolor{yellow!25}&\cellcolor{red!25}&\cellcolor{red!25}\\[0.1em]
\hline
$\lambda$4489&\cellcolor{green!25}&\cellcolor{red!25}&\cellcolor{red!25}&\cellcolor{red!25}&\cellcolor{red!25}&\cellcolor{yellow!25}& \cellcolor{red!25}\\[0.1em]

\hline
$\lambda$4508&\cellcolor{green!25}& \cellcolor{red!25} &\cellcolor{red!25}&\cellcolor{red!25}&\cellcolor{red!25}&\cellcolor{red!25}&\cellcolor{red!25}\\[0.1em]

\hline
$\lambda$4515&\cellcolor{green!25}&\cellcolor{red!25} &\cellcolor{red!25}&\cellcolor{red!25}&\cellcolor{red!25}&\cellcolor{red!25}&\cellcolor{red!25}\\[0.1em]

\hline
$\lambda$4520&\cellcolor{red!25}&\cellcolor{red!25} &\cellcolor{red!25}&\cellcolor{red!25}&\cellcolor{red!25}&\cellcolor{yellow!25}&\cellcolor{red!25}\\[0.1em]

\hline
$\lambda$4522&\cellcolor{green!25}& \cellcolor{yellow!25}&\cellcolor{red!25}&\cellcolor{yellow!25}&\cellcolor{yellow!25}&\cellcolor{yellow!25}&\cellcolor{green!25}\\[0.1em]

\hline
$\lambda$4555&\cellcolor{green!25}&\cellcolor{red!25} &\cellcolor{red!25}&\cellcolor{red!25}&\cellcolor{red!25}&\cellcolor{red!25}&\cellcolor{red!25}\\[0.1em]

\hline
$\lambda$4583&\cellcolor{green!25}&\cellcolor{red!25} &\cellcolor{green!25}&\cellcolor{red!25}&\cellcolor{green!25}&\cellcolor{green!25}&\cellcolor{red!25}\\[0.1em]

\hline
$\lambda$4629&\cellcolor{red!25}&\cellcolor{red!25} &\cellcolor{red!25}&\cellcolor{red!25}&\cellcolor{red!25}&\cellcolor{red!25}&\cellcolor{red!25}\\[0.1em]

\hline
$\lambda$4923&\cellcolor{green!25}&\cellcolor{green!25}&\cellcolor{green!25}&\cellcolor{red!25}&\cellcolor{green!25}&\cellcolor{green!25}&\cellcolor{green!25}\\[0.1em]

\hline
$\lambda$5018&\cellcolor{green!25}&\cellcolor{yellow!25}&\cellcolor{green!25}&\cellcolor{green!25}&\cellcolor{green!25}&\cellcolor{green!25}&\cellcolor{green!25}\\[0.1em]

\hline
$\lambda$5169&\cellcolor{green!25}& \cellcolor{green!25}&\cellcolor{green!25}&\cellcolor{green!25}&\cellcolor{green!25}&\cellcolor{green!25}&\cellcolor{green!25}\\[0.1em]

\hline
$\lambda$5197&\cellcolor{green!25}&\cellcolor{yellow!25} &\cellcolor{red!25}&\cellcolor{red!25}&\cellcolor{yellow!25}&\cellcolor{red!25}&\cellcolor{red!25}\\[0.1em]

\hline
$\lambda$5234&\cellcolor{green!25}&\cellcolor{red!25} &\cellcolor{red!25}&\cellcolor{red!25}&\cellcolor{red!25}&\cellcolor{yellow!25}&\cellcolor{red!25}\\[0.1em]

\hline
$\lambda$5276&\cellcolor{green!25}&\cellcolor{yellow!25} &\cellcolor{yellow!25}&\cellcolor{green!25}&\cellcolor{green!25}&\cellcolor{yellow!25}&\cellcolor{red!25}\\[0.1em]

\hline
$\lambda$5316&\cellcolor{green!25}& \cellcolor{yellow!25}&\cellcolor{green!25}&\cellcolor{yellow!25}&\cellcolor{green!25}&\cellcolor{green!25}&\cellcolor{red!25}\\[0.1em]

\hline
$\lambda$5362&\cellcolor{green!25}&\cellcolor{red!25} &\cellcolor{yellow!25}&\cellcolor{red!25}&\cellcolor{green!25}&\cellcolor{yellow!25}&\cellcolor{red!25}\\[0.1em]

\hline

\end{tabular}\\
\label{tab:detection_separated_nights_M2}

\end{table}

%--------------------------------------------------------------------
\section{Detected lines - KELT-9b}\label{Section:K9-detected_lines}

\begin{figure*}[h!]
\centering
\includegraphics[width=0.95\textwidth]{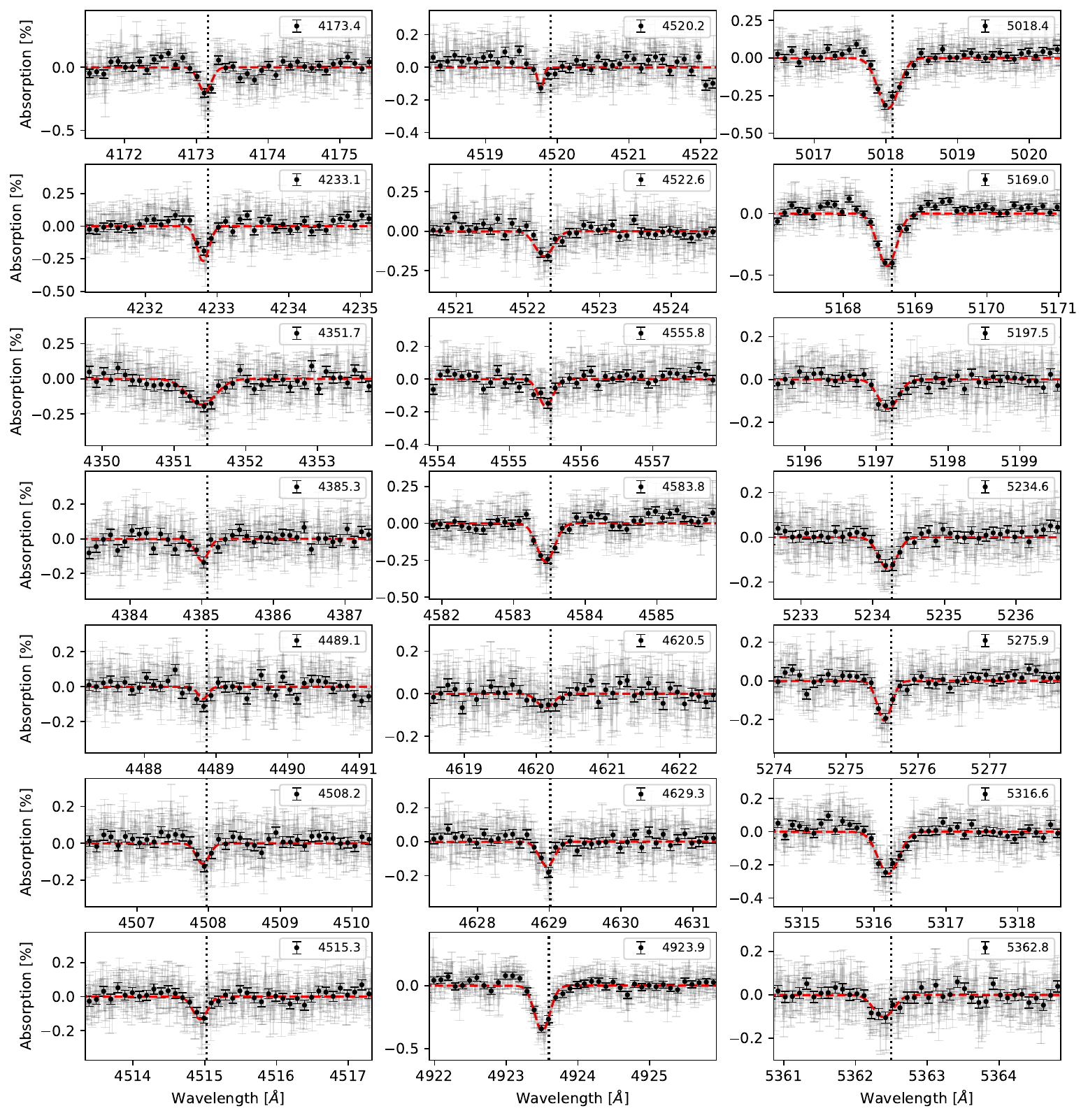}
\caption{Transmission spectra for all lines detected in KELT-9b (gray dots). Black dots indicate the binned transmission spectrum with a step of 0.1 $\AA$. The red dashed lines represent the best Gaussian fit of the planetary signal from the MCMC analysis.  }
\label{Fig:TS_ALL_K9}
\end{figure*}

Our analysis led to the detection of 21 \ion{Fe}{ii} single lines in the atmosphere of the UHJ KELT-9b. The list of detected lines and the fitted parameters are presented in Table \ref{tab:K9_detection}, and all transmission spectra of the considered lines are plotted in Fig. \ref{Fig:TS_ALL_K9}. All detected lines are simultaneously presented by \citealt{DArpa_2024} using the same data set, but using significantly different methodology. In our work, we focus on the analysis using the MCMC method, with several parameters  fitted simultaneously. A detection is reported only when all the parameters have been converged in the MCMC analysis. %It is important to recognize that the two works are distinct and contribute equally to the advancement of the field. 
In our analysis, we are focusing exclusively on the unblended lines of \ion{Fe}{ii}.  Four lines detected by \citealt{DArpa_2024} were excluded from our analysis because, for the \ion{Fe}{ii} $\lambda$4296 $\AA$ line, the MCMC did not converge for both $K_{\rm p}$ and amplitude, the \ion{Fe}{ii} $\lambda$4549 $\AA$ line is blended with a strong \ion{Ti}{ii} line, the \ion{Fe}{ii} $\lambda$4303 $\AA$ line blends with a \ion{Fe}{i} line, and the \ion{Fe}{ii} $\lambda$6456 $\AA$ line may potentially blend with two additional lines to the extent that their RM effect ends up covering the \ion{Fe}{ii} line.

\begin{figure}[h!]
\centering
\includegraphics[width=0.9\hsize]{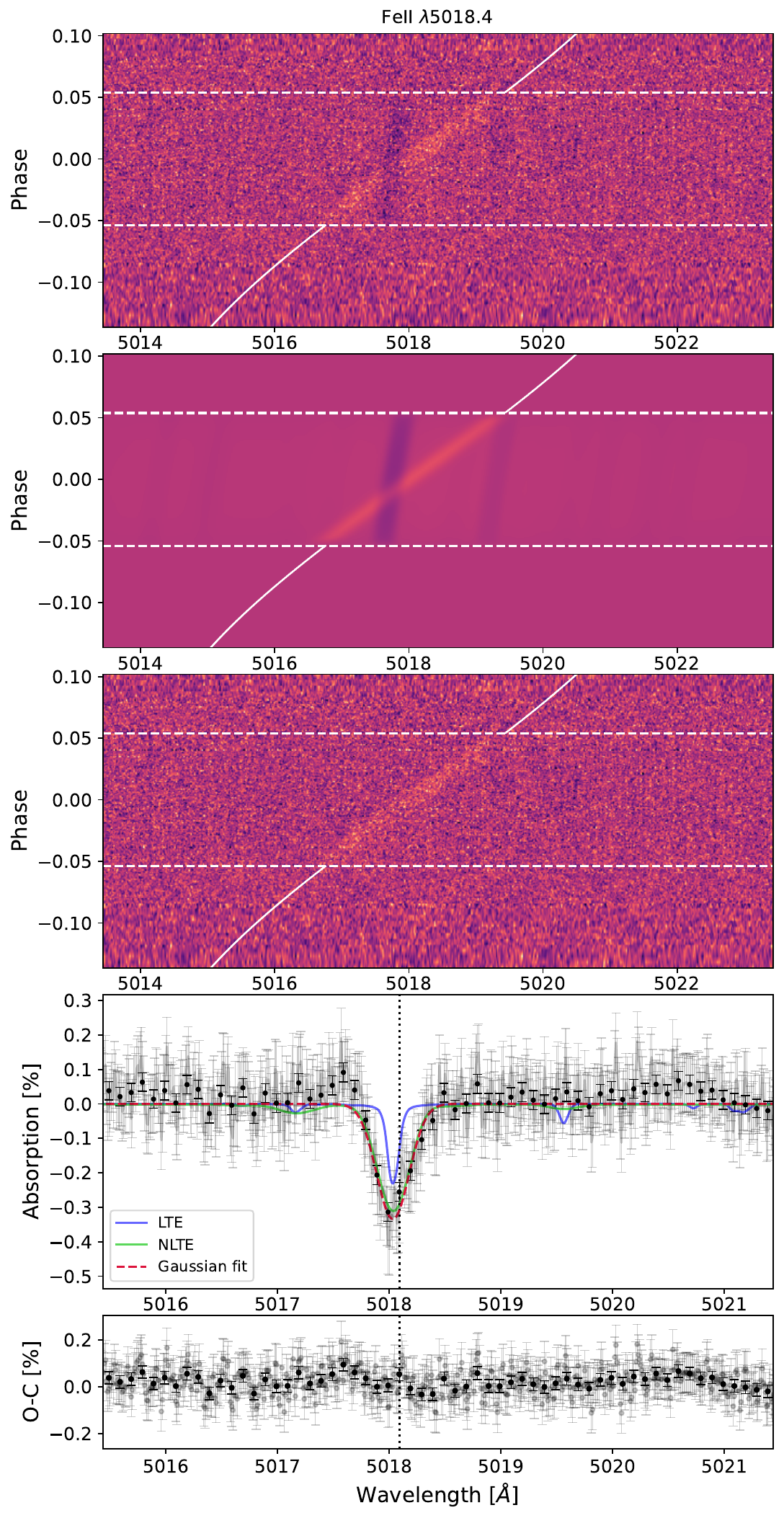}
%\caption{\textit{Top panel:} Residual map for line $\lambda$5018.4 $\AA$ for KELT-9b, shown in the stellar rest frame. A light-tiled signal is the atmospheric signal of the \ion{Fe}{ii} and the dark almost vertical signal is the RM effect. The white horizontal lines indicate the start and end of the transit and the tilted white line indicates the expected velocities for the signal coming from the planet's atmosphere, assuming $v_{sys}$ as the mean $v_{sys}$ from our analysis. \textit{Second panel:} The best-fit model of the planetary signal and the RM and CLV effects. \textit{Third panel:} Same as the top panel, but with the RM and CLV effects corrected. \textit{Forth panel:} Transmission spectrum for the detected line (gray dots). The black dots indicate the binned transmission spectrum with a step of 0.1 $\AA$. The red plot is the best fit of the planetary signal from the MCMC analysis, \textbf{the blue plot is an LTE}, and the green plot is the NLTE model for the atmosphere of KELT-9b. \textit{Bottom panel:} Residuals.}
\caption{The results of the analysis of the \ion{Fe}{ii} line at $\lambda$5018.4 $\AA$ for KELT-9b. \textit{Top panel}: Residual map in the stellar rest frame. The light-tilted signal is the atmospheric signal of \ion{Fe}{ii}, while the dark almost vertical signal is the RM effect. The white horizontal lines indicate the start and end of the transit, while the tilted white line indicates the expected trace of the \ion{Fe}{ii} line considering velocities coming from the planet's atmosphere, assuming $v_{\rm sys}$ = $-$21.61 km~s$^{-1}$ as the average $v_{\rm sys}$ from our analysis. \textit{Second panel:} Best-fit model of the planetary signal and of the RM and CLV effects. \textit{Third panel:} Same as the top panel, but with the RM and CLV effects corrected. \textit{Fourth panel:} Transmission spectrum of the detected line (gray dots). The black dots indicate the binned transmission spectrum with a step of 0.1 $\AA$. The red line is the best Gaussian fit of the planetary signal derived from the MCMC analysis, the blue line shows the LTE model, and the green line indicates the NLTE model. \textit{Bottom panel:} Residuals after removing Gaussian fit from the TS.}
\label{Fig:TS_steps_K9}
\end{figure}

In Fig. \ref{Fig:TS_steps_K9}, we present an example of the MCMC and TS analysis carried out for the \ion{Fe}{ii} $\lambda$5018.4 $\AA$ line. The figure consists of a residual map in the stellar rest frame, with a visible signal coming from the planetary atmosphere (light-tilted signal) and the RM effect (dark vertical signal). In the second panel, we show the best-fitted model of the planetary signal and RM and CLV effect using the MCMC analysis described in Sect. \ref{Section:TS}. The residual map after correcting for the RM and CLV effect is presented in the third panel. The transmission spectrum compared to the best-fitting Gaussian model, synthetic atmospheric spectrum computed assuming local-thermodynamic equilibrium (LTE) and accounting for non-local thermodynamic equilibrium (NLTE) effects (see Sect.~\ref{SEC:TS_models}) is shown in the fourth panel. The corner plots of the MCMC analysis and the transmission spectrum for all the studied lines can be found in the appendix (Figs. \ref{Fig:TS_K9_1}, \ref{Fig:TS_K9_2}, \ref{Fig:TS_K9_3}, \ref{Fig:TS_K9_4}, \ref{Fig:TS_K9_5}, and \ref{Fig:TS_K9_6} for TS, and Figs. \ref{fig:corner_k9_1}, \ref{fig:corner_k9_2},  \ref{fig:corner_k9_3},  \ref{fig:corner_k9_4},  \ref{fig:corner_k9_5},  \ref{fig:corner_k9_6},  \ref{fig:corner_k9_7}, \ref{fig:corner_k9_8},  \ref{fig:corner_k9_9},  \ref{fig:corner_k9_10}, and  \ref{fig:corner_k9_11} for corner plots).

\begin{figure}[h]
\centering
\includegraphics[width=0.98\hsize]{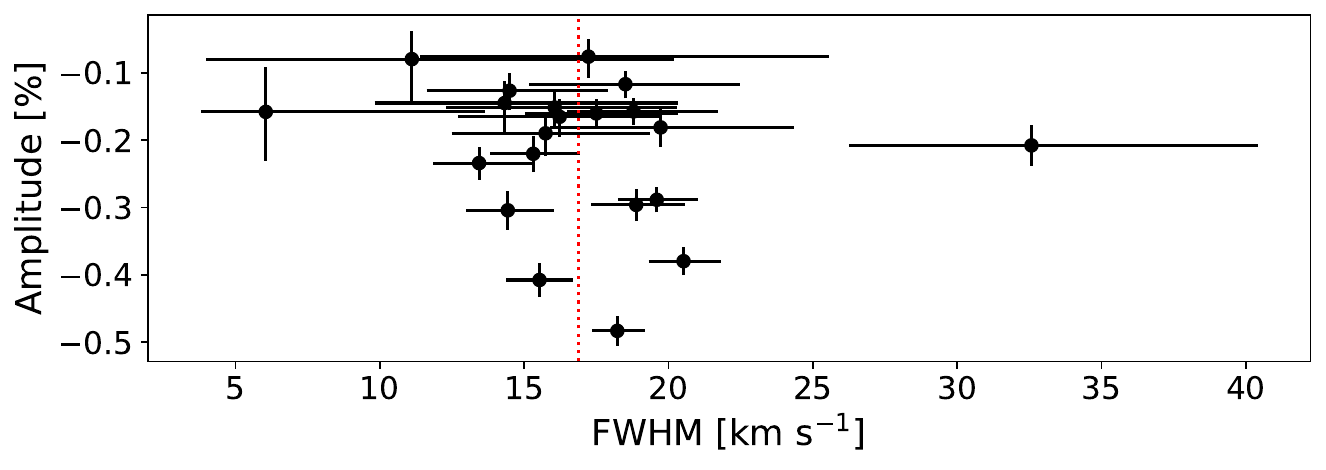}
\includegraphics[width=0.98\hsize]{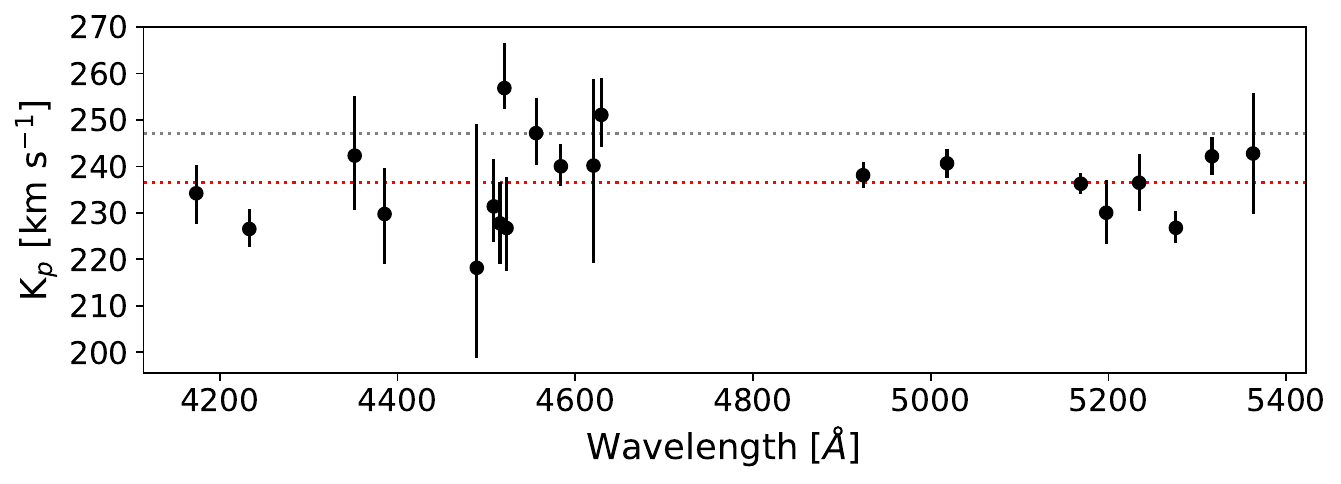}   
\caption{Amplitude, K$_p$, and FWHM of the \ion{Fe}{ii} lines detected in atmosphere of KELT-9b. \textit{Top panel}: amplitude of the transmission signal detected from the MCMC analysis for each of the detected lines as the function of their FWHM. The red vertical dashed line indicates the mean value of FWHM. \textit{Bottom panel:} $K_{\rm p}$ as a function of wavelength of the detected lines. The red horizontal dashed line indicates the mean $K_{\rm p}$ value, while the gray horizontal dashed line indicates the theoretical value. }
\label{Fig:RES_K9_AMP_FWHM}
\end{figure}

In Fig. \ref{Fig:RES_K9_AMP_FWHM} we present the amplitude as the function of the FWHM (upper panel) and $K_{\rm p}$ as the function of wavelength (lower panel) for all of the detected \ion{Fe}{ii} lines. The mean values of FWHM and $K_{\rm p}$ are 16.87 $\pm$ 4.75 km\,s$^{-1}$ and 236.4 $\pm$ 9.0 km\,s$^{-1}$, respectively, where the latter is consistent with the $K_{\rm p}$ obtained from the orbital parameters, that is $K_{\rm p}$ = 246.99 $\pm$ 5.72 km\,s$^{-1}$.

%In Figure \ref{Fig:RES_K9_AMP_FWHM} we present the amplitude, FWHM, and in Figure \ref{Fig:RES_K9_KP_RP} in  $K_p$, $R_{\lambda}$, and $R_{eff}$ for each of the detected lines, with the mean value for \textbf{FWHM = 16.87 $\pm$ 4.75} km\,s$^{-1}$, \textbf{$R_{eff}$ = 1.15 $\pm$ 0.07} $R_p$, and \textbf{$K_p$ = 236.4 $\pm$ 9.0} km\,s$^{-1}$, which is consistent with the $K_p$ assuming the orbital parameters, namely $K_p$ = 246.99 \textbf{$\pm$ 5.72} km\,s$^{-1}$. We observed that the detected radius of the planet $R_{\lambda}$ with a method using the RM effect, grows with the wavelength of the studied line. We discuss this effect in the Section \ref{SEC:Radius}.

\begin{figure}[h]
\centering
\includegraphics[width=0.98\hsize]{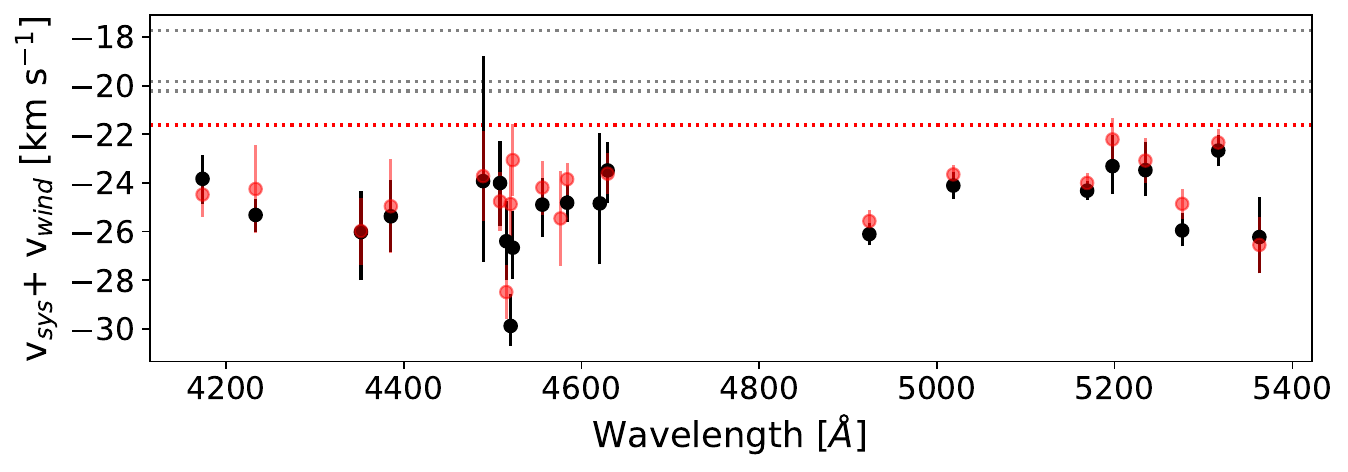}
\includegraphics[width=0.98\hsize]{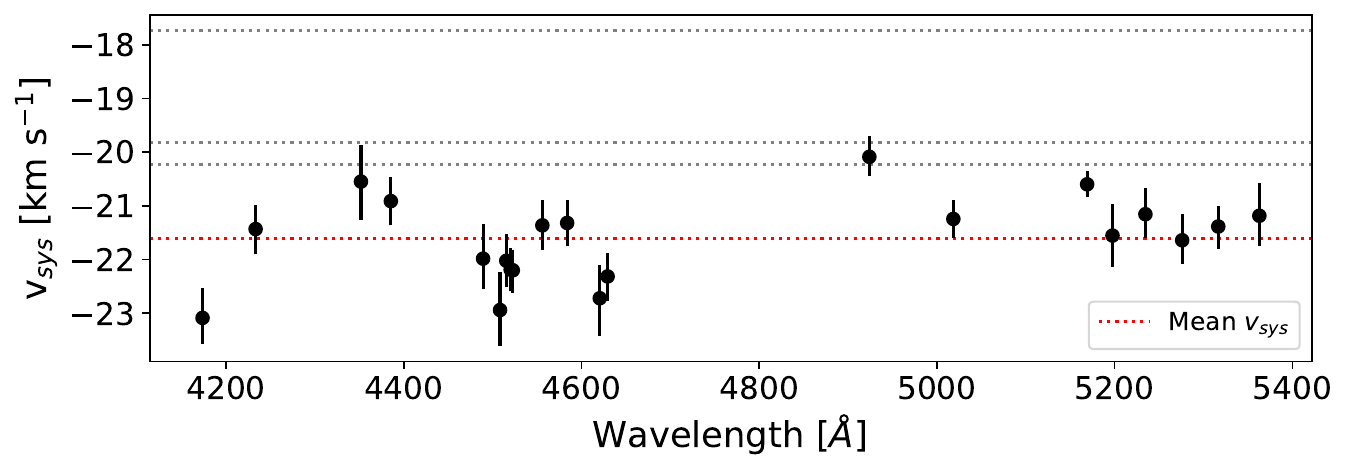}
\includegraphics[width=0.98\hsize]{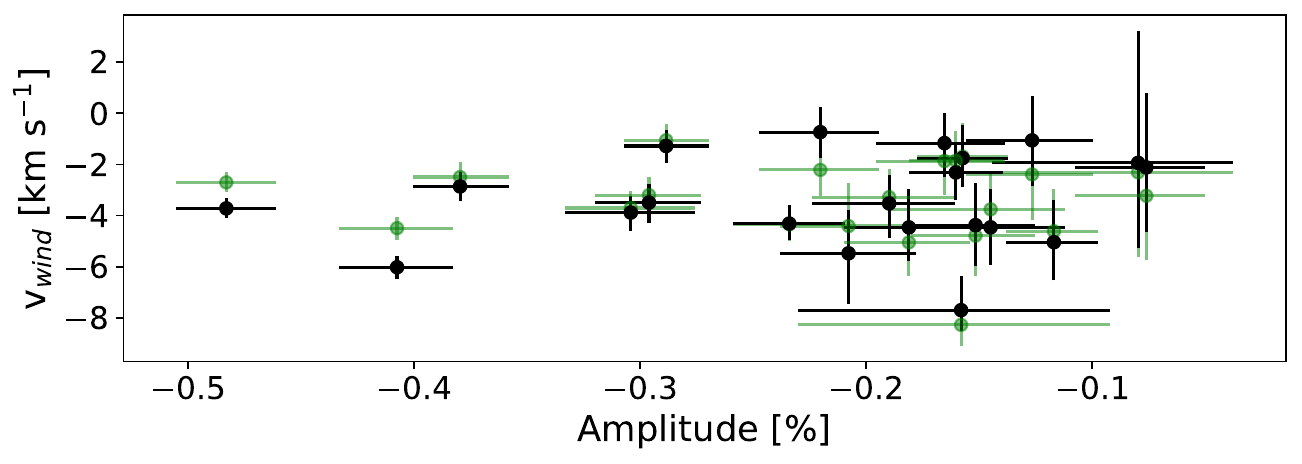}
\caption{$v_{\rm sys+wind}$ and $v_{\rm sys}$ of the \ion{Fe}{ii} lines detected in atmosphere of KELT-9b.  \textit{Top panel}: fitted $v_{\rm sys+wind}$ for each of the detected lines. The red dots represent the results of \citealt{DArpa_2024} \textit{Middle panel:} fitted $v_{\rm sys}$ for each of the detected lines. The red dashed horizontal lines indicate the mean value of the $v_{\rm sys}$ = $-$21.61 $\pm$ 0.77 km\,s$^{-1}$, while the grey horizontal dashed lines indicate literature $v_{\rm sys}$ values: $-$17.74 $\pm$ 0.11 km\,s$^{-1}$ \citep{Hoeijmakers_K9_2}, $-$19.819 $\pm$ 0.024 km\,s$^{-1}$ \citep{Borsa_K9}, and $-$20.22 $\pm$ 0.49 km\,s$^{-1}$ (\textit{Gaia}). \textit{Bottom panel:} $v_{\rm wind}$ versus amplitude plot, where green points represent the values calculated by correcting the fitted $v_{\rm sys+wind}$ by the mean $v_{\rm sys}$ and the black points represent the values corrected by $v_{\rm sys}$ fitted for each of the lines separately.}
\label{Fig:RES_K9_2}
\end{figure}

In Fig. \ref{Fig:RES_K9_2}, we present the values of $v_{\rm sys+wind}$ and $v_{\rm sys}$ as a function of the wavelength of the detected line, and $v_{\rm wind}$ as a function of the amplitude of the detected signal. The mean measured $v_{\rm sys}$ = $-$21.61 $\pm$ 0.77 km\,s$^{-1}$ is slightly larger than those reported in the literature by \citet[][$-$17.74 $\pm$ 0.11 km\,s$^{-1}$]{Hoeijmakers_K9_2}, \citet[][$-$19.819 $\pm$ 0.024 km\,s$^{-1}$]{Borsa_K9}, and \textit{Gaia} ($-$20.22 $\pm$ 0.49 km\,s$^{-1}$). By calculating the velocity of the atmospheric winds and considering the $v_{\rm sys}$ value from our MCMC analysis, we were able to detect atmospheric winds in the atmosphere of KELT-9b of $v_{\rm wind}$ = $-$3.41 $\pm$ 1.56 km\,s$^{-1}$. Given that the $v_{\rm sys}$ calculated in our work is the lowest among those presented in the literature, it can be stated with confidence that the KELT-9b has strong atmospheric blue-shifted day-to-night side winds with a minimum velocity of $-$3.41 km\,s$^{-1}$.

%\begin{figure}[h]
%%\centering
%\includegraphics[width=\hsize]{Figures/kelt-9/vel_all_K9_Mattia_N_B.jpg}
%\caption{The $v_{sys} + v_{wind}$ in compared to the literature measurements for detected lines. The red dots represent the results of D'Arpa et al. (submitted, priv. comm.). The red dashed horizontal lines indicate the mean value of the $v_{sys}$ = -21.61 $\pm$ 0.77 km\,s$^{-1}$, and the gray dashed horizontal lines indicate the literature $v_{sys}$ : -17.74 $\pm$ 0.11 km\,s$^{-1}$ \citep{Hoeijmakers_K9_2}, -19.819 $\pm$ 0.024 km\,s$^{-1}$ \citep{Borsa_K9}, and -20.22  $\pm$ 0.49 km\,s$^{-1}$ (GAIA).}
%\label{Fig:RES_K9_3}
%\end{figure}

%%   \begin{figure}
 %  \centering
 %  \includegraphics[width=\hsize]{Figures/FeII 5018.436.jpg}
 %     \caption{}
 %        \label{FigVibStab}
  % \end{figure}
   
%\begin{landscape}

\begin{table*}[]
\centering
\caption{Detected \ion{Fe}{ii} lines in the atmosphere of KELT-9b. }
\tiny
\setlength{\tabcolsep}{7pt}
\begin{tabular}{c|cccccc|cc|c}
\hline\hline
\\[-0.8em]
Line &   h    &FWHM  & K$_p$   &$v_{sys}+v_{wind}$  & $v_{sys}$  &  R$_\lambda$ &$\nu_{\rm mic}$ & $\nu_{\rm mac}$ &  R$_{eff}$ \\[0.2em]
\\[-1.1em]
 &   [$\%$]    &[km\,s$^{-1}$]&   [km\,s$^{-1}$]  &  [km\,s$^{-1}$]&  [km\,s$^{-1}$]&  [R$_p$]& [km\,s$^{-1}$]&  [km\,s$^{-1}$]&  [R$_p$]\\[0.2em]
\hline
\hline
\\[-0.8em]
%$\lambda$4173.45& -0.22$\pm$ 0.3 & 15.27$^{+1.73}_{-1.48}$ & 234.53$^{+6.24}_{-6.36}$ & -23.87$^{+1.00}_{-1.07}$ & -22.27$^{+0.57}_{-0.58}$&1.02$\pm$0.4 & 1.77$^{+1.01}_{-1.06}$&2.60$^{+2.37}_{-1.68}$&1.15 $\pm$ 0.01

$\lambda$4173.45& -0.22 $\pm$ 0.3 & 15.31$^{+1.61}_{-1.47}$ & 234.21$^{+5.99}_{-6.48}$ & -23.83$^{+0.98}_{-1.02}$ & -23.09$^{+0.55}_{-0.49}$&0.91$\pm$0.3 & 1.73 $^{+1.01}_{-0.99}$&2.64$^{+2.36}_{1.71}$& 1.15 $\pm$ 0.02
\\[0.2em]
\hline
\\[-0.8em]
%$\lambda$4233.16 & -0.30$\pm$0.03 & 14.39$^{+1.60}_{-1.41}$ & 226.71$^{+4.30}_{-4.04}$ & -25.26$^{+0.67}_{-0.73}$ & -20.13$^{+0.48}_{-0.49}$ & 0.92$\pm$0.03 &  1.26$^{+0.63}_{-0.64}$ & 6.06$^{+2.95}_{-3.20}$&1.20$\pm$ 0.01\\[0.2em]
$\lambda$4233.16 & -0.30$\pm$0.03 & 14.43$^{+1.59}_{-1.44}$ & 226.53$^{+4.35}_{-3.86}$ & -25.32$^{+0.63}_{-0.71}$ & -21.43$^{+0.45}_{-0.46}$ & 0.88$\pm$0.03 & 1.32$^{+0.65}_{-0.62}$  & 6.18$^{+2.76}_{-3.24}$& 1.20 $\pm$ 0.02\\[0.2em]\hline

\\[-0.8em]
%$\lambda$4351.76&-0.21$\pm$0.03&28.76$^{+6.39}_{-5.02}$& 243.67$^{+11.19}_{-10.59}$&-25.80$^{+1.63}_{-1.76}$&-18.81$^{+0.72}_{-0.75}$&1.19$^{+0.05}_{-0.06}$& 4.78$^{+2.36}_{-1.81}$ & 15.34$^{+4.97}_{-5.55}$ & 1.14$\pm$ 0.01 \\[0.2em]
$\lambda$4351.76&-0.21$\pm$0.03&32.57$^{+7.82}_{-6.26}$& 242.31$^{+12.71}_{-11.68}$&-26.02$^{+1.68}_{-1.96}$&-20.55$^{+0.68}_{-0.72}$&1.08$^{+0.05}_{-0.04}$& 5.98 $^{+2.67}_{-2.20}$ & 14.96$^{+5.12}_{-6.64}$& 1.14 $\pm$ 0.02 \\[0.2em]
\hline
\\[-0.8em]
%$\lambda$4385.37&-0.15$^{+0.03}_{-0.04}$&14.56$^{+5.87}_{-4.24}$&229.33$^{+10.61}_{-10.38}$&-25.34$^{+1.54}_{-1.58}$&-19.96$^{+0.48}_{-0.51}$&0.94$\pm$0.04 & 1.83$^{+1.69}_{-1.03}$ & 9.18$^{+8.78}_{-6.39}$& 1.10$\pm$ 0.01 \\[0.2em]
$\lambda$4385.37&-0.14$^{+0.03}_{-0.04}$&14.32$^{+5.98}_{-4.47}$&229.74$^{+9.83}_{-10.68}$&-25.37$^{+1.48}_{-1.45}$&-20.91$^{+0.45}_{-0.45}$&0.86$^{+0.04}_{-0.03}$ & 1.18$^{+1.70}_{-1.08}$ & *9.08$^{+8.66}_{-6.45}$ & 1.10 $\pm$ 0.03 \\[0.2em]
\hline
\\[-0.8em]
%$\lambda$4489.17 & -0.08$^{+0.04}_{-0.05}$&11.90$^{+8.73}_{-6.76}$ &215.00$^{+22.71}_{-16.58}$ &-24.09$^{+5.48}_{-3.07}$&-21.49$^{+0.77}_{-0.79}$&1.03$\pm$0.05 & 0.88$^{+0.80}_{-0.56}$ & 5.58$^{+4.48}_{-3.69}$ & 1.06$\pm$ 0.01\\[0.2em]
$\lambda$4489.17& -0.08$^{+0.04}_{-0.06}$&11.11$^{+8.90}_{-7.09}$ &218.17$^{+30.38}_{-19.13}$ &-23.93$^{+5.15}_{-3.31}$&-21.99$^{+0.64}_{-0.55}$&0.96$^{+0.04}_{-0.05}$ & 0.93$^{+0.77}_{-0.58}$ & 5.40$^{+4.08}_{-3.61}$& 1.06 $\pm$ 0.04 \\[0.2em]

\hline
\\[-0.8em]
%$\lambda$4508.28&-0.13$\pm$ 0.03 & 14.49$^{+3.36}_{-2.74}$&232.37$^{+10.82}_{-7.84}$&-24.17$^{+1.76}_{-1.90}$ & -22.17$^{+0.73}_{-0.69}$ &0.96$\pm$0.05 & *0.67$^{+0.67}_{-0.45}$ & 7.36$^{+3.78}_{-4.12}$ &1.09$\pm$ 0.01 \\[0.2em]
$\lambda$4508.28&-0.13$\pm$ 0.03 & 14.50$^{+3.37}_{-2.87}$&231.38$^{+10.23}_{-7.68}$&-24.01$^{+1.72}_{-1.76}$ & -22.94$^{+0.70}_{-0.66}$ &0.31 $\pm$ 0.04 & *0.65$^{+0.64}_{-0.45}$& 7.09$^{+3.99}_{-4.03}$ & 1.09 $\pm$ 0.02 \\[0.2em]

\hline
\\[-0.8em]
%$\lambda$4515.76 & -0.15$\pm$0.03 &16.01$^{+4.29}_{-3.51}$&227.71$^{+8.84}_{-8.83}$ & -26.38$^{+1.60}_{-1.61}$ & -21.14$^{+0.63}_{-0.57}$ & 0.99$\pm$0.04 & 1.55$^{+1.01}_{-0.93}$ & 9.93$^{+5.73}_{-6.05}$&1.08$\pm$ 0.01\\[0.2em]
$\lambda$4515.76 & -0.15$\pm$0.03 &16.06$^{+4.20}_{-3.76}$&227.73$^{+8.93}_{-8.63}$ & -26.40$^{+1.64}_{-1.57}$ & -22.02$^{+0.50}_{-0.50}$ & 0.92$\pm$0.04 & 1.50$^{+1.00}_{-0.88}$ & 9.78$^{+5.84}_{-6.09}$& 1.11 $\pm$ 0.02\\[0.2em]

\hline
\\[-0.8em]
%$\lambda$4520.21 & 0.16$^{+0.07}_{-0.08}$ & 6.18$^{+7.48}_{-2.43}$ & 256.61$^{+10.43}_{-4.64}$&-29.83$^{+1.41}_{-0.84}$ & -21.18$^{+0.43}_{-0.40}$ & 1.19$\pm$0.03 & *0.63$^{+0.87}_{-0.47}$ & 10.10$^{+5.26}_{-5.84}$ & 1.11$\pm$ 0.01\\[0.2em]
$\lambda$4520.21 & 0.16$^{+0.07}_{-0.07}$ & 6.05$^{+7.56}_{-2.24}$ & 256.84$^{+9.70}_{-4.55}$&-29.88$^{+1.32}_{-0.81}$ & -22.19$^{+0.41}_{-0.40}$ & 1.11$\pm$0.03 & *0.46$^{+0.66}_{-0.34}$& 8.85$^{+5.63}_{-5.21}$& 1.11 $\pm$ 0.04\\[0.2em]

\hline
\\[-0.8em]
%$\lambda$4522.63&-0.18$\pm$0.03 & 19.89$^{+4.62}_{-3.82}$& 226.61$^{+10.02}_{-8.23}$& -26.64$^{+1.46}_{-1.25}$& -21.22$^{+0.52}_{-0.40}$ & 1.16$\pm$0.03 & 2.08$^{+0.89}_{-0.80}$ & 9.36$^{+3.85}_{-3.90}$& 1.29$\pm$ 0.01\\[0.2em]
$\lambda$4522.63&-0.18$\pm$0.03 & 19.73$^{+4.61}_{-3.81}$& 226.73$^{+10.87}_{-9.09}$& -26.66$^{+1.49}_{-1.29}$& -22.20$^{+0.38}_{-0.42}$ & 1.08$\pm$0.03 & 1.97$^{+0.83}_{-0.81}$ & 9.36$^{+3.76}_{-3.62}$& 1.13 $\pm$ 0.02\\[0.2em]

\hline
\\[-0.8em]
%$\lambda$4555.89 & -0.20$\pm$0.03&15.52$^{+3.72}_{-3.12}$&246.41$^{+6.82}_{-7.26}$ & -24.80$^{+1.15}_{-1.20}$ & -20.48 $\pm$ 0.49& 1.03$\pm$0.04 & 2.49$^{+1.32}_{-1.04}$ & 13.70$^{+4.27}_{-3.49}$& 1.14$\pm$ 0.01\\[0.2em]
$\lambda$4555.89 & -0.19$\pm$0.03&15.74$^{+3.56}_{-3.22}$&247.15$^{+7.56}_{-6.89}$ & -24.89$^{+1.10}_{-1.33}$ & -21.36 $^{+0.47}_{-0.46}$ & 0.97$\pm$0.03 & 2.27$^{+1.33}_{-1.00}$ & 13.20$^{+4.36}_{-3.62}$ & 1.13 $\pm$ 0.02\\[0.2em]
\hline
\\[-0.8em]
%$\lambda$4583.83 &  -0.30$^{+0.03}_{-0.02}$ &  18.94$^{+1.74}_{-1.58}$ &  239.57$^{+4.40}_{-4.51}$  & -24.76$^{+0.74}_{-0.76}$ &-20.34$^{+0.45}_{-0.46}$& 1.00$\pm$0.04 & 3.94$^{+0.90}_{-0.87}$ &8.34$^{+2.08}_{-2.84}$& 1.20$\pm$ 0.01\\[0.2em]
$\lambda$4583.83 &  -0.30 $\pm$ 0.02 &  18.88$^{+1.69}_{-1.55}$ &  240.01$^{+4.73}_{-4.29}$  & -24.81$^{+0.71}_{-0.80}$ &-21.32$^{+0.44}_{-0.43}$& 0.92$\pm$0.03 & 3.88$^{+0.89}_{-0.88}$&8.49$^{+2.10}_{-2.56}$& 1.20 $\pm$ 0.01 \\[0.2em]
\hline
\\[-0.8em]

%$\lambda$4620.51 &-0.08$\pm$ 0.03 & 17.29$^{+8.11}_{-5.95}$ & 240.01$^{+17.27}_{-20.03}$ &-24.79$^{+2.66}_{-2.65}$&-21.82$^{+0.71}_{-0.73}$& 1.00$\pm$0.06  & 6.49$^{+4.24}_{-3.45}$ & 13.79$^{+6.47}_{-7.25}$& 1.06$\pm$ 0.02 \\[0.2em]
$\lambda$4620.51 &-0.08$\pm$ 0.03 & 17.23$^{+8.31}_{-5.82}$ & 240.17$^{+18.69}_{-20.80}$ &-24.84$^{+2.89}_{-2.50}$&-22.72$^{+0.62}_{-0.70}$& 0.93$\pm$0.05  & 6.42$^{+4.19}_{-3.48}$ & 13.94$^{+6.25}_{-7.39}$ & 1.05 $\pm$ 0.02 \\[0.2em]

\hline
\\[-0.8em]
%$\lambda$4629.33 &  -0.17$\pm$0.03 & 15.84$^{+3.72}_{-3.24}$ & 251.48$^{+7.57}_{-6.36}$ & -23.40$^{+1.25}_{-1.26}$ &-21.45$^{+0.48}_{-0.47}$& 1.09$\pm$0.04& 2.11$^{+0.99}_{-0.88}$& 6.63$^{+3.38}_{-3.79}$ & 1.12$\pm$ 0.01\\[0.2em]
$\lambda$4629.33 &  -0.17$\pm$0.03 & 16.23$^{+3.42}_{-3.50}$ & 251.08$^{+7.90}_{-6.95}$ & -23.49$^{+1.16}_{-1.33}$ &-22.32$^{+0.44}_{-0.45}$& 1.02$^{+0.03}_{-0.04}$& 2.09$^{+0.94}_{-0.87}$& 6.75$^{+3.45}_{-3.73}$& 1.12 $\pm$ 0.02\\[0.2em]

\hline
\\[-0.8em]
%$\lambda$4923.93 & -0.41$\pm$ 0.03 & 15.57$^{+1.17}_{-1.14}$& 237.72$^{+2.71}_{-2.63}$ &-26.13$^{+0.45}_{-0.46}$ & -19.36$^{+0.43}_{-0.38}$ &1.23$\pm$ 0.03 &3.98$^{+0.45}_{-0.48}$ & 3.30$^{+2.47}_{-2.18}$& 1.27$\pm$ 0.01\\[0.2em]
$\lambda$4923.93& -0.41$^{+0.02}_{-0.03}$ & 15.53$^{+1.15}_{-1.15}$& 238.10$^{+2.90}_{-2.79}$ &-26.11$^{+0.45}_{-0.46}$ & -20.09$^{+0.39}_{-0.36}$ &1.12$\pm$ 0.03 & 3.95$^{+0.42}_{-0.47}$& 3.53$^{+2.42}_{-2.28}$& 1.27 $\pm$ 0.01\\[0.2em]
\hline
\\[-0.8em]
%$\lambda$5018.43 & -0.39$\pm$0.02 & 20.57$^{+1.31}_{-1.23}$ & 239.68$^{+3.23}_{-3.16}$ & -24.12$^{+0.56}_{-0.54}$ & -20.39$^{+0.39}_{-0.40}$ & 1.24 $\pm$0.03 & 4.34$^{+0.52}_{-0.49}$ & 9.59$^{+1.56}_{-1.76}$& 1.25$\pm$ 0.01\\[0.2em]
$\lambda$5018.43 & -0.38$\pm$0.02 & 20.52$^{+1.29}_{-1.21}$ & 240.65$^{+2.98}_{-3.07}$ & -24.11$^{+0.57}_{-0.56}$ & -21.25$^{+0.35}_{-0.36}$ & 1.14 $\pm$0.03 & 4.23$^{+0.54}_{-0.52}$ & 9.84$^{+1.51}_{-1.68}$ & 1.25 $\pm$ 0.01 \\[0.2em]
\hline
\\[-0.8em]
%$\lambda$5169.03 & -0.49$\pm$0.02 & 18.83$^{+0.93}_{-0.88}$ & 236.12 $^{+2.03}_{-2.22}$ & -24.29$^{+0.41}_{-0.38}$ & -19.66$^{+0.26}_{-0.27}$ &1.46$\pm$0.02 & 3.97$^{+0.35}_{-0.43}$ & 3.02$^{+2.26}_{-1.93}$& 1.31$\pm$ 0.01\\[0.2em]
$\lambda$5169.03& -0.48$\pm$0.02 & 18.23$^{+0.95}_{-0.86}$ & 236.25 $^{+2.38}_{-2.24}$ & -24.32$^{+0.41}_{-0.37}$ & -20.60$^{+0.24}_{-0.23}$ &1.25$\pm$0.02 & 3.70$^{+0.43}_{-0.29}$& 3.07$^{+2.12}_{-1.99}$ & 1.31 $\pm$ 0.01 \\[0.2em]
\hline
\\[-0.8em]
%$\lambda$5197.57 & -0.16 $\pm$ 0.02 & 19.09$^{+2.66}_{-2.50}$& 230.28$^{+7.63}_{-6.99}$ & -23.12$^{+1.25}_{-1.17}$& -20.83$^{+0.63}_{-0.60}$& 1.16$^{+0.06}_{-0.05}$ & 2.02$^{+0.96}_{-1.00}$ & 11.49$^{+3.10}_{-2.93}$& 1.11$\pm$ 0.01\\[0.2em]
$\lambda$5197.57 & -0.16 $\pm$ 0.02 & 18.79$^{+2.91}_{-2.29}$& 230.02$^{+7.03}_{-6.65}$ & -23.31$^{+1.28}_{-1.14}$& -21.55$^{+0.59}_{-0.59}$& 1.06$^{+0.04}_{-0.05}$ & 1.88$^{+0.96}_{-1.03}$ &11.52$^{+3.29}_{-3.21}$&1.11 $\pm$ 0.01\\[0.2em]
\hline
\\[-0.8em]
%$\lambda$5234.62 & -0.16$\pm$ 0.02 & 18.07$^{+2.76}_{-2.29}$ & 236.89$^{+6.48}_{-5.92}$ &-23.36$^{+1.18}_{-1.07}$ & -20.68$^{+0.53}_{-0.50}$ & 1.44$\pm$0.04 & 1.16$^{+0.52}_{-0.48}$ & 9.21$^{+2.97}_{-2.95}$ & 1.11$\pm$ 0.01\\[0.2em]
$\lambda$5234.62  & -0.16$\pm$ 0.02 & 17.50$^{+2.80}_{-2.46}$ & 236.48$^{+6.23}_{-6.16}$ &-23.47$^{+1.17}_{-1.06}$ & -21.16$^{+0.48}_{-0.44}$ & 1.23$\pm$0.04 & 1.51$^{+0.48}_{-0.49}$ & 8.86$^{+2.97}_{-2.94}$ &1.11 $\pm$0.01\\[0.2em]
\hline
\\[-0.8em]
%$\lambda$5275.99 & -0.24 $\pm$ 0.02 & 13.70$^{+1.70}_{-1.55}$ & 226.38$^{+3.50}_{-3.51}$ &-25.92$^{+0.69}_{-0.64}$ & -21.21$^{+0.50}_{-0.49}$ &1.26$\pm$0.05 & 3.46$^{+0.81}_{-0.73}$ & *3.28$^{+3.00}_{-2.26}$& 1.16$\pm$ 0.01 \\[0.2em]
$\lambda$5275.99 & -0.23 $\pm$ 0.02 & 13.44$^{+1.92}_{-1.58}$ & 226.77$^{+3.53}_{-3.36}$ &-25.95$^{+0.71}_{-0.65}$ & -21.64$^{+0.48}_{-0.44}$ &1.12$\pm$0.04 & 3.28$^{+0.76}_{-0.75}$& *3.30$^{-2.90}_{2.26}$& 1.16 $\pm$ 0.02 \\[0.2em]
\hline
\\[-0.8em]
%$\lambda$5316.6 &-0.29$\pm$0.02 & 19.79$^{+1.51}_{-1.32}$ & 241.62$^{+4.04}_{-4.13}$ &-22.73$^{+0.68}_{-0.62}$ & -20.28$^{+0.46}_{-0.44}$ & 1.43$\pm$0.03 & 3.94$^{+0.78}_{-0.64}$ & *2.43$^{+2.53}_{-1.68}$ & 1.19$\pm$ 0.01\\[0.2em]
$\lambda$5316.6&-0.29$\pm$0.02 & 19.59$^{+1.43}_{-1.34}$ & 242.16$^{+4.05}_{-4.04}$ &-22.67$^{+0.64}_{-0.65}$ & -21.39$^{+0.38}_{-0.41}$ & 1.26$\pm$0.03 & 3.79$^{+0.77}_{-0.69}$& *2.61$^{+2.57}_{-1.78}$ & 1.19 $\pm$ 0.01\\[0.2em]
\hline
\\[-0.8em]
%$\lambda$5362.86 & -0.12$\pm$0.02& 18.67$^{+3.91}_{-3.37}$& 242.52$^{+14.48}_{-12.99}$ & -26.18$^{+1.72}_{-1.55}$ & -20.35$\pm$0.77 &1.44$^{+0.04}_{-0.05}$& 3.64$^{+1.47}_{-1.19}$ & 12.87$^{+3.76}_{-3.60}$ & 1.08$\pm$ 0.02\\[0.2em]
$\lambda$5362.86& -0.12$\pm$0.02& 18.51$^{+3.90}_{-3.25}$& 242.76$^{+12.99}_{-12.79}$ & -26.23$^{+1.65}_{-1.48}$ & -21.19$^{+0.60}_{-0.57}$ &1.31 $\pm$ 0.05& 3.26$^{+1.47}_{-1.12}$& 12.73$^{+3.87}_{-3.65}$ & 1.08 $\pm$ 0.01\\[0.2em]
\hline

\end{tabular}\\
%\tablefoot{}
\tablefoot{The list includes the parameters  fitted in the MCMC analysis to detect the planetary signal, as well as the MCMC analysis to fit the micro- and macroturbulence velocity to the observed transmission spectrum based on the NLTE synthetic transmission spectra. The values marked with an asterisk ``*'' are consistent with 0 km s$^{-1}$.}
\label{tab:K9_detection}
\end{table*}

%\end{landscape}

\begin{figure}[h]
\centering
\includegraphics[width=0.95\hsize]{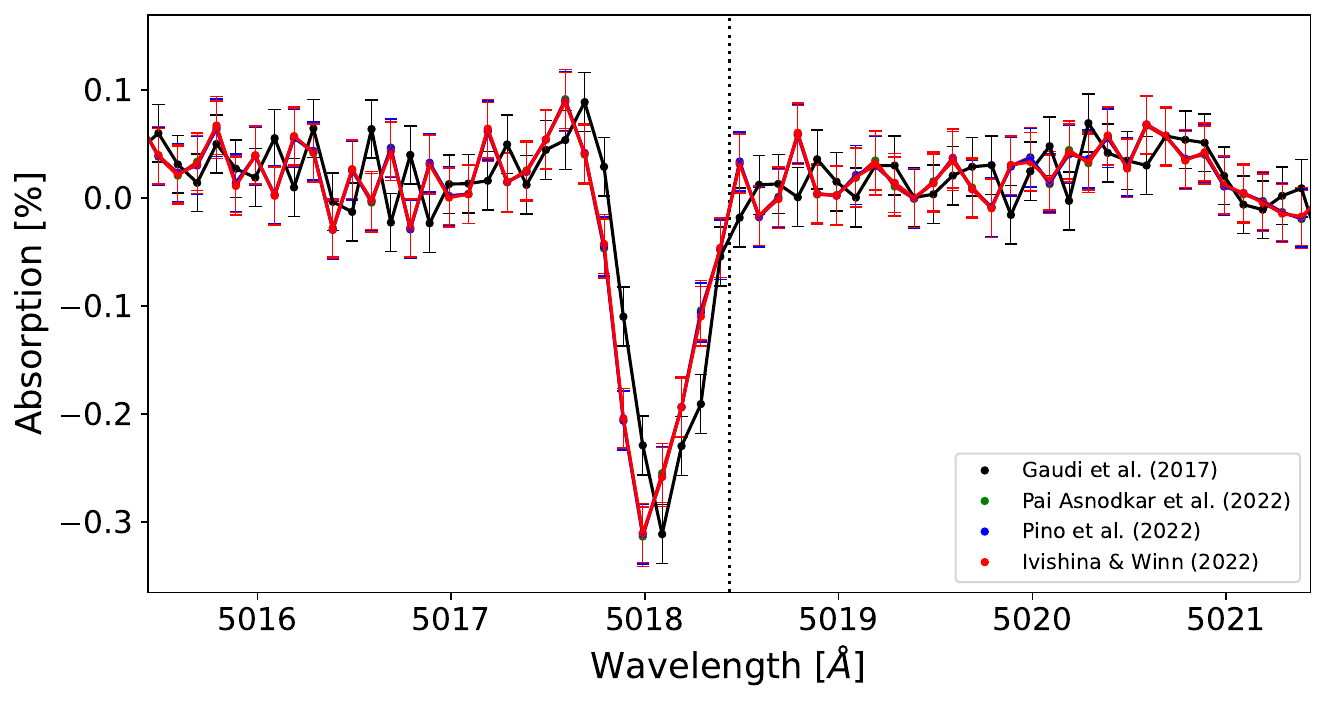}   
\caption{Transmission spectrum around the \ion{Fe}{ii} 5018.4 $\AA$ line in the atmosphere of KELT-9b for T0 and period from \citealt{Gaudi_K9_2017} (black dots), \citealt{Pai_K9_FeII} (green dots), \citealt{Pino_K9_2022} (blue points), and \citealt{Ivshina_Winn_2022_K9} (red points). The black vertical line represents the theoretical position of the studied line. Blue and green points are indistinguishable from the red and black points.   } 
\label{Fig:ts_all_5018}
\end{figure}

As previously stated, the final velocity of the wind is dependent on the different literature values of T0 and the period. We repeated the calculation around \ion{Fe}{ii} 5018.4 \AA,  using the values presented in \citealt{Gaudi_K9_2017}, \citealt{Pai_K9_FeII}, \citealt{Pino_K9_2022}, and \citealt{Ivshina_Winn_2022_K9}. The transmission spectrum can is shown in Fig. \ref{Fig:ts_all_5018}. The only discrepancy is observed in the values derived from \citealt{Gaudi_K9_2017}, where the position of the line is approximately 6 km s$^{-1}$ away. However, this discrepancy is not reflected in the line position for values derived from the remaining literature. As the values obtained using TESS data are in alignment with each other, we have determined that these are the most precise.

%%%%%%%%%%%%%%%%%%%%%%%%%%%%%%%%%%%%%%%%%%%%%%%%%%%%%%%%%%%%%%%%%%%%%%%%%%%%%%%%%%%%%%%%%%%%%%%%%%%%%%%%%%%%%%%%%%%%%%%%%%%%%%%%%%%%%%%%%%%%%%%%%%%%%%%%%%%%%%%%%%%%%%%%%%%%%%%%%%%%%%%%%%%%%%%%%%%%%%%%%%%%%%%%%%%%%%%%%%%%%%%%%%%%%%%%%%%%%%%%%%%%%%%%%%%%%%%%%%%%%%%%%%%%%%%%%%%%

\section{Detected lines:\ KELT-20b}\label{Section:M2-detected_lines}

\begin{figure*}[h]
\centering
\includegraphics[width=0.95\textwidth]{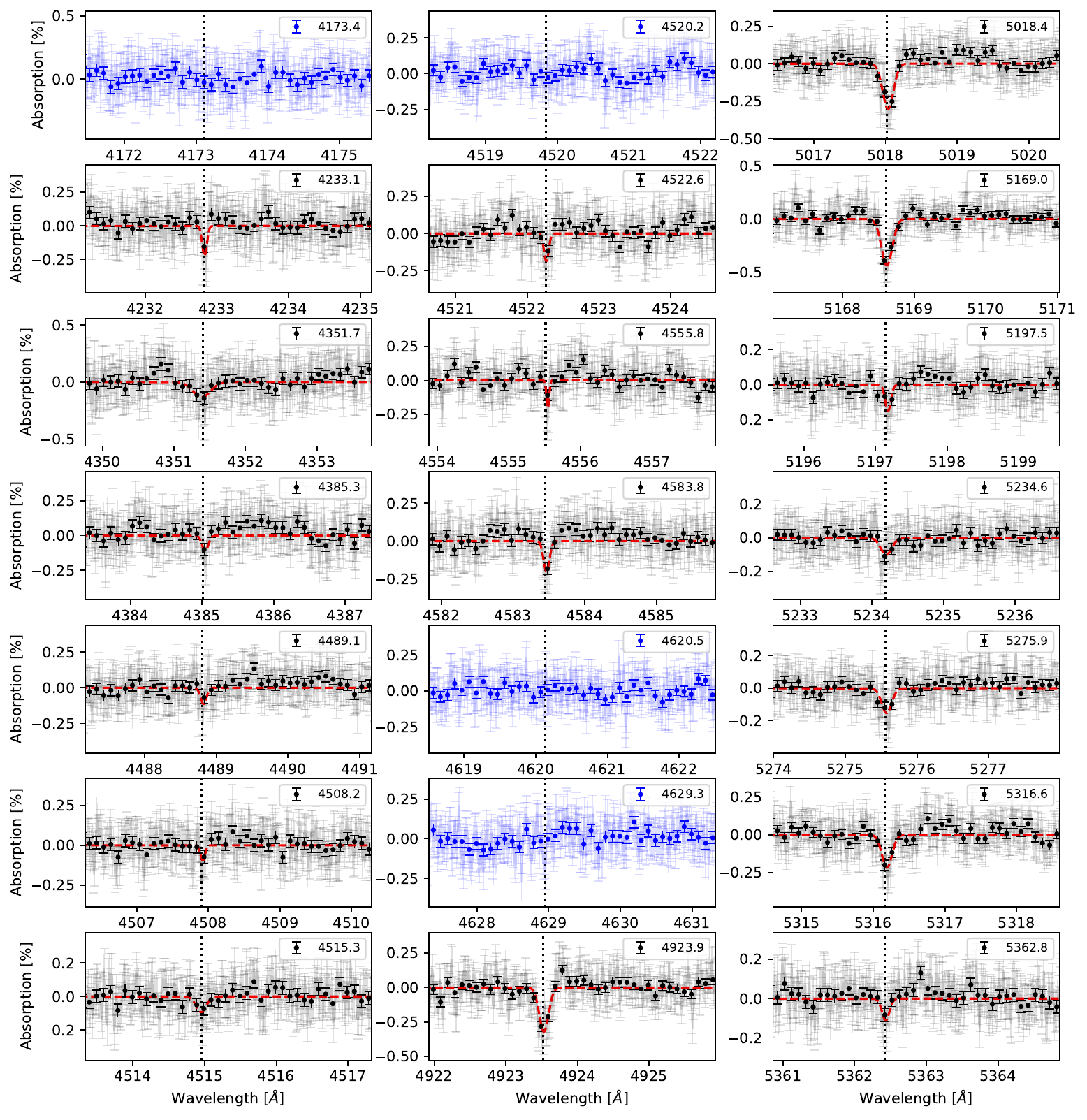}
\caption{Transmission spectra for all the lines detected in KELT-20b (gray dots) and the transmission spectra around the non-detected lines are marked with blue dots. Dark dots indicate the binned transmission spectrum with a step of 0.1 $\AA$. The red dashed lines represent the best fit of the planetary signal from the MCMC analysis. }
\label{Fig:TS_ALL_K20}
\end{figure*}

The analysis of six transits of KELT-20b led to the detection of 17 single lines of \ion{Fe}{ii}. The list of detected lines and the fitted parameters are presented in Table \ref{tab:K20_detection} and the transmission spectra of the studied lines are presented in Fig. \ref{Fig:TS_ALL_K20}. The lines $\lambda$5018 $\AA$, $\lambda$5169 $\AA$, and $\lambda$5316 $\AA$ were previously detected by \citet{M2_Casasayas_2019}, while the detection of the remaining lines is presented here for the first time. 

\begin{figure}[h]
\centering
\includegraphics[width=0.9\hsize]{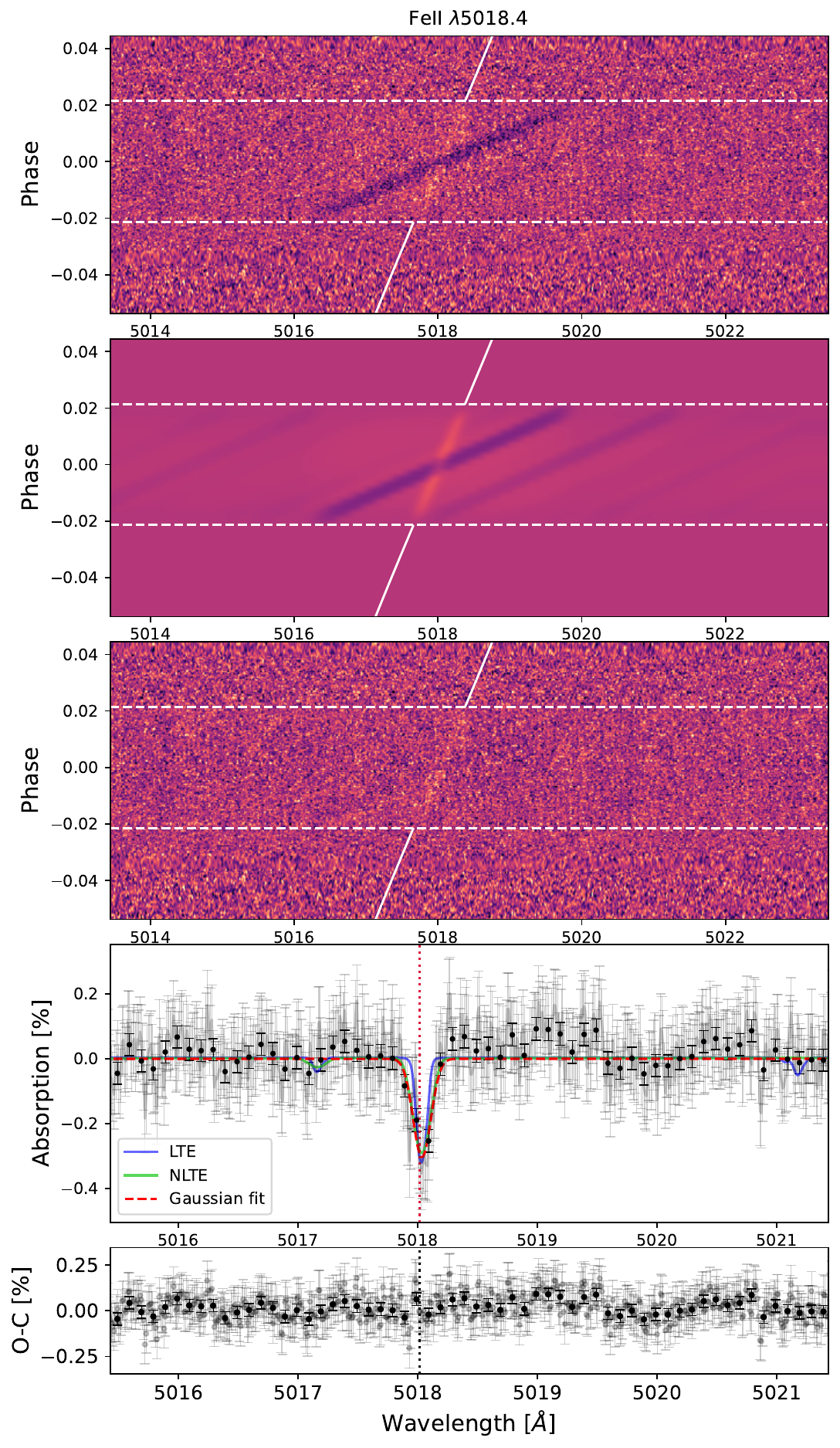}
\caption{Same as Fig. \ref{Fig:TS_steps_K9}, but for KELT-20b. In the top panel, the RM effect is a tilted dark signal and the planetary signal from \ion{Fe}{ii} is a bright vertical signal.}
\label{Fig:TS_steps_M2}
\end{figure}

Similarly to Fig. \ref{Fig:TS_steps_K9}, in Fig. \ref{Fig:TS_steps_M2} we present the MCMC and TS analysis around the \ion{Fe}{ii} $\lambda$5018.4 $\AA$ line. The RM effect is visible as a dark tilted signal, while the atmospheric detection is a light slightly tilted signal. The corner plots of the MCMC analysis and the transmission spectrum for all the studied lines can be found in the appendix (Figs. \ref{Fig:TS_M2_1}, \ref{Fig:TS_M2_2}, \ref{Fig:TS_M2_3}, \ref{Fig:TS_M2_4}, \ref{Fig:TS_M2_5}, and \ref{Fig:TS_M2_6} for TS, and Figs. \ref{fig:corner_k20_1}, \ref{fig:corner_k20_2},  \ref{fig:corner_k20_3},  \ref{fig:corner_k20_4},  \ref{fig:corner_k20_5},  \ref{fig:corner_k20_6},  \ref{fig:corner_k20_7}, \ref{fig:corner_k20_8},  \ref{fig:corner_k20_9},  \ref{fig:corner_k20_10}, and  \ref{fig:corner_k20_11} for corner plots.)  

\begin{figure}[h]
\centering
\includegraphics[width=0.98\hsize]{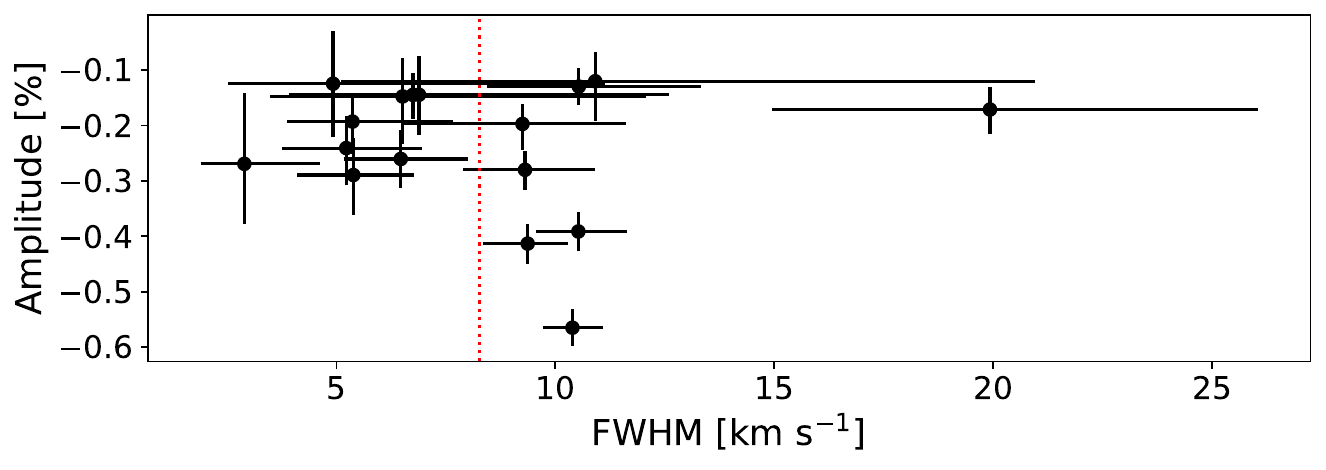}
\includegraphics[width=0.98\hsize]{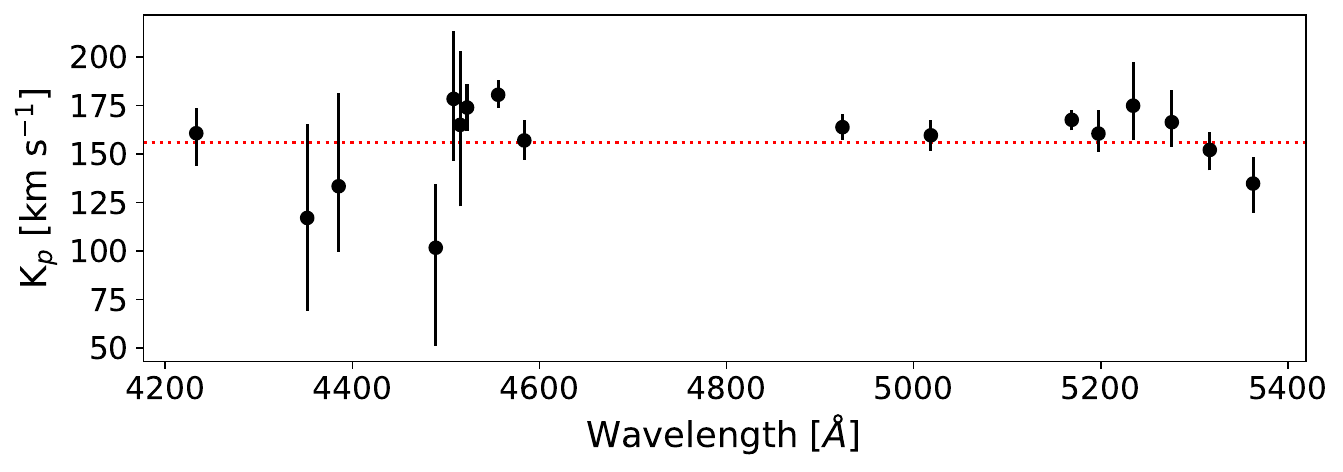}
\caption{Same as Fig. \ref{Fig:RES_K9_AMP_FWHM}, but for KELT-20b.}
\label{Fig:RES_M2_AMP_FWHM}
\end{figure}

In Fig. \ref{Fig:RES_M2_AMP_FWHM} we present the amplitude of the absorption signal as a function of FWHM and $K_{\rm p}$ as a function of the wavelength of the detected lines. We find an average FWHM value of 8.27 $\pm$ 3.74 km\,s$^{-1}$ and $K_{\rm p}$ of 155.8 $\pm$ 21.2 km\,s$^{-1}$, where the latter is consistent with the $K_{\rm p}$ obtained considering the orbital parameters, namely, $K_{\rm p}$ = 169.34 $\pm$ 6.56 km\,s$^{-1}$. 

%observe that the detected radius of the planet (R$_\lambda$) grows with the wavelength of the studied line. We discuss this effect in Section \ref{SEC:Radius}. 

\begin{figure}[h]
\centering
\includegraphics[width=0.98\hsize]{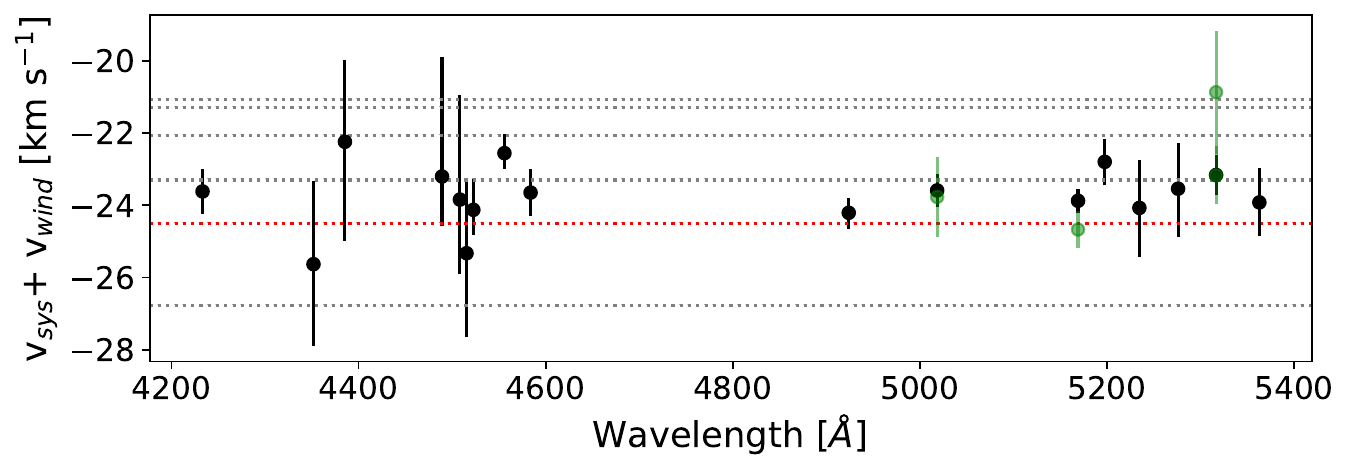}
\includegraphics[width=0.98\hsize]{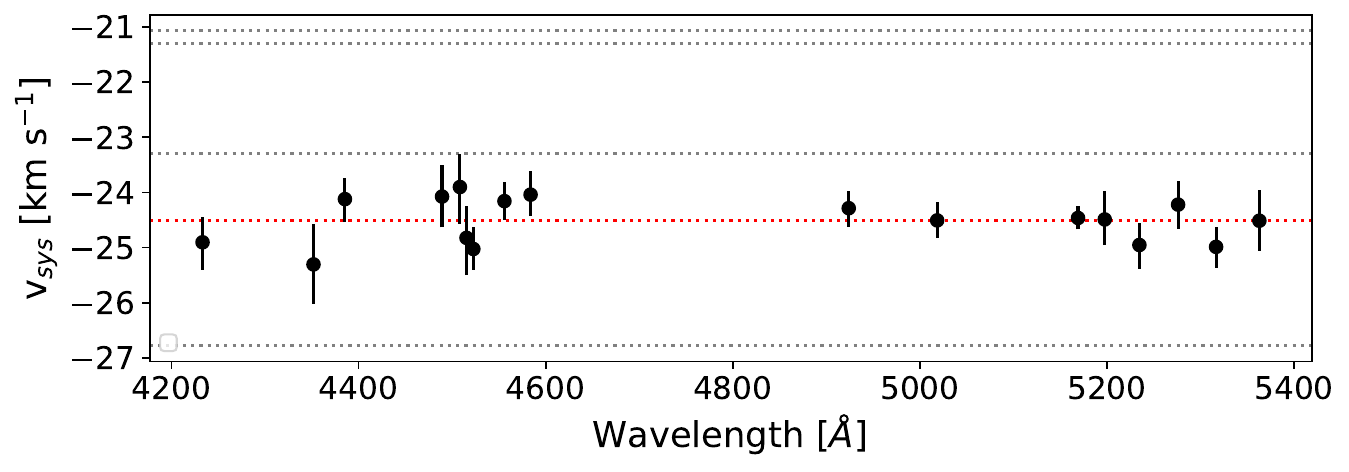}
\includegraphics[width=0.98\hsize]{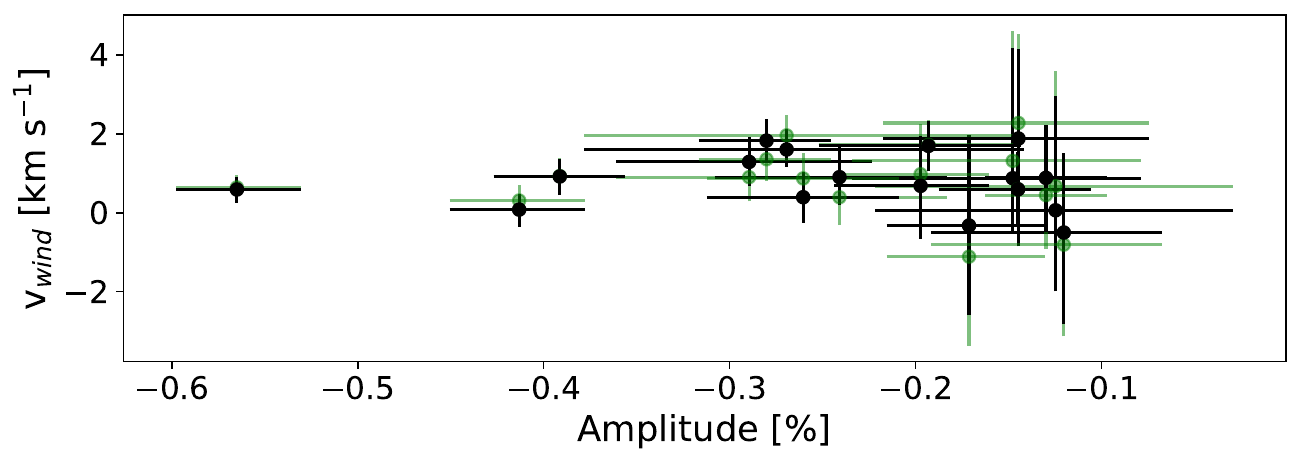}
\caption{Same as Fig. \ref{Fig:RES_K9_2}, but for KELT-20b. In the upper panel, the green points represent the measurement obtained by \citet{M2_Casasayas_2019}. In the middle panel, we plot several literature systemic velocities with gray horizontal lines: $-$23.3 $\pm$ 0.3 km\,s$^{-1}$ \citep{Lund_M2}, $-$21.3 $\pm$ 0.4  \citep{Talens_M2}, $-$22.06 $\pm$ 0.35 km\,s$^{-1}$ \citep{Nugroho_M2}, $-$24.48 $\pm$ 0.04 km\,s$^{-1}$ \citep{Rainer_M2}, and $-$26.78 $\pm$ 0.71 km\,s$^{-1}$ (\textit{Gaia}). The red horizontal line indicates the mean $v_{\rm sys}$ value of $-$24.52 km\,s$^{-1}$ derived in this work.}
\label{Fig:RES_M2_2}
\end{figure}

In Fig. \ref{Fig:RES_M2_2} we present the obtained $v_{\rm sys+wind}$ and $v_{\rm sys}$ values as a function of the wavelength of the detected lines, and $v_{\rm wind}$ as a function of the amplitude of the detected signal. We obtain an average $v_{\rm sys}$ value of $-$24.52 $\pm$ 0.40 km\,s$^{-1}$. Due to the discrepancy among $v_{\rm sys}$ values present in the literature, namely $-$23.3 $\pm$ 0.3 km\,s$^{-1}$ \citep{Lund_M2}, $-$21.3 $\pm$ 0.4 and $-$21.07 $\pm$ 0.03 km\,s$^{-1}$ \citep{Talens_M2}, $-$22.06 $\pm$ 0.35 and $-$22.02 $\pm$ 0.47 km\,s$^{-1}$ \citep{Nugroho_M2}, $-$24.48 $\pm$ 0.04 km\,s$^{-1}$ \citep{Rainer_M2}, and $-$26.78 $\pm$ 0.71 km\,s$^{-1}$ (\textit{Gaia}), it is difficult to determine the optimal value and correct for $v_{\rm sys}$. Therefore, the existence or non-existence of winds in the atmosphere of this planet cannot be reliably determined. The mean value of $v_{\rm sys+winds}$ obtained from the MCMC analysis is $-$23.73 $\pm$ 0.84 km\,s$^{-1}$. Assuming the value of $v_{\rm sys}$ calculated in this work, we obtain $v_{\rm wind}$ = 0.8 $\pm$ 0.8 km\,s$^{-1}$, which is consistent with the non-detection of atmospheric winds.

%Due to the discrepancy between $v_{sys}$ from the literature -23.3\textbf{$\pm$ 0.3} km\,s$^{-1}$ \citep{Lund_M2}, -21.3\textbf{$\pm$ 0.4} and \textbf{-21.07$\pm$ 0.03} km\,s$^{-1}$ \citep{Talens_M2}, -22.06\textbf{$\pm$ 0.35} and -22.02\textbf{$\pm$ 0.47}  km\,s$^{-1}$ \citep{Nugroho_M2}, -24.48 \textbf{$\pm$ 0.04} km\,s$^{-1}$ \citep{Rainer_M2}, and -26.78 \textbf{$\pm$ 0.71} km\,s$^{-1}$ from GAIA, it is difficult to rule out the best and correct $v_{sys}$ of this system. As a result of this, the existence or non-existence of the winds in the atmosphere of this planet can not be determined. \textbf{The mean value of $v_{sys+winds}$  obtain with MCMC analysis is -23.73 $\pm$ 0.84 km\,s$^{-1}$. By assuming the $v_{sys}$ calculated in this work, $v_{wind}$ = 0.8 $\pm$ 0.8 km\,s$^{-1}$, which is consistent with no-detection of atmospheric winds.}%{\color{red}Assuming the $v_{sys}$ calculated in our work $v_{wind}$ = 0.9 $\pm$ 0.8 km\,s$^{-1}$,} showing slightly red-shifted signal, as a consequence of possible night-to-day side winds. \textit{-23.64 0.7979}

%\begin{figure}[h]
%%\centering
%\includegraphics[width=\hsize]{Figures/mascara-2/all_results/vel_all_M2.jpg}
%\caption{\textbf{New Figure:}The $v_{sys}+v_{wind}$ for each of the detected lines compared with the literature values. The green points represent the measurement obtained by \citet{M2_Casasayas_2019}.}
%%\label{Fig:VEL_M2_2}
%\end{figure}

%\begin{landscape}
\begin{table*}[]
\centering
\tiny
\caption{Detected \ion{Fe}{ii} lines in the transmission spectrum of KELT-20b.}
\setlength{\tabcolsep}{5pt}
\begin{tabular}{c|ccccccc|cc|c}
\hline\hline
\\[-0.8em]
Line &   h    &FWHM    & K$_p$ &$v_{sys}+v_{wind}$  & $v_{sys}$ &  R$_\lambda$&Angle&$\nu_{\rm mic}$ & $\nu_{\rm mac}$& R$_{eff}$\\[0.1em]
\\[-1.1em]
&    [$\%$]    &[km\,s$^{-1}$]& [km\,s$^{-1}$]  & [km\,s$^{-1}$]&  [km\,s$^{-1}$]&  [R$_p$]&[deg]&[km\,s$^{-1}$]& [km\,s$^{-1}$] & [R$_p$]\\[0.1em]
\hline\hline
\\[-0.9em]
$\lambda$4233.16& -0.29 $\pm$ 0.07& 5.39$^{+1.38}_{-1.27}$ & 160.76$^{+13.01}_{-16.78}$ & -23.62$^{+0.63}_{-0.61}$ & -24.90 $^{+0.45}_{-0.50}$ & 1.00$\pm$ 0.02 & 29.00$^{+7.63}_{-8.09}$ & 0.46$^{+0.40}_{-0.33}$ &*1.98$^{+2.66}_{-1.38}$ & 1.1 $\pm$ 0.02 \\
%\\[-0.8em]
% & -0.37 $\pm$ 0.07 & 9.84$^{+2.48}_{-1.87}$ & 127.33 $^{+19.47}_{-15.05}$ & -22.89 $^{+0.9}_{-1.04}$ & -24.41$^{+0.57}_{-0.62}$ & 1.00 $\pm$ 0.03 & 21.79$^{+9.23}_{-9.83}$ & 3.55$^{+2.26}_{-1.64}$ & *6.42$^{+6.06}_{-4.33}$  \\
 \\[-1em]

\hline
\\[-0.9em]
$\lambda$4351.7 &-0.17 $\pm$ 0.04&19.93$^{+6.07}_{-4.96}$ & 117.07$^{+47.69}_{-47.48}$&-25.63$^{+2.28}_{-2.25}$ &-25.30$^{+0.74}_{-0.70}$ &0.92$\pm$0.03& 21.82$^{+10.91}_{-10.88}$ & *0.80$^{+1.18}_{-0.60}$ & 6.34$^{+6.74}_{-3.73}$ &1.06$^{+0.01}_{-0.02}$\\
\\[-1em]

\hline
\\[-0.9em]
$\lambda$4385.3 &-0.14 $\pm$ 0.07&6.89 $^{+5.64}_{-2.95}$ & 133.42$^{+48.04}_{-33.82}$&-22.24$^{+2.26}_{-2.73}$ &-24.12$^{+0.39}_{-0.41}$ &1.11$\pm$0.02& 27.36$^{+6.18}_{-6.71}$ & 0.13$^{+0.09}_{-0.09}$ & *4.49$^{+4.28}_{-3.21}$  & 1.05$^{+0.02}_{-0.03}$\\
\\[-1em]

\hline
\\[-0.9em]
$\lambda$4489.1 &-0.15$^{+0.07}_{-0.09}$&6.51$^{+5.52}_{-3.02}$ & 101.71$^{+32.82}_{-50.26}$&-23.20$^{+3.26}_{-1.36}$ &-24.07$^{+0.58}_{-0.56}$ &0.91 $\pm$ 0.03& 45.27$^{+10.02}_{-9.95}$ & *0.44$^{+0.52}_{-0.32}$  & **14.73$^{+7.38}_{-8.06}$  & 1.05$^{+0.02}_{-0.03}$\\
\\[-1em]

\hline
\\[-0.9em]
$\lambda$4508.2 &-0.12 $\pm$ 0.10&4.92$^{+6.20}_{-2.38}$ & 178.47$^{+34.87}_{-30.94}$&-23.84$^{+2.87}_{-2.05}$ &-23.90$^{+0.60}_{-0.66}$ &0.94 $\pm$ 0.03& 34.12$^{+9.78}_{-9.72}$ & *0.70$^{+1.18}_{-0.52}$  & **11.40$^{+8.51}_{-6.47}$  & 1.04 $\pm$ 0.03 \\
\\[-1em]

\hline
\\[-0.9em]
$\lambda$4515.33 &-0.12$^{+0.05}_{-0.07}$&10.91$^{10.00}_{-5.79}$ & 165.06$^{+37.92}_{-41.46}$&-25.33$^{+2.01}_{-2.32}$ &-24.82$^{+0.57}_{-0.67}$ &0.99 $\pm$ 0.03& 29.17$^{+9.93}_{-9.29}$ & *1.39$^{+1.69}_{-0.96}$ & 11.02$^{+8.52}_{-6.37}$ & 1.04 $\pm$ 0.02\\
\\[-1em]

\hline
\\[-0.9em]
$\lambda$4522.63 & -0.24$^{+0.06}_{-0.07}$& 5.23$^{+1.72}_{-1.47}$ & 174.00$^{+12.01}_{-12.19}$ &-24.13$^{+0.85}_{-0.70}$ & -25.02$^{+0.39}_{-0.37}$ & 1.07 $\pm$ 0.02 &49.33$^{+6.61}_{-7.02}$ & 0.24$^{+0.47}_{-0.17}$ & 3.24$^{+2.76}_{-2.05}$ & 1.08 $\pm$ 0.02
\\
%\\[-0.8em]
% & -0.19$^{+0.06}_{-0.09}$ & 10.52$^{+6.67}_{-4.78}$ & 190.03$^{+18.58}_{-34.44}$ &  -24.90$^{+1.42}_{-2.48}$ & -25.37$^{+0.52}_{-0.55}$ & 1.05$^{+0.02}_{-0.03}$ &37.93$^{+9.09}_{-8.55}$ &  \\
\\[-1em]

\hline
\\[-0.9em]
$\lambda$4555.89& -0.27$^{+0.13}_{-0.11}$& 2.90$^{+1.71}_{-0.99}$ & 180.59$^{+7.60}_{-6.57}$ &-22.56$^{+0.51}_{-0.43}$ & -24.16$^{+0.35}_{-0.34}$ & 1.14 $\pm$ 0.02 & 37.37$^{+5.41}_{-6.44}$ & *0.15$^{+0.10}_{-0.10}$ & *1.65$^{+3.07}_{-1.14}$ & 1.09 $\pm$ 0.04 \\%\\[-0.8em]
%&  -0.29$^{+0.08}_{-0.09}$& 5.13$^{+2.29}_{-1.57}$ & 187.12$^{+12.12}_{-11.00}$ &-22.39$^{+0.67}_{-0.69}$ & -24.00$^{+0.42}_{-0.46}$ & 1.14$^{+0.02}_{-0.03}$ & 29.50$^{+7.54}_{-7.23}$ & *0.22$^{+0.49}_{-0.15}$ & *1.64$^{+1.98}_{-1.18}$ \\
\\[-1em]
\hline
\\[-0.9em]
$\lambda$4583.83 &-0.26 $\pm$ 0.05&6.47$^{+1.51}_{-1.30}$ & 157.03$^{+10.37}_{-9.81}$&-24.65$^{+0.65}_{-0.64}$ &-24.04 $^{+0.42}_{-0.39}$ &1.09$\pm$0.02& 44.18$^{+7.05}_{-6.50}$ &0.16$^{+0.13}_{-0.16}$ & 2.50$^{+1.88}_{-1.60}$ & 1.09 $\pm$ 0.02\\
%\\[-0.8em]
%&-0.27 $\pm$ 0.07&6.98$^{+2.02}_{-1.52}$ & 166.08$^{+12.85}_{-11.04}$&-24.41$^{+0.76}_{-0.86}$&-23.66$^{+0.44}_{-0.51}$&1.11$^{+0.03}_{-0.02}$& 41.87$^{+8.06}_{-8.09}$ & *0.25$^{+0.46}_{-0.17}$ & 2.85$^{+2.15}_{-1.79}$ \\
\\[-1em]

 \hline
 \\[-0.9em]
$\lambda$4923.93 & -0.41 $\pm$ 0.04 & 9.37$^{+0.91}_{-1.00}$& 163.88$^{+6.74}_{-6.69}$ &-24.21$^{+0.40}_{-0.44}$ & -24.24$^{+0.32}_{-0.33}$&1.23 $\pm$ 0.02&48.32$^{+6.07}_{-5.42}$ & 3.03$^{+0.52}_{-0.53}$  & 3.45$^{+2.10}_{-2.22}$& 1.14 $\pm$ 0.01\\
%\\[-0.8em]
%& -0.43 $\pm$ 0.045 & 9.17$^{+1.10}_{-1.00}$& 158.94$^{+6.85}_{-7.37}$ &-24.09$^{+0.47}_{-0.52}$ & -24.44$^{+0.41}_{-0.40}$&1.23$^{+0.02}_{-0.03}$&43.57$^{+6.91}_{-6.78}$ & 2.45$^{+0.57}_{-0.56}$ & 2.73$^{+2.08}_{-1.81}$ \\
\\[-1em]

 \hline
 \\[-0.9em]
$\lambda$5018.43&-0.39 $\pm$ 0.04 & 10.53 $^{+1.09}_{-0.96}$ &159.68$^{+7.86}_{-7.86}$&-23.59$^{+0.45}_{-0.47}$& -24.51$^{+0.32}_{-0.32}$ & 1.22 $\pm$ 0.02 &38.03$^{+5.13}_{-5.69}$ & 1.80$^{+0.61}_{-0.66}$ & 5.33$^{+1.57}_{-1.88}$ & 1.13 $\pm$ 0.11\\
%\\[-0.8em]
%&-0.39$\pm$0.04 & 10.73$^{+1.28}_{-1.19}$ &160.06$^{+10.36}_{-10.14}$&-24.04$^{+0.56}_{-0.61}$& -24.53$^{+0.38}_{-0.39}$&1.23$^{+0.03}_{-0.02}$ &30.73$^{+6.56}_{-6.44}$ & 2.11$^{+0.74}_{-0.73}$ & 4.06$^{+1.89}_{-2.33}$\\
\\[-1em]

\hline
\\[-0.9em]
$\lambda$5169.03 & -0.56 $\pm$ 0.03 & 11.19$^{+0.88}_{-0.83}$ & 167.66$^{+4.87}_{-5.39}$ & -23.88$^{+0.32}_{-0.33}$ &-24.46$^{+0.22}_{-0.20}$ & 1.20 $\pm$ 0.01 &48.47$^{+3.85}_{-4.09}$ & 2.27$^{+0.44}_{-0.36}$& 1.97$^{+1.65}_{-1.32}$ & 1.19 $\pm$ 0.01\\
%\\[-0.8em]
%& -0.58 $\pm$ 0.04 & 11.35$^{+0.95}_{-0.84}$ & 166.13$^{+6.27}_{-6.39}$ & -24.13 $\pm$ 0.41 & -24.55$^{+0.25}_{-0.24}$ & 1.21 $\pm$ 0.01 &48.37$^{+3.62}_{-4.45}$ & 3.06$^{+0.42}_{-0.49}$ & 2.79$^{+2.03}_{-1.87}$\\
\\[-1em]

%\hline
%\\[-0.9em]
%$\lambda$5197.57&-0.16 $\pm$ 0.07& 6.00$^{+4.42}_{-2.16}$& 163.79$^{-19.48}_{-14.12}$&-22.84$^{+0.99}_{-1.23}$& -24.62$^{+0.64}_{-0.65}$ &1.15 $\pm$ 0.03 & 36.30$^{+10.19}_{-9.14}$ & *0.17$^{+0.29}_{-0.11}$ & *4.19$^{+7.41}_{-3.40}$ \\
%%\\[-0.8em]
%&-0.19 $\pm$ 0.07& 5.12$^{+3.49}_{-1.73}$& 159.18$^{-15.24}_{-11.86}$&-22.81$^{+0.80}_{-1.16}$& -24.67$^{+0.70}_{-0.63}$ &1.15$^{+0.04}_{-0.03}$ & 38.39$^{+9.14}_{-9.32}$ & *0.21$^{+0.40}_{-0.14}$ & *3.49$^{+4.60}_{-2.64}$ \\
%\\[-1em]
\hline
\\[-0.9em]
$\lambda$5197.57 &-0.19$^{+0.05}_{-0.06}$& 5.37$^{+2.30}_{-1.50}$& 160.58$^{+12.06}_{-9.15}$&-22.80$^{+0.62}_{-0.64}$ &-24.49$^{+0.52}_{-0.47}$ &1.17$\pm$0.07& 38.53$^{+7.44}_{-7.36}$ &*0.14$^{+0.11}_{-0.10}$ & *4.51$^{+5.40}_{-3.42}$  & 1.07 $\pm$ 0.02\\
\\[-1em]

\hline
\\[-0.9em]
$\lambda$5234.62 &-0.13 $\pm$ 0.03& 10.54$^{+2.77}_{-2.09}$& 174.96$^{+22.51}_{-17.88}$&-24.07$^{+1.34}_{-1.37}$& -24.95$^{+0.40}_{-0.43}$ & 1.24$\pm$0.02 & 39.50$^{+6.42}_{-6.45}$ & 3.03$^{+2.43}_{-1.58}$& 6.07$^{+6.85}_{-4.11}$ & 1.05 $\pm$ 0.01\\
%\\[-0.8em]
%&-0.13 $\pm$ 0.05& 9.00$^{+4.67}_{-2.63}$& 160.20$^{+41.08}_{-31.87}$&-22.86$^{+1.87}_{-2.56}$& -25.07$^{+0.48}_{-0.49}$ &1.24 $\pm$ 0.03 & 33.36$^{+8.68}_{-7.61}$ & 0.68$^{+0.62}_{-0.41}$ & 8.30$^{+10.36}_{-4.78}$\\
\\[-1em]

\hline
\\[-0.9em]
$\lambda$5275.99 &-0.20 $^{+0.04}_{-0.05}$& 9.25$^{+2.37}_{-2.71}$& 166.43$^{+16.27}_{-12.49}$&-23.54$^{+1.26}_{-1.33}$& -24.22$^{+0.43}_{-0.43}$ &1.22 $\pm$ 0.03 & 48.08$^{+7.17}_{-7.33}$ & 4.31$^{+2.00}_{-1.67}$ & *3.51$^{3.00}_{-2.40}$ & 1.07 $\pm$ 0.02\\
%\\[-0.8em]
%&-0.28 $\pm$ 0.06& 5.72$^{+1.76}_{-1.16}$& 162.12$^{+8.24}_{-8.58}$&-22.25$^{+0.57}_{-0.68}$& -24.33$^{+0.56}_{-0.50}$ &1.23 $\pm$ 0.03 & 47.24$^{+8.95}_{-8.84}$ & 2.01$^{+1.68}_{-1.08}$ & *3.06$^{+3.44}_{-2.10}$\\
\\[-1em]

\hline
\\[-0.9em]
$\lambda$5316.16 &-0.28 $^{+0.03}_{-0.04}$ &9.31$^{+1.60}_{-1.40}$& 152.15$^{+9.40}_{-10.53}$&-23.16$\pm$0.55& -24.99$^{+0.36}_{-0.38}$ &1.24 $\pm$ 0.02 & 47.20$^{+5.64}_{-6.07}$ &2.52$^{+1.20}_{-0.92}$ &3.25$^{+2.78}_{-2.12}$ & 1.10 $\pm$ 0.01\\
%\\[-0.8em]
%&-0.28 $\pm$ 0.04& 11.67$^{+2.64}_{-1.94}$& 154.28$^{+15.35}_{-12.28}$&-23.02$^{+0.77}_{-0.75}$& -25.46 $\pm$ 0.44 &1.25 $\pm$ 0.02 & 45.33$^{+6.71}_{-7.18}$ & 4.59$^{+1.85}_{-1.43}$  & *3.43$^{+3.17}_{-2.28}$\\
\\[-1em]

\hline
\\[-0.9em]
$\lambda$5362.86 &-0.14 $\pm$ 0.04& 6.75$^{+2.54}_{-1.82}$& 134.80$^{+13.65}_{-15.07}$&-23.92$\pm$ 0.93& -24.51 $^{+0.55}_{-0.54}$ &1.19 $\pm$ 0.03 & 43.13$^{+9.18}_{-7.46}$ &*3.42$^{+4.30}_{-2.35}$ & *5.37$^{+7.17}_{-3.74}$ & 1.05 $\pm$ 0.01\\
%\\[-0.8em]
% &-0.20$^{+0.05}_{-0.06}$& 6.60$^{+2.66}_{-1.74}$& 139.64$^{+10.73}_{-14.71}$&-23.70$^{+0.79}_{-0.87}$& -24.69$^{+0.61}_{-0.64}$ &1.20 $\pm$ 0.04 & 51.11$^{+9.48}_{-9.95}$ & *0.98$^{+1.19}_{-0.66}$ & *4.80$^{+7.86}_{-3.49}$\\
%\\[-1em]
\hline
\end{tabular}\\
\tablefoot{The values marked with an asterisk ``*'' are consistent with 0 km s$^{-1}$, while a double asterisk ``**'' indicates values that could not be determined through the MCMC analysis.}
\label{tab:K20_detection}
\end{table*}
%\end{landscape}

\section{NLTE transmission spectroscopic models}\label{SEC:TS_models}

% In order to compare our results with the models, we computed synthetic transmission spectrum for both planets with xxxx. The models were computed assuming local thermodynamic equilibrium (LTE) and non-LTE for several macroturbulences.

\citet{Fossati_NLTE_K9,Fossati_2023_M2} showed that NLTE effects impact significantly the temperature-pressure (TP) profile and transmission spectrum of both KELT-9b and KELT-20b. In particular, NLTE effects increase by more than 3000\,K the temperature in the middle and upper atmosphere, compared to TP profiles computed assuming LTE. Generally NLTE effects lead to stronger absorption in transmission spectra, except for some typically weak lines mostly of neutral iron and calcium.

We compared the results of our observations with the synthetic NLTE transmission spectra of \citet{Fossati_NLTE_K9,Fossati_2023_M2}.  The considered TP profiles are the same as those presented by \citet{Fossati_NLTE_K9} for KELT-9b and by \citet{Fossati_2023_M2} for KELT-20b. Each of the two TP profiles has been computed employing the {\sc helios} code \citep[][see also \citealt{Fossati_NLTE_K9} for the additionally considered opacities]{malik2017,malik2019} in the lower atmosphere (pressures higher than about 0.1\,mbar) and the {\sc Cloudy} NLTE radiative transfer code \citep[version 17.03;][]{cloudy_2013,cloudy_2017}, through the {\sc Cloudy} for Exoplanets (CfE) interface \citep{Fossati_NLTE_K9,W121_Young}, at lower pressures, and assuming solar abundances. This separation was applied because {\sc helios} does not account for NLTE effects that are relevant in the middle and upper atmosphere; whereas {\sc Cloudy}, which considers NLTE effects, is unreliable at densities greater than 10$^{15}$\,cm$^{-3}$ \citep[see][for more details]{cloudy_2017}. \citet{Fossati_NLTE_K9,Fossati_2023_M2} presented also LTE TP profiles that have been computed in the same way, but assuming LTE for the {\sc Cloudy} calculations. Transmission spectra have been computed on the basis of the TP and abundance profiles obtained from the LTE and NLTE calculations following the procedure described by \citet{young2020} and \citet{fossati2020_KELT9datadriven}. We remark that the NLTE transmission spectra were computed by employing the NLTE TP profiles and enabling NLTE in the {\sc Cloudy} calculations. The LTE transmission spectra have been computed using the LTE TP profiles and imposing the LTE assumption for the {\sc Cloudy} transmission spectra calculation. Furthermore, these are forward self-consistent models that do not employ free parameters that are tweaked to fit the observations (i.e., the computation of the TP profiles and of the synthetic spectra is agnostic of the observations and employs exclusively the system parameters and the stellar spectral energy distribution as input). All details of the atmospheric structure (i.e., TP and abundance profiles) and transmission spectra calculations for KELT-9b and KELT-20b can be found in \citet{Fossati_NLTE_K9,Fossati_2023_M2}, respectively.

As in previous works \citep[][]{Borsa_K9_O}, we broadened the NLTE synthetic spectra accounting for microturbulence velocity ($\nu_{\rm mic}$; added in quadrature to the thermal velocity), macroturbulence velocity ($\nu_{\rm mac}$), and planetary rotation. This was done with the assumption a tidally locked planet (i.e., $v_{\rm rot} = 6.64$ km\,s$^{-1}$ for KELT-9b and $v_{\rm rot}= 2.73$ km\,s$^{-1}$ for KELT-20b), using \texttt{fastRotBroad} from \texttt{PyAstronomy}. In particular, the synthetic spectra were computed for $\nu_{\rm mic}$ values of 1, 2, 3, 4, 6, 8, 10, 12, 14, and 18 km\,s$^{-1}$, with the transmission spectra for in-between $\nu_{\rm mic}$ values obtained through interpolation using a fifth-degree polynomial. 

We compared the observed and NLTE synthetic line profiles employing an MCMC algorithm to finally extract for each detected line the best fitting $\nu_{\rm mic}$--$\nu_{\rm mac}$ pair. We ran the MCMC employing 10000 steps and 20 walkers, assuming uniform priors on $\nu_{\rm mic}$ and $\nu_{\rm mac}$ in the 0--14 and 0--25\,km\,s$^{-1}$ ranges, respectively, and leaving the line center as a free parameter (see \citealt{DArpa_2024}, for more details). %\citep[see][for more details] {D'Arpa et al 2024}. 
The best fitting $\nu_{\rm mic}$ and $\nu_{\rm mac}$ values are listed in Tables~\ref{tab:K9_detection} and \ref{tab:K20_detection}, while the relative corner plots obtained from the MCMC fitting are displayed in the appendix (Fig. \ref{fig:corner_vv_K9}, and Fig. \ref{fig:corner_vv_K9_1}, for KELT-9b, and Fig. \ref{fig:M2_MCMC_VV_1}, and Fig. \ref{fig:M2_MCMC_VV_2} for MASCARA-2b).
%We run the MCMC for 10000 steps and 20 walkers, assuming that all priors have uniform probability and range of $$\nu_{\rm mic}$$: 0 - 14 km\,s$^{-1}$, $$\nu_{\rm mac}$$: 0 - 25 km\,s$^{-1}$, and $v_{shift}$: -50 - 50 km\,s$^{-1}$. The fitting was performed for the best transmission spectrum presented in Section \ref{Section:TS} with the range of $\pm 3$ $\AA$ from the studied line.

\begin{figure}[h]
\centering
\includegraphics[width=0.9\hsize]{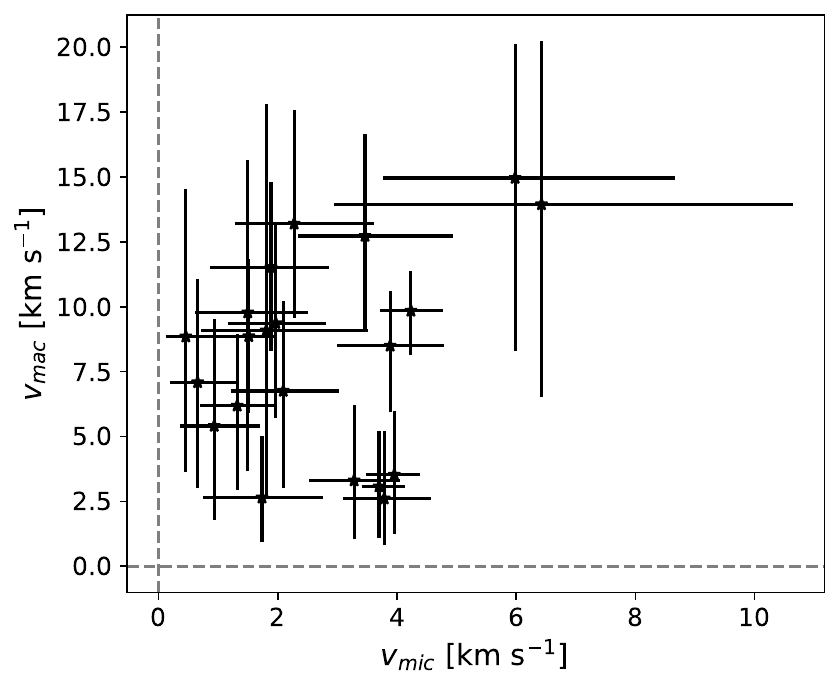}
\caption{Best-fitting $\nu_{\rm mic}$ and $\nu_{\rm mac}$ values obtained for each of the \ion{Fe}{ii} lines detected in the atmosphere of KELT-9b.}
\label{FIG:vmic_mac_K9}
\end{figure}

In the case of KELT-9b, we find average $\nu_{\rm mic}$ and $\nu_{\rm mac}$ values of 2.7 $\pm$ 1.6 km\,s$^{-1}$ and 8.2 $\pm$ 3.7 km\,s$^{-1}$, respectively, which agree within 1$\sigma$ with what independently obtained in previous studies by \citet[][$\nu_{\rm mic}$ = 3.0 $\pm$ 0.7 km\,s$^{-1}$, $\nu_{\rm mac}$ = 13 $\pm$ 5 km\,s$^{-1}$]{Borsa_K9_O}, and \citealt{DArpa_2024} ( $\nu_{\rm mic}$ = 3.0 $\pm$ 0.3 km\,s$^{-1}$, $\nu_{\rm mac}$ = 6.8 $\pm$ 1.2 km\,s$^{-1}$)
%\citet[][$\nu_{\rm mic}$ = 3.0 $\pm$ 0.3 km\,s$^{-1}$, $\nu_{\rm mac}$ = 6.8 $\pm$ 1.2 km\,s$^{-1}$]{D'Arpa et al 2024} 
employing the same models, but considering different spectral lines and/or different data analysis methods. In Fig. \ref{FIG:vmic_mac_K9}, we present the fitted values of $\nu_{\rm mic}$ and $\nu_{\rm mac}$ for each of the detected lines. 

\begin{figure*}[h]
\centering
\includegraphics[width=0.45\textwidth]{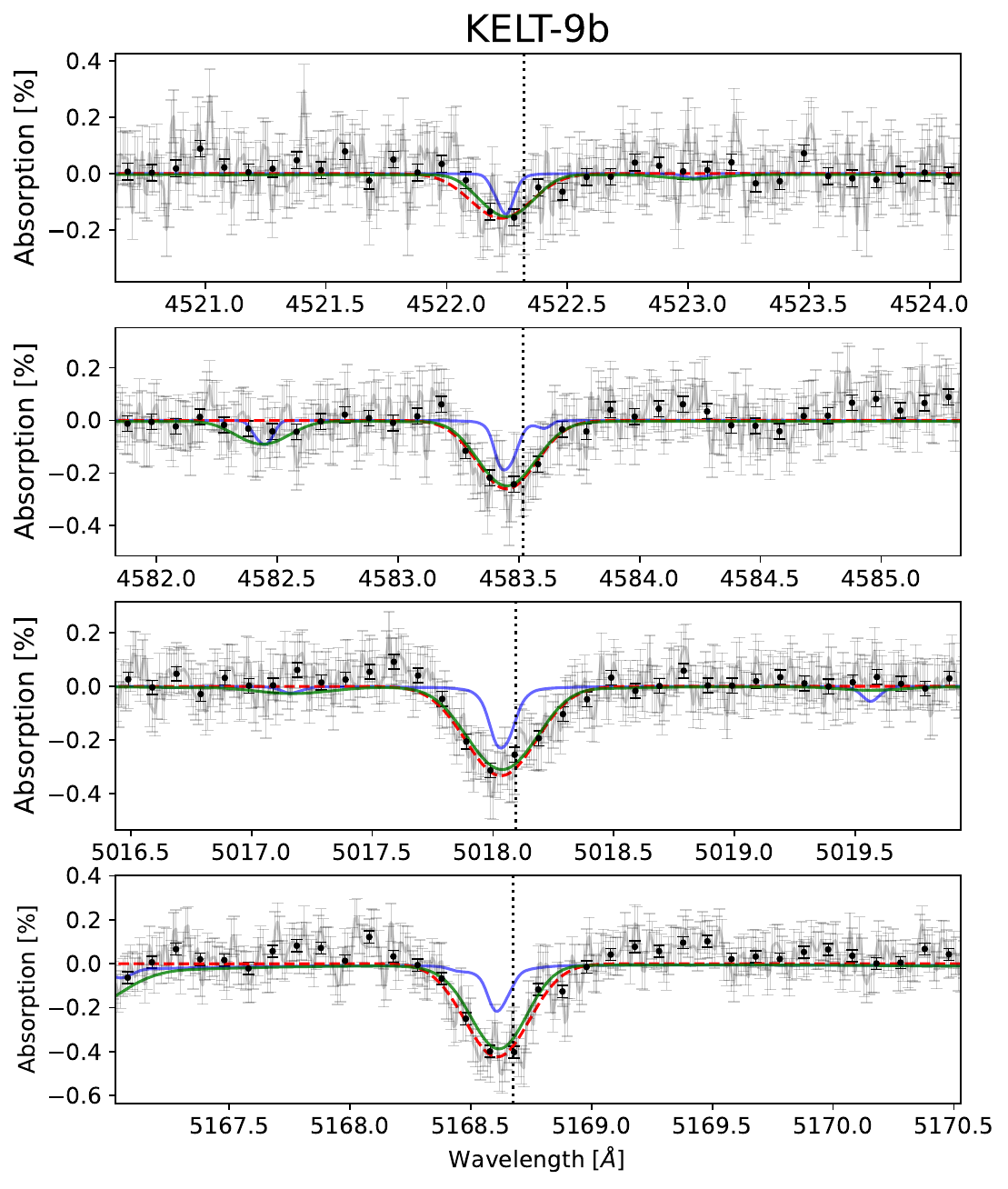}
\includegraphics[width=0.45\textwidth]{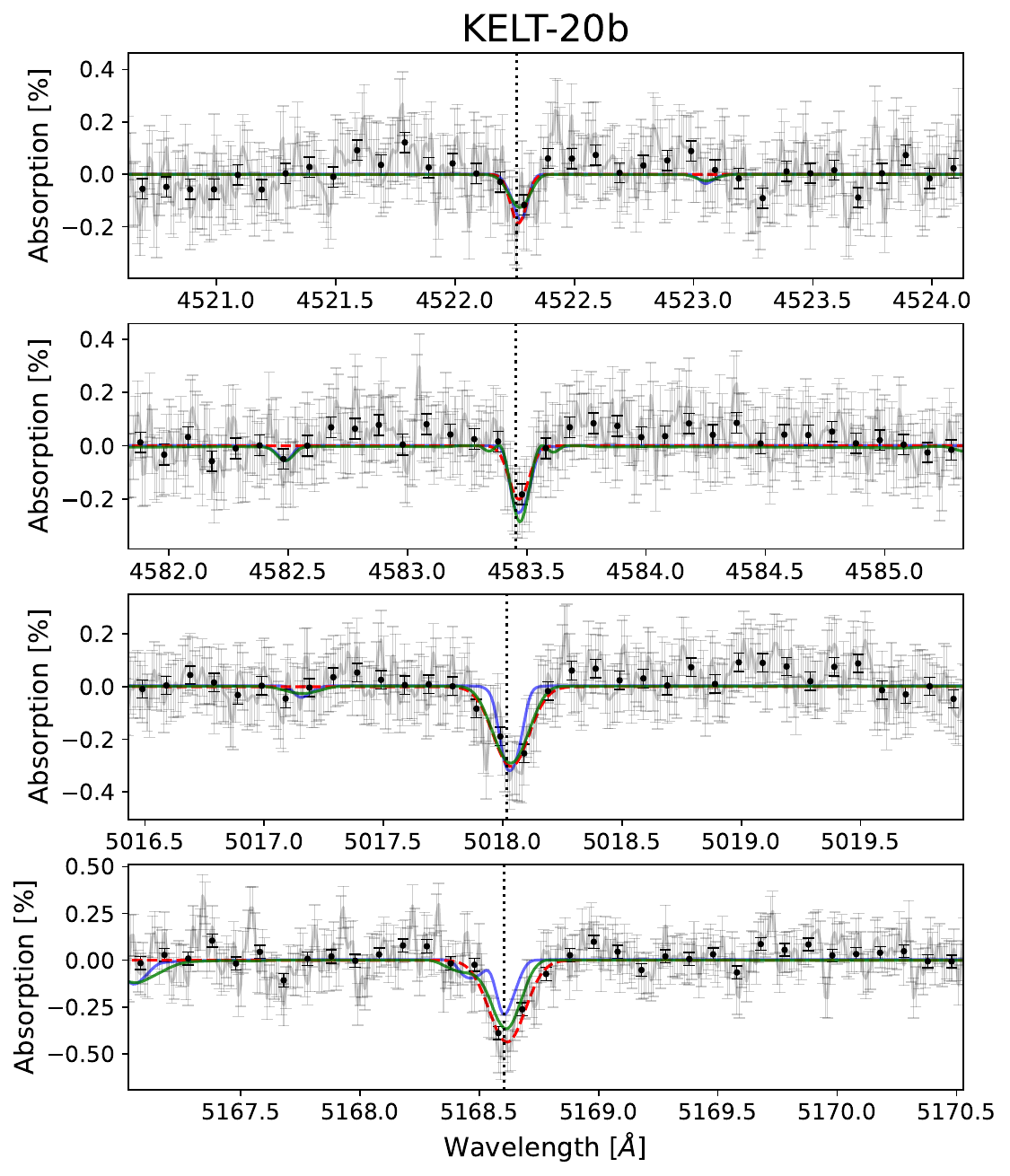}

\caption{Transmission spectra of the \ion{Fe}{ii} lines at $\lambda$4522 $\AA$, $\lambda$4583 $\AA$, $\lambda$5018 $\AA$, and $\lambda$5169 $\AA$ for KELT-9b (left column) and KELT-20b (right column) in comparison to the LTE (blue) and NLTE (green) synthetic profiles computed for the best fitting $\nu_{\rm mic}$ and $\nu_{\rm mac}$ values. The red line is the best Gaussian fit of the planetary signal. }
\label{FIG:4VV}
\end{figure*}

\begin{figure}[h]
\centering
\includegraphics[width=0.9\hsize]{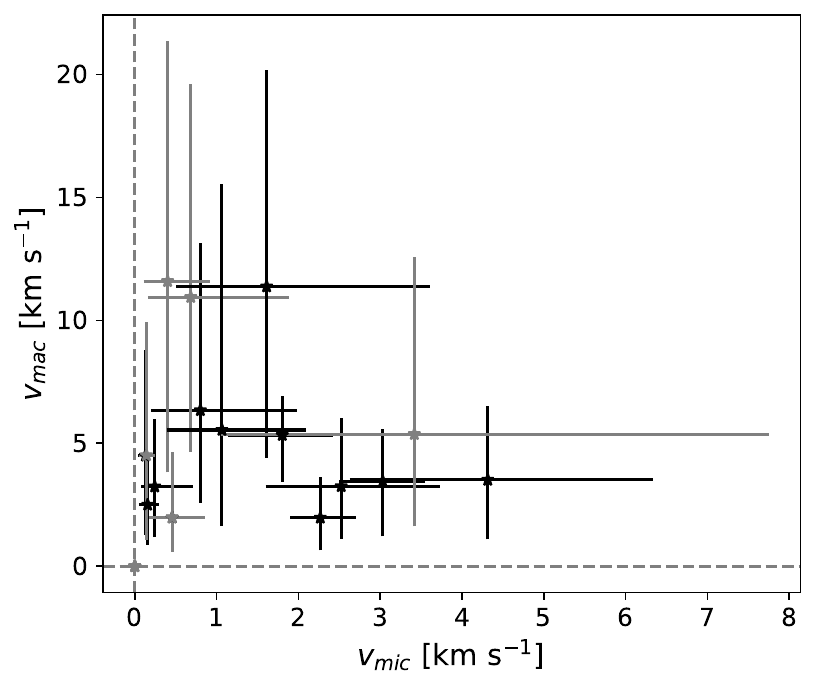}
\caption{Same as Fig. \ref{FIG:vmic_mac_K9}, but for KELT-20b. The values for which both $\nu_{\rm mic}$ and $\nu_{\rm mac}$ are consistent with 0 km~s$^{-1}$ or could not be determined are plotted with grey color}. 
\label{FIG:vmic_mac_M2}
\end{figure}

Also, in the case of KELT-20b, we find that the best fitting NLTE models are for non-zero microturbulence and/or macroturbulence values and on average we find $\nu_{\rm mic}$ =1.4  $\pm$ 1.3 km\,s$^{-1}$ and $\nu_{\rm mac}$ = 5.1 $\pm$ 3.1 km\,s$^{-1}$. Figure~\ref{FIG:vmic_mac_M2} shows the fitted values of $\nu_{\rm mic}$ and $\nu_{\rm mac}$ for each of the detected lines.

In Fig. \ref{FIG:4VV}, we show as an example the observed transmission spectra compared to the LTE and NLTE synthetic profiles for KELT-9b and KELT-20b. In the case of KELT-9b, the NLTE profiles reproduce significantly better the observed features compared to the LTE profiles. In the case of KELT-20b, NLTE effects seem to have less of an impact on the line profiles, particularly for weak lines; whereas the difference between the LTE and NLTE features is larger for stronger lines, for which the NLTE synthetic spectra allow for a better ft to the observations.

 These results support those of \citet{Fossati_NLTE_K9,Fossati_2023_M2}, who found that transmission spectra computed accounting for NLTE effects fit significantly better the hydrogen Balmer line profiles compared to those obtained assuming LTE. Altogether, these results indicate that the higher middle and upper atmospheric temperature obtained when accounting for NLTE effects is a better representation of the planetary atmosphere. The good fit obtained by the NLTE transmission spectra of the metal lines presented here is a further indication that the NLTE model is capable to adequately reproduce not only the TP profile, but also the \ion{Fe}{ii} ionisation fraction and level population across the atmosphere. In turn, this supports the conclusion that in KELT-9b and KELT-20b, the temperature inversion is driven by metal-line heating; thus, NLTE effects play the role of over-populating the lower energy levels of \ion{Fe}{ii} and under-populating \ion{Mg}{ii}, which are responsible for driving most of the heating and cooling, respectively.

\section{Cross-correlation analysis of the \ion{Fe}{ii} lines}\label{SEC:CC}
In addition to transmission spectroscopy around single lines, for both planets, we applied the cross-correlation (CC) method to the available data sets. Following the data reduction steps described in Sect. \ref{Section:TS}, we cross-correlated the residuals with synthetic models of \ion{Fe}{ii}. The CC was performed with \texttt{crosscorrRV} from \texttt{PyAstronomy} for each order and each orbital phase separately from $-$200 to 200 km\,s$^{-1}$ with a step size of 0.8 km\,s$^{-1}$.

\begin{figure}[h]
\centering
\includegraphics[width=0.98\hsize]{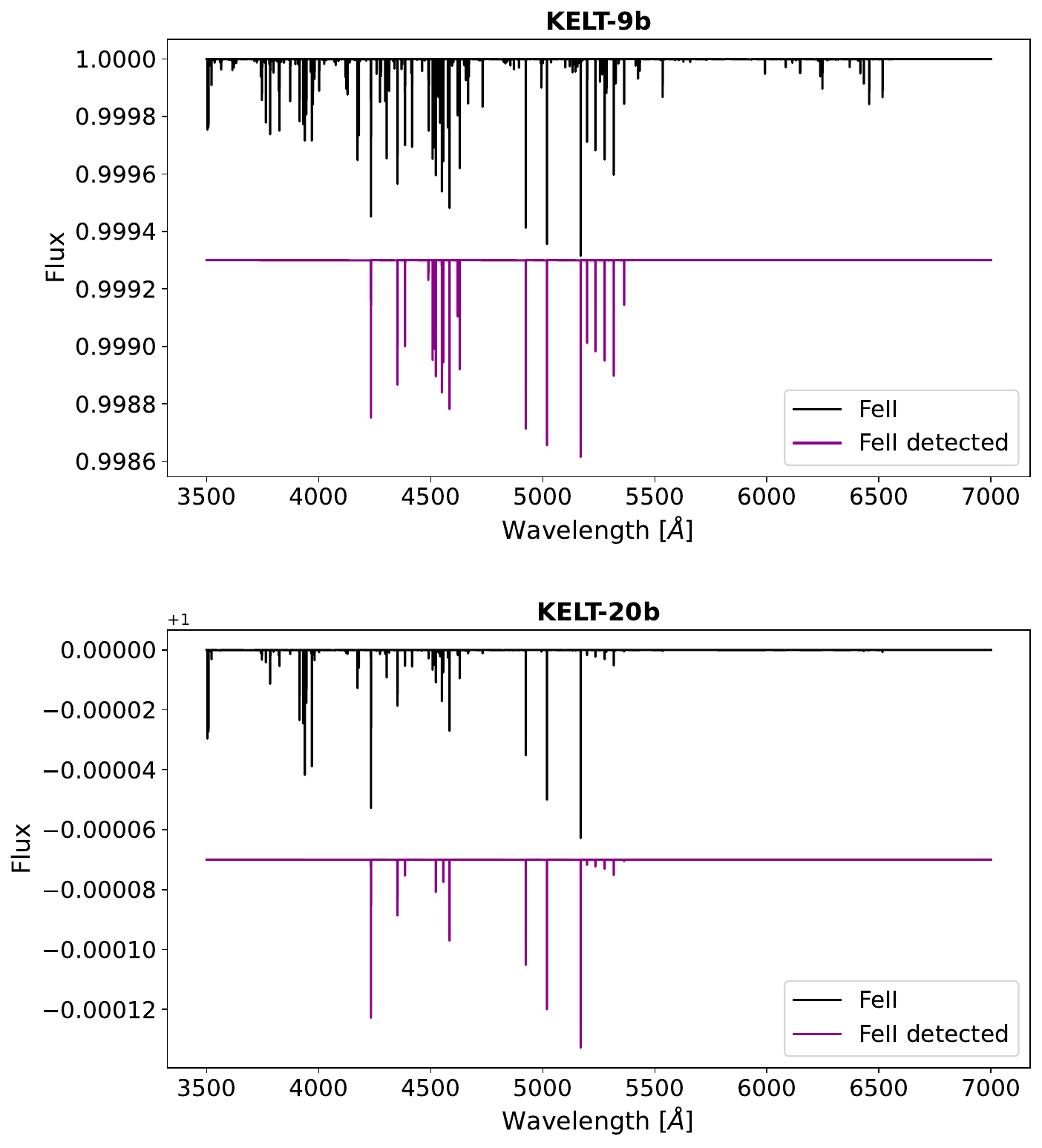}
\caption{Synthetic transmission spectra computed considering only \ion{Fe}{ii} for KELT-9b (top panel) and KELT-20b (bottom panel). The black lines are for models created with \texttt{PetitRadTrans} considering all lines, while the purple lines are for models computed considering only the lines detected through the single-line analysis. }
\label{FIG:FeII_model}
\end{figure}

Synthetic LTE models of \ion{Fe}{ii} were created with \texttt{PetitRadTrans} \citep{PetitRadTrans} assuming an isothermal pressure-temperature profile at $T= T_{\rm eq}$ (see Table \ref{tab:params_K9_K20}), solar abundances \citep{Asplund_sun}, and a cloud layer at P$_0$ = 1 mbar to simulate the continuum opacity produced by H$^-$. (e.g., \citealt{Hoeijmakers_K9_2}, \citealt{Stangret_6plan}, \citealt{Borsa_w121_RM}).  The CC was performed for three different synthetic models, one consisting of all the lines from \texttt{PetitRadTrans}, the second consisting only of lines detected by single-line analysis, and a third one consisting only the lines not detected by single-line analysis. The models computed for KELT-9b and KELT-20b are shown in Fig. \ref{FIG:FeII_model} (right panel).

%Due to the fact of the strong influence of the RM effect on each of the \ion{Fe}{ii} lines, it was necessary to correct the CC maps from this effect. In the first step, we cross-correlated the models of RM and CLV effects with the synthetic models of \ion{Fe}{ii}. This method turned out to be insufficient in our case, and we decided to correct the RM and CLV effects by fitting gaussian to the visible by-eye structures coming from these effects. Since the RM effect for both planets creates in the CC maps different structures,  in the case of KELT-9b we removed RM signal around -100, -40, 20, and 90 km\,s$^{-1}$, and in the case of KELT-20b we removed only a main RM signal from -150 - 100 km\,s$^{-1}$. 

Due to the strong influence of the RM effect on each of the \ion{Fe}{ii} lines, it was necessary to correct the CC maps from this effect. In the first step, we cross-correlated the models of RM and CLV effects with the synthetic models of \ion{Fe}{ii}. This method turned out to be insufficient in our case, and thus we decided to correct the RM and CLV effects by fitting a Gaussian to the visible structures coming from these effects. Given that the RM effect produces different structures in the CC maps for the two planets, in the case of KELT-9b, we removed the RM signal around $-$100, $-$40, 20, and 90 km\,s$^{-1}$. Then,  in the case of KELT-20b, we removed  the RM signal only from $-$150 to  100 km\,s$^{-1}$. 

In the next step, we moved the residuals to the planetary rest frame assuming $v_{\rm sys} = 0$ km\,s$^{-1}$ and $K_{\rm p}$ in the range of 0--300 km\,s$^{-1}$, where the planetary signal is expected to have its maximum near the theoretical $K_{\rm p}$ and $v_{\rm sys}$. Then, after co-adding the in-transit residuals (T1 -- T4), we find the best $K_{\rm p}$ and calculate the S/N plot by dividing the results by the standard deviation calculated away from the expected planetary signal ($-$200 to $-$100 km\,s$^{-1}$ and 100 to 200 km\,s$^{-1}$). 

%The synthetic models of \ion{Fe}{ii} were CC with the models of CLV and RM effect for different radii of the planet from 0.7 to 1.5 R$_p$ which plays the role of the scaling factor of the CLV and RM effect in the CC maps. 

%We used the MCMC algorithm to fit the CLV and RM effect together with the planetary signal assuming the Gaussian shape of the signal scaled by the transit model from \texttt{PyTransit}. 

%We used the MCMC algorithm to fit the RM and CLV effects, and planetary signal assuming the Gaussian shape of the signal (amplitude $a_{CC}$, FWHM$_{CC}$), which moves in time with the velocity of the planet on its orbit. In the analysis, we fitted $K_p$$_{CC}$,  $a_{CC}$, FWHM$_{CC}$, $v_{sys}$$_{CC}$ and $v_{CC}$, which is the combination of the $v_{sys}$ and the $v_{wind}$ in case of the signal from the planet and amplitude, FWHM, position, and angle of the RM and CLV effects, while assuming that the angle for each of the structure in the effect is the same. The analysis was performed on each night separately as well as the combined CC maps. The fitted values for both planets are collected in Table XXX.

\begin{figure*}
\centering
\includegraphics[width=0.98\hsize]{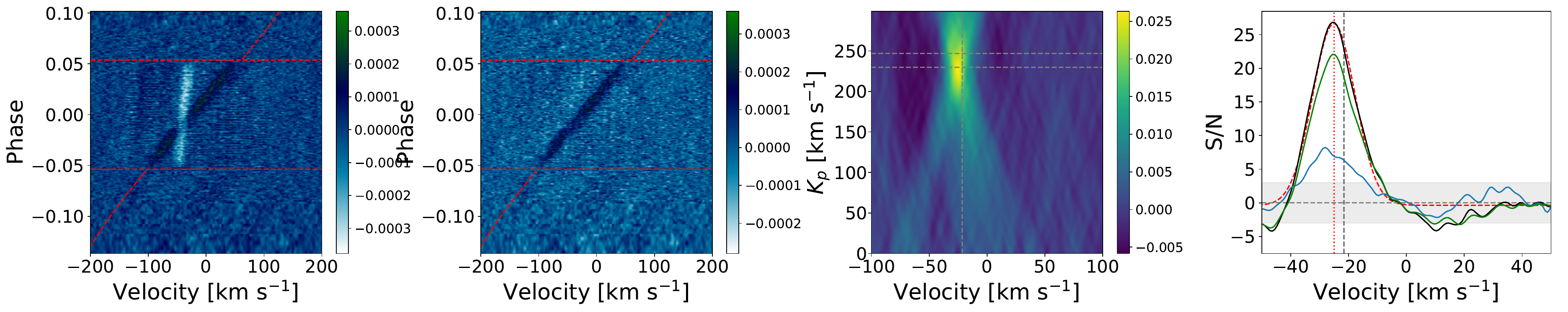}
\includegraphics[width=0.98\hsize]{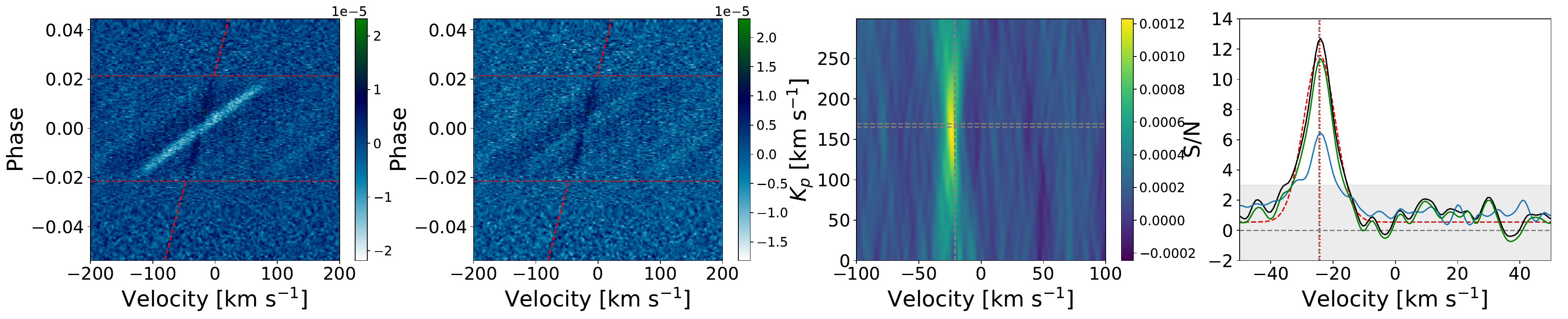}
\caption{Cross-correlation results for KELT-9b (top panels) and KELT-20b (bottom panels) using all \ion{Fe}{ii} lines. \textit{Left column:} Cross-correlation residual maps, where the lbright tilted signal is the RM effect, the dark signal is the \ion{Fe}{ii} atmospheric signal, the red horizontal dashed lines indicate the T1 and T4, and the dashed tilted line represents the expected velocity of the planet where for visualization reasons we assumed $v_{\rm sys}$ equal to the average $v_{\rm sys}$ from the single line analysis. \textit{Center-left column:} Same as the maps in the first column, but after correcting for the RM effect. \textit{Center-right column:} $K_{\rm p}$ maps in the range of $K_{\rm p}$ from 0 to 300 km s$^{-1}$, where the two horizontal gray dashed lines indicate the theoretical $K_{\rm p}$ and the best $K_{\rm p}$ values, respectively, and the vertical gray dashed line indicates the mean $v_{\rm sys}$ value from the single line analysis. \textit{Right column:} S/N for the best $K_{\rm p}$ value (black line), fitted Gaussian function (red dashed line), S/N obtained using only the lines detected in the single-line analysis (green line), and S/N obtained using only the lines not detected in the single-line analysis (blue line). The black dashed vertical line indicates the mean $v_{\rm sys}$ value and the red vertical line indicates the velocity of the signal from the Gaussian fit.}
\label{Fig:CC_K9_K20}
\end{figure*}

In the top panels of Fig. \ref{Fig:CC_K9_K20}, we present the CC analysis for KELT-9b. The left panel shows the CC residual maps, where the vertical bright signals around $-$50 km\,s$^{-1}$ and 90 km\,s$^{-1}$, and the darker signal around $-$120 km\,s$^{-1}$ are the CLV and RM effects. The planetary signal is visible as a tilted dark signal following a red tilted dashed line, which indicates the theoretical radial velocity of the planetary signal. The two horizontal red dashed lines indicate T1 and T4. In the second panel, we present the residual maps after correcting for the CLV and RM effects. In the third panel, we present the $K_{\rm p}$ map calculated for $K_{\rm p}$ ranging from 0 to 300 km\,s$^{-1}$. The \ion{Fe}{ii} signal was detected with S/N = 27.02 $\pm$ 0.92 and $v_{\rm sys+wind}=$ $-$25.06 $\pm$ 0.10 km\,s$^{-1}$ for the CC computed considering all lines, with S/N = 21.97 $\pm$ 0.74 and $v_{\rm sys+wind}=$ $-$24.88 $\pm$ 0.11 km\,s$^{-1}$ considering the lines detected in the single-line analysis, and with S/N = 8.06 $\pm$ 0.27 and $v_{\rm sys+wind}=$ $-$25.72 $\pm$ 0.44 km\,s$^{-1}$ considering the lines not detected in the single-line analysis. We successfully identified the signal from the lines not detected in the single-line analysis using the cross-correlation method, although the signal was much weaker than the one coming from the detected lines. This indicates that the main contributors to the cross-correlation analysis were the lines that were detected also through single-line analysis.

%Both values are almost exactly the same, suggesting that the detected single lines are the main contributors to the cross-correlation analysis, and the remaining lines, are too weak to be significant.

In the bottom row of Fig. \ref{Fig:CC_K9_K20}, we present the cross-correlation results for KELT-20b. The signal was detected with S/N = 11.51 $\pm$ 0.46 and $v_{\rm sys+wind}=$ $-$24.27 $\pm$ 0.13 km\,s$^{-1}$ for the CC computed considering all lines, with S/N = 11.00 $\pm$ 0.44 and $v_{\rm sys+wind}= $ $-$24.32 $\pm$ 0.12 km\,s$^{-1}$, considering the lines detected in the single-line analysis, and with S/N = 6.98 $\pm$ 0.55 and $v_{\rm sys+wind}=$ $-$25.19 $\pm$ 1.04 km\,s$^{-1}$ considering the lines not detected in the single-line analysis. Similarly to KELT-9b, the main contributors to the CC signal are the lines detected with the single-line analysis. 

\section{Planetary radius}\label{SEC:Radius}

\begin{figure}[h]
\centering

\includegraphics[width=0.98\hsize]{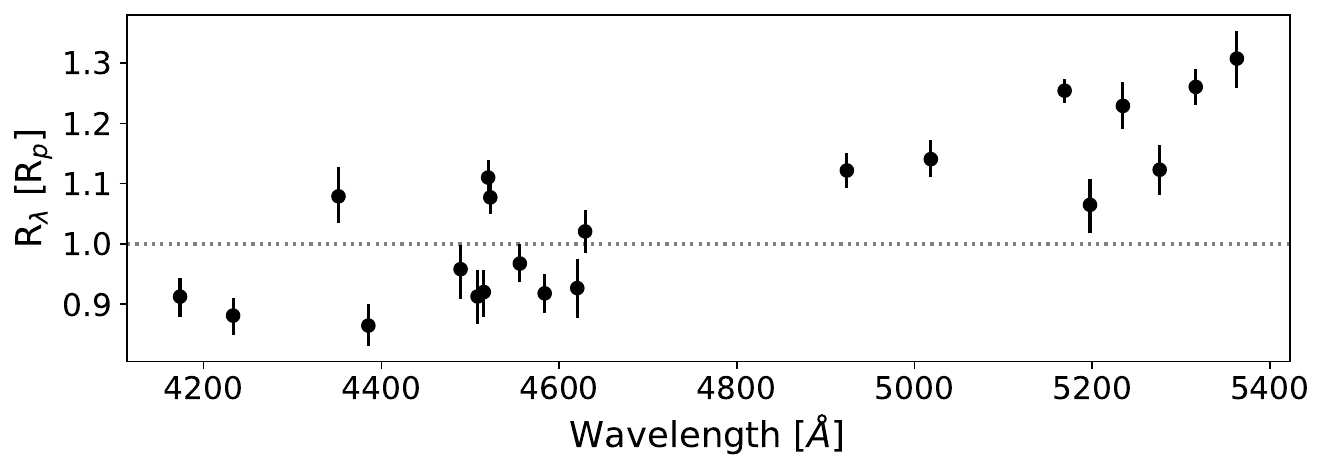}
\includegraphics[width=0.98\hsize]{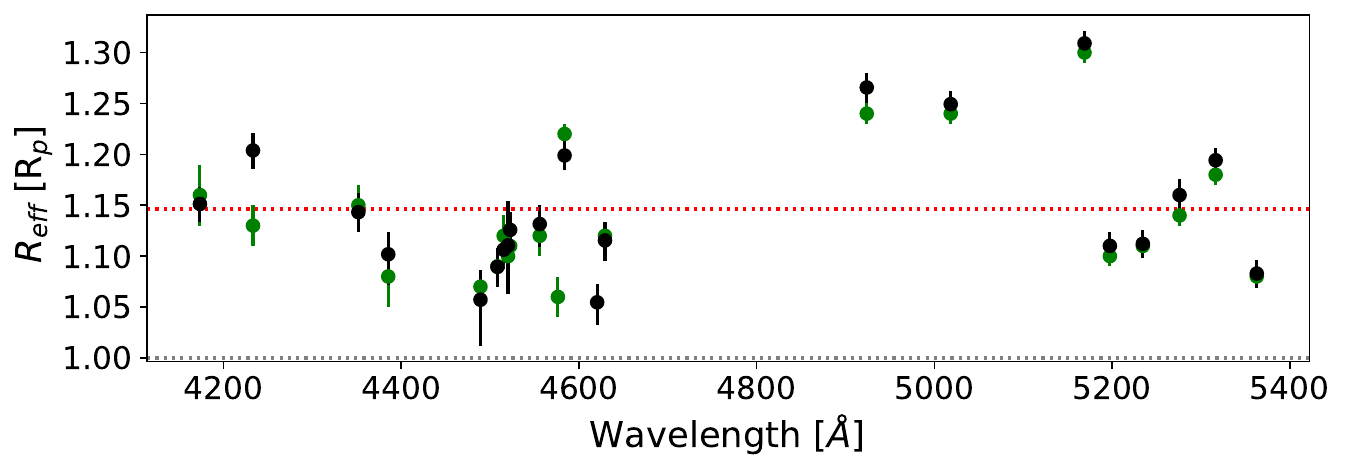}
\caption{$R_{\lambda}$ and $R_{\rm eff}$ of the \ion{Fe}{ii} lines detected in atmosphere of KELT-9b. \textit{Top panel}: Fitted $R_{\lambda}$ for each of the lines. The dashed horizontal line indicates $R_\lambda = R_{\rm p}$. \textit{Bottom panel:} Calculated $R_{\rm eff}$ values using the amplitude of the detected signal for each of the lines (black dots) and the values obtained by \citealt{DArpa_2024} (green dots).}
\label{Fig:RES_K9_KP_RP}
\end{figure}

\begin{figure}[h]
\centering
\includegraphics[width=0.98\hsize]{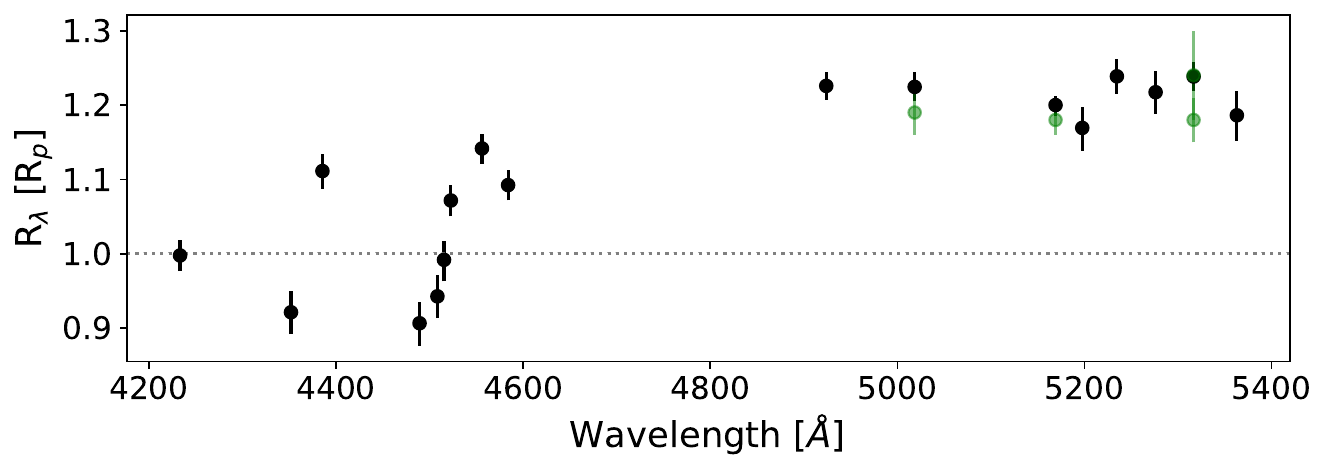}
\includegraphics[width=0.98\hsize]{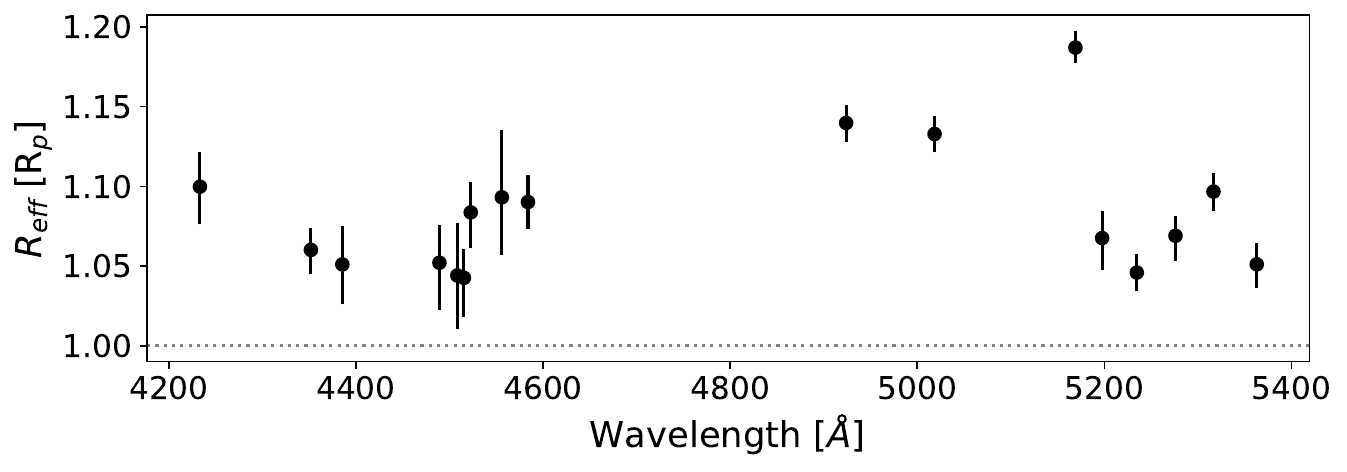}

\caption{Same as Fig. \ref{Fig:RES_K9_KP_RP}, but for KELT-20b. The green dots in the top panel represent the measurements obtained by \citet{M2_Casasayas_2019} through fitting the strength of the RM effect. }
\label{Fig:RES_M2_KP_RP}
\end{figure}

We employ two distinct methodologies for determining the radius of the planet from the detected lines. In the first methodology, we concentrate on the detected signal from the planet, following the approach of \citet{Chen_RM} to calculate the effective radius of the planet (hereafter, $R_{\rm eff}$). Using the amplitude of the detected signal, we considered $R_{\rm eff}/R_{\rm p} = (\delta +h)/ \delta$, where h is the line contrast and $\delta$ is $(R_{\rm p}/R_*)^2$ from Table \ref{tab:params_K9_K20}. The measured average values are $R_{\rm eff}$ = 1.15 $\pm$ 0.07 $R_{\rm p}$ for KELT-9b and $R_{\rm eff}$ = 1.08 $\pm$ 0.04 $R_{\rm p}$ for KELT-20b. In the bottom panels of Figs. \ref{Fig:RES_K9_KP_RP} and \ref{Fig:RES_M2_KP_RP}, we present the obtained $R_{\rm eff}$ values as a function of wavelength for all lines detected through the single-line analysis.

In the second methodology, we focused on the detected RM and CLV effect, disregarding the signal coming from the planet. Here, the radius of the planet was calculated by fitting the modeled RM and CLV effect, which were calculated assuming different radii of the planet (hereafter $R_\lambda$). The top plots of Figs. \ref{Fig:RES_K9_KP_RP} and \ref{Fig:RES_M2_KP_RP} present the $R_{\lambda}$ as a function of wavelength for all detected lines. 

This method allowed us to find the optimal strength of the RM and CLV effect to clear the map from the effects and, thus, detect weaker lines. For both planets, we find that the radius of the planet ($R_\lambda$), as measured using the RM effect, exhibits a linear trend with wavelength. Furthermore, some of the calculated radii were found to be smaller than 1 R$_p$. The same analysis was repeated using stellar models computed with \texttt{TurboSpectrum} \citep{Turbospectrum_2012}, presenting a similar behavior (see Fig. A.21 for KELT-9b). As discussed by \citet{M2_Casasayas_2019}, where the radius was fitted using a similar methodology focusing on the RM and CLV models, the discrepancy can be attributed to a number of factors. Firstly, the models depend on several parameters, such as $T_{eff}$, $\log{g}$, and [Fe/H], where their uncertainties could lead to significant differences among the stellar models. We expect that NLTE effects in both KELT-9 and KELT-20 do not play a significant role in shaping the stellar spectra \citep[e.g.,][]{Fossati2009}. However, the models used in the analysis, which assumed an LTE atmosphere of those stars, could potentially introduce a minor additional uncertainty to the overall strength of the effect. Consequently, the collective impact of all uncertainties could result discrepancies in the planetary radius. Furthermore, as seen for example in Fig. \ref{fig:rad_TS_SME}, the observed trend appears to be linear. %However, due to the lack of detected \ion{Fe}{ii} lines in the wavelength range  4700 - 4850 $\AA$, we can group them into lines <4700 $\AA$  and lines >4700 $\AA$.

Here, it is crucial to stress that the obtained R$_\lambda$ values should not be interpreted as the physical parameter of the planet, but as a scaling factor for the RM and CLV effect. Furthermore, we want to show that because the scaling factor is different for different lines, the RM and CLV effects cannot be corrected for the CC maps without considering an additional scaling factor and/or scaling the strength depending on wavelength, which would require further investigation. Also, we strongly encourage future analyses of the planetary signal to fit the RM and CLV models in the 2D maps. This will ensure that the effects are not under- or overcorrected. %It will also allow one to perform statistical studies to identify the main causes of this behavior. These include stellar parameters, NLTE effects, and the spectrograph used in the analysis etc. 

\section{Conclusions}\label{Section:conclusions}

In this work, we focused on the two ultra-hot Jupiters KELT-9b and KELT-20b, presenting single-line detections and cross-correlation analysis of \ion{Fe}{ii}. We combined six transit observations for each of the planets obtained with the HARPS-N high-resolution spectrograph.

In the case of KELT-9b, using single-line analysis, we detected 21 separated lines of \ion{Fe}{ii}, with a mean $v_{\rm wind}$ of $-$3.41  $\pm$ 1.56 km s$^{-1}$. The blue-shifted lines can indicate the existence of day-to-night winds. In addition, we measured a $v_{\rm sys}$ of $-$21.61 $\pm$ 0.77 km s$^{-1}$. With the cross-correlation method, we detected the \ion{Fe}{ii} signal at the level of S/N = 27.02. Additionally, we show that atmospheric models accounting for NLTE effects fit the observations significantly better, as compared to models computed assuming LTE. On the basis of the NLTE models, we obtained $\nu_{\rm mic}$ = 2.7 $\pm$ 1.6 km\,s$^{-1}$ and $\nu_{\rm mac}$ = 8.2 $\pm$ 3.7 km\,s$^{-1}$, in agreement with previous works.

In the transmission spectrum of KELT-20b, we identified 17 single lines of \ion{Fe}{ii}. With the measurements of $v_{\rm sys}$ using MCMC fitting ($v_{\rm sys}$ = $-$24.52 $\pm$ 0.40 km s$^{-1}$) and comparing our results with the literature, it is not clear if the atmospheric signal can conclusively lead to the detection of atmospheric winds. Further studies are required to achieve more precise measurements of $v_{\rm sys}$. The cross-correlation analysis has revealed a strong detection of \ion{Fe}{ii} in the atmosphere of KELT-20b with an S/N of 11.51. Similarly to KELT-9b, the observations support models computed considering NLTE effects, from which we obtained $\nu_{\rm mic}$ = 1.4 $\pm$ 1.3 km\,s$^{-1}$ and $\nu_{\rm mac}$ = 5.1 $\pm$ 3.1 km\,s$^{-1}$.

The analysis of the systemic velocity focused on the analysis of the RM effect. As mentioned in Sect. \ref{Section:TS}, we modeled the RM and CLV effects assuming $v_{\rm sys}$ = 0.0 km s$^{-1}$, and during the MCMC analysis we found the best $v_{\rm sys}$ for each of the lines. In the case of KELT-9b, we find an average $v_{\rm sys}$ value of $-$21.61 $\pm$ 0.77 km s$^{-1}$. The literature values presented by \citet{Hoeijmakers_K9_2}, \citet{Borsa_K9}, and \textit{Gaia} are $-$17.74 $\pm$ 0.11 km s$^{-1}$, $-$19.819 $\pm$ 0.024 km s$^{-1}$, and $-$20.22 $\pm$ 0.49 km s$^{-1}$, respectively. Depending on which $v_{\rm sys}$ is considered for the correction of the wind velocity, we obtain average wind velocities ranging between $-$7.2 km s$^{-1}$ and $-$4.2 km s$^{-1}$. 

For KELT-20b the obtained mean value of $v_{\rm sys}$ is $-$24.63 km s$^{-1}$. The literature measurements presented in the past were $-$23.3 km\,s$^{-1}$ \citep{Lund_M2}, $-$21.3 \citep{Talens_M2}, $-$22.06 km\,s$^{-1}$ \citep{Nugroho_M2}, $-$24.48 km\,s$^{-1}$ \citep{Rainer_M2}, and $-$26.78 km\,s$^{-1}$ (\textit{Gaia}). These values are inconsistent among each other. Our measurements are comparable to the values obtained by \citet{Rainer_M2} using the same dataset from HARPS-N, which leads to the detection of red-shifted winds of 1.12 km s$^{-1}$. When considering all the literature $v_{\rm sys}$ values, $v_{\rm winds}$ varies from $-$2.21 km s$^{-1}$ to 3.27 km s$^{-1}$. In the case of KELT-20b, precise measurements of $v_{\rm sys}$ are necessary to confirm the existence of atmospheric winds.  

%KELT-9b and KELT-20b are ultra-hot Jupiters orbiting bright A-type stars.  Due to the fact that KELT-20b's $T_{\rm eq}$ is about half that of KELT-9b, we were unable to detect the weakest \ion{Fe}{ii} in the former, which are instead clearly detectable in the case of KELT-9b. Furthermore, we were unable to detect winds in the atmosphere of KELT-20b, whereas, in the case of KELT-9b, the detected wind showed a high blue-shifted velocity. Additionally, we demonstrate that the role of the NLTE effects for KELT-9b is considerably more significant than in the case of the much colder planet KELT-20b. 

\begin{acknowledgements}
The authors and, in particular, M.S. acknowledge the support of the PRIN INAF 2019 through the project ``HOT-ATMOS". 
This work made use of PyAstronomy (Czesla et al. 2019) and the VALD database, operated at Uppsala University, the Institute of Astronomy RAS in Moscow,
We acknowledge the use of the ExoAtmospheres database during the preparation of this work
The authors also acknowledge financial contribution from INAF GO Large Grant 2023 GAPS-2 and from the European Union - Next Generation EU RRF M4C2 1.1 PRIN MUR 2022 project 2022CERJ49 (ESPLORA).
We acknowledge the Italian Center for Astronomical Archive https://www.ia2.inaf.it/ for providing technical assistance, services and supporting activities of the GAPS collaboration.
L.M. acknowledges financial contribution from PRIN MUR 2022 project 2022J4H55R.
TZi acknowledges support from CHEOPS ASI-INAF agreement n. 2019-29-HH.0, NVIDIA Academic Hardware Grant Program for the use of the Titan V GPU card and the Italian MUR Departments of Excellence grant 2023-2027 “Quantum Frontiers”.
Part of the research activities described in this paper were carried out with contribution of the Next Generation EU funds within the National Recovery and Resilience Plan (PNRR), Mission 4 - Education and Research, Component 2 - From Research to Business (M4C2), Investment Line 3.1 - Strengthening and creation of Research Infrastructures, Project IR0000034 – “STILES - Strengthening the Italian Leadership in ELT and SKA”.\\

\end{acknowledgements}

\bibliographystyle{aa} % style aa.bst
\bibliography{biblio} % your references Yourfile.bib


\onecolumn
\begin{appendix} %First appendix

\section{Additional figures for KELT-9b} \label{app:add_fig_k9}

\begin{figure}[h]
\centering
\includegraphics[width=0.49\textwidth]{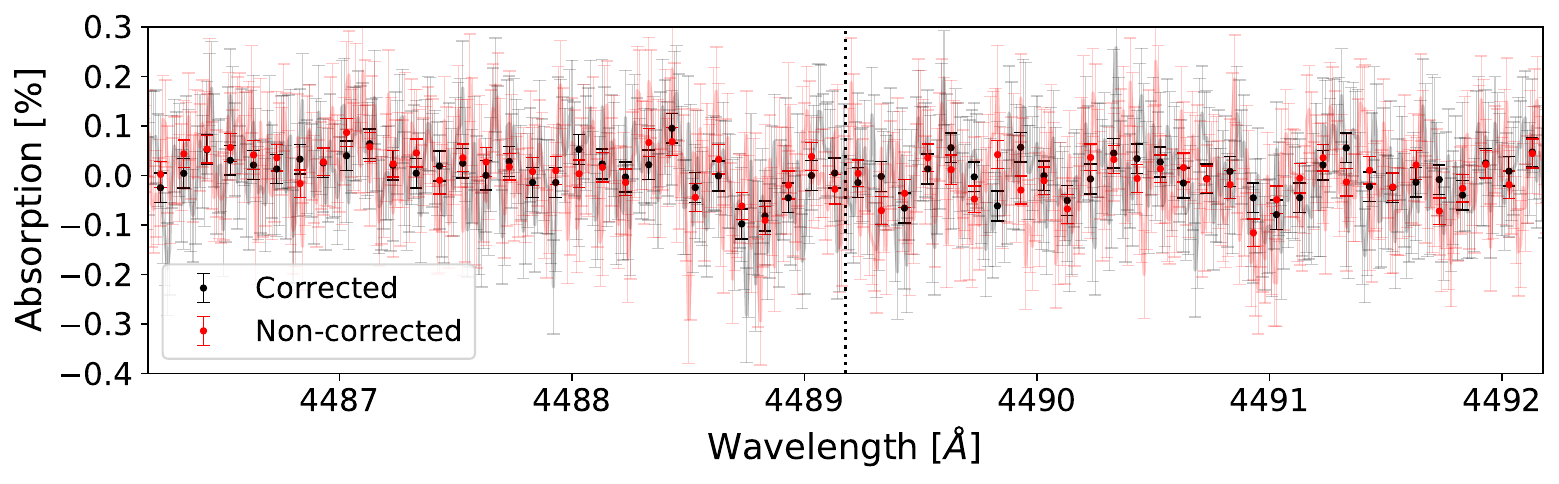}
\includegraphics[width=0.49\textwidth]{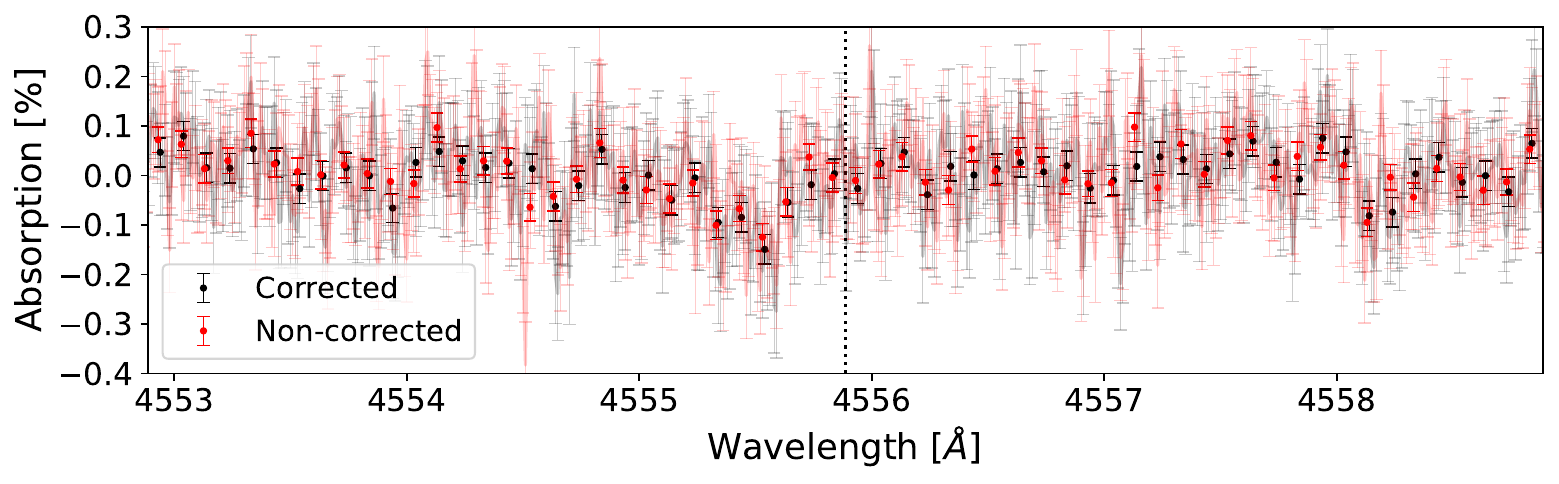}
\includegraphics[width=0.49\textwidth]{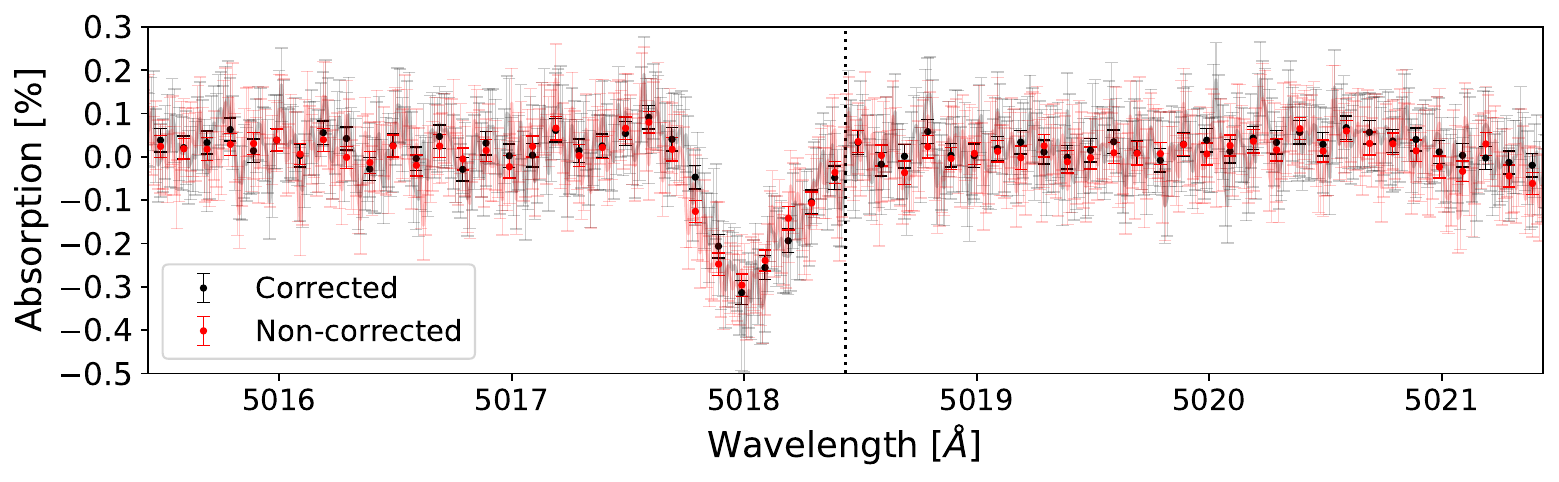}
\includegraphics[width=0.49\textwidth]{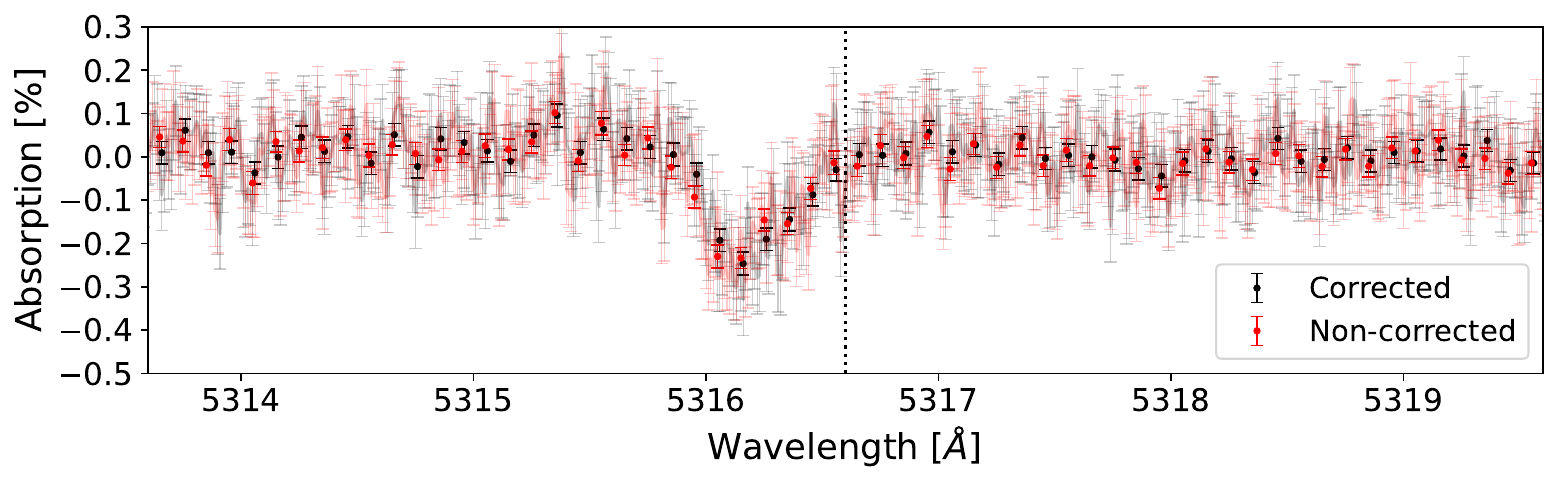}

\caption{Example of the transmission spectrum around lines  $\lambda$4489 $\AA$, $\lambda$4555 $\AA$, $\lambda$5018 $\AA$, and $\lambda$ 5316 $\AA$ for KELT-9b before (red) and after (black) correcting the telluric lines using \texttt{Molecfit}. The black vertical line indicates the theoretical position of the line. }
\label{Fig:molec_tell}
\end{figure}

\subsection{TS}

\begin{figure}[h]
\centering
\includegraphics[width=0.49\textwidth]{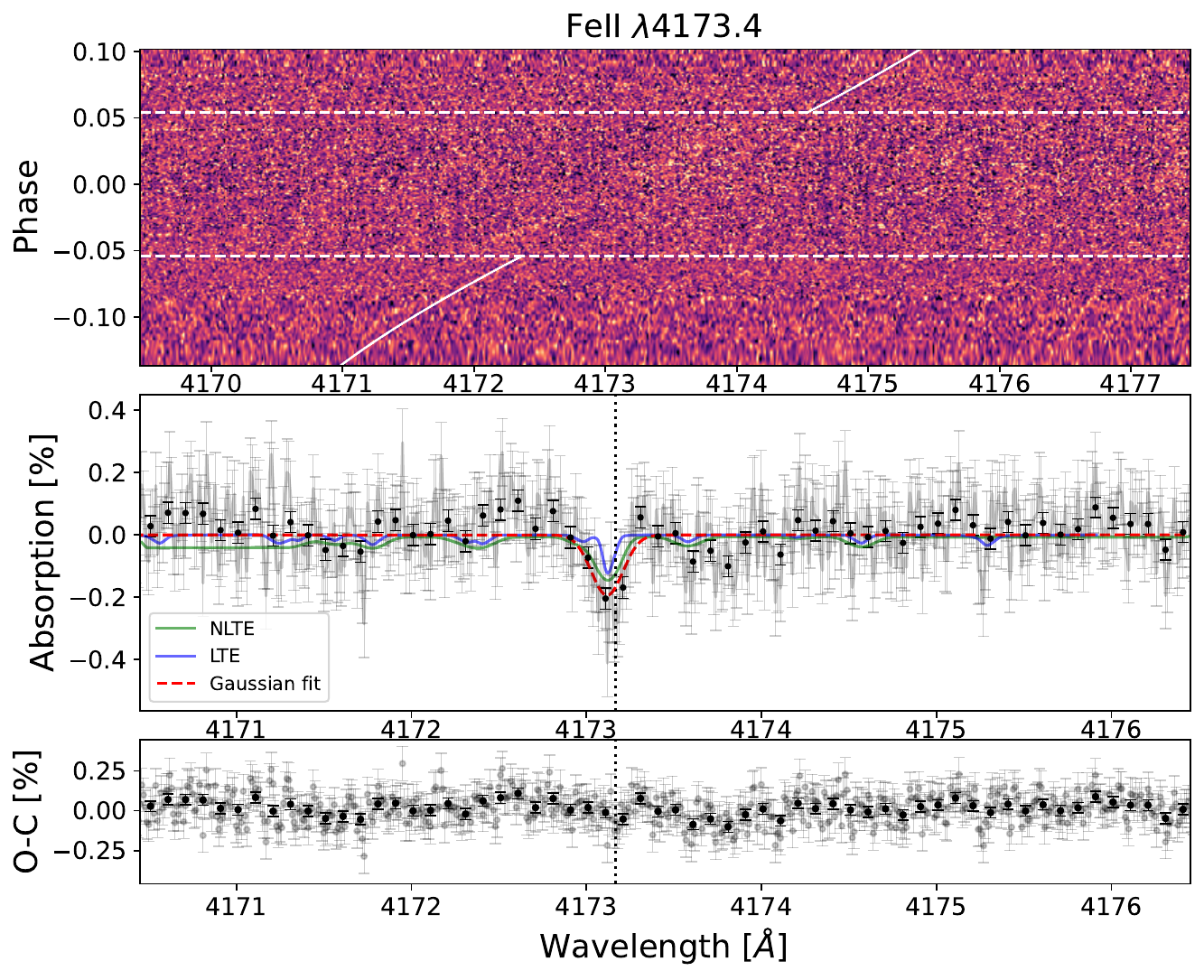}
\includegraphics[width=0.49\textwidth]{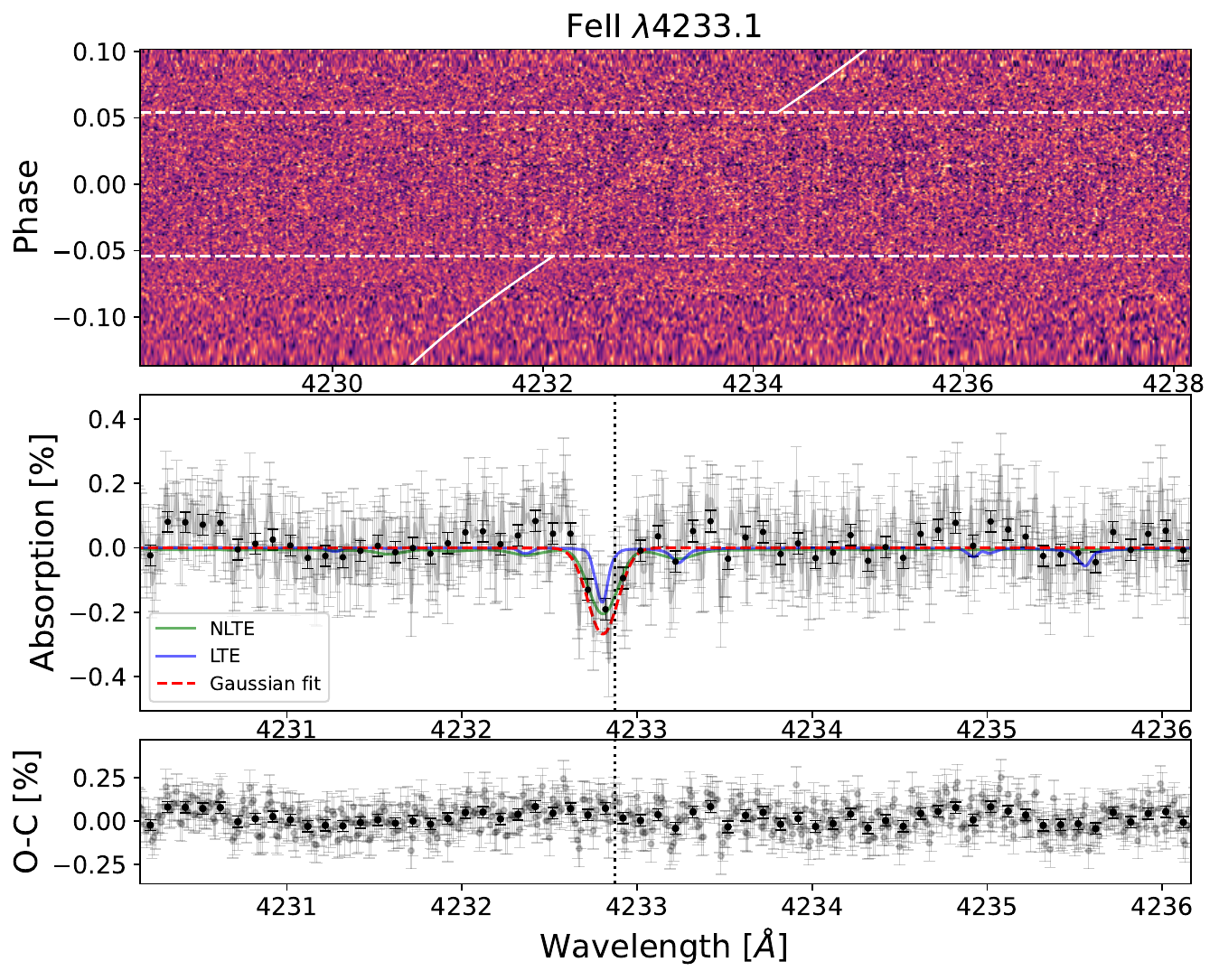}

\includegraphics[width=0.49\textwidth]{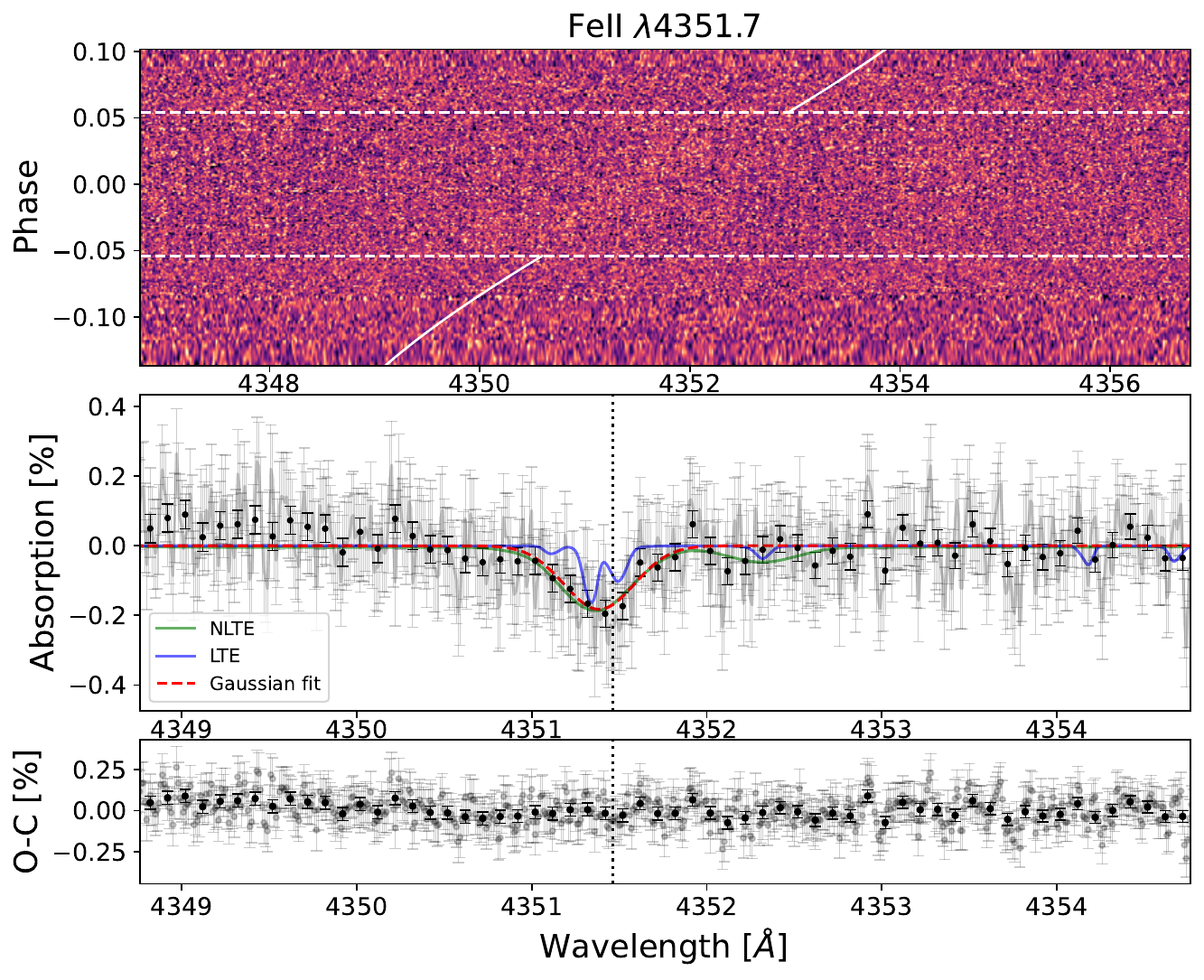}
\includegraphics[width=0.49\textwidth]{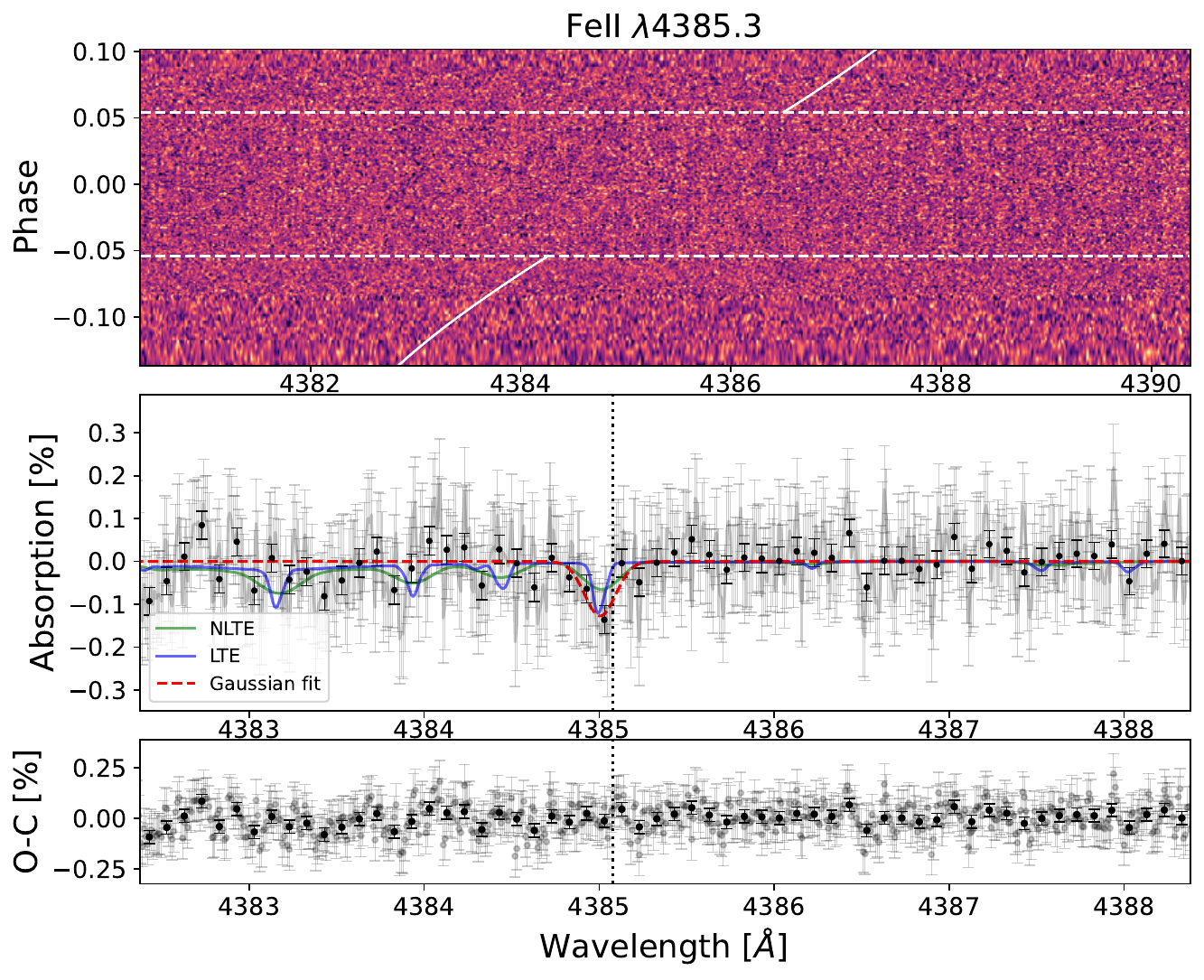}

\caption{\textit{Top panel:} Residual map for lines $\lambda$4173 $\AA$, $\lambda$4233 $\AA$, $\lambda$4351 $\AA$, and $\lambda$4385 $\AA$ for KELT-9b shown in the stellar rest frame. The white horizontal lines indicate the start and end of the transit and the tilted white line indicates the expected velocities for the signal coming from the atmosphere of the planet while assuming $v_{sys}$ as the mean $v_{sys}$ from our analysis. \textit{Middle panel:} Transmission spectrum for the detected line (gray dots). The black dots indicate the binned transmission spectrum with a step of 0.1 $\AA$, plotted for visualization only. The red plot is the best fit of the planetary signal from the MCMC analysis, the blue plot is the LTE model, and the green plot is the NLTE model for the atmosphere of KELT-9b. \textit{Bottom panel:} Residuals.  }
\label{Fig:TS_K9_1}
\end{figure}

\begin{figure}[h]
\centering
\includegraphics[width=0.49\textwidth]{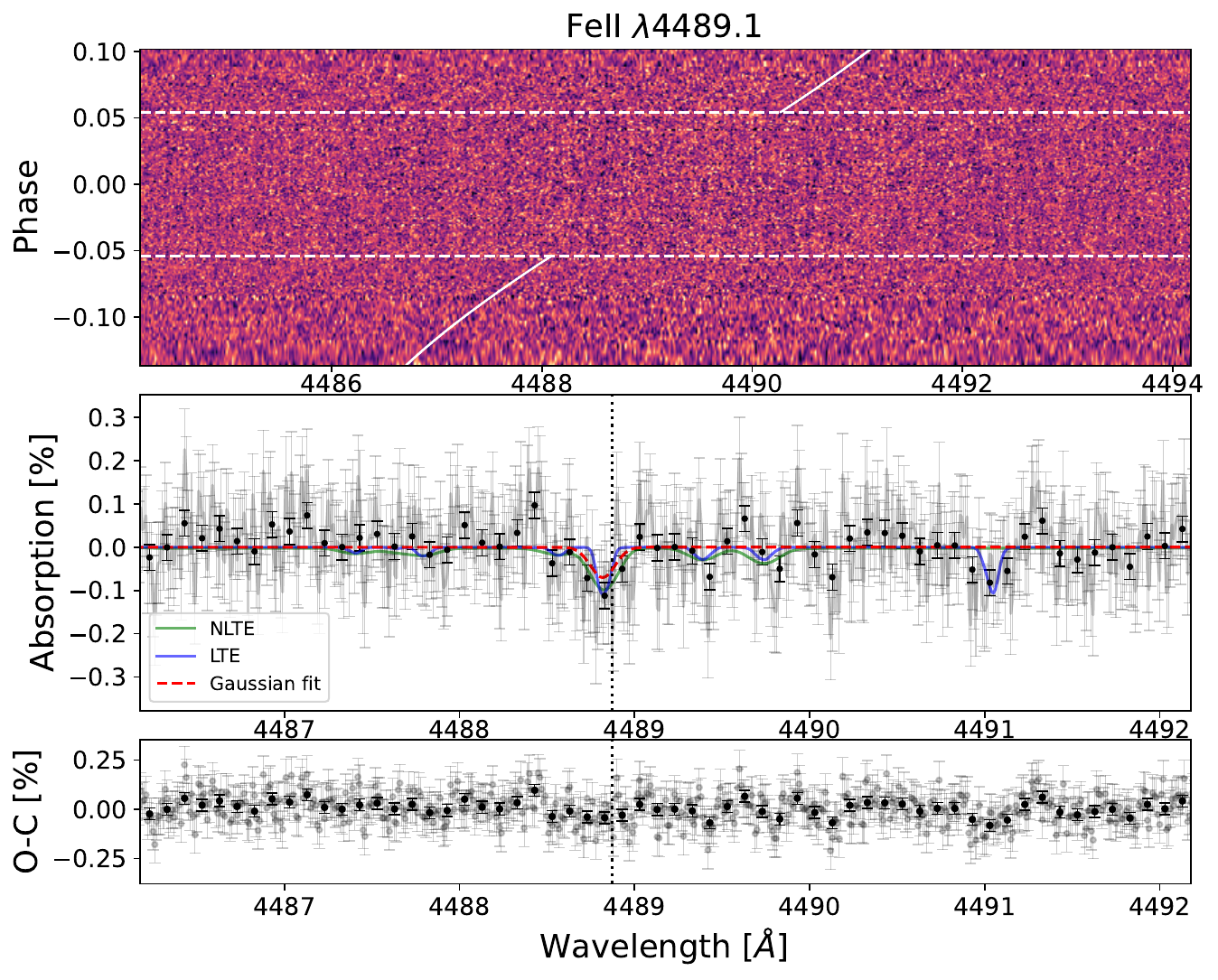}
\includegraphics[width=0.49\textwidth]{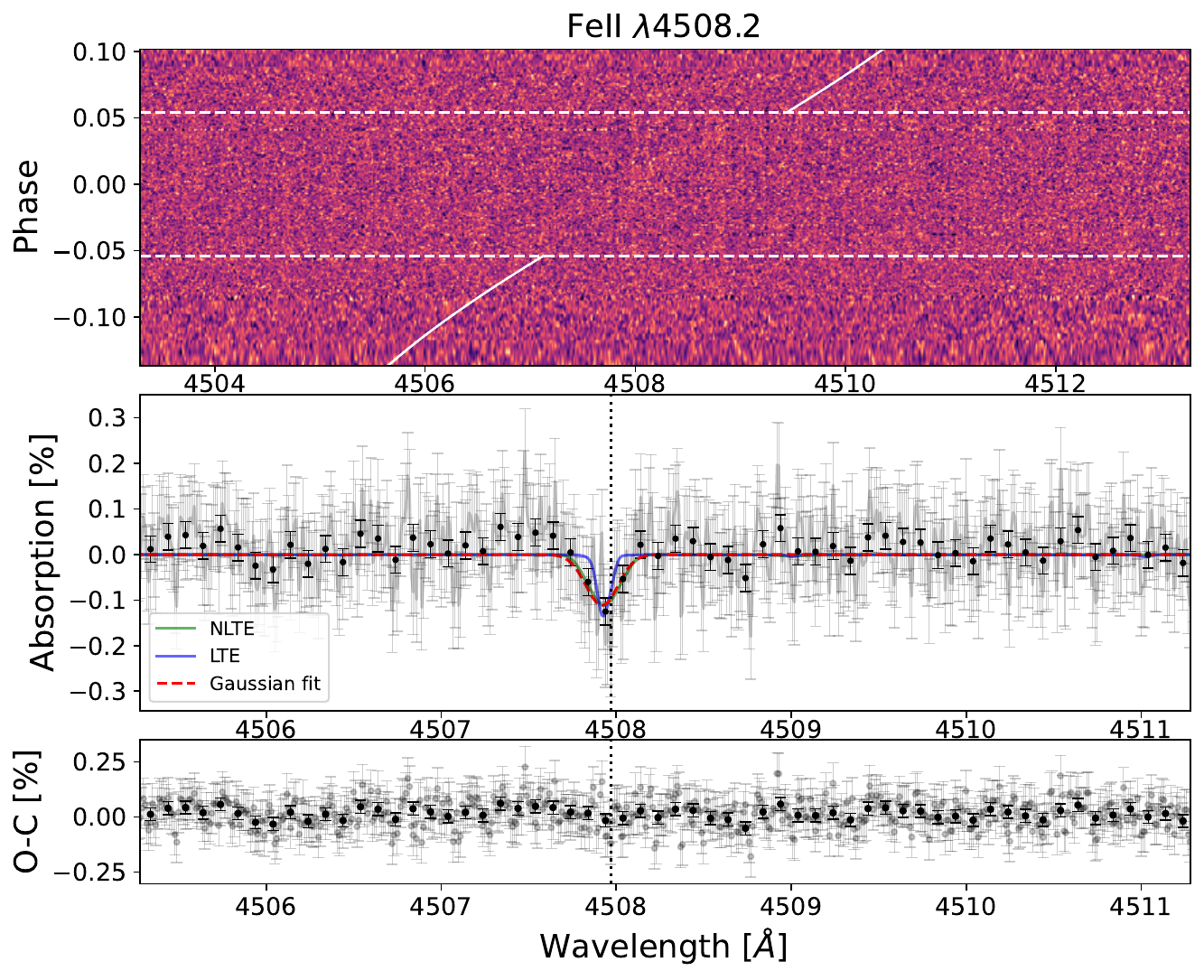}

\includegraphics[width=0.49\textwidth]{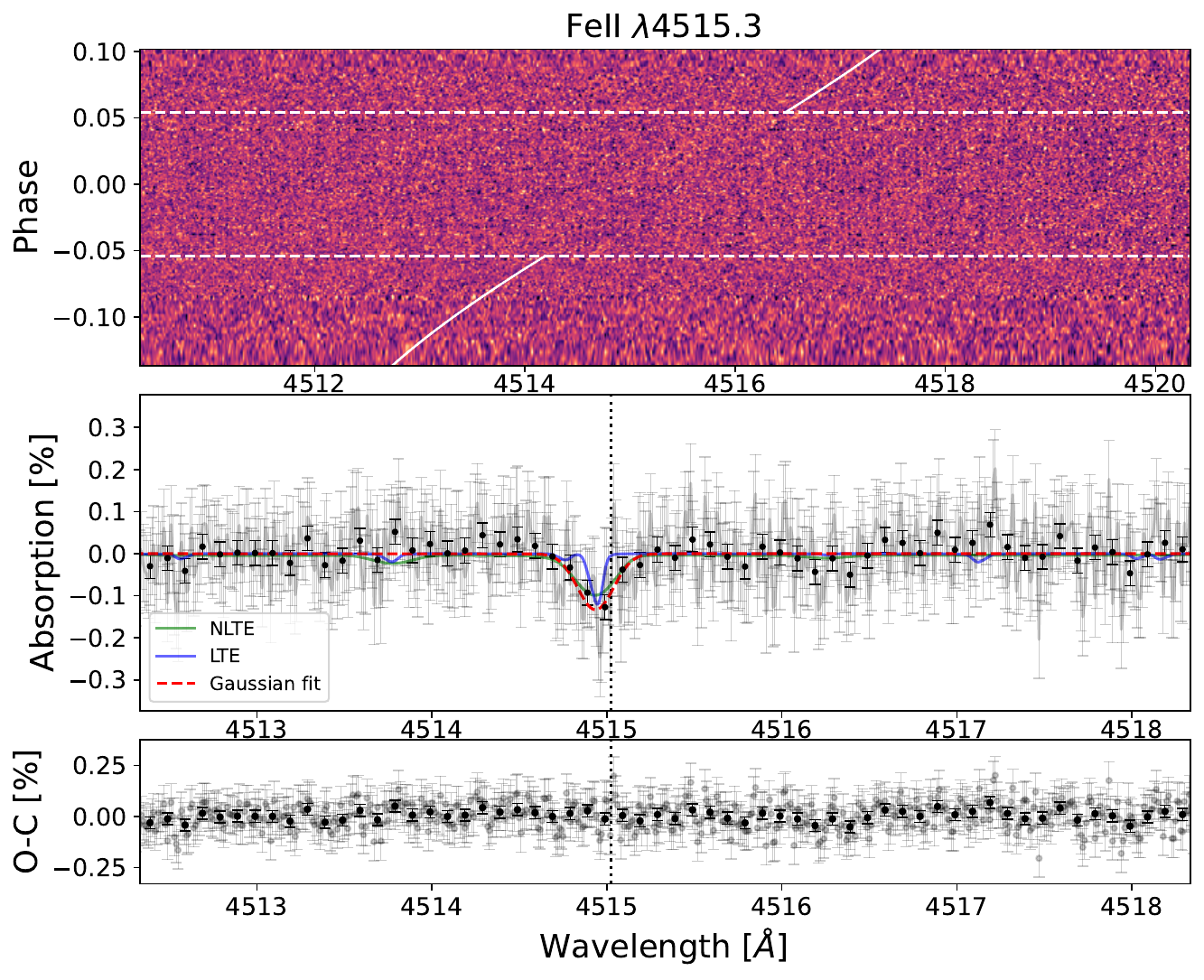}
\includegraphics[width=0.49\textwidth]{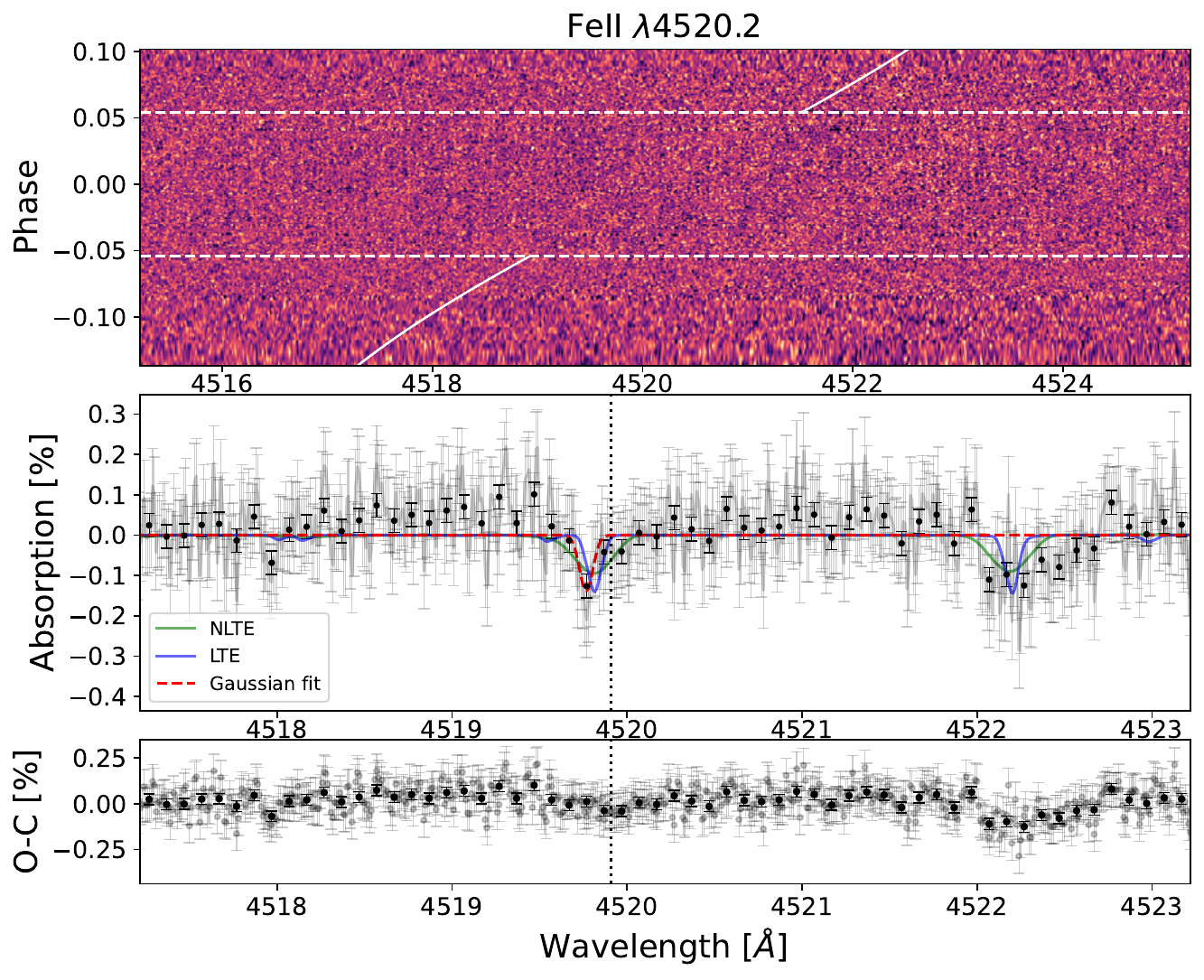}

\caption{Same as Figure \ref{Fig:TS_K9_1}, but for lines $\lambda$4489 $\AA$, $\lambda$4508 $\AA$, $\lambda$4515 $\AA$, and $\lambda$4520 $\AA$.}
\label{Fig:TS_K9_2}
\end{figure}

\begin{figure}[h]
\centering
\includegraphics[width=0.5\textwidth]{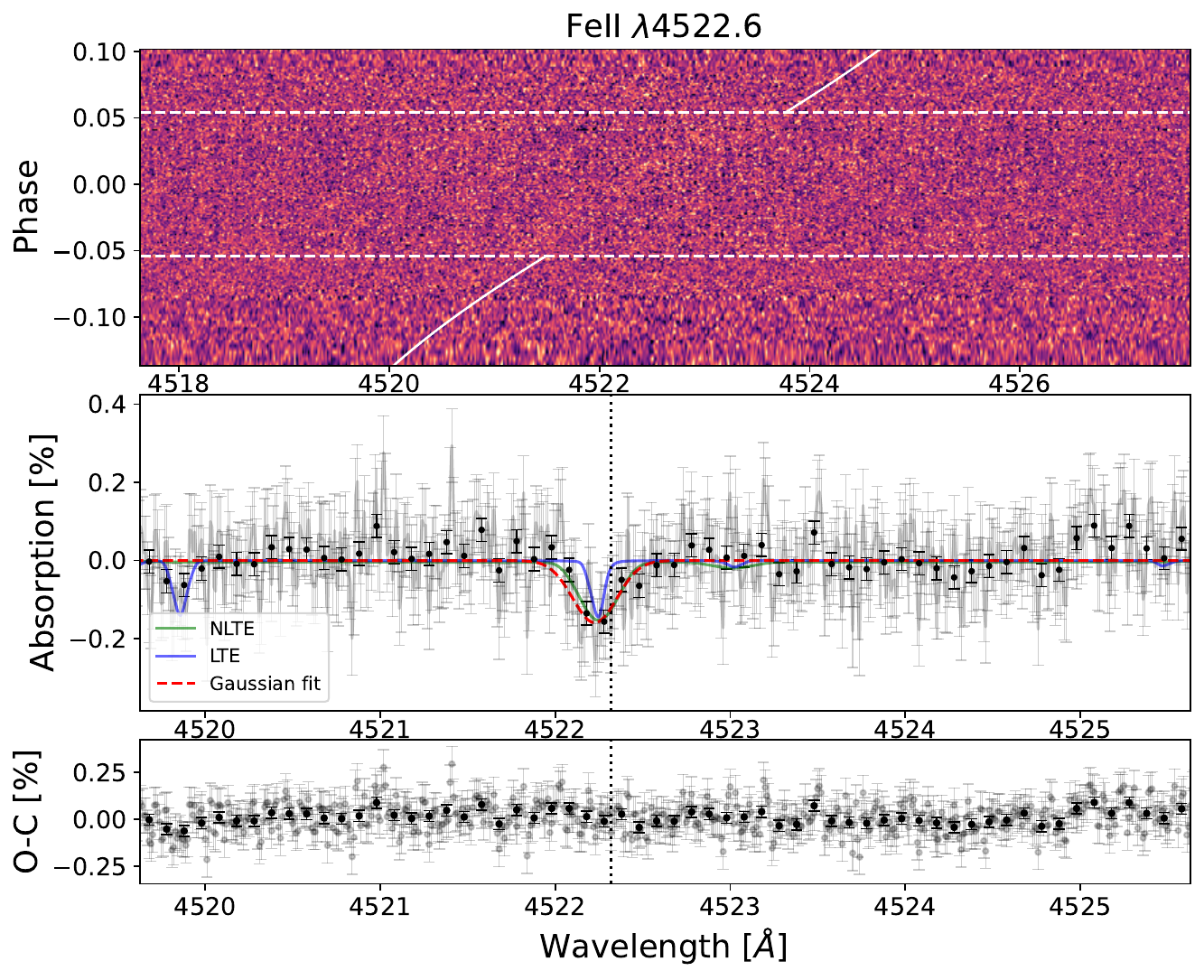}
\includegraphics[width=0.49\textwidth]{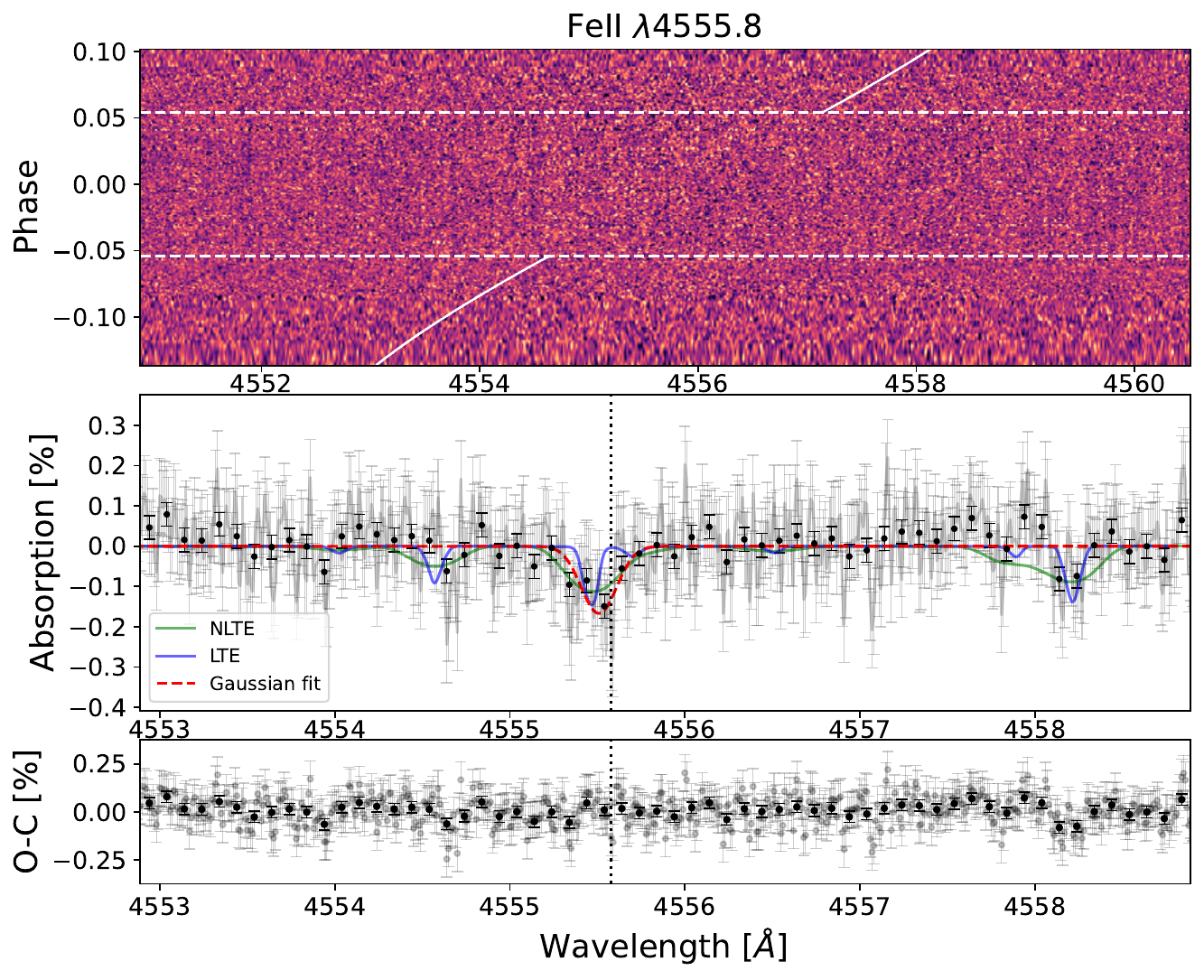}

\includegraphics[width=0.49\textwidth]{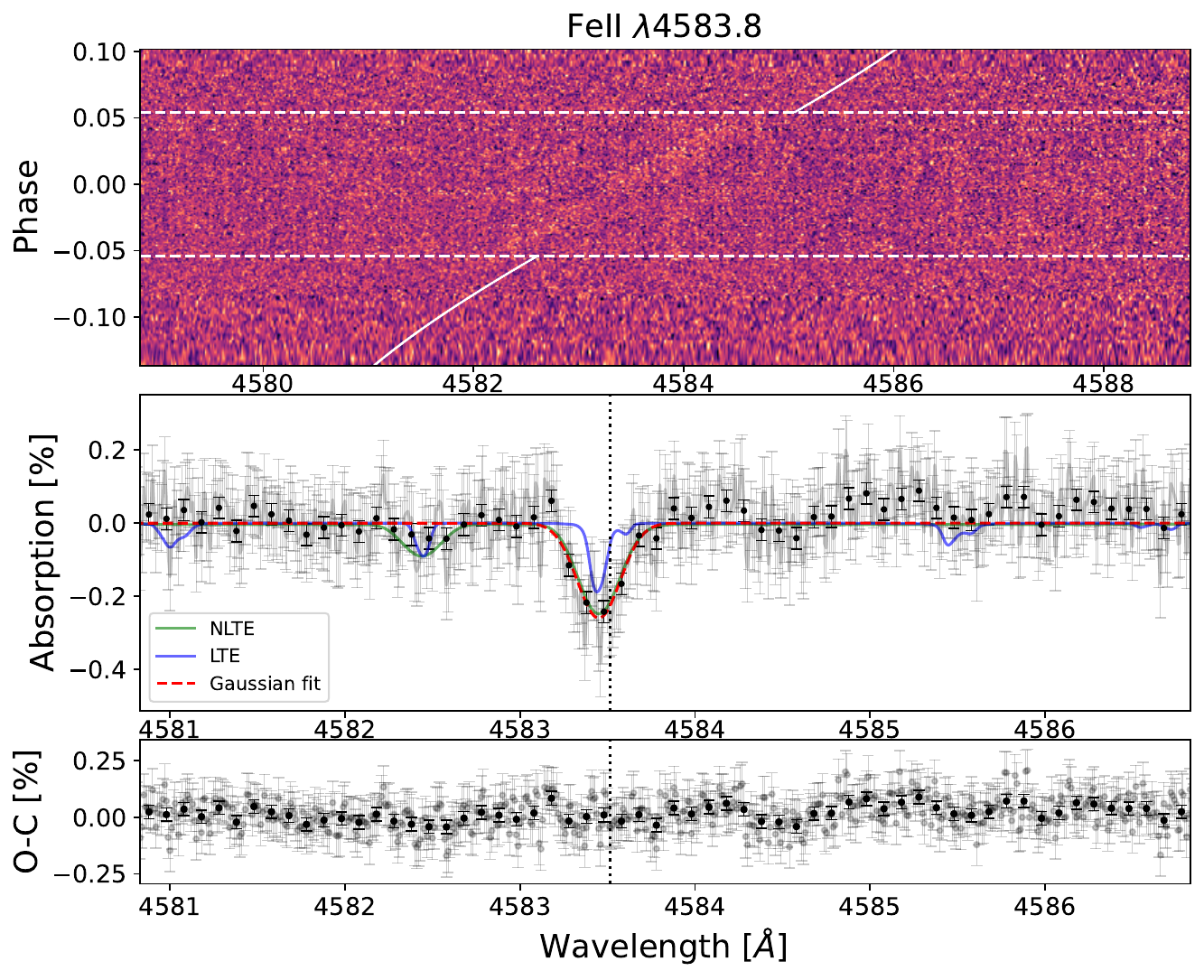}
\includegraphics[width=0.49\textwidth]{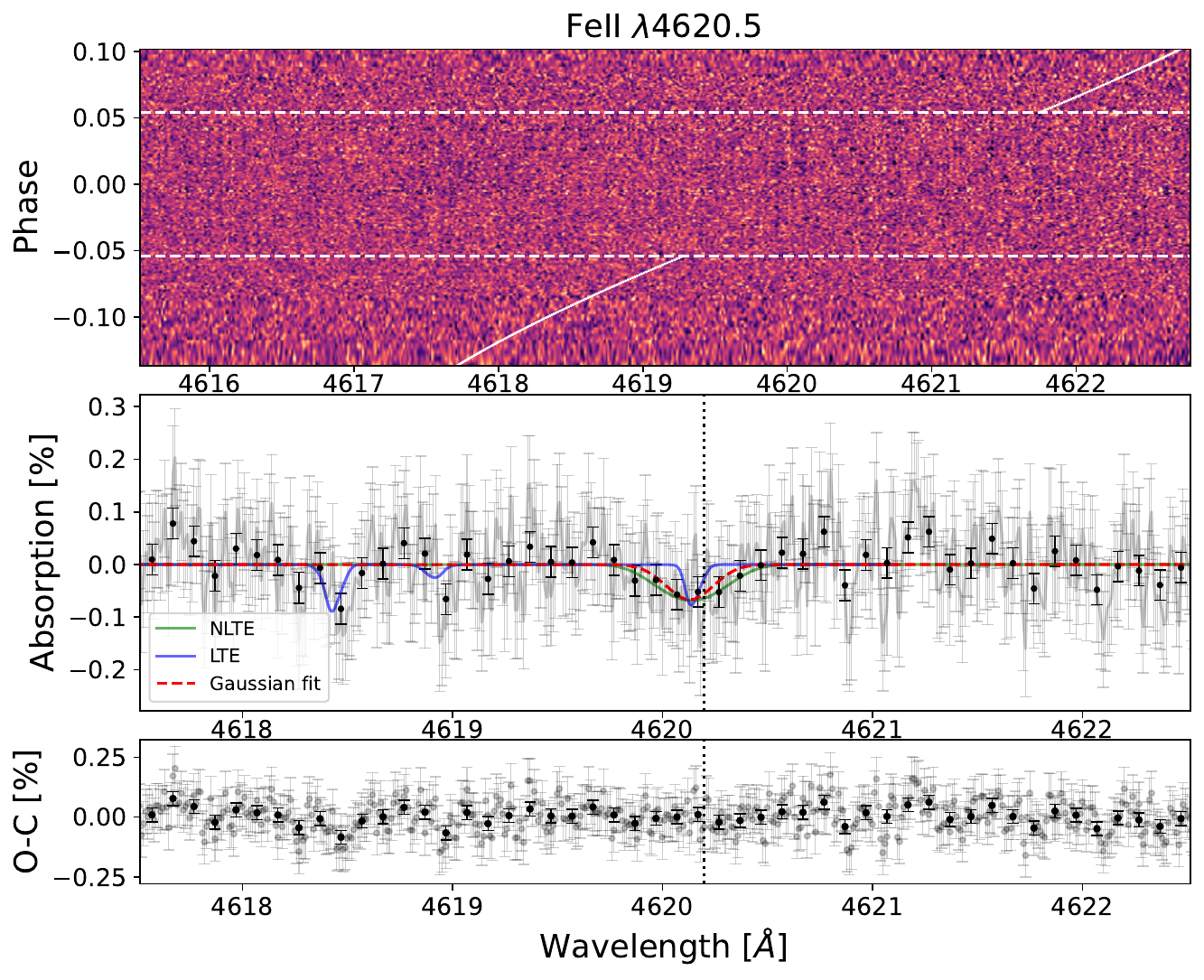}

\caption{Same as Figure \ref{Fig:TS_K9_1}, but for lines $\lambda$4522 $\AA$, $\lambda$4555 $\AA$, $\lambda$4583 $\AA$, and $\lambda$4620 $\AA$.}
\label{Fig:TS_K9_3}
\end{figure}

\begin{figure}[h]
\centering
\includegraphics[width=0.49\textwidth]{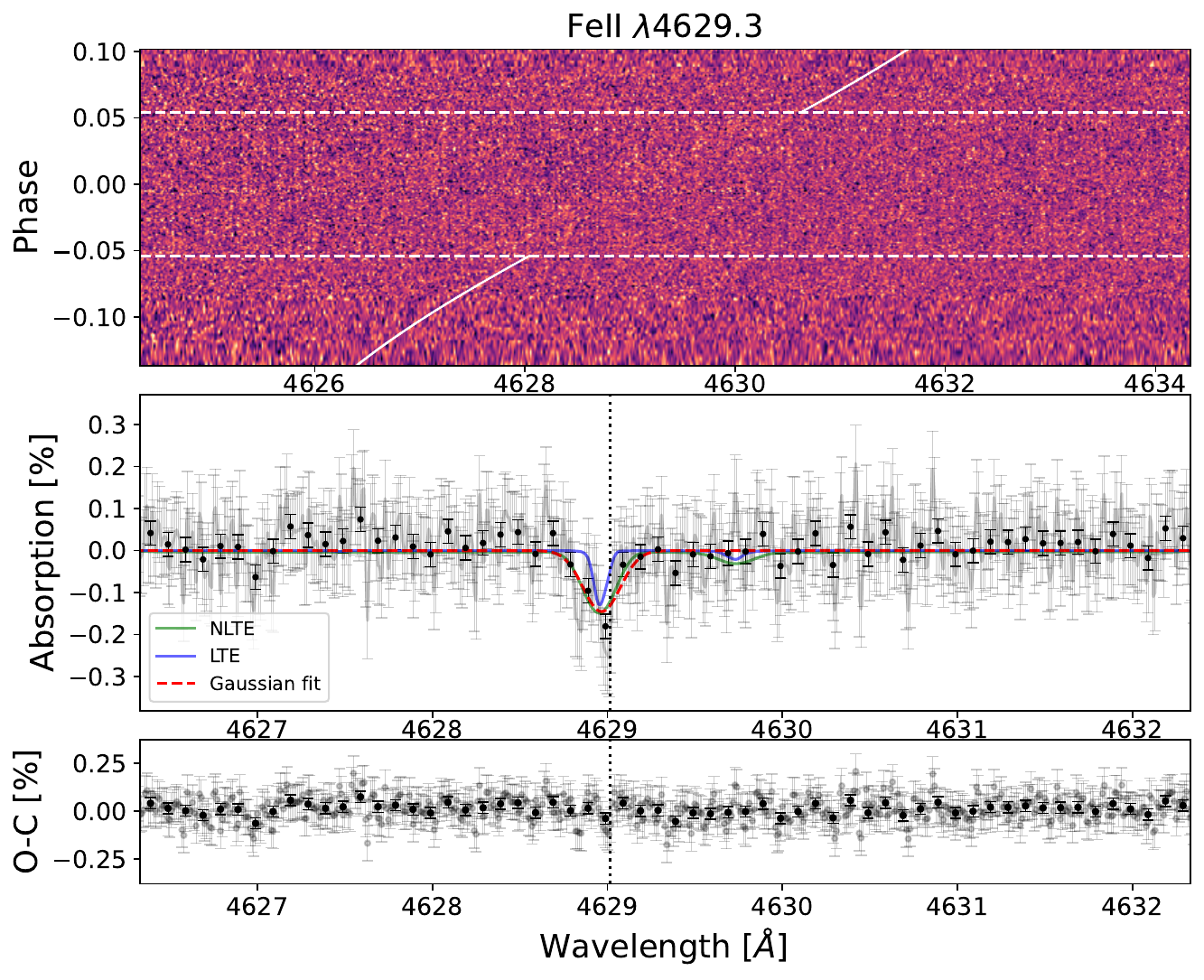}
\includegraphics[width=0.49\textwidth]{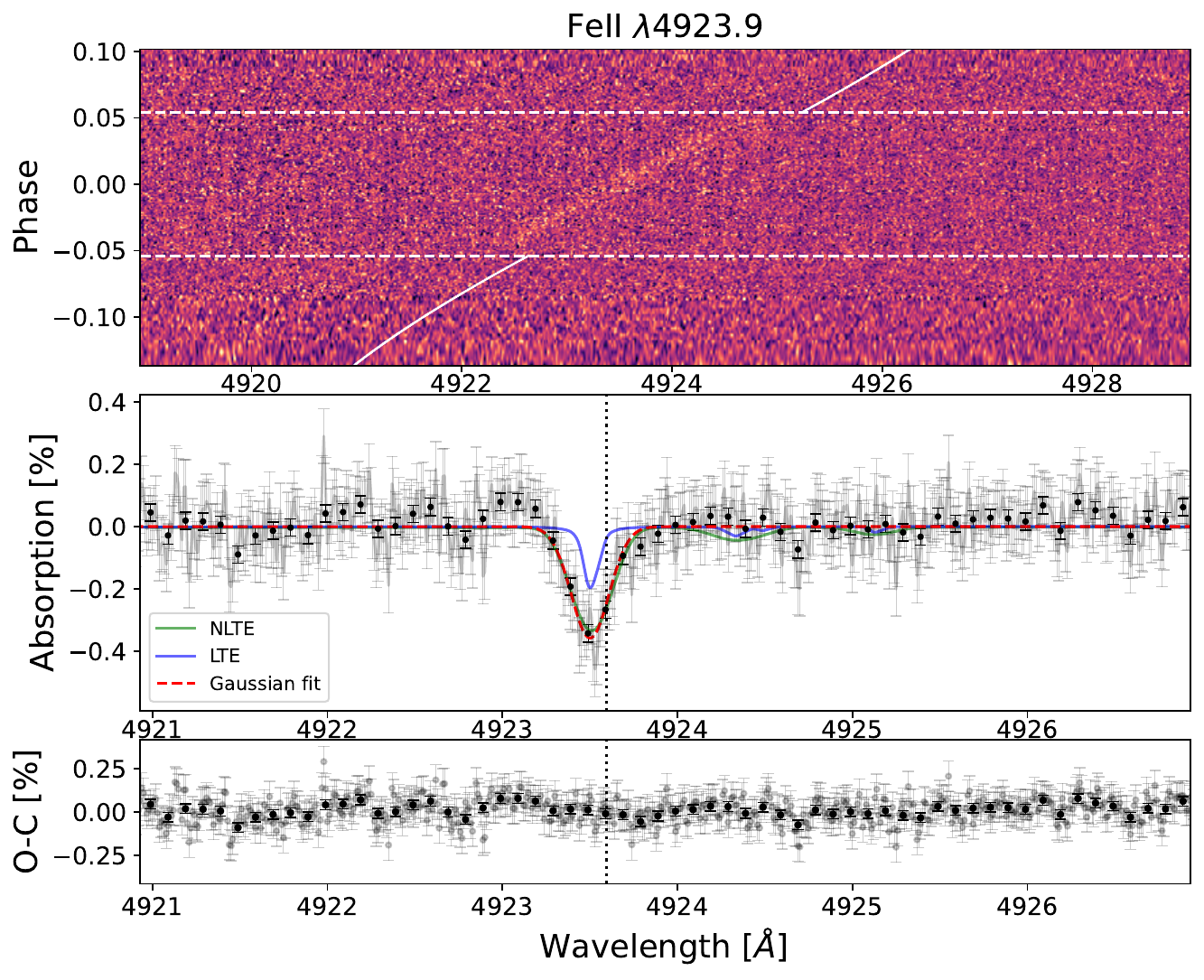}

\includegraphics[width=0.49\textwidth]{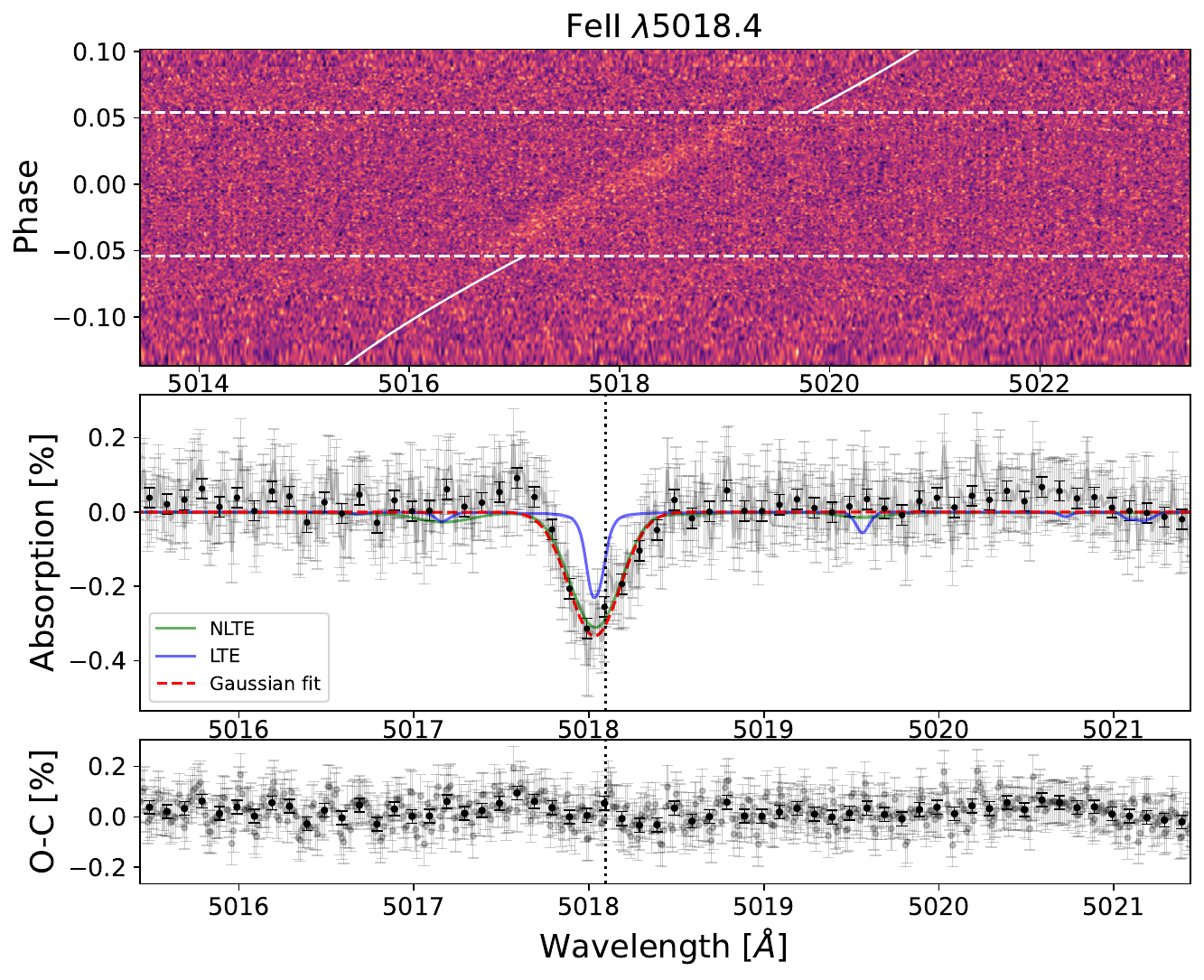}
\includegraphics[width=0.49\textwidth]{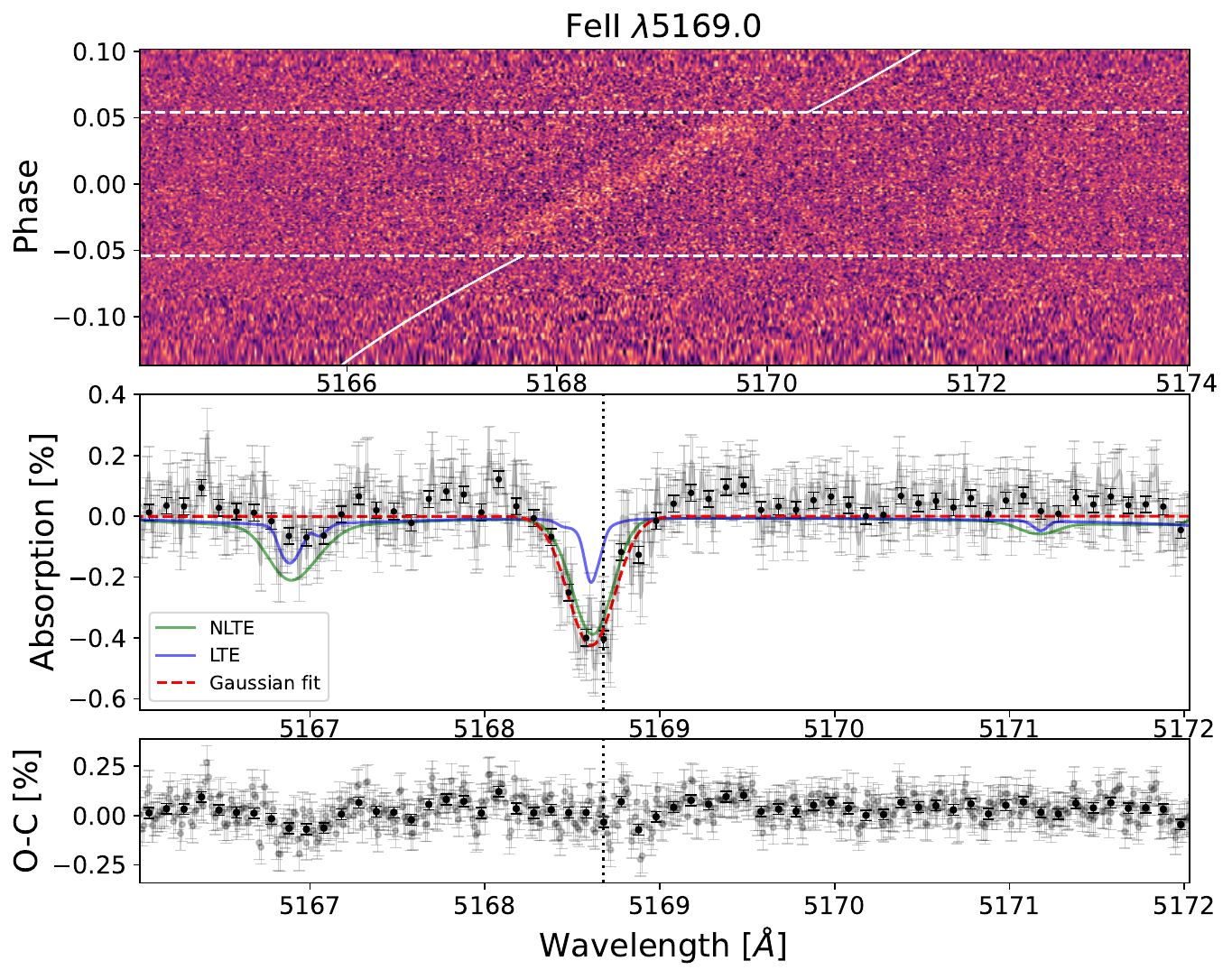}

\caption{Same as Figure \ref{Fig:TS_K9_1}, but for lines $\lambda$4629 $\AA$, $\lambda$4923 $\AA$, $\lambda$5018 $\AA$, and $\lambda$5169 $\AA$.}
\label{Fig:TS_K9_4}
\end{figure}

\begin{figure}[h]
\centering
\includegraphics[width=0.49\textwidth]{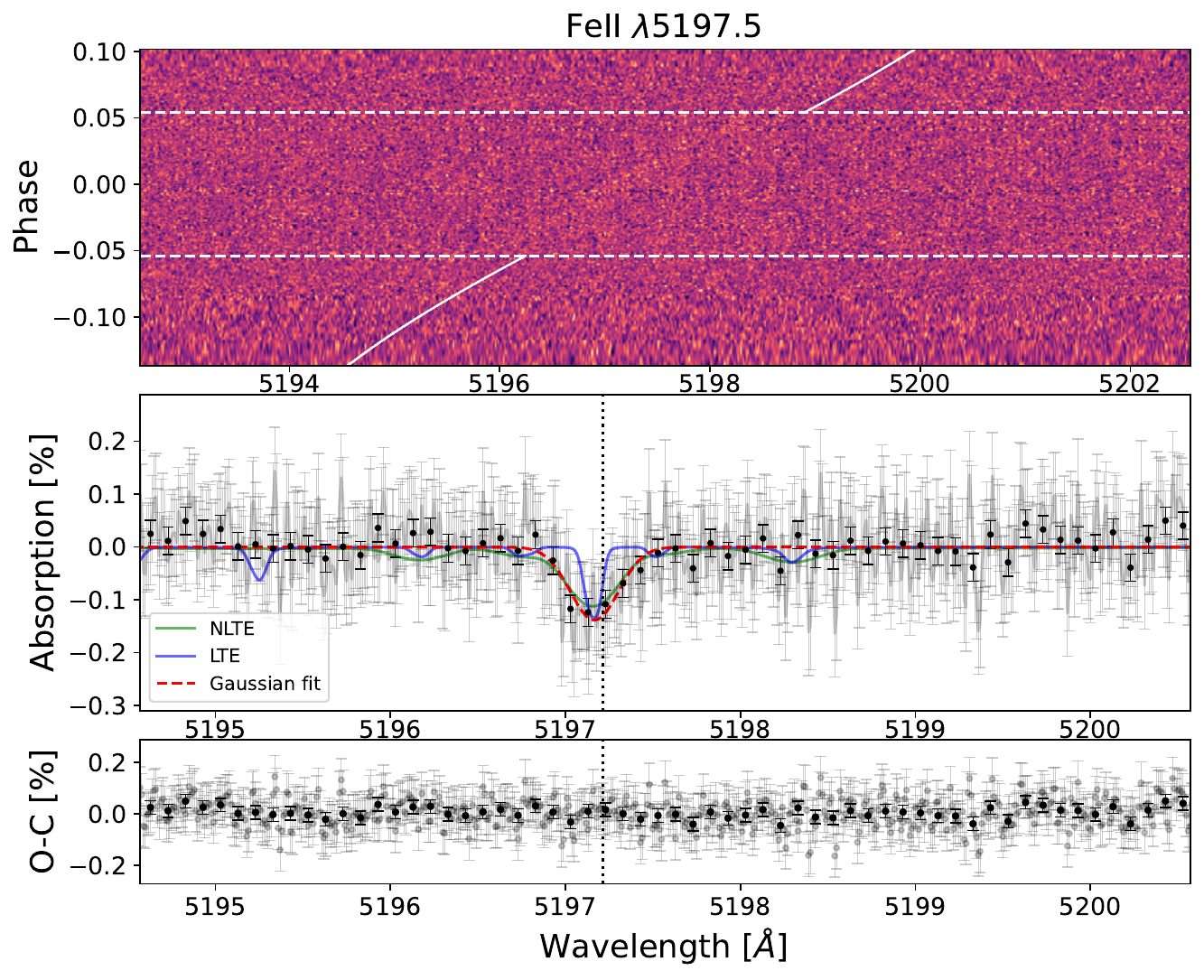}
\includegraphics[width=0.49\textwidth]{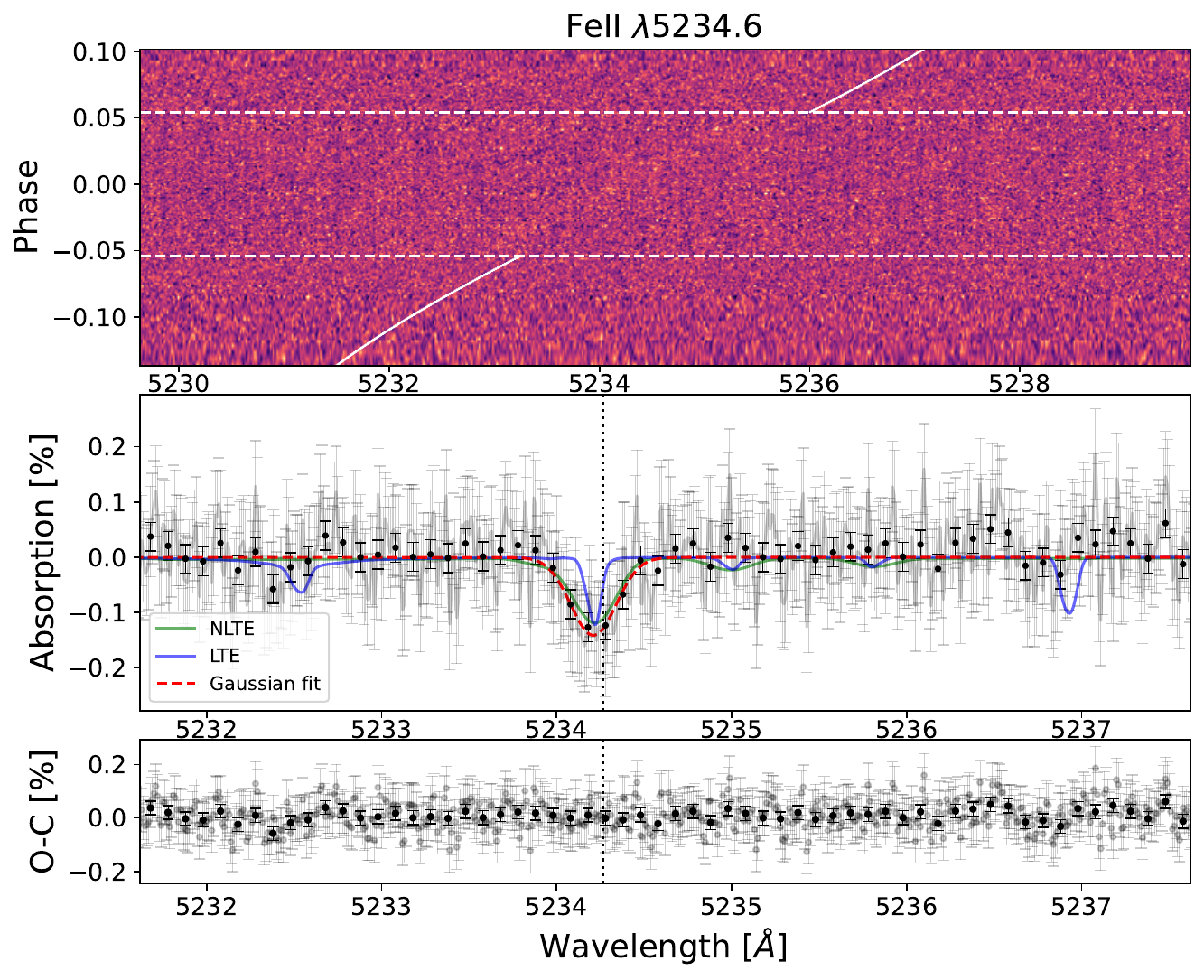}

\includegraphics[width=0.49\textwidth]{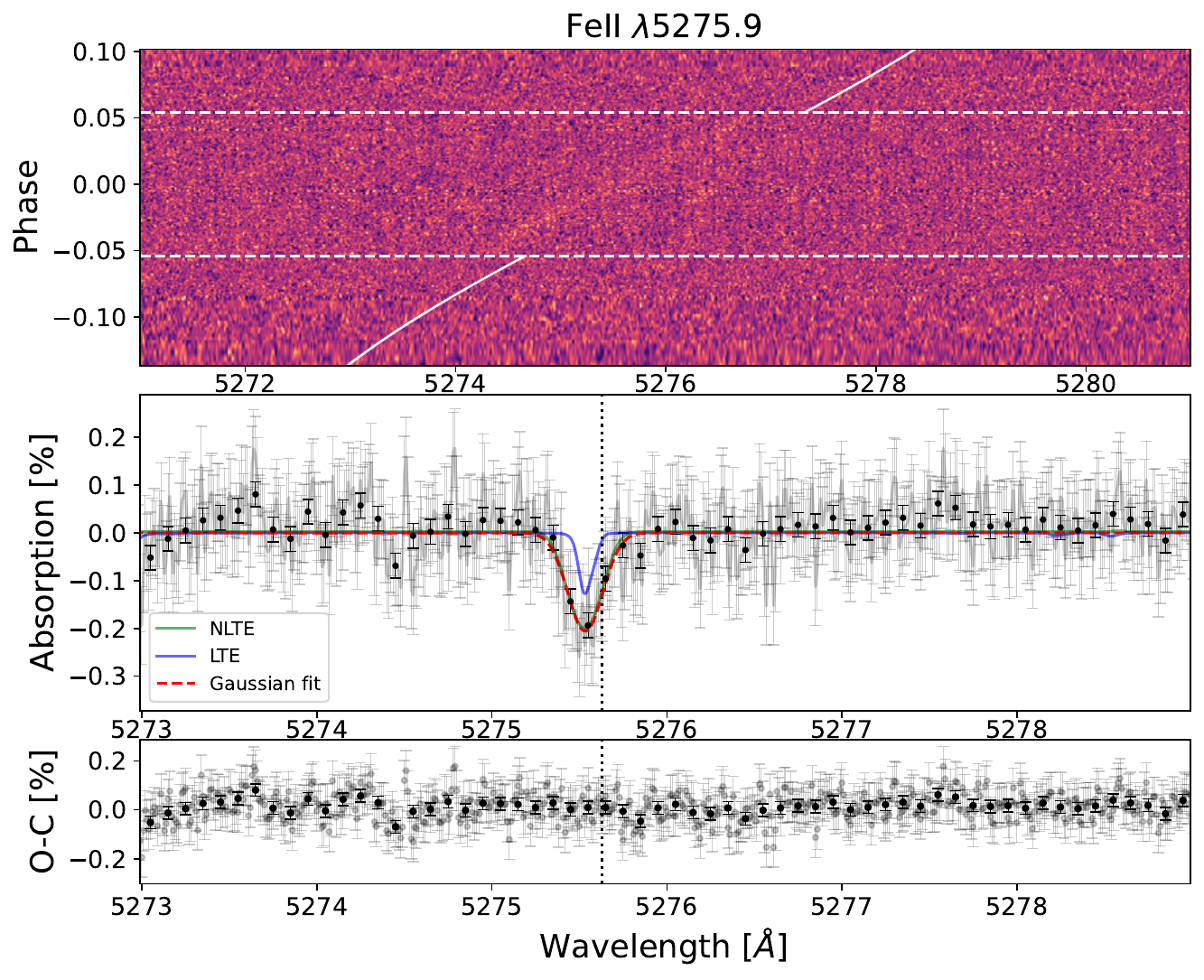}
\includegraphics[width=0.49\textwidth]{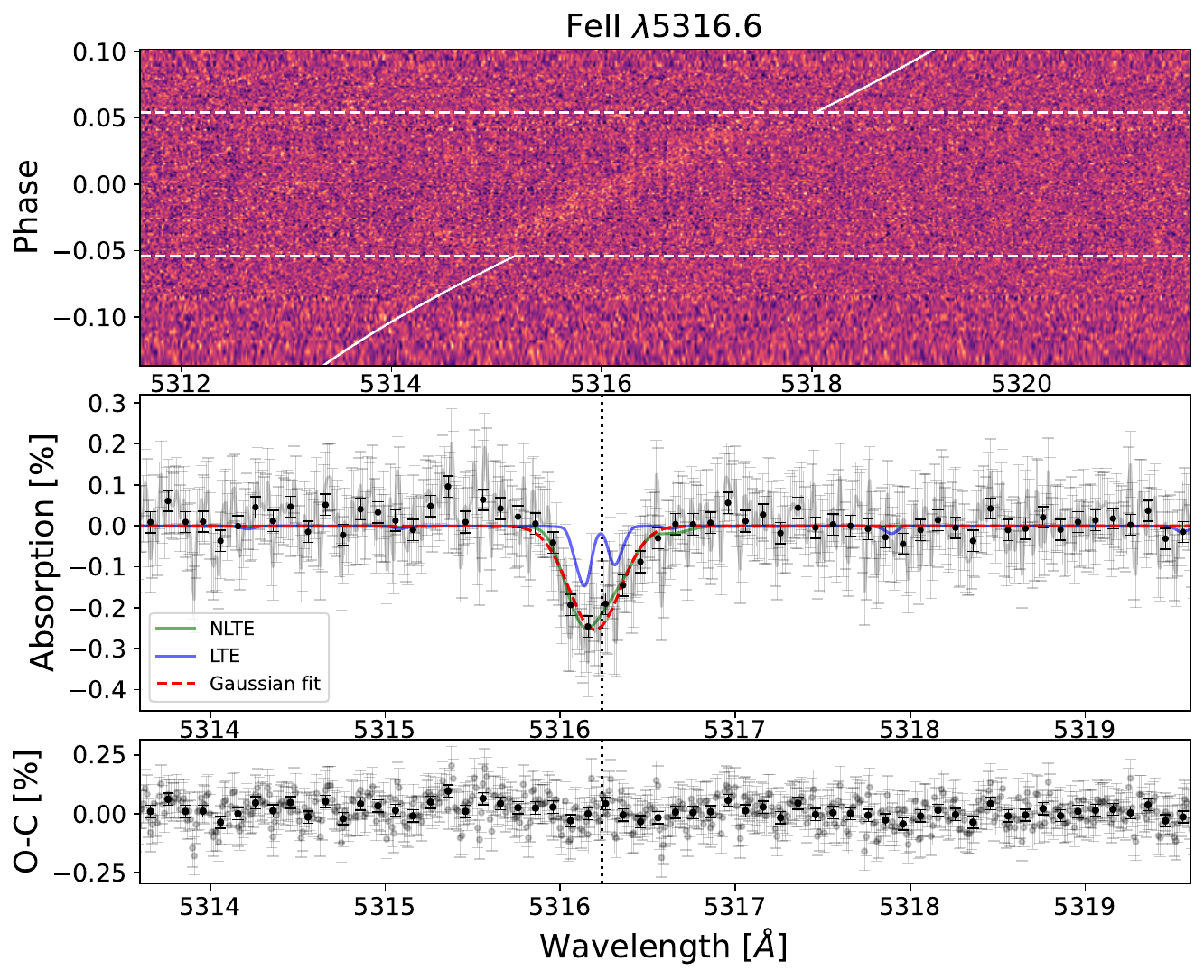}

\caption{Same as Figure \ref{Fig:TS_K9_1}, but for lines $\lambda$5197 $\AA$, $\lambda$5234 $\AA$, $\lambda$5276 $\AA$, and $\lambda$5316 $\AA$.}
\label{Fig:TS_K9_5}
\end{figure}

\begin{figure}[h]
\centering
\includegraphics[width=0.49\textwidth]{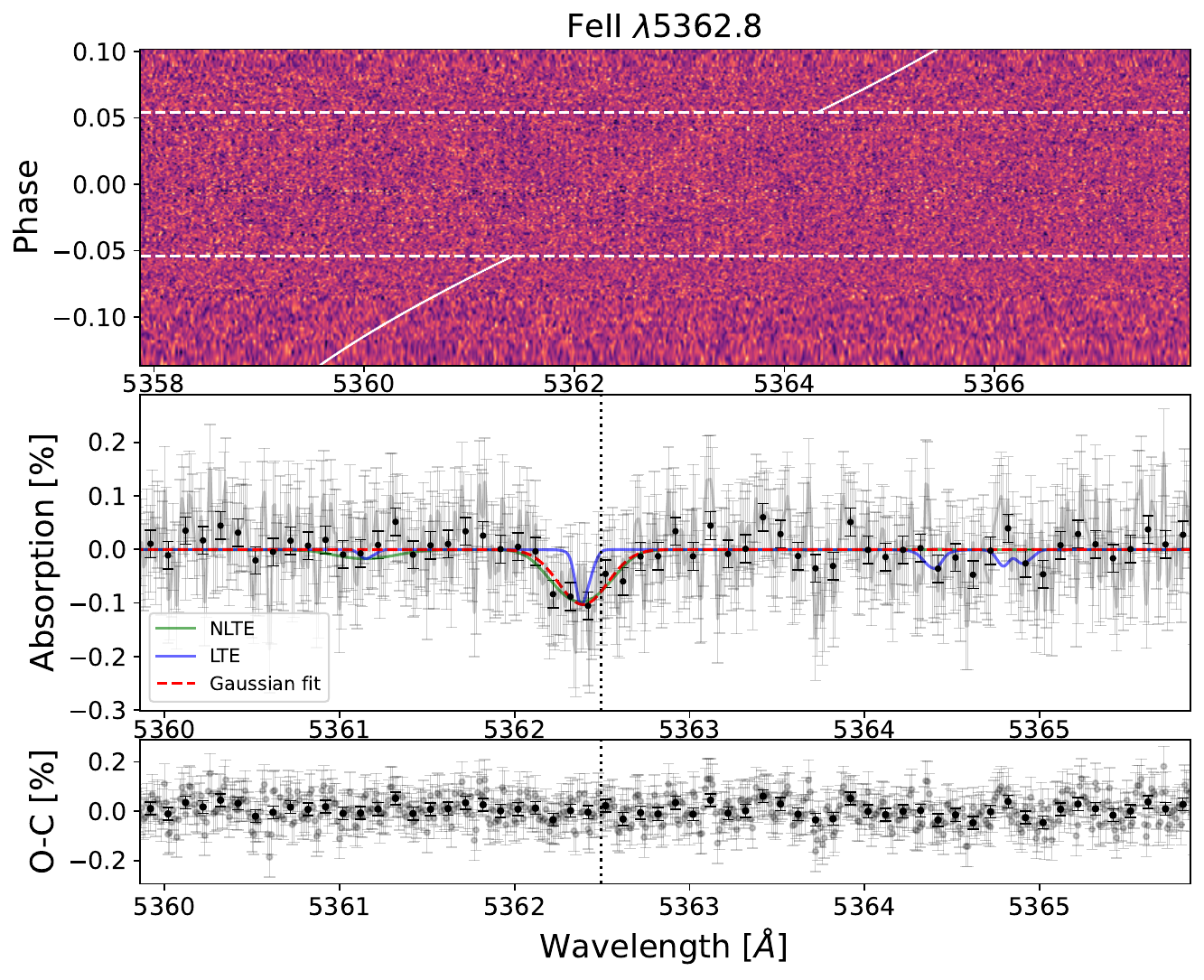}

\caption{Same as Figure \ref{Fig:TS_K9_1}, but for lines $\lambda$5362 $\AA$.}
\label{Fig:TS_K9_6}
\end{figure}

\begin{figure}[h]
\centering
\includegraphics[width=0.38\textwidth]{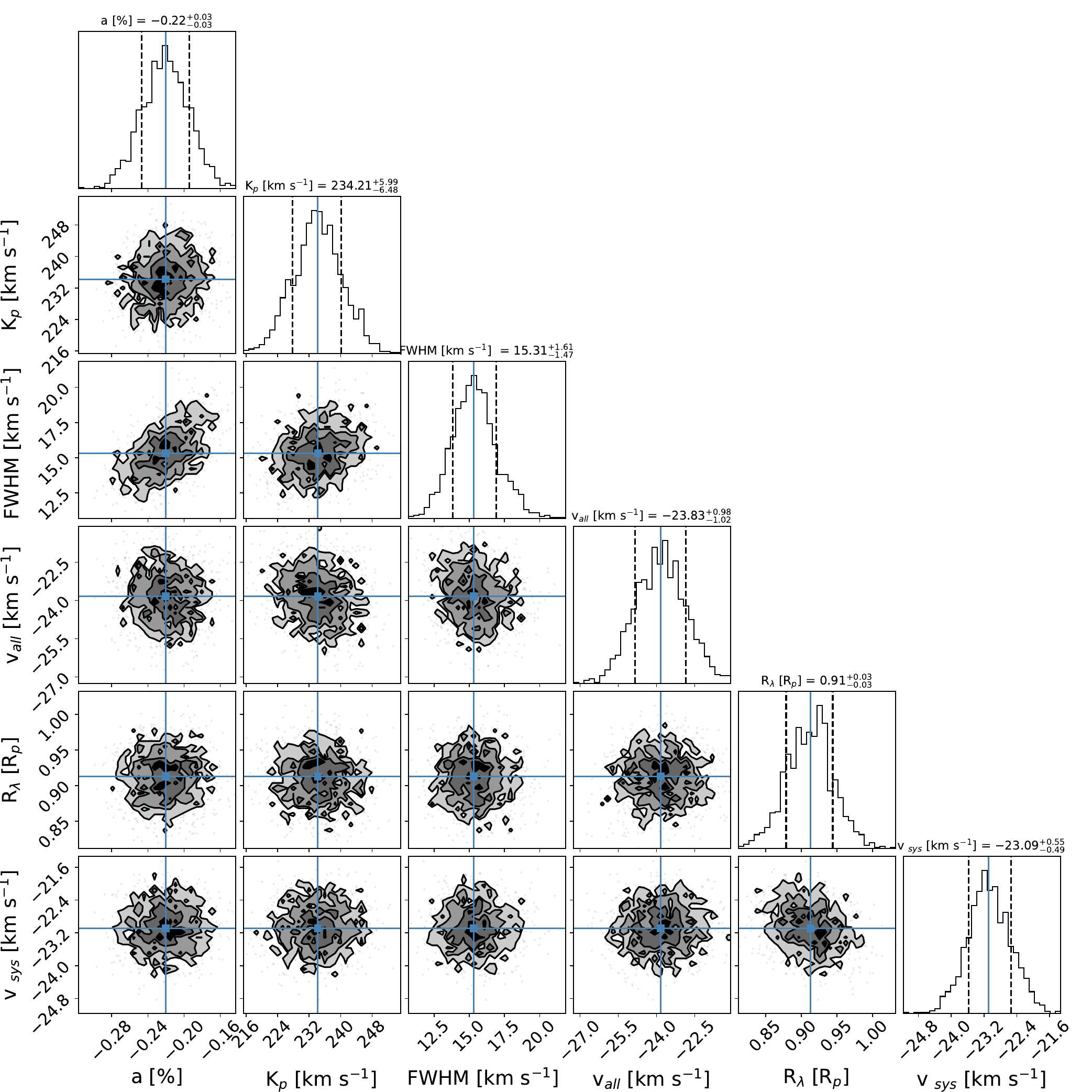}
\includegraphics[width=0.38\textwidth]{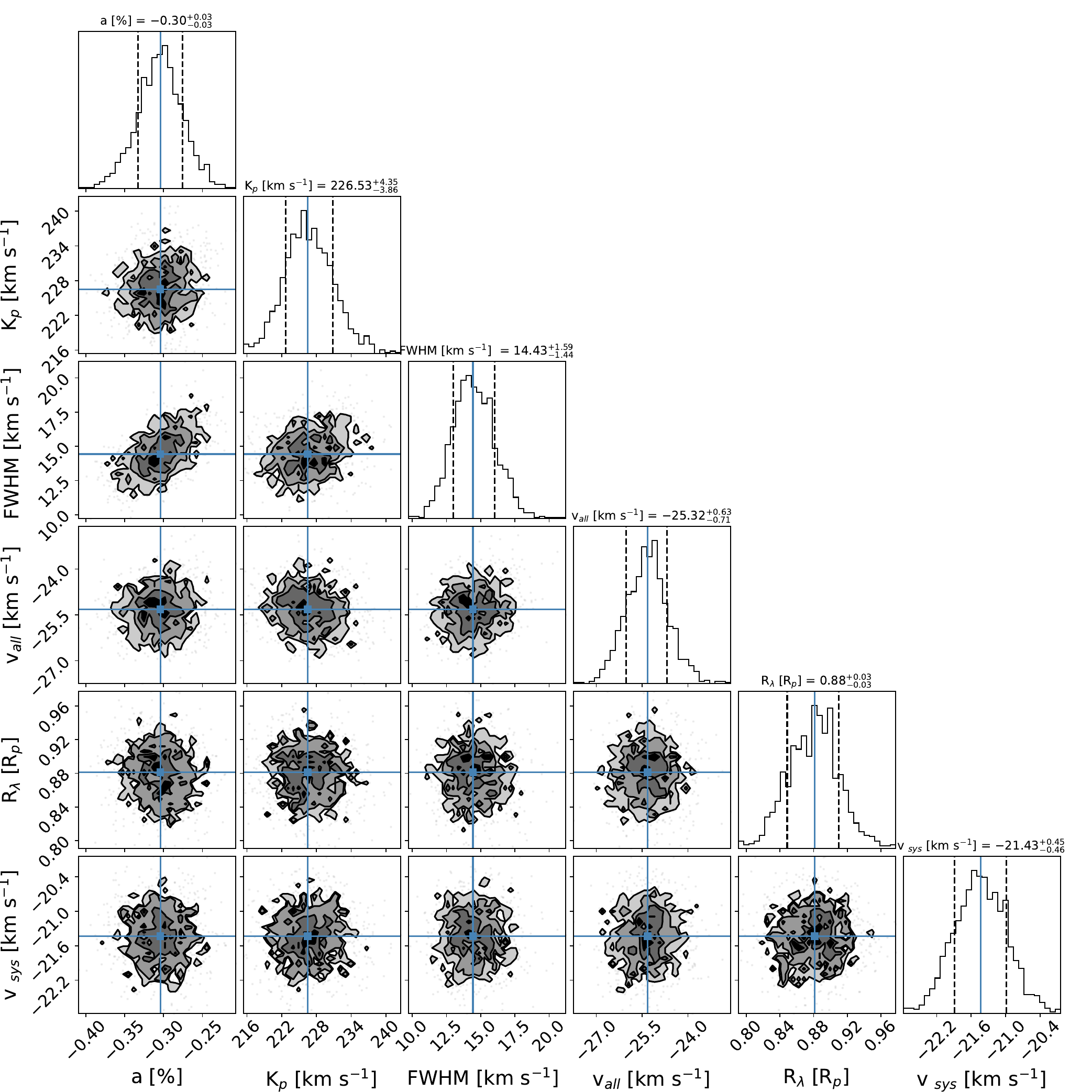}
\caption{The corner plots of the MCMC analysis for KELT-9b of the lines \ion{Fe}{ii} $\lambda$4173.4 $\AA$, (left panel) and \ion{Fe}{ii} $\lambda$4233.16 $\AA$ (left panel). }
\label{fig:corner_k9_1}
\end{figure}

\begin{figure}[h]
\centering
\includegraphics[width=0.38\textwidth]{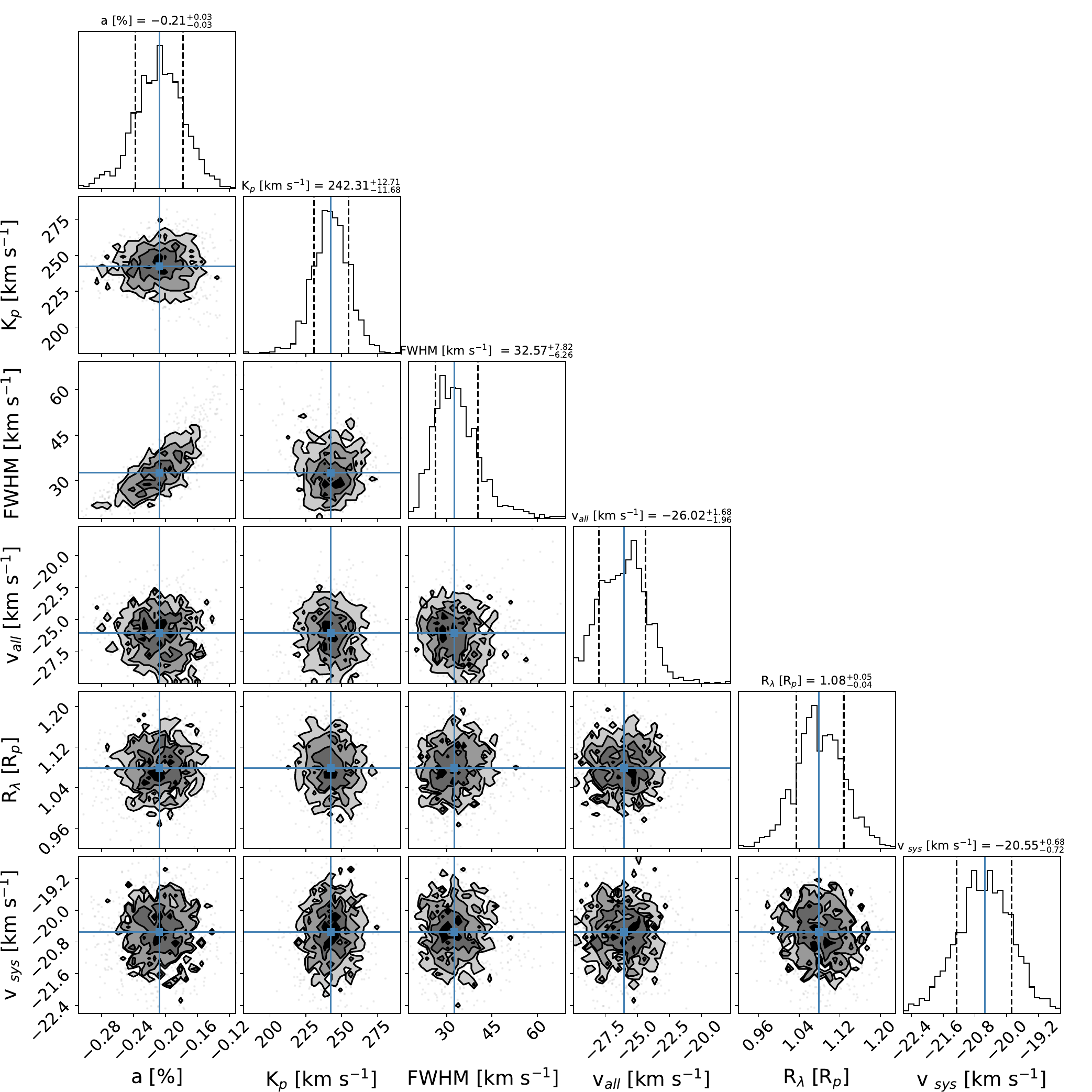}
\includegraphics[width=0.38\textwidth]{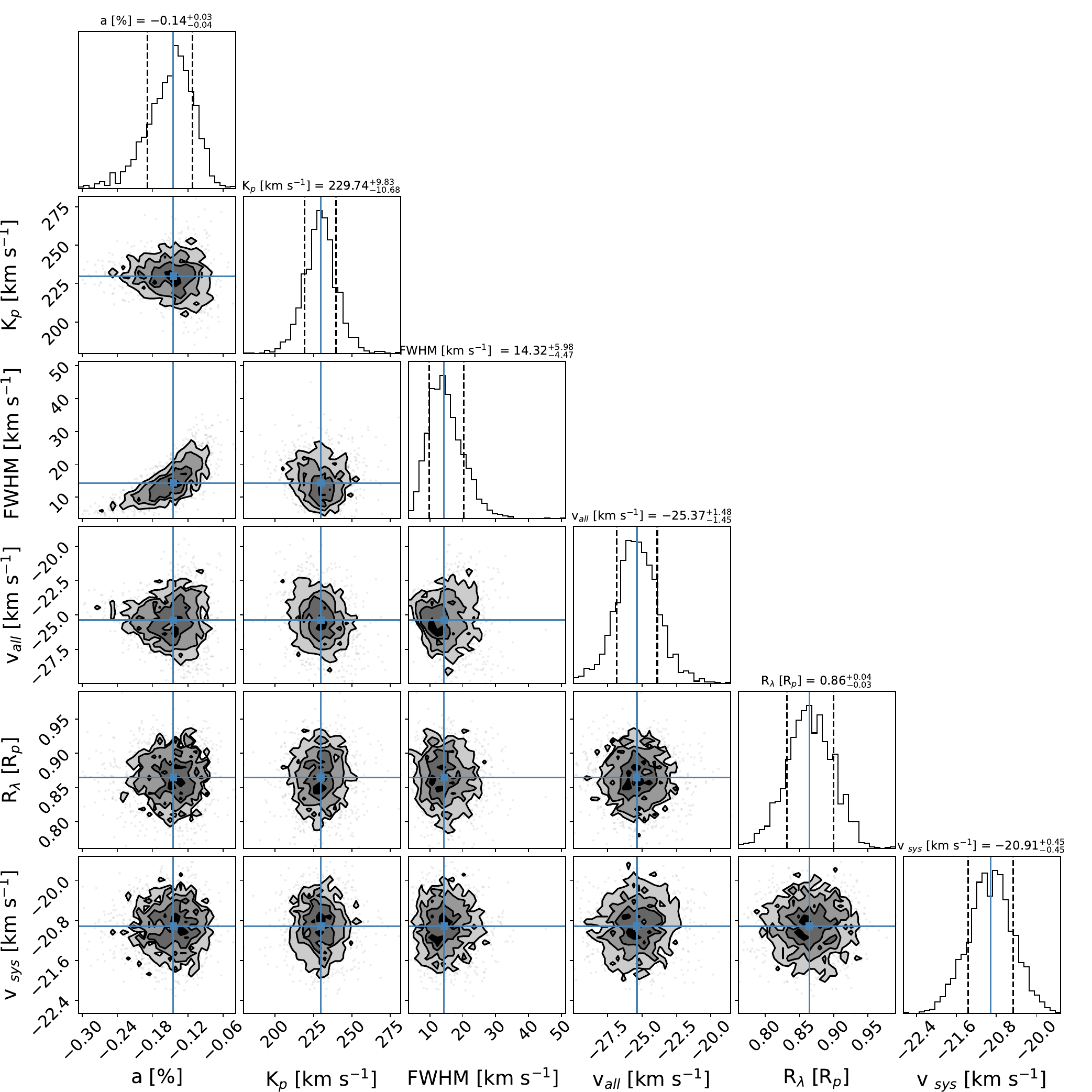}
\caption{Same as Figure \ref{fig:corner_k9_1} but for \ion{Fe}{ii} $\lambda$4351 $\AA$ (left) and \ion{Fe}{ii} $\lambda$4385 $\AA$ (right).}
\label{fig:corner_k9_2}
\end{figure}

\begin{figure}[h]
\centering
\includegraphics[width=0.38\textwidth]{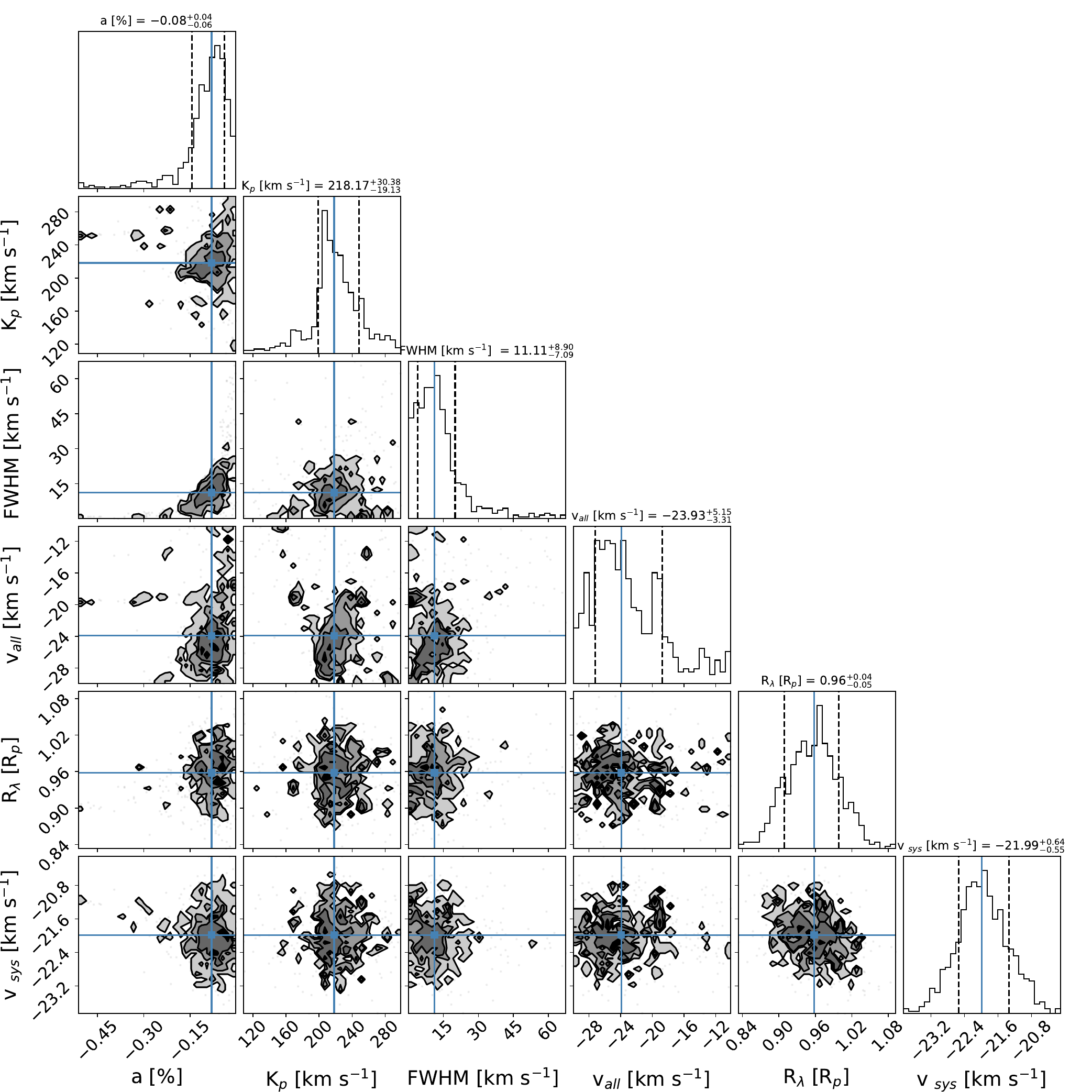}
\includegraphics[width=0.38\textwidth]{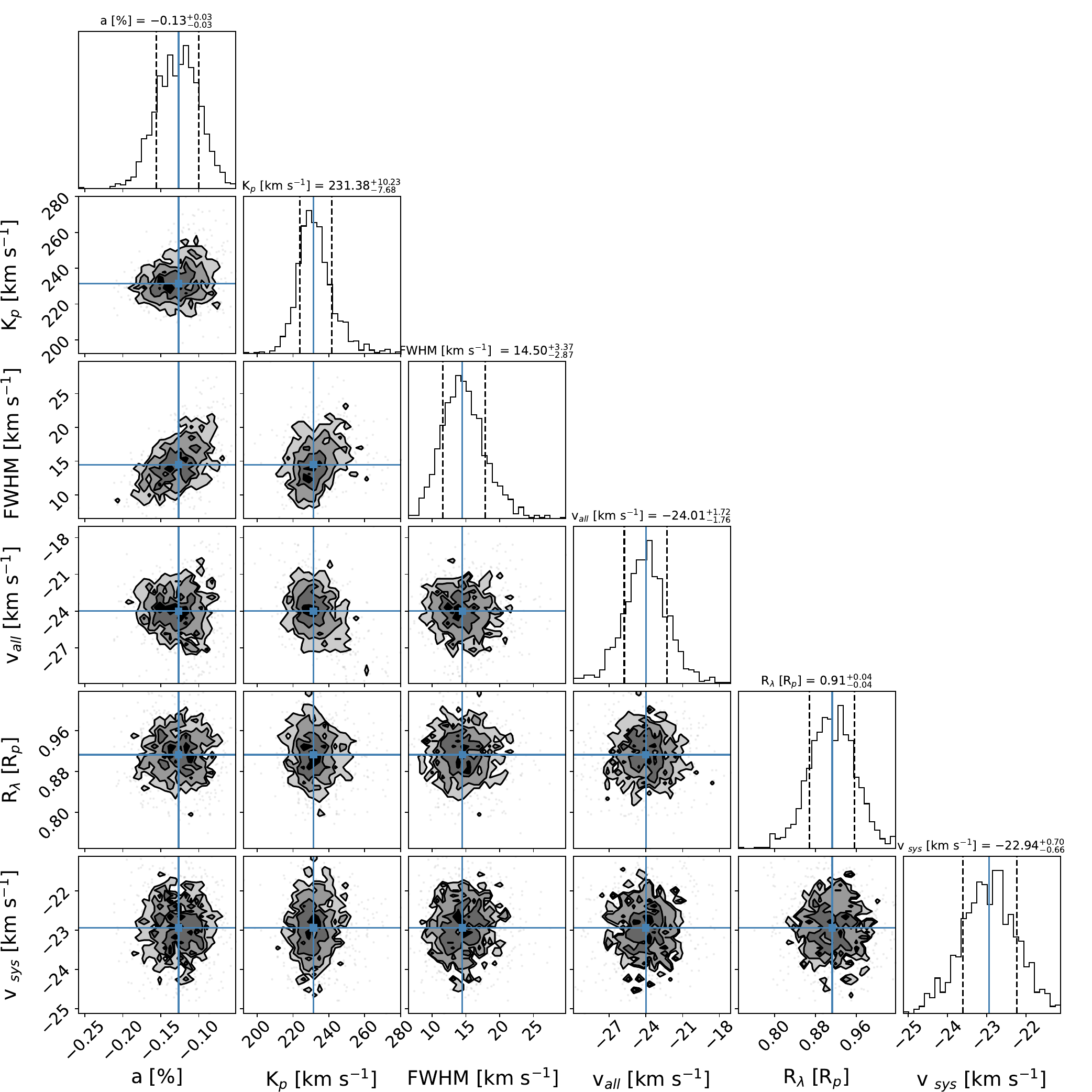}
\caption{Same as Figure \ref{fig:corner_k9_1} but for \ion{Fe}{ii} $\lambda$4489 $\AA$ (left) and \ion{Fe}{ii} $\lambda$4508 $\AA$ (right).}
\label{fig:corner_k9_3}
\end{figure}

\begin{figure}[h]
\centering
\includegraphics[width=0.38\textwidth]{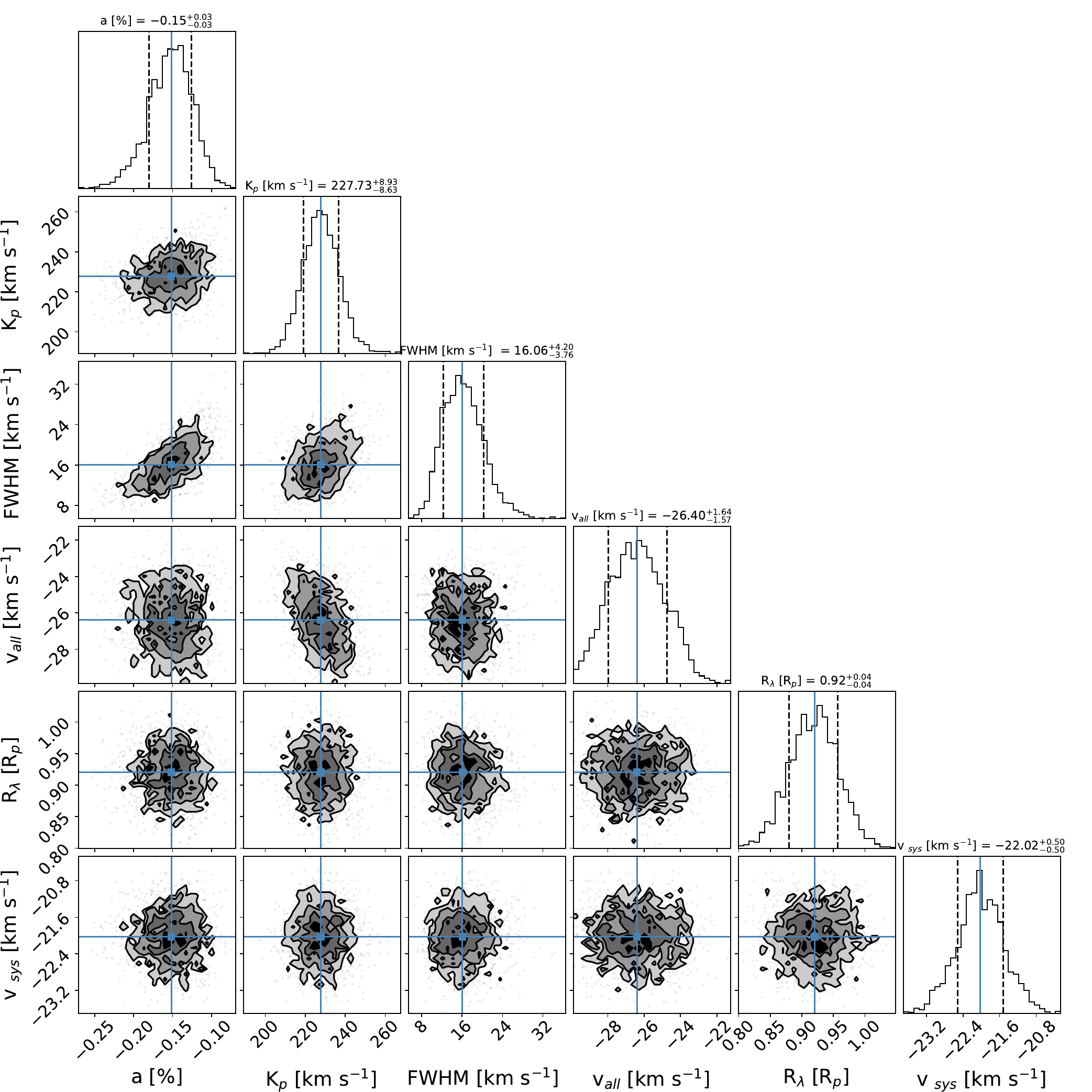}
\includegraphics[width=0.38\textwidth]{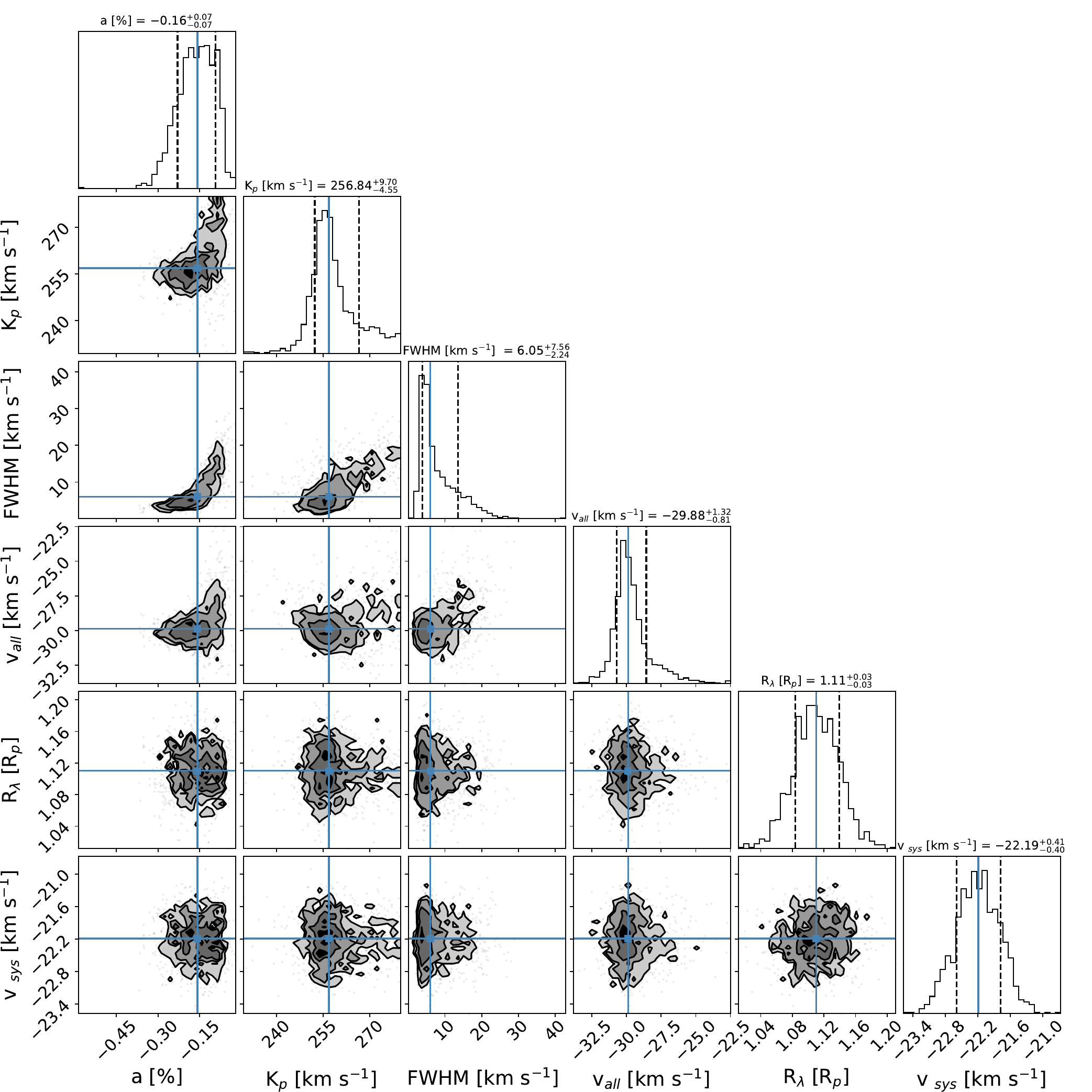}
\caption{Same as Figure \ref{fig:corner_k9_1} but for \ion{Fe}{ii} $\lambda$4515 $\AA$ (left) and \ion{Fe}{ii} $\lambda$4520 $\AA$ (right).}
\label{fig:corner_k9_4}
\end{figure}

\begin{figure}[h]
\centering
\includegraphics[width=0.38\textwidth]{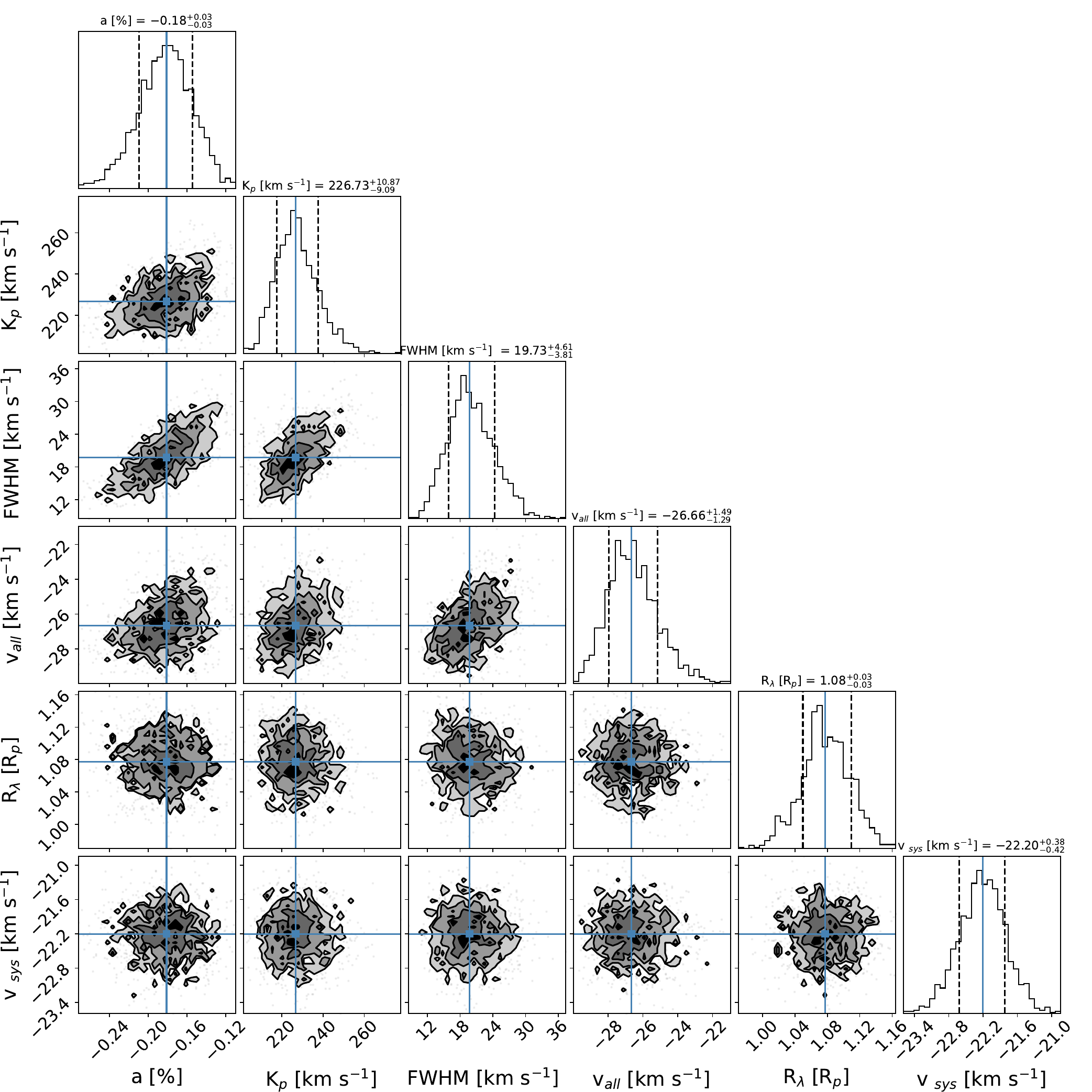}
\includegraphics[width=0.38\textwidth]{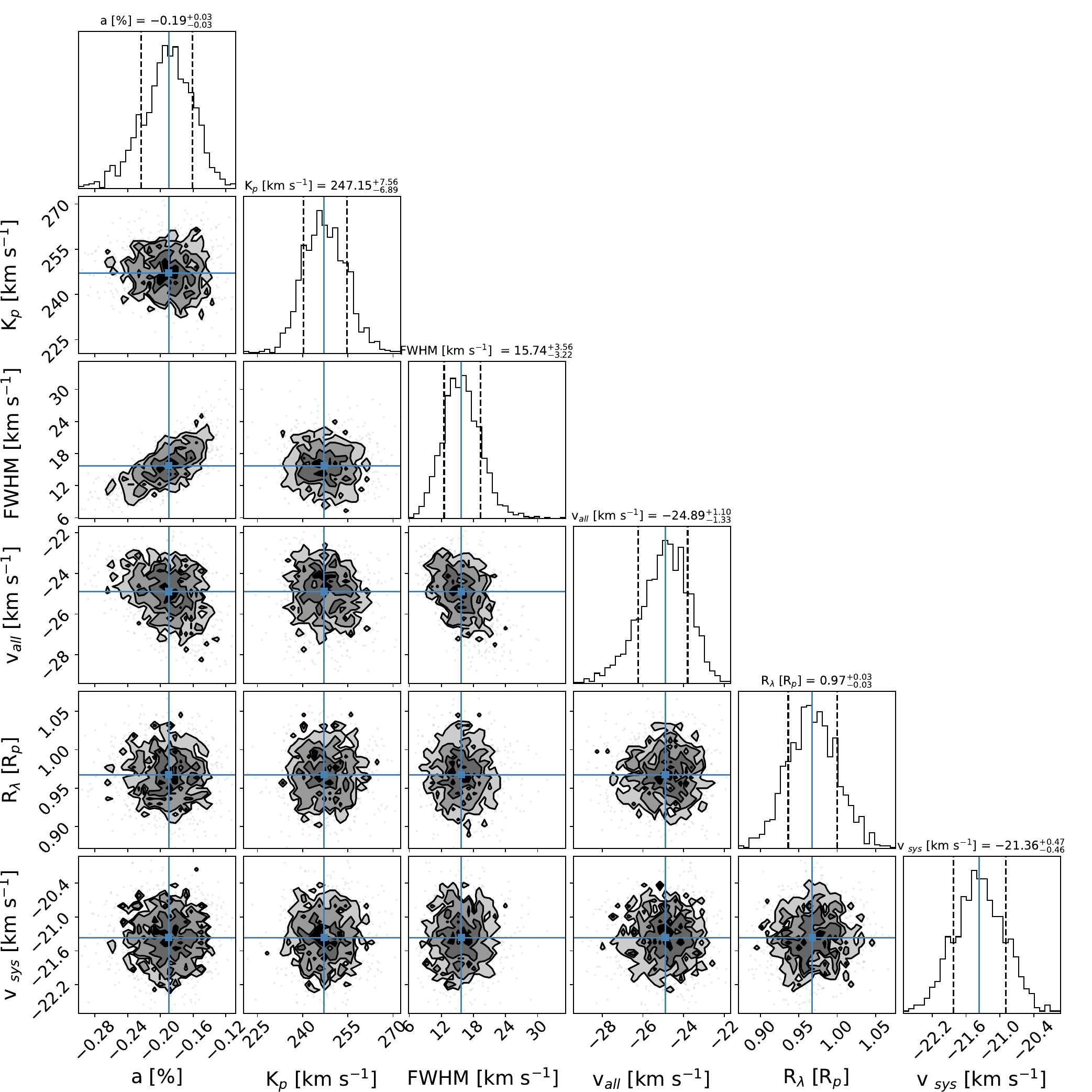}
\caption{Same as Figure \ref{fig:corner_k9_1} but for \ion{Fe}{ii} $\lambda$4522 $\AA$ (left) and \ion{Fe}{ii} $\lambda$4555 $\AA$ (right).}
\label{fig:corner_k9_5}
\end{figure}

\begin{figure}[h]
\centering
\includegraphics[width=0.38\textwidth]{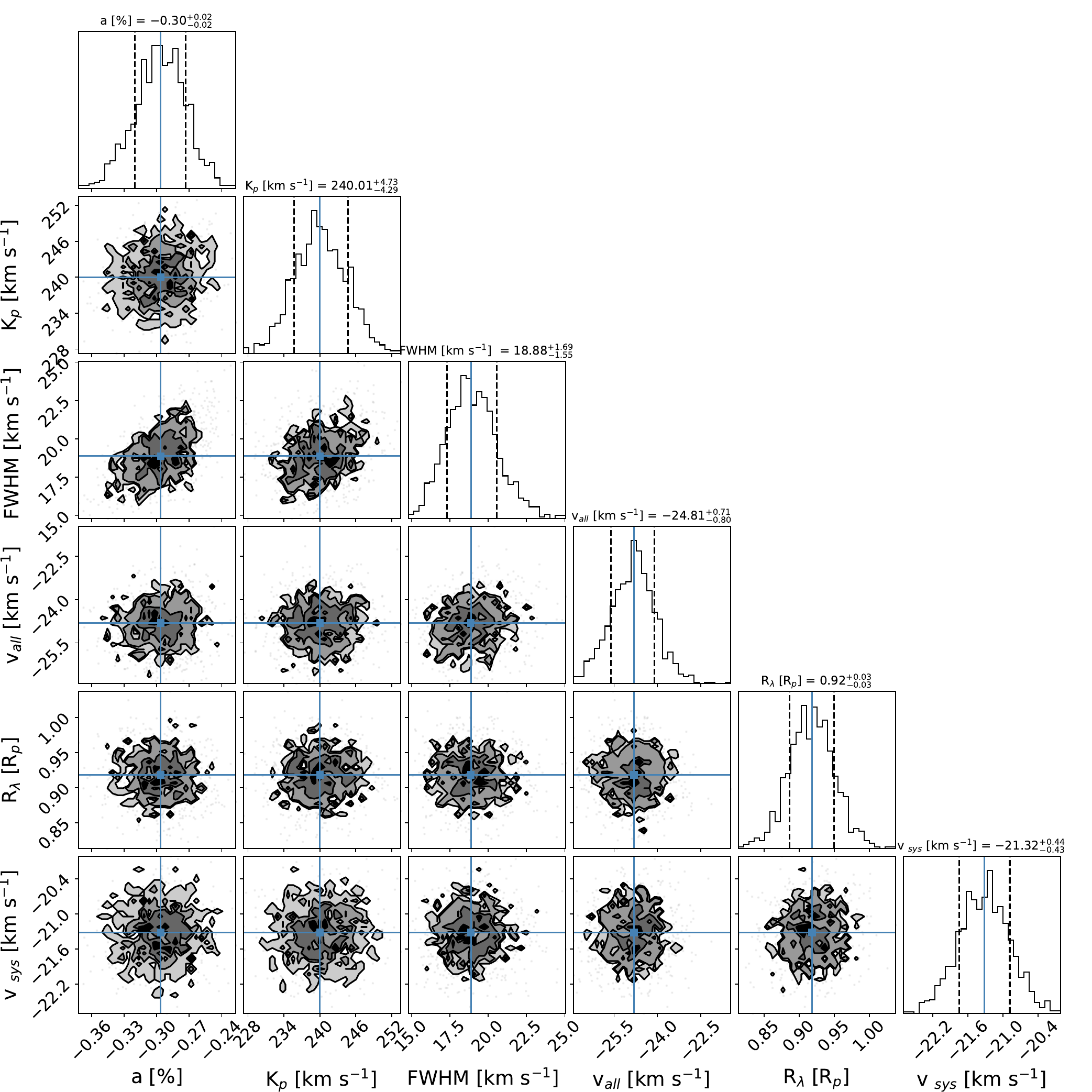}
\includegraphics[width=0.38\textwidth]{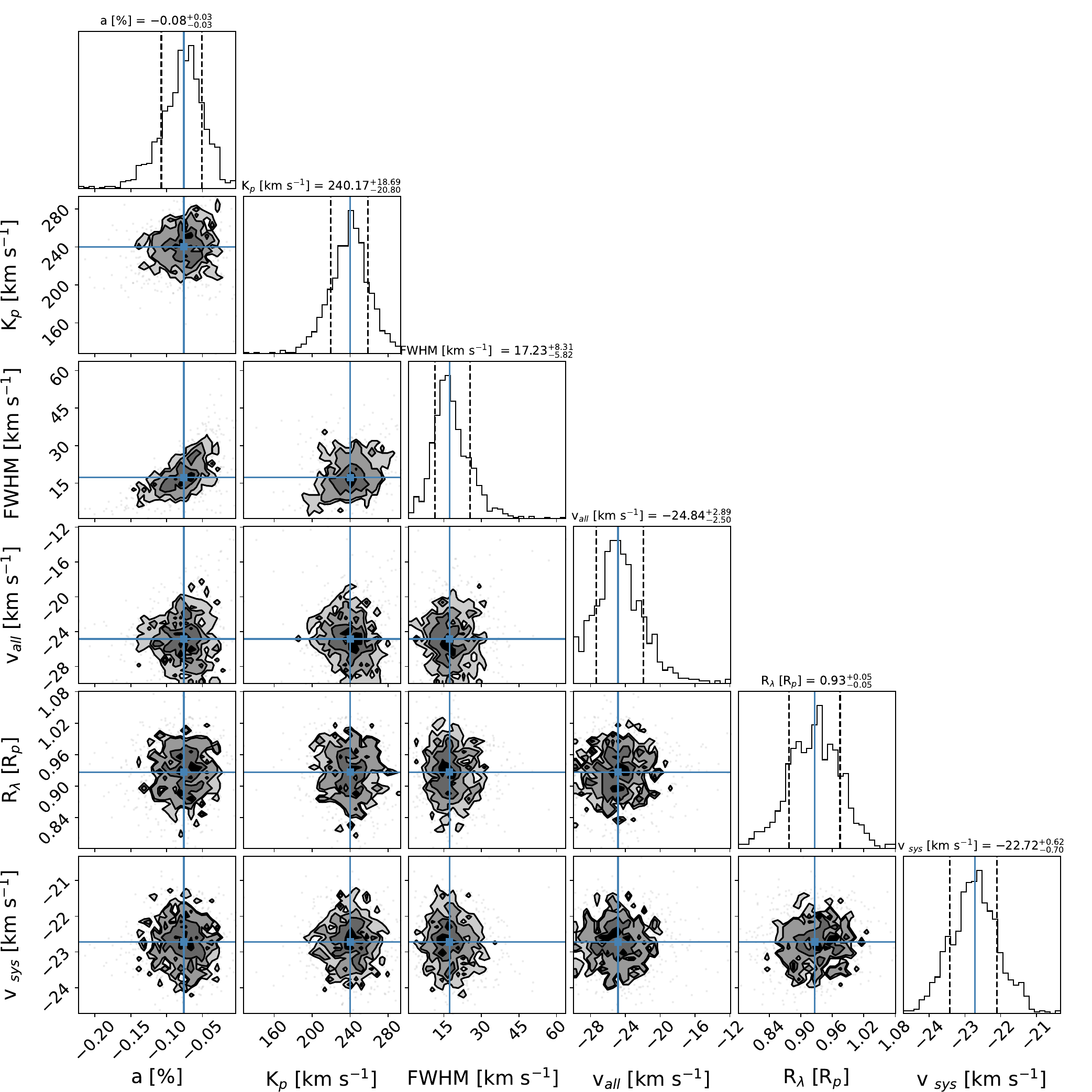}

\caption{Same as Figure \ref{fig:corner_k9_1} but for \ion{Fe}{ii} $\lambda$4583 $\AA$ (left) and \ion{Fe}{ii} $\lambda$4620 $\AA$(right).}
\label{fig:corner_k9_6}
\end{figure}

\begin{figure}[h]
\centering
\includegraphics[width=0.38\textwidth]{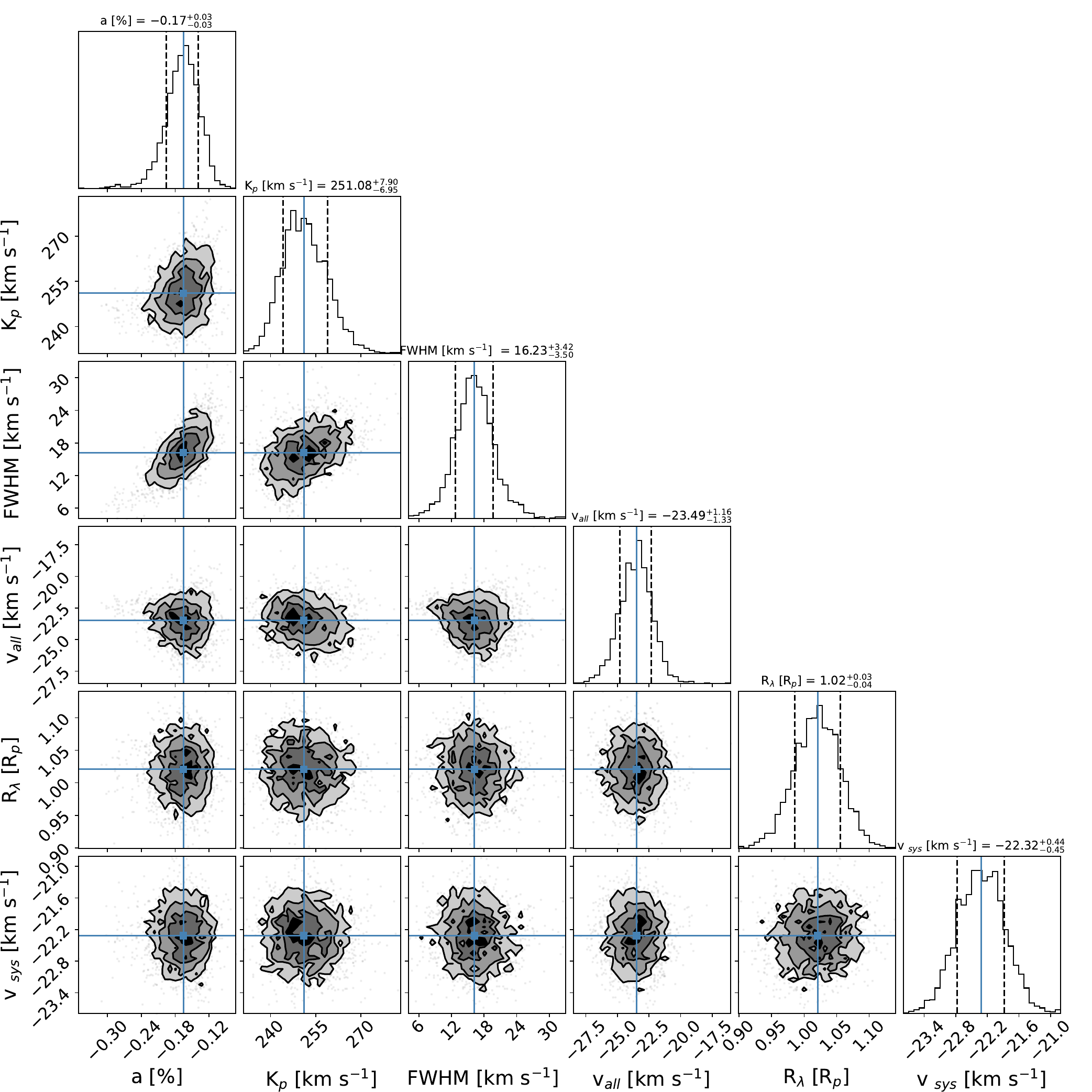}
\includegraphics[width=0.38\textwidth]{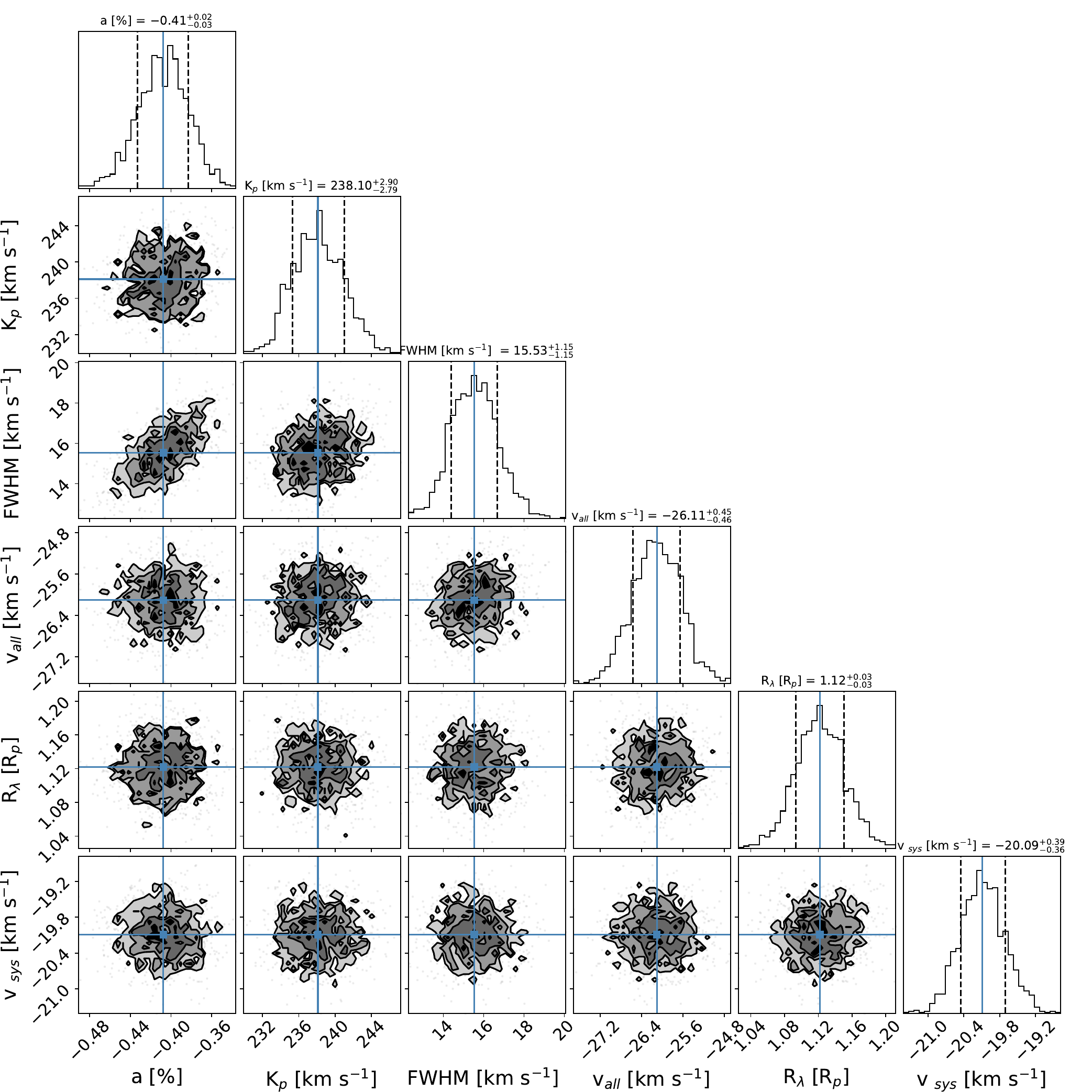}

\caption{Same as Figure \ref{fig:corner_k9_1} but for \ion{Fe}{ii} $\lambda$4629 $\AA$ (left) and \ion{Fe}{ii} $\lambda$4923 $\AA$ (right).}
\label{fig:corner_k9_7}
\end{figure}

\begin{figure}[h]
\centering
\includegraphics[width=0.38\textwidth]{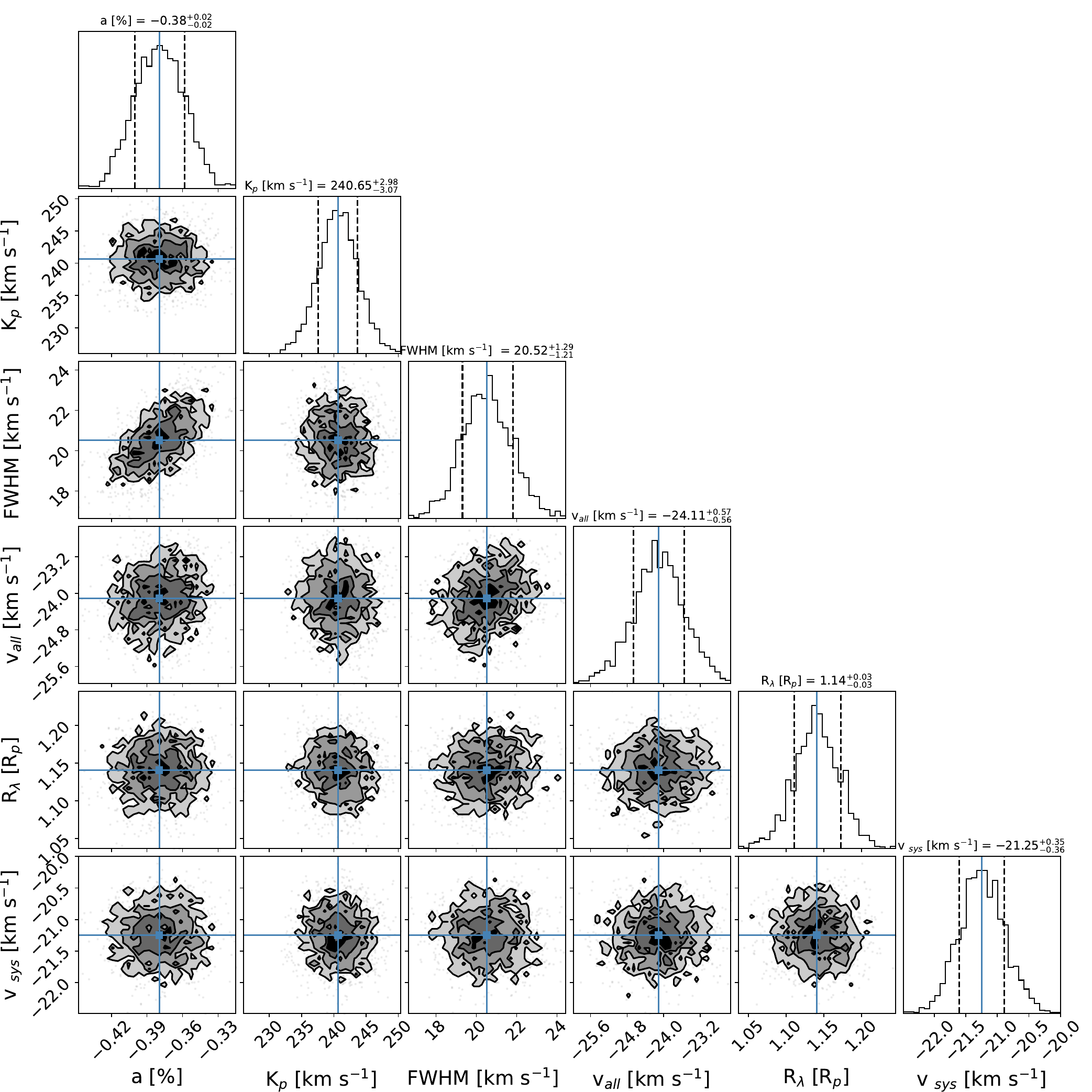}
\includegraphics[width=0.38\textwidth]{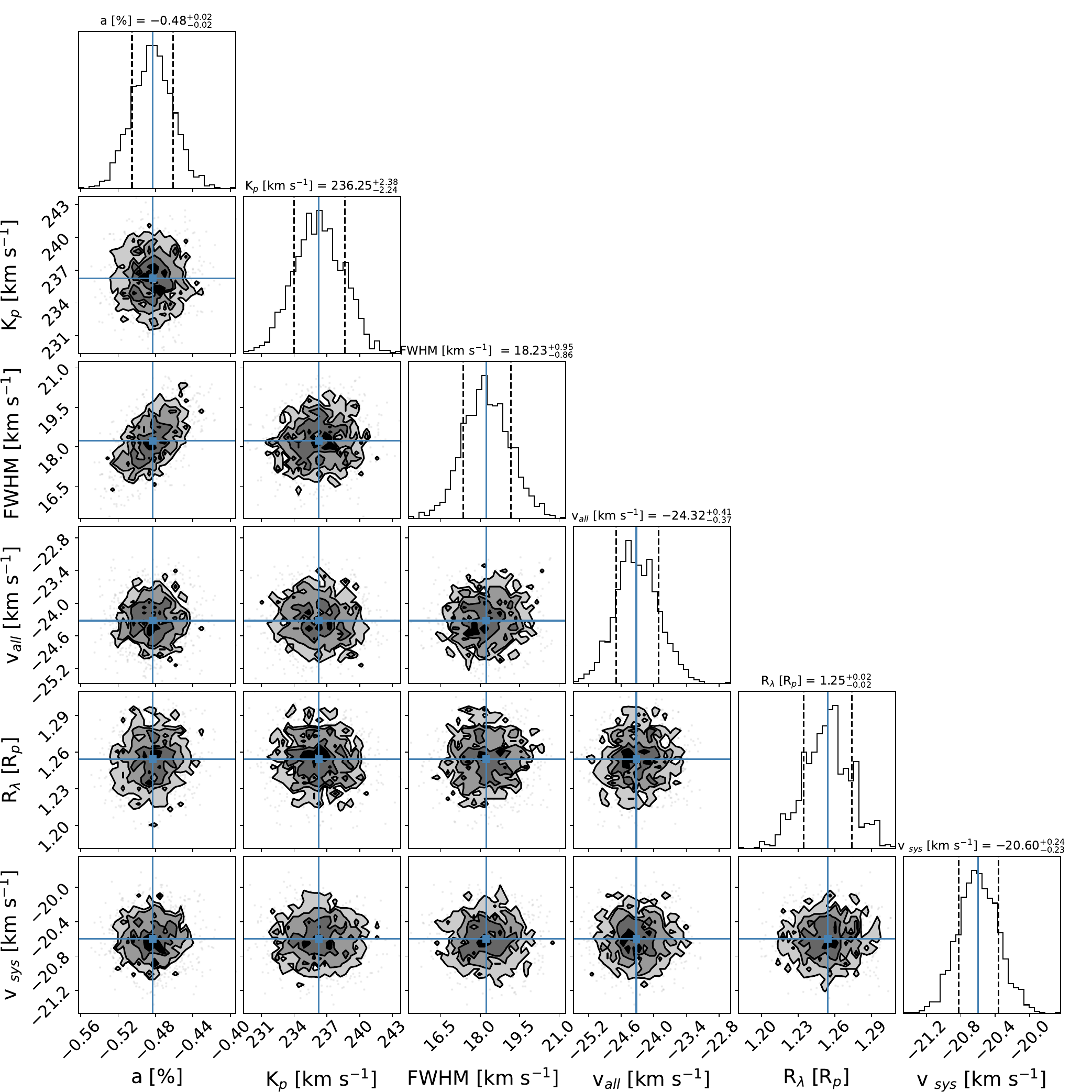}

\caption{Same as Figure \ref{fig:corner_k9_1} but for \ion{Fe}{ii} $\lambda$5018 $\AA$ (left) and \ion{Fe}{ii} $\lambda$5169 $\AA$ (right).}
\label{fig:corner_k9_8}
\end{figure}

\begin{figure}[h]
\centering
\includegraphics[width=0.38\textwidth]{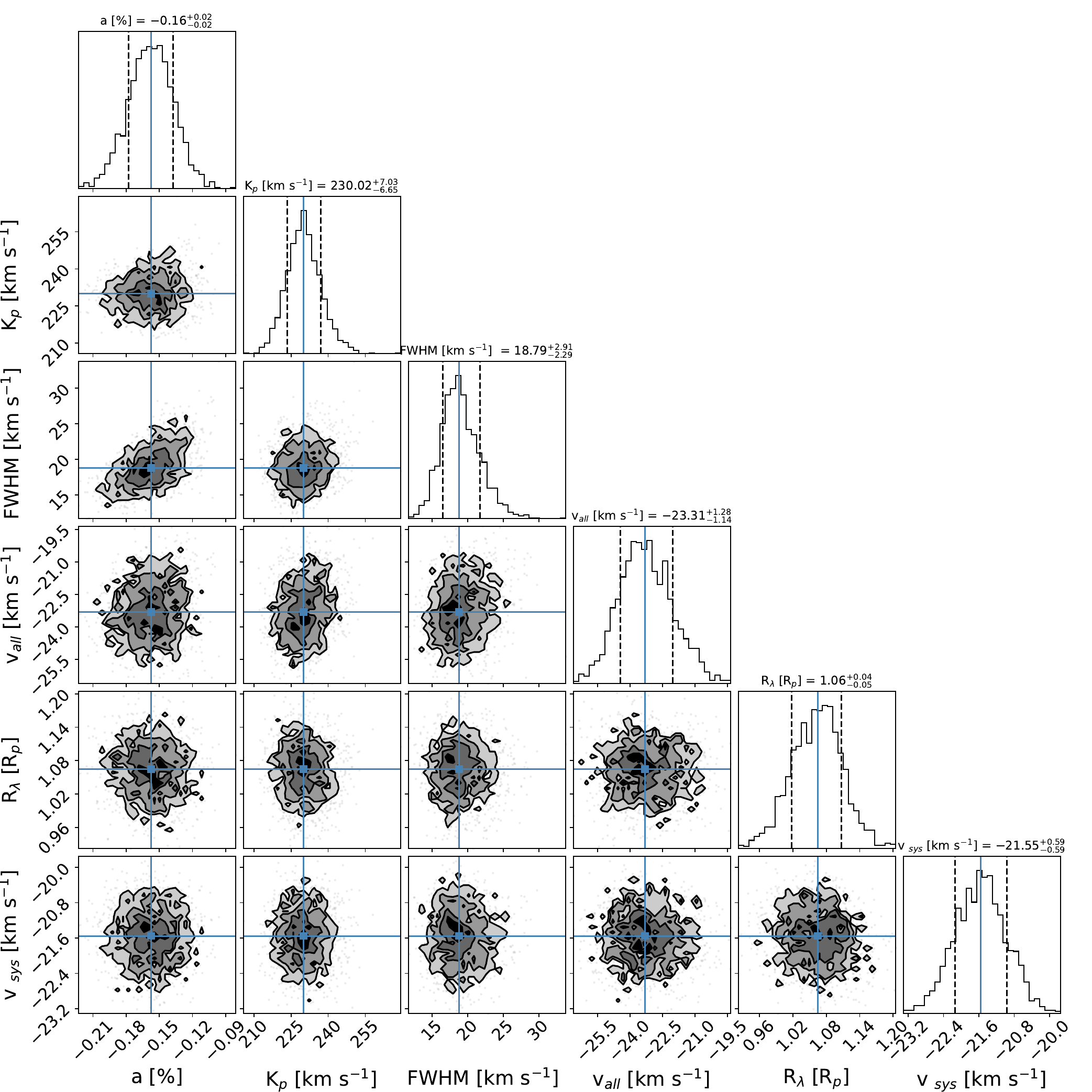}
\includegraphics[width=0.38\textwidth]{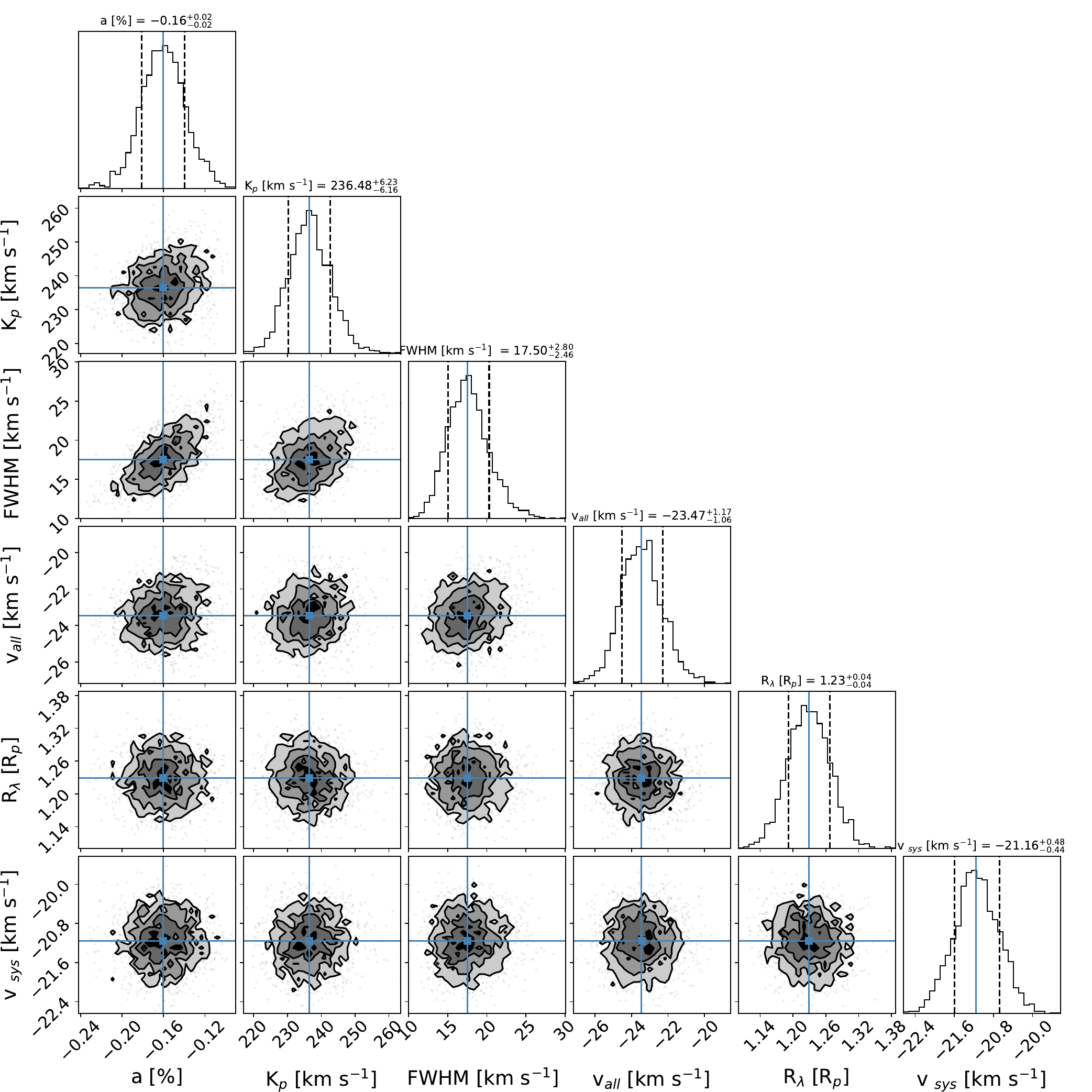}

\caption{Same as Figure \ref{fig:corner_k9_1} but for \ion{Fe}{ii} $\lambda$5197 $\AA$ (left) and \ion{Fe}{ii} $\lambda$5234 $\AA$ (right).}
\label{fig:corner_k9_9}
\end{figure}

\begin{figure}[h]
\centering
\includegraphics[width=0.38\textwidth]{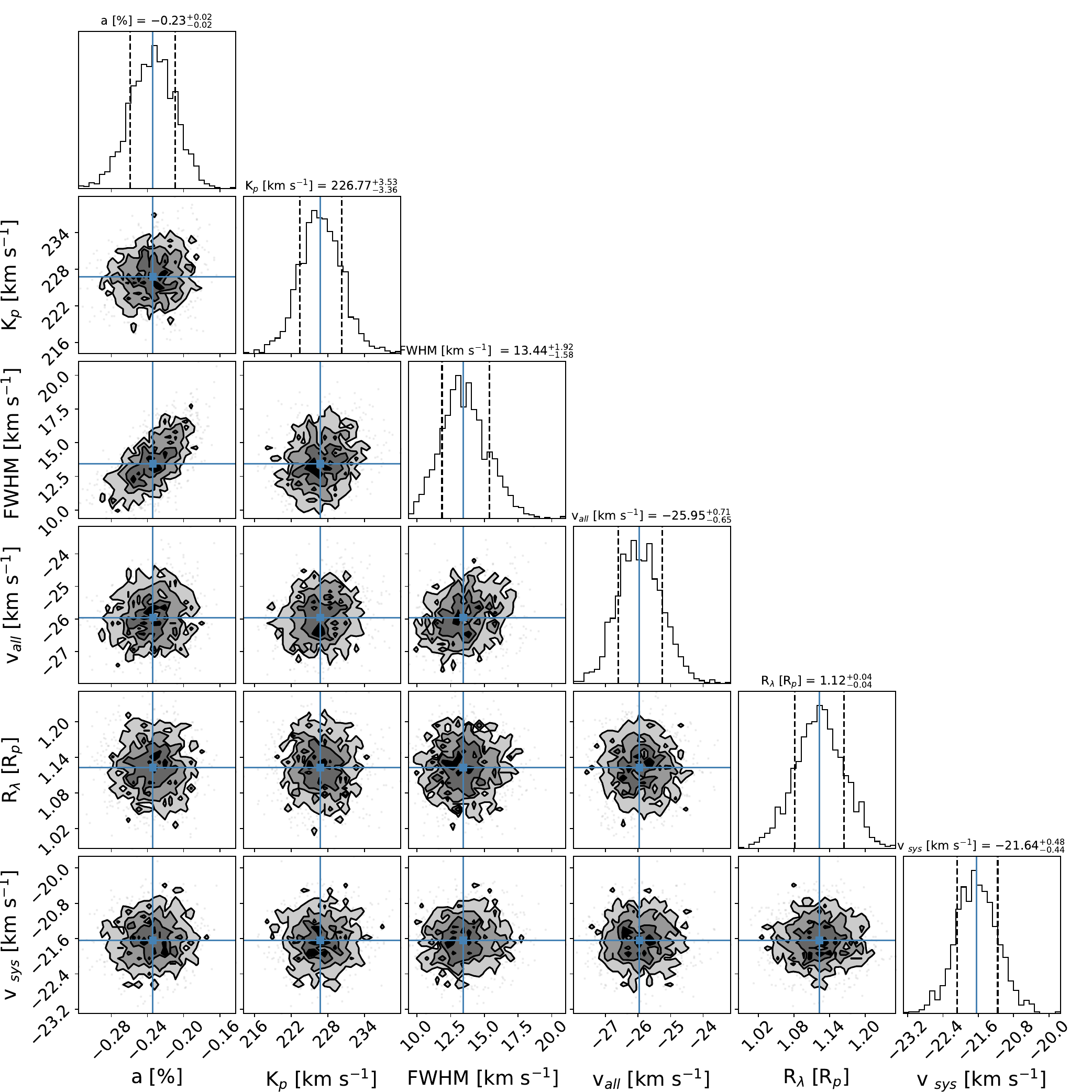}
\includegraphics[width=0.38\textwidth]{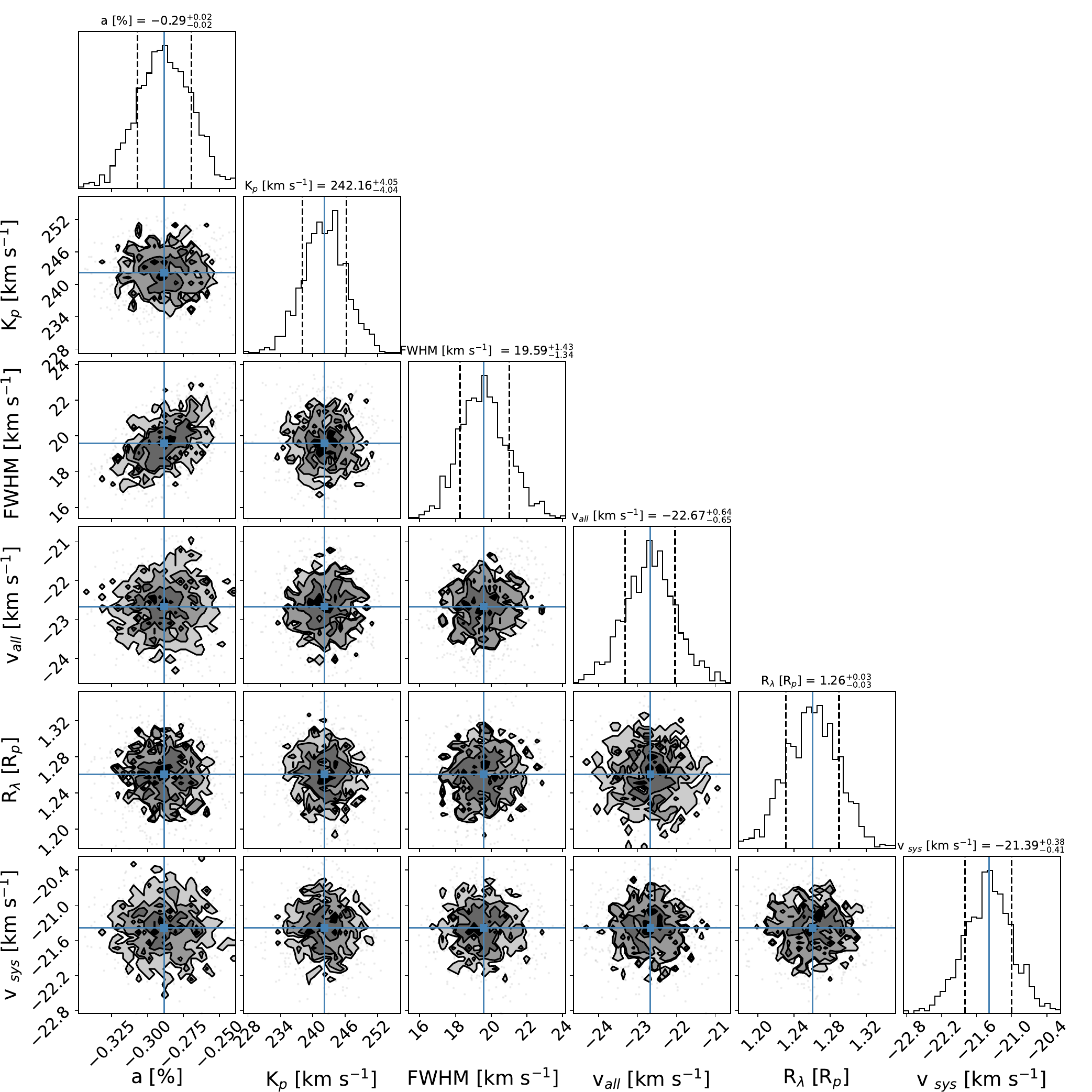}
\caption{Same as Figure \ref{fig:corner_k9_1} but for \ion{Fe}{ii} $\lambda$5275 $\AA$ (left) and \ion{Fe}{ii} $\lambda$5316 $\AA$ (right).}
\label{fig:corner_k9_10}
\end{figure}

\begin{figure}[h]
\centering
\includegraphics[width=0.38\textwidth]{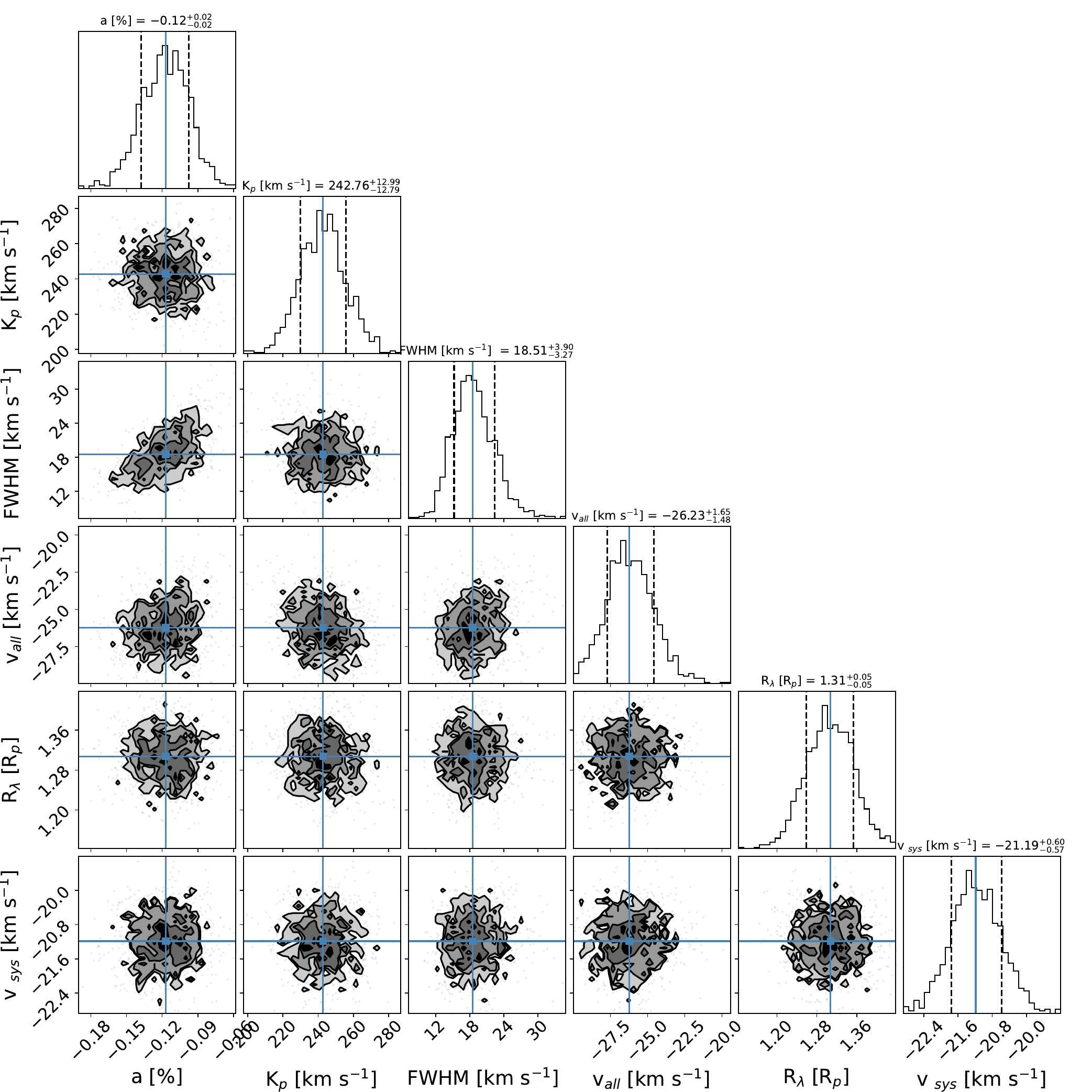}
\caption{Same as Figure \ref{fig:corner_k9_1} but for \ion{Fe}{ii} $\lambda$5362 $\AA$.}
\label{fig:corner_k9_11}
\end{figure}

\begin{figure}[h]
\centering
\includegraphics[width=0.195\textwidth]{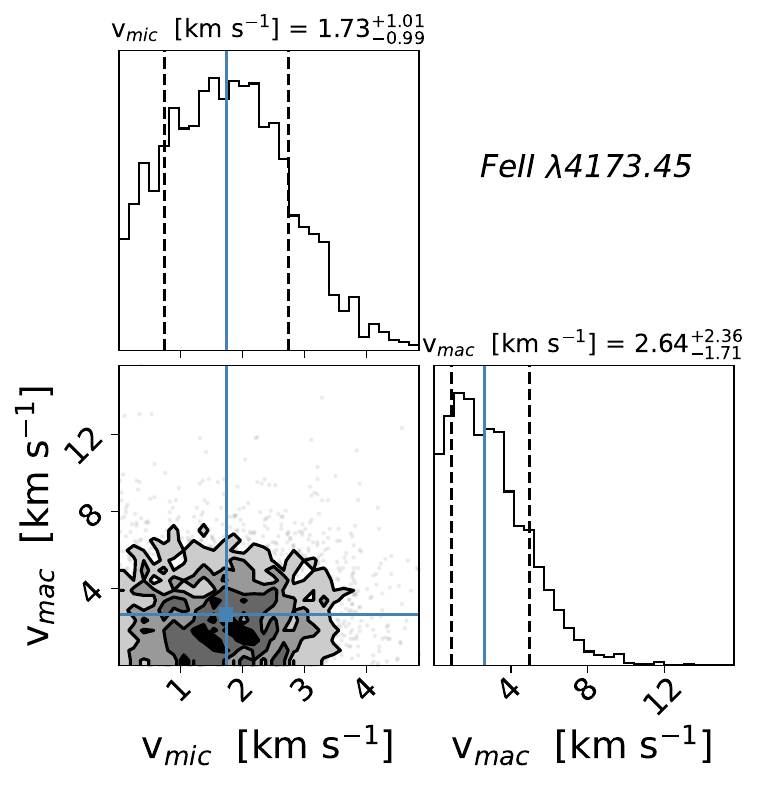}
\includegraphics[width=0.195\textwidth]{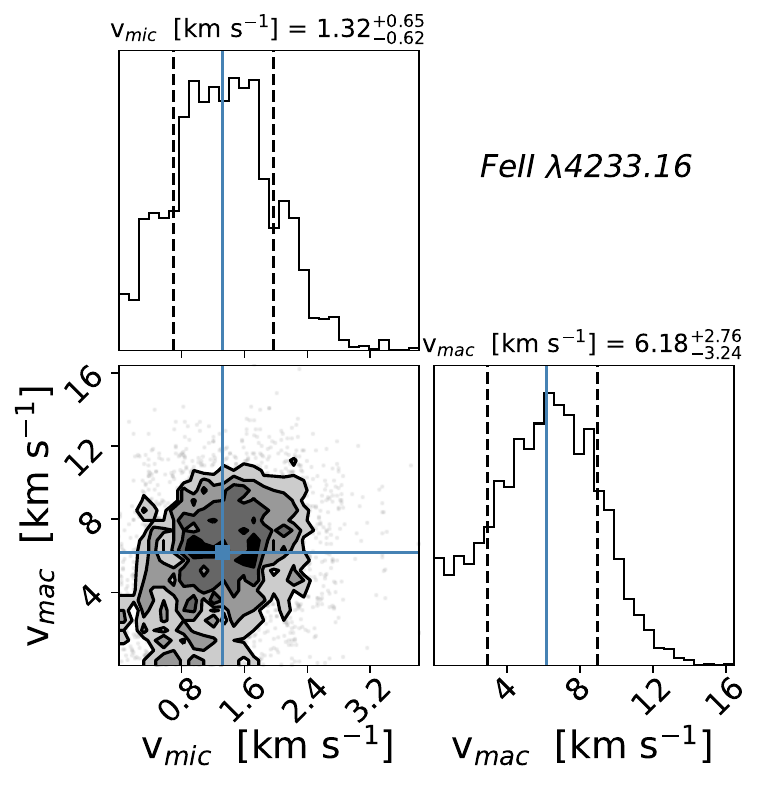}
\includegraphics[width=0.195\textwidth]{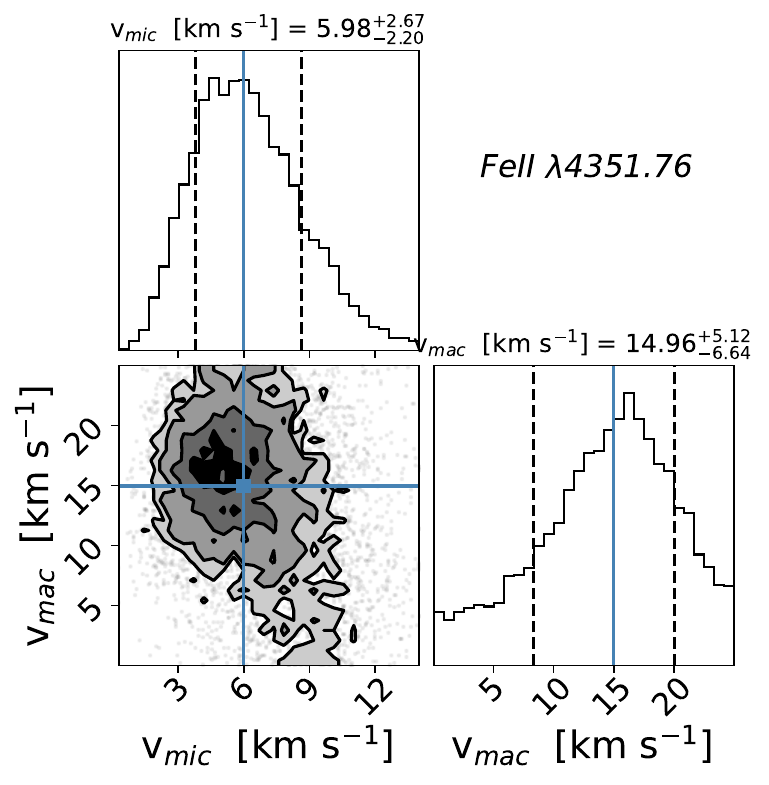}
\includegraphics[width=0.195\textwidth]{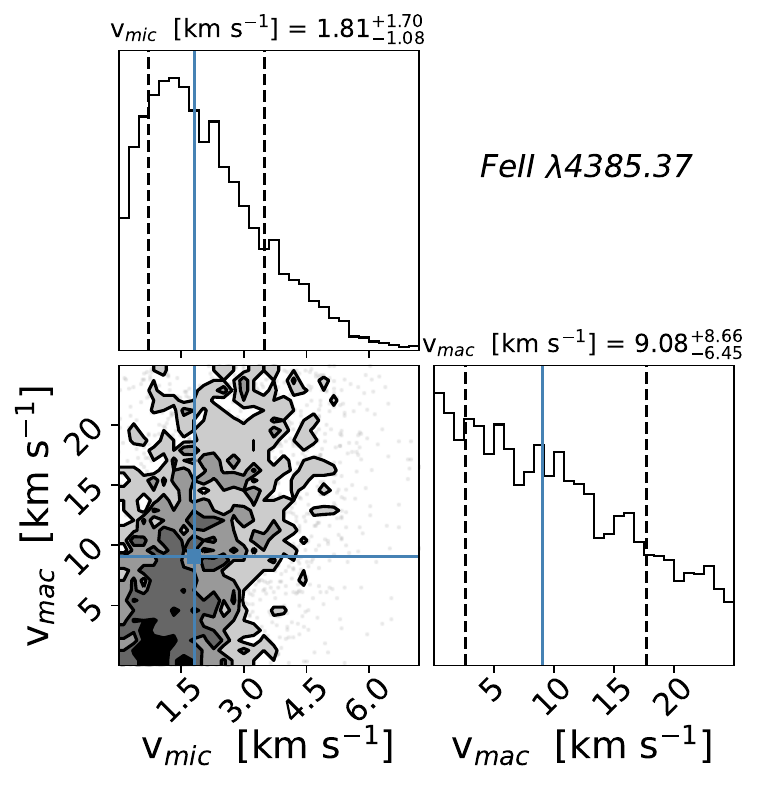}
\includegraphics[width=0.195\textwidth]{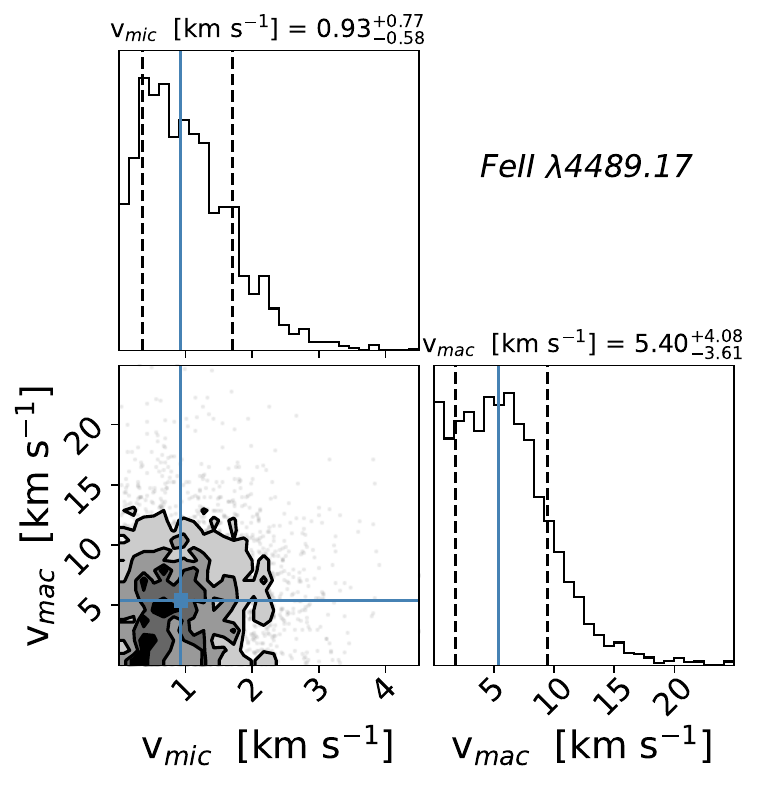}

\includegraphics[width=0.195\textwidth]{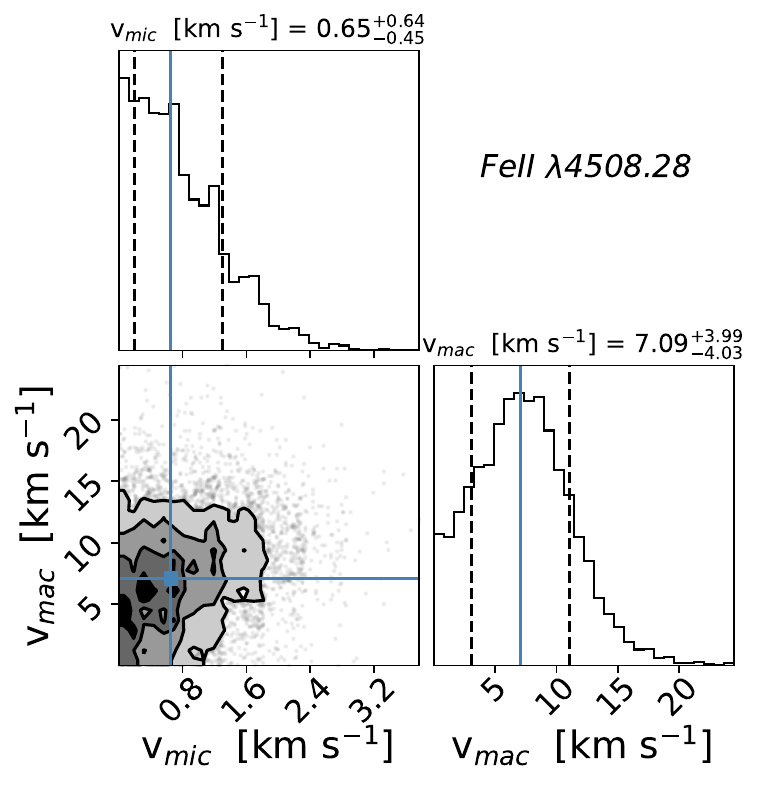}
\includegraphics[width=0.195\textwidth]{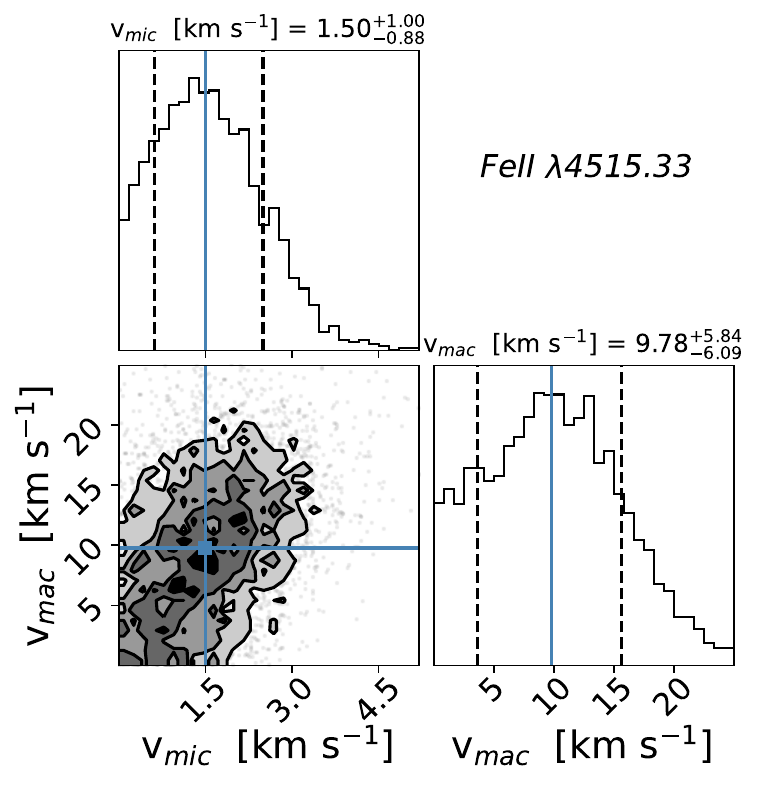}
\includegraphics[width=0.195\textwidth]{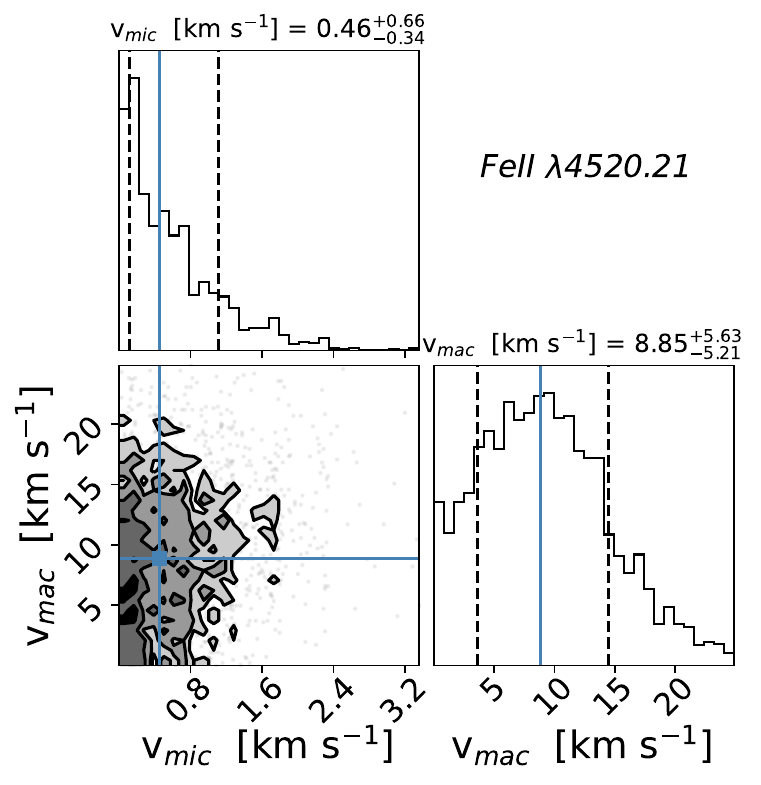}
\includegraphics[width=0.195\textwidth]{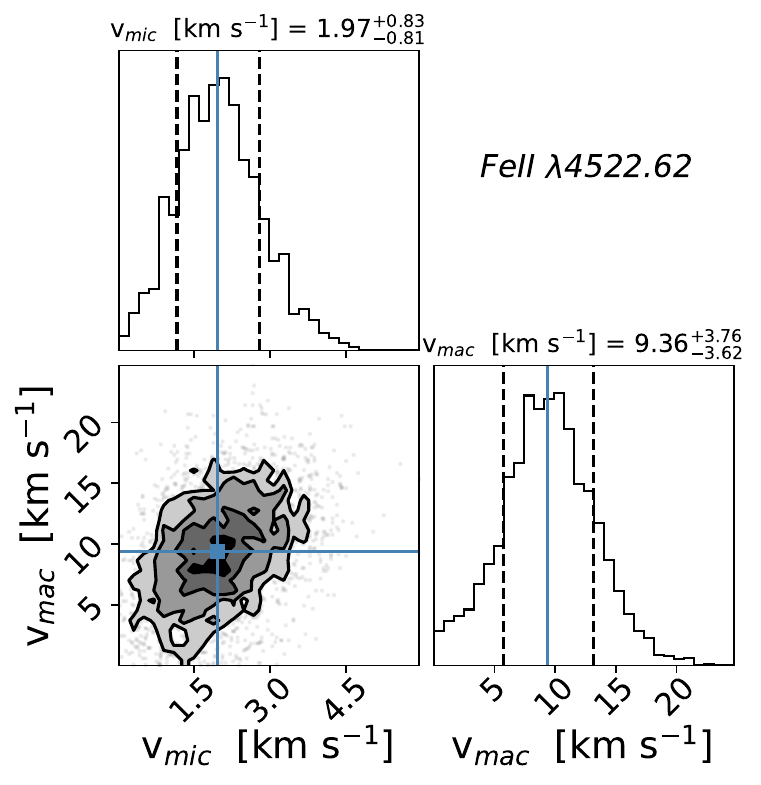}
\includegraphics[width=0.195\textwidth]{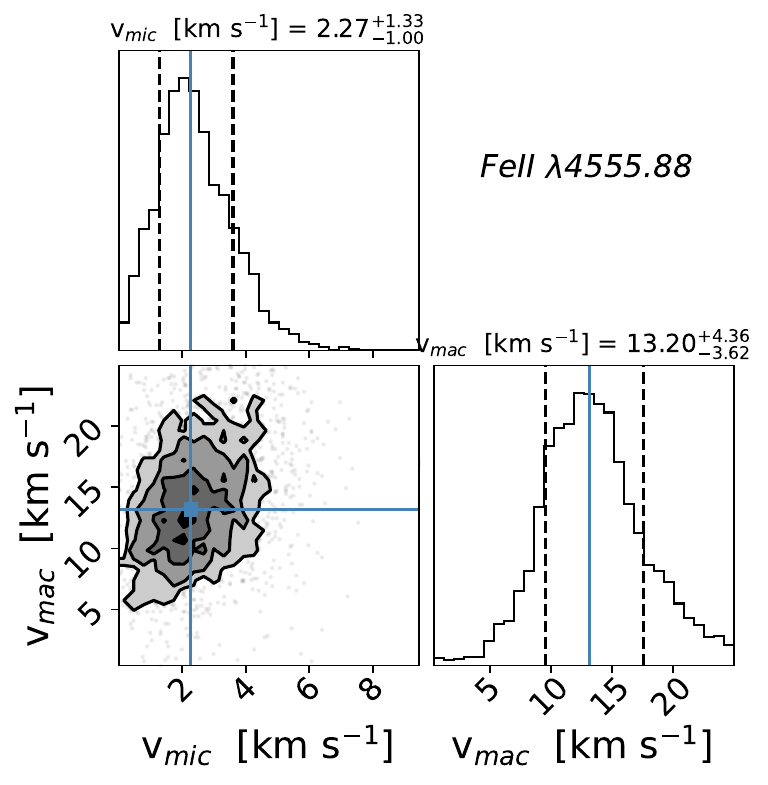}

\includegraphics[width=0.195\textwidth]{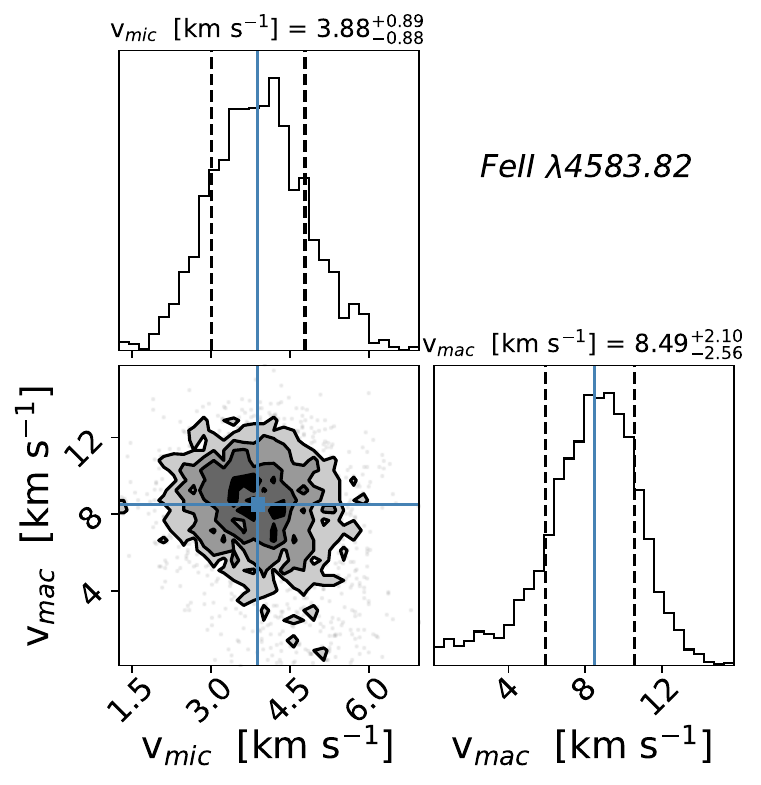}
\includegraphics[width=0.195\textwidth]{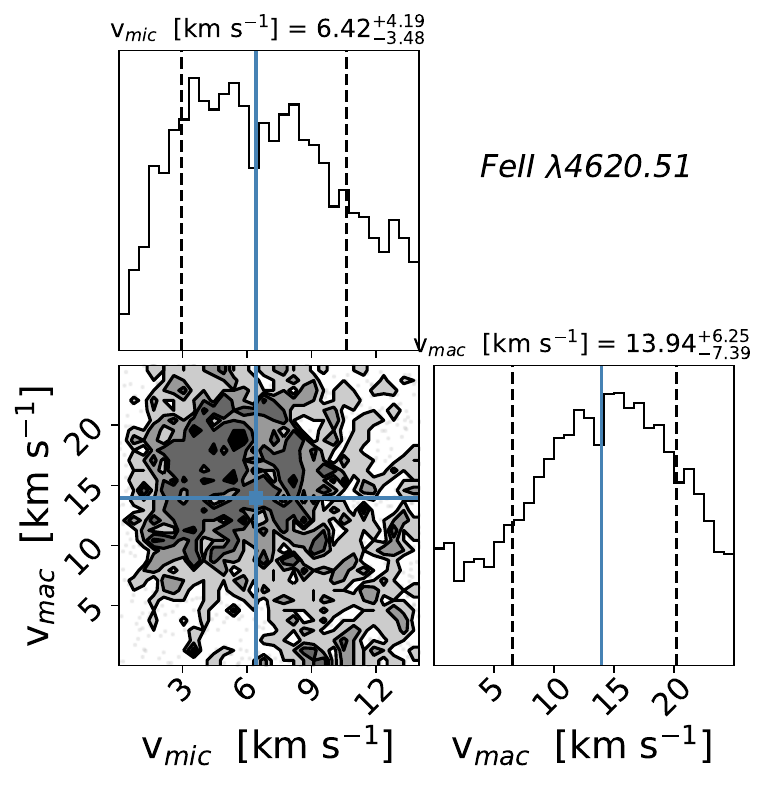}
\includegraphics[width=0.195\textwidth]{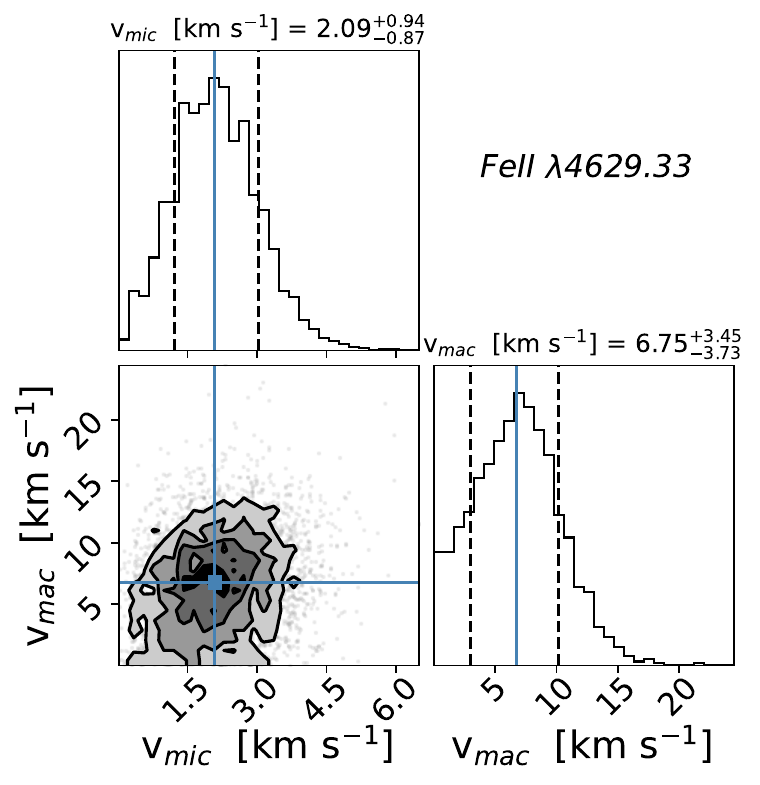}
\includegraphics[width=0.195\textwidth]{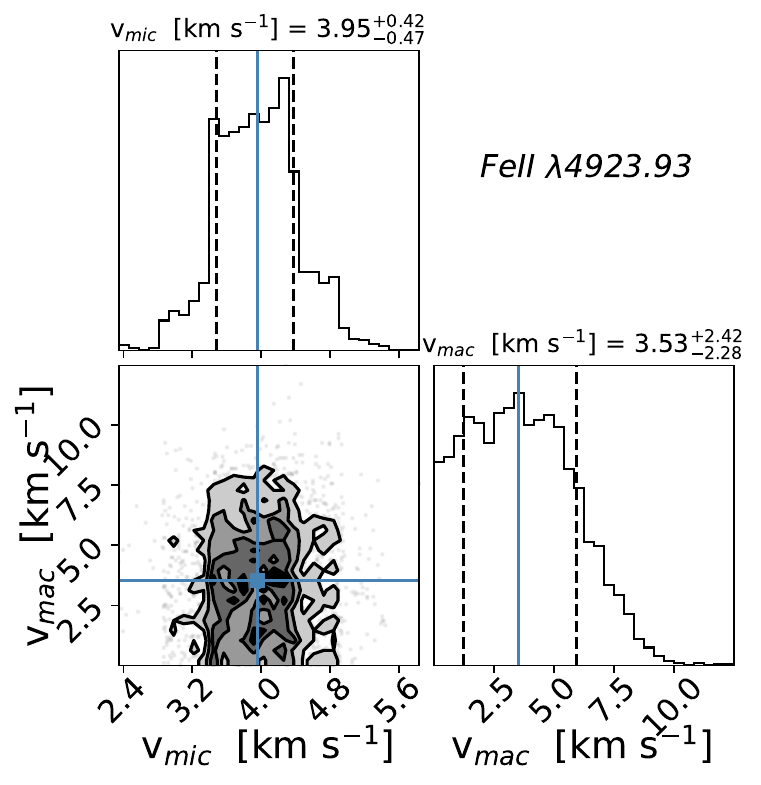}
\includegraphics[width=0.195\textwidth]{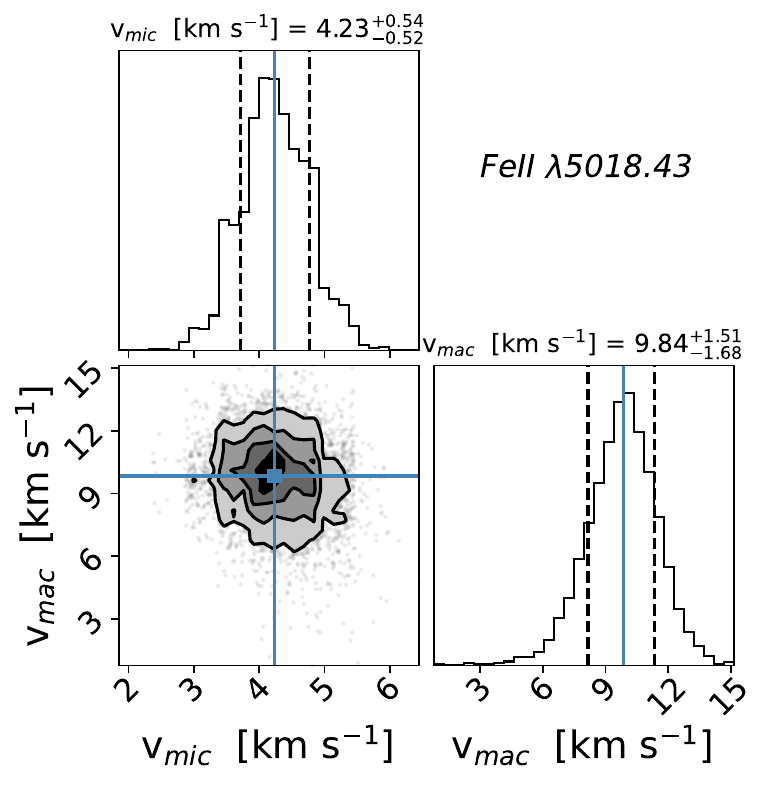}

\caption{The corner plots of the MCMC analysis for KELT-9b focusing on finding the best $\nu_{\rm mic}$ and $\nu_{\rm mac}$ for lines $\lambda$4173 $\AA$, $\lambda$4233 $\AA$, $\lambda$4351 $\AA$, $\lambda$4385 $\AA$, $\lambda$4489 $\AA$, $\lambda$4508 $\AA$, $\lambda$4515 $\AA$, $\lambda$4520. $\AA$,
$\lambda$4522 $\AA$, $\lambda$4555 $\AA$,
$\lambda$4583 $\AA$, $\lambda$4620 $\AA$, $\lambda$4629 $\AA$, $\lambda$4923 $\AA$, and $\lambda$5018 $\AA$. }
\label{fig:corner_vv_K9}
\end{figure}

%%%%%%%%%%%%%%%%%%%%%%%%%%%%%%%%%%%%%%%%%%%%%%%%%%%%%%%%%%%%%%%%%%%%%%%%%%%%%%%%%%%%%%%%%%%%%%%%%%%%%%%%%%%%%%%%%

\begin{figure}[h]
\centering

\includegraphics[width=0.195\textwidth]{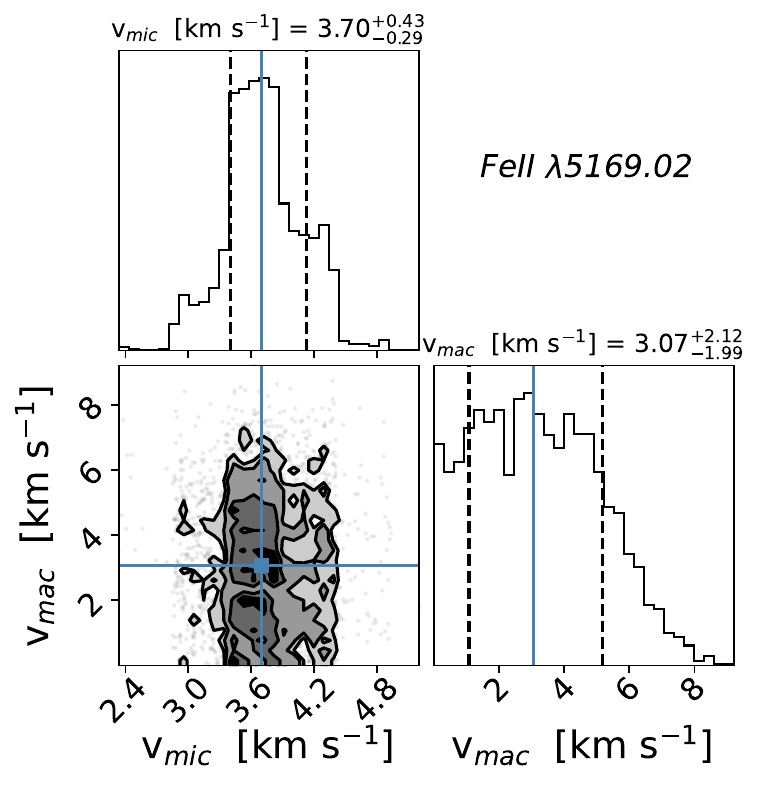}
\includegraphics[width=0.195\textwidth]{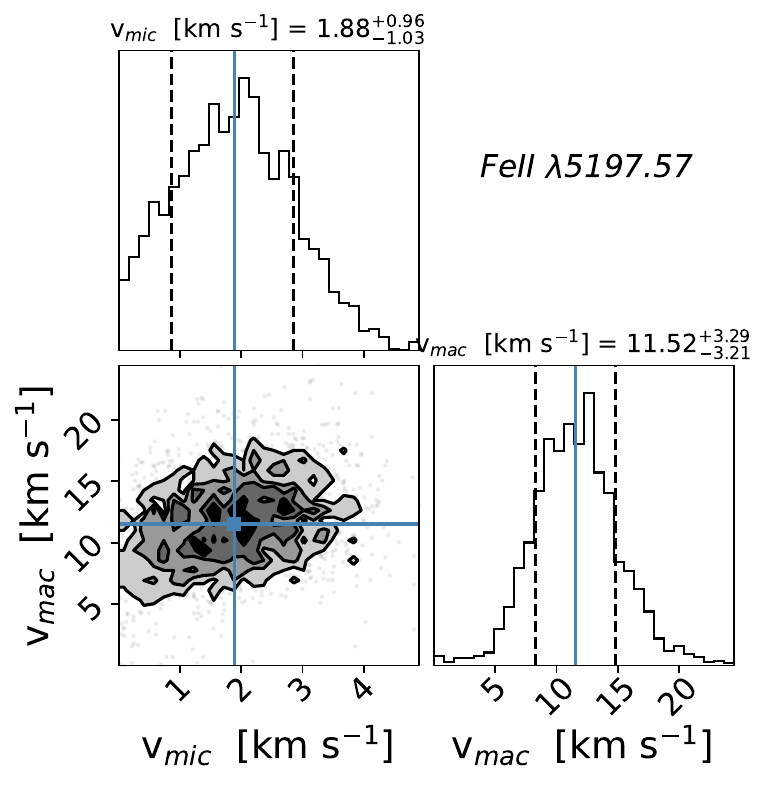}
\includegraphics[width=0.195\textwidth]{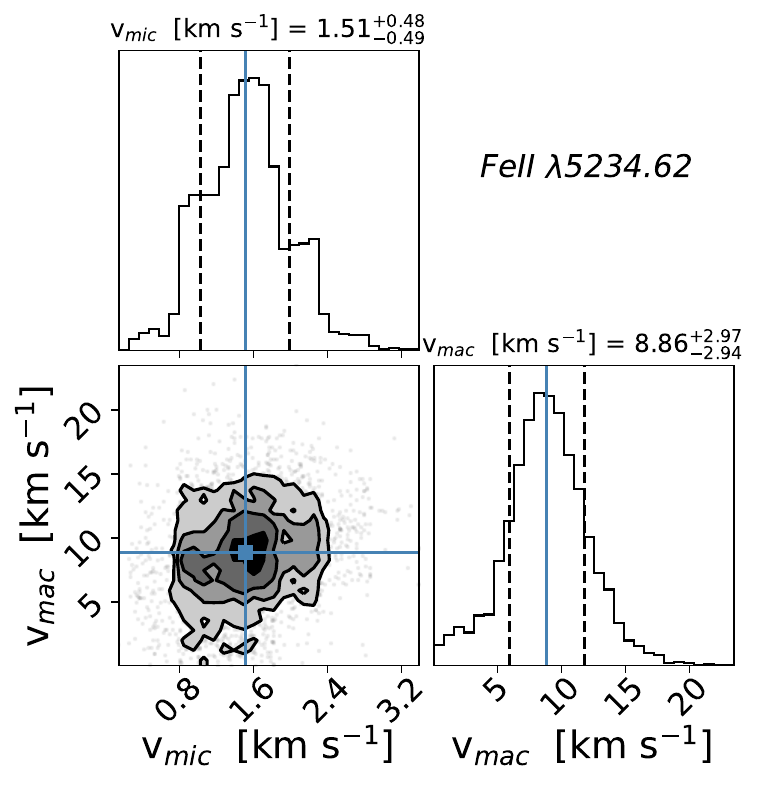}
\includegraphics[width=0.195\textwidth]{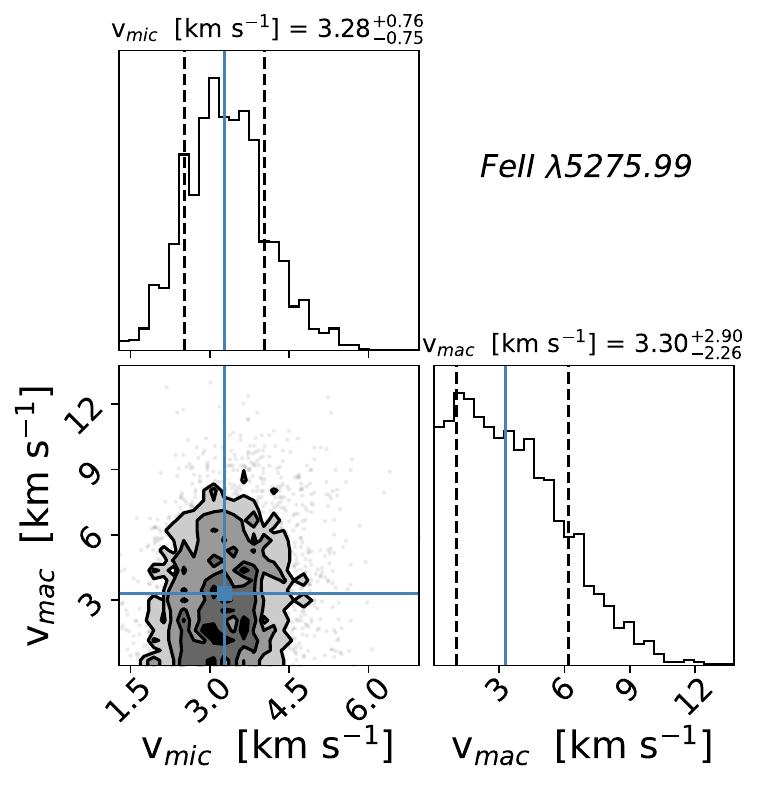}
\includegraphics[width=0.195\textwidth]{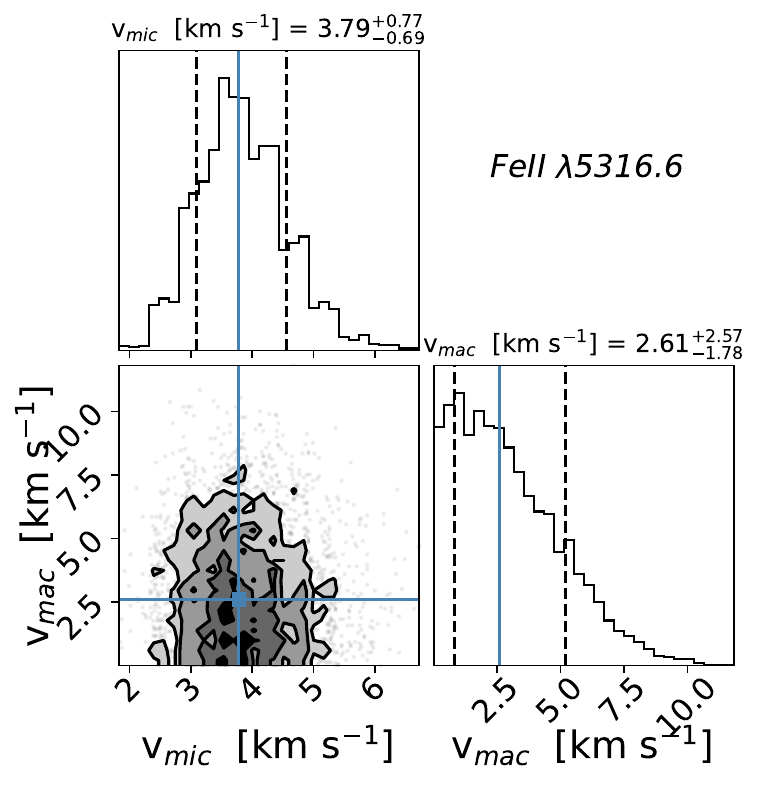}

\includegraphics[width=0.2\textwidth]{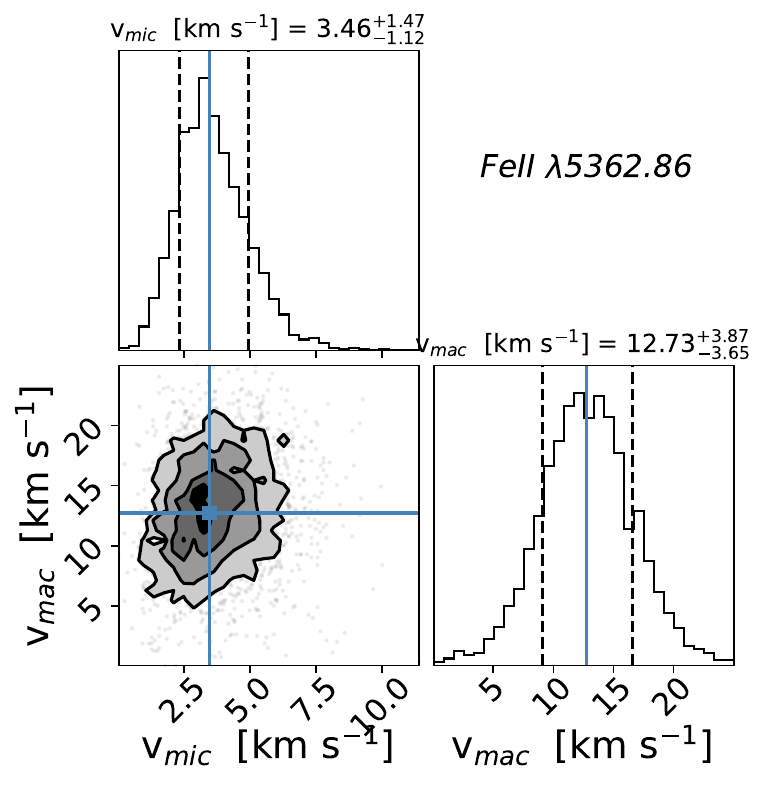}

\caption{Same as Figure \ref{fig:corner_vv_K9}, but for $\lambda$5169 $\AA$,
$\lambda$5197 $\AA$, $\lambda$5234 $\AA$, $\lambda$5275.99 $\AA$, $\lambda$5316 $\AA$, and $\lambda$5362 $\AA$   }
\label{fig:corner_vv_K9_1}
\end{figure}

\begin{figure}[h]
\centering
\includegraphics[width=\textwidth]{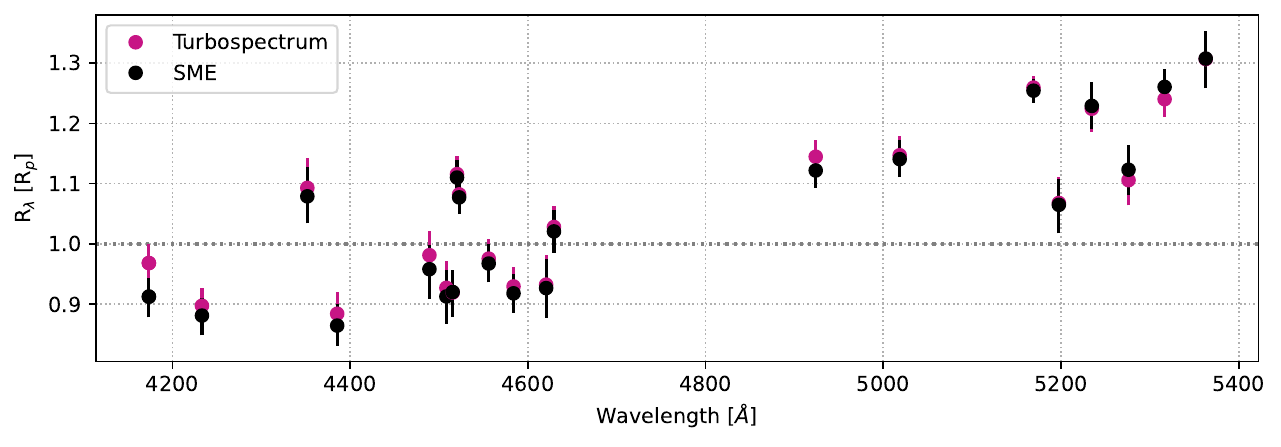}
\caption{Fitted R$_\lambda$ using MCMC for each of the \ion{Fe}{ii} lines in the atmosphere of KELT-9b versus the wavelength of the line. The black points represent fitting using RM and CLV models calculated using SME and the pink dots using \texttt{Turbospectrum}.  }
\label{fig:rad_TS_SME}
\end{figure}

%%%%%%%%%%%%%%%%%%%%%%%%%%%%%%%%%%%%%%%%%%%%%%%%%%%%%%%%%%%%%%%%%%%%%%%%%%%%%%%%%%%%%%%%%%%%%%%%%%%%%%%%%%%%%%%%%%%%%%%%%%%%%%%%%%%%%%%%%%%%%%%%%%%%%%%%%%%%%%%%%%%%%%%%%%%%%%%%%%%%%%%%%%%%%%%%%%%%%%%%%%%%%%%%%%%%%%%%%%%%%%%%%%%%%%%%%%%%%%%%%%%%%%%%%%%%%%%%%%%%%%%%%%%%%%%%%%%%%%%%%%%%%%%%%%%%%%%%%%%
\section{Additional figures for KELT-20b} \label{app:add_fig_k20}

%4173,4233,4351,4385
%4489,4508,4515,452

\begin{figure}[h]
\centering
\includegraphics[width=0.49\textwidth]{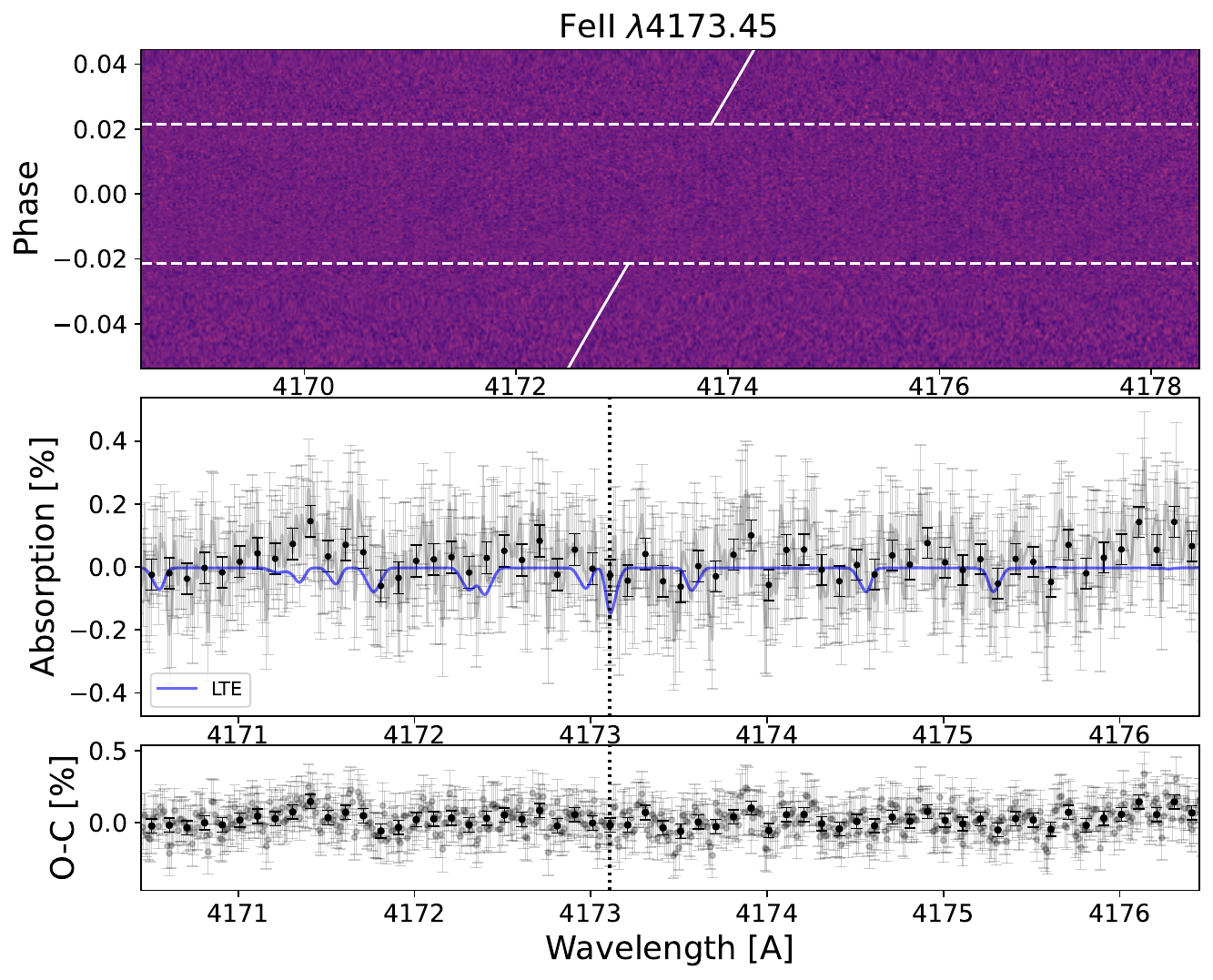}
\includegraphics[width=0.49\textwidth]{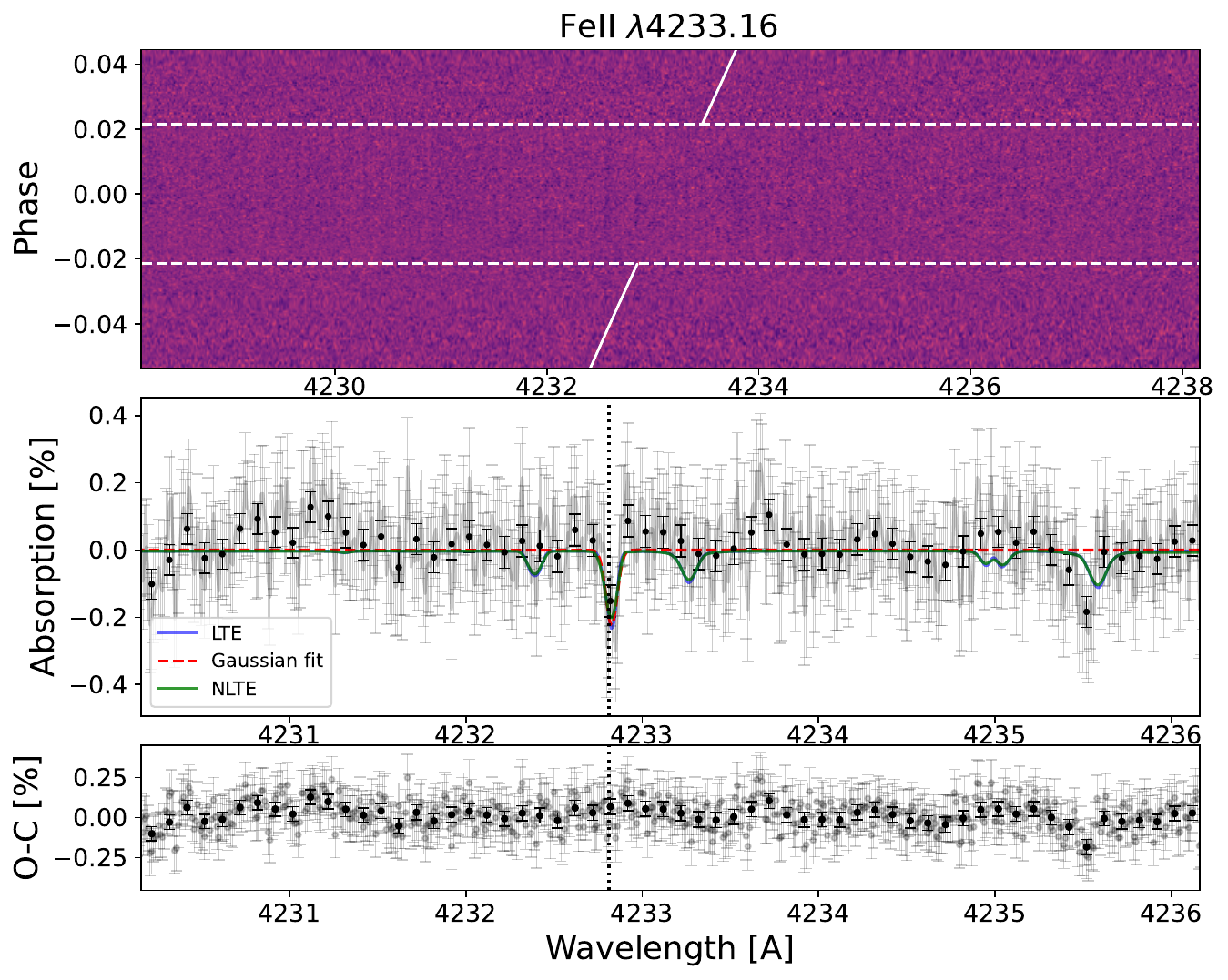}

\includegraphics[width=0.49\textwidth]{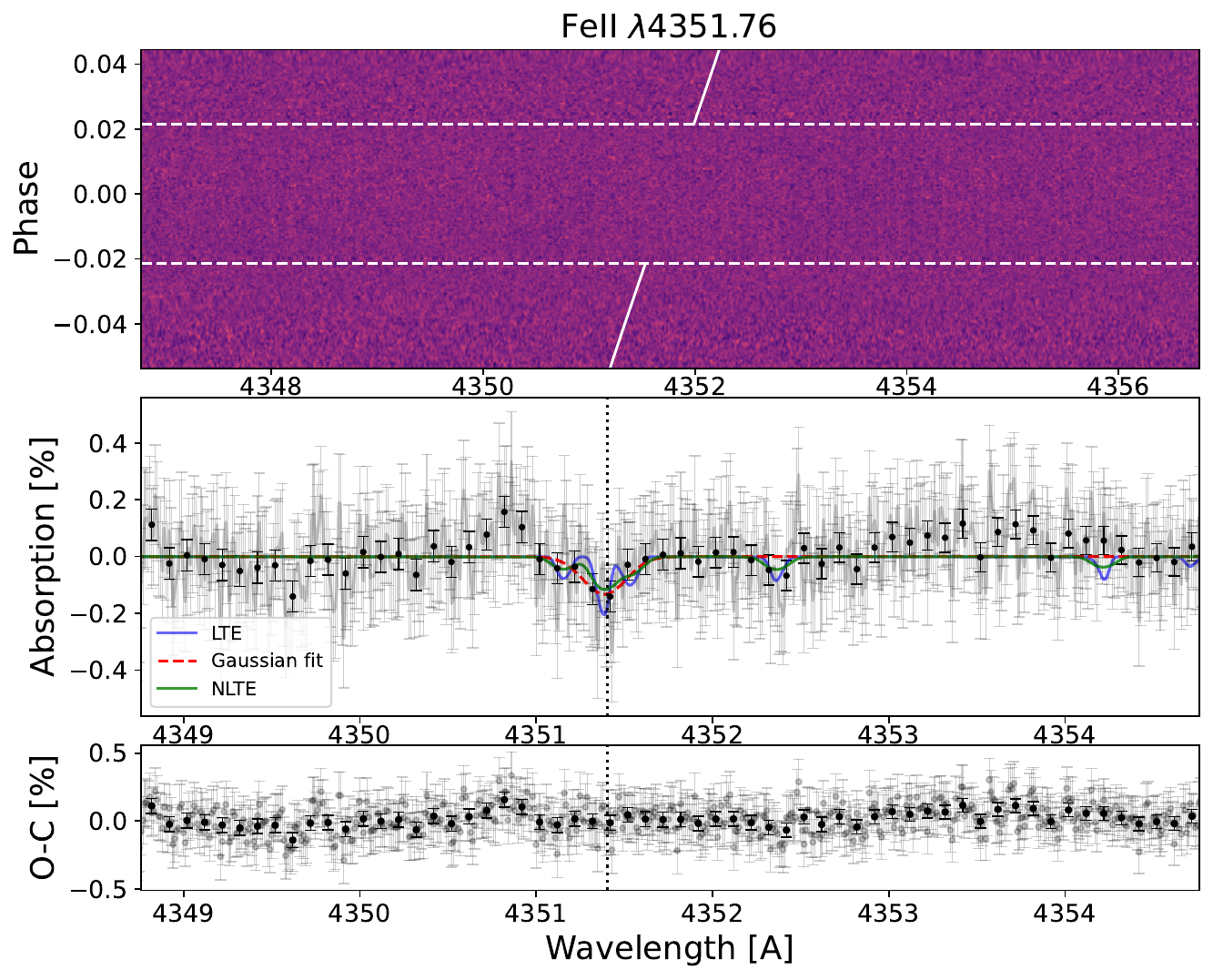}
\includegraphics[width=0.49\textwidth]{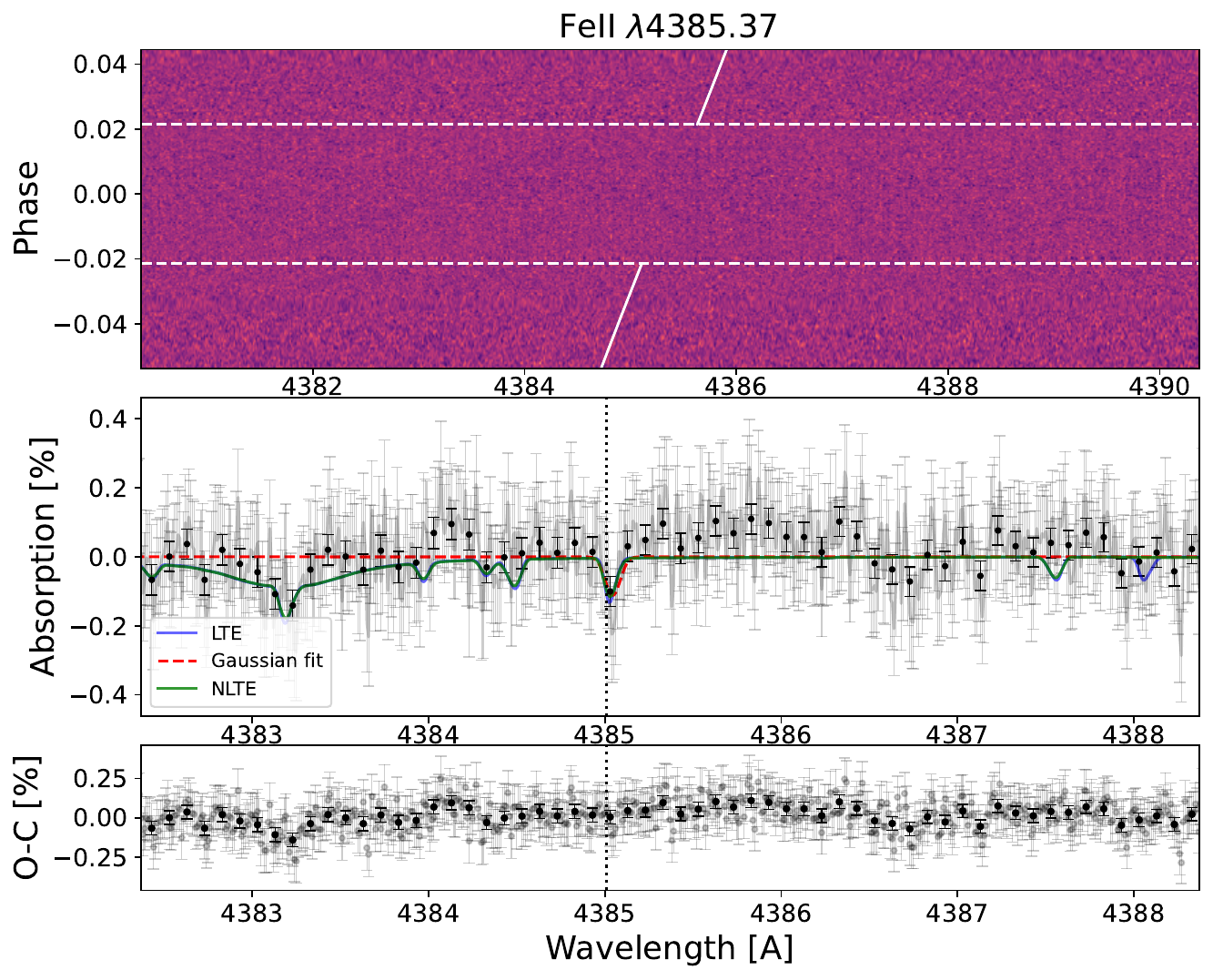}

\caption{\textit{Top panel:} Residual map for lines $\lambda$4173 $\AA$, $\lambda$4233 $\AA$, $\lambda$4351 $\AA$, and $\lambda$4385 $\AA$ for KELT-20b presented in the stellar rest frame. White horizontal lines indicate the beginning and end of the transit and the tilted white line indicates the expected velocities for the signal coming from the atmosphere of the planet while assuming $v_{sys}$ as the mean $v_{sys}$ from our analysis. \textit{Middle panel:} Transmission spectrum for the detected line (grey dots). The black dots indicate the binned transmission spectrum with the step of 0.1 $\AA$, the plot only for visualization. The red plot is the best fit of the planetary signal from the MCMC analysis, the blue plot is the NLE and the green is the NLTE model for the atmosphere of KELT-20b. In the case of models for which v$_{mic}$ or/and $v_{mac}$ was consistent with 0 km s$^{-1}$ or not determined we assumed 0 km s$^{-1}$ while plotting the models. \textit{Bottom panel:} Residuals.  }
\label{Fig:TS_M2_1}
\end{figure}

\begin{figure}[h]
\centering
\includegraphics[width=0.49\textwidth]{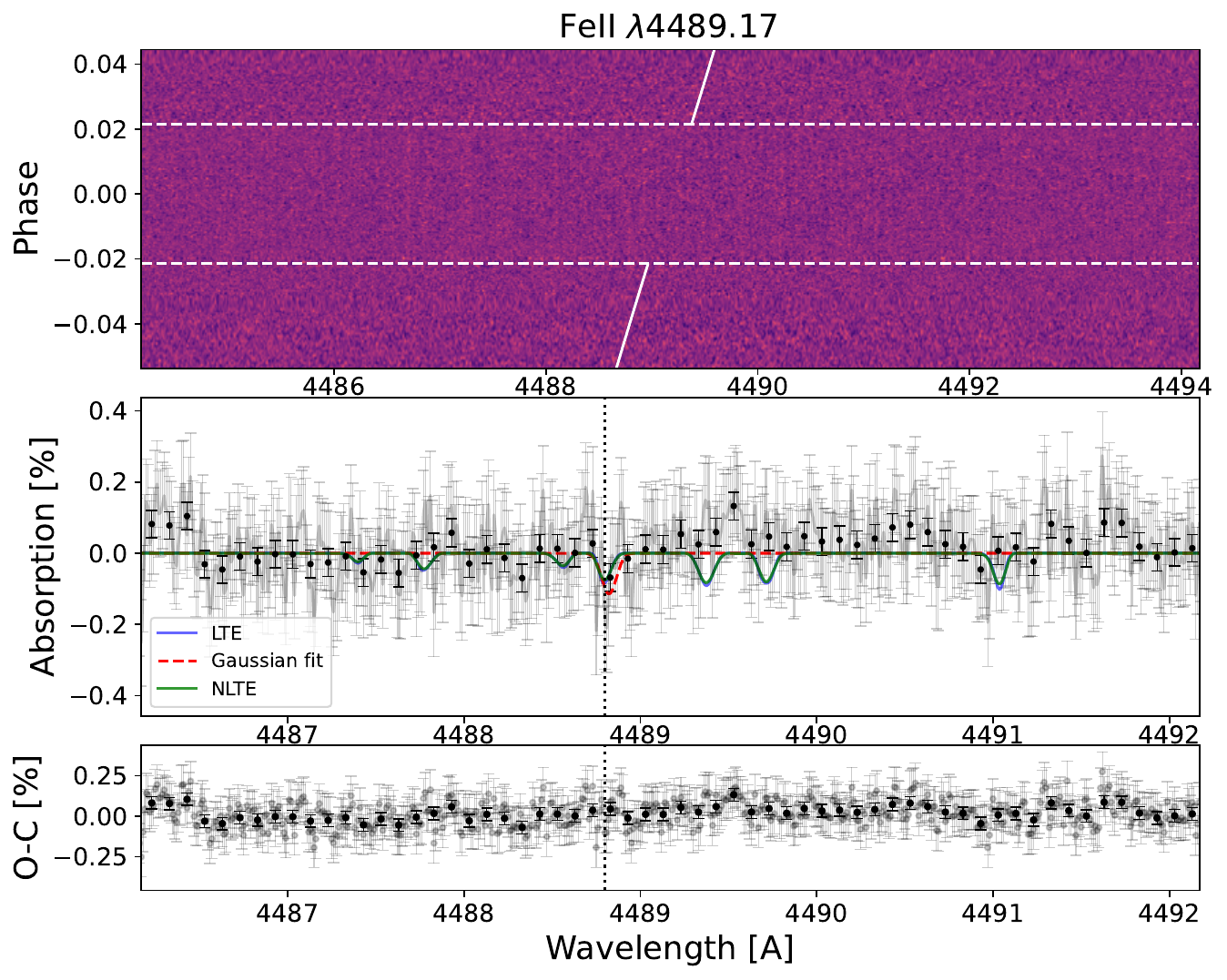}
\includegraphics[width=0.49\textwidth]{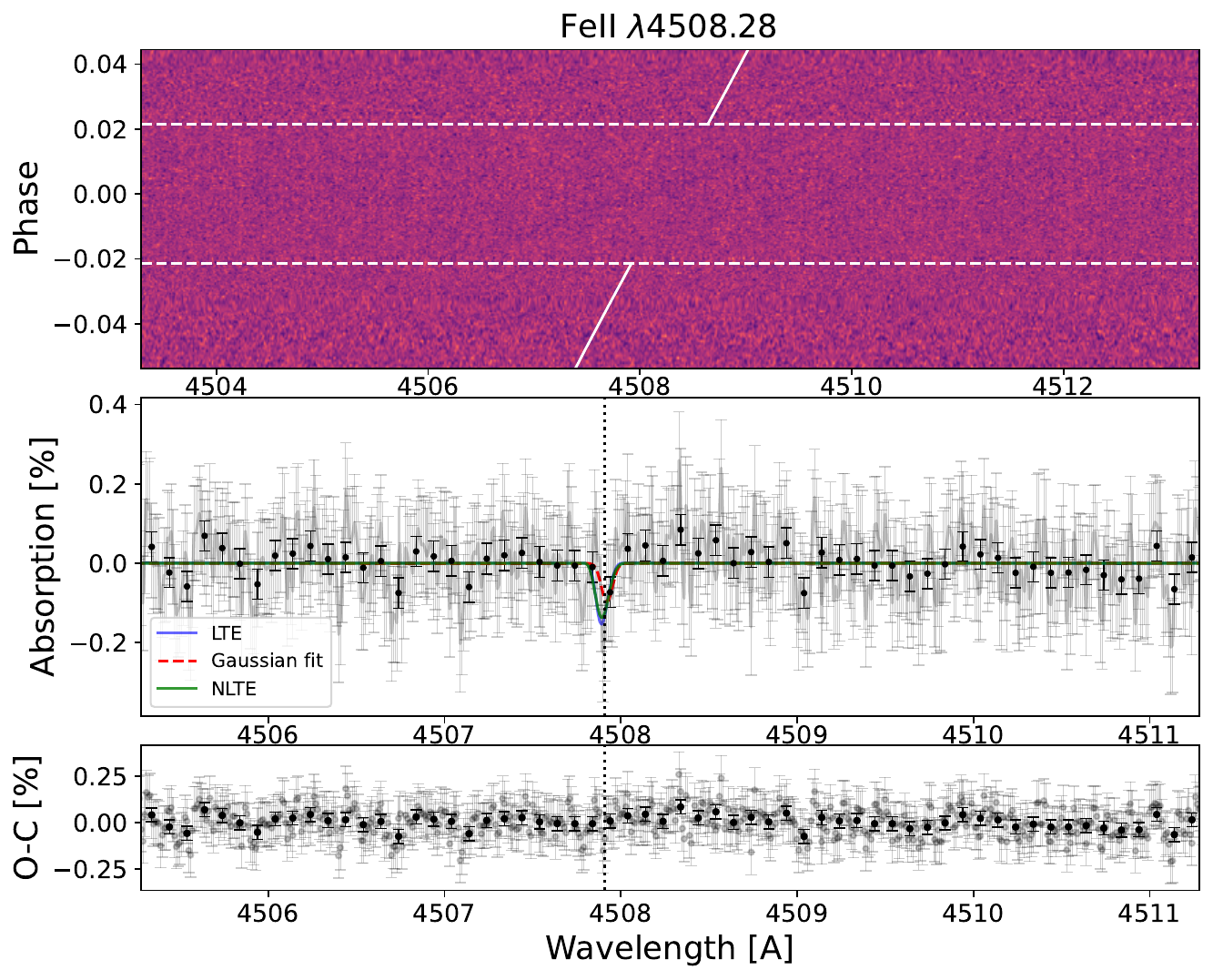}

\includegraphics[width=0.49\textwidth]{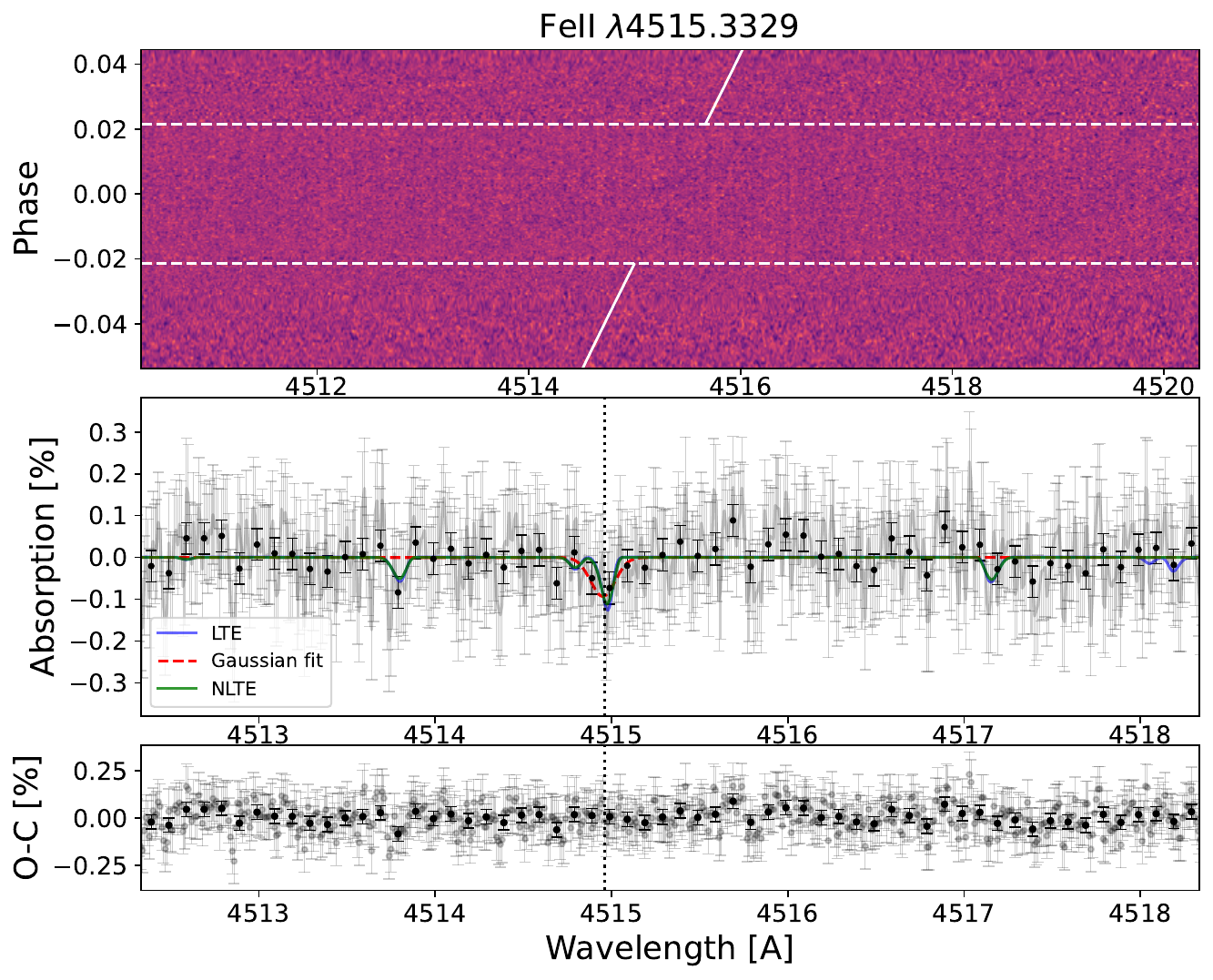}
\includegraphics[width=0.49\textwidth]{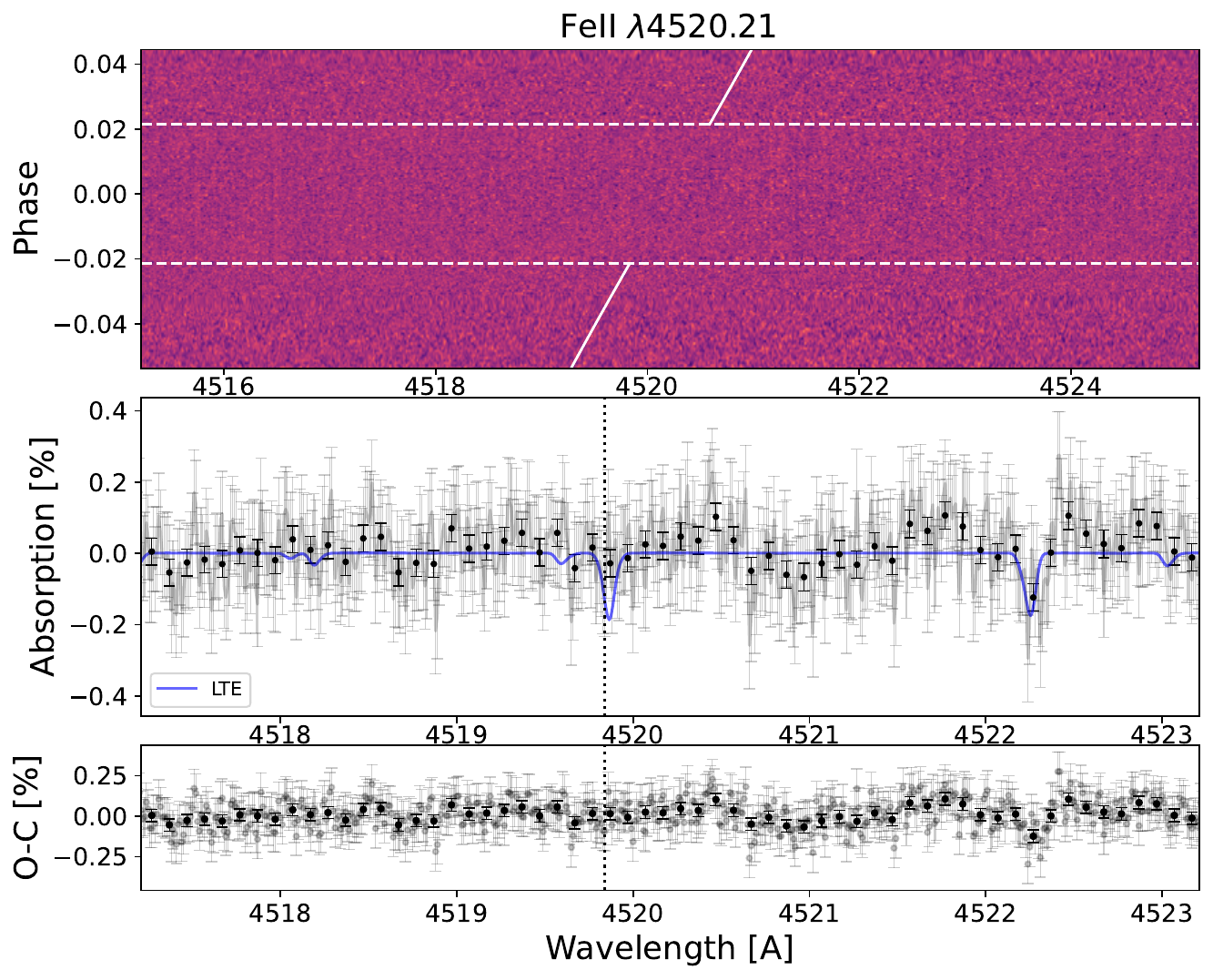}

\caption{Same as Figure \ref{Fig:TS_M2_1}, but for $\lambda$4489, $\lambda$4508, $\lambda$4515, and $\lambda$4520.}
\label{Fig:TS_M2_2}
\end{figure}

\begin{figure}[h]
\centering
\includegraphics[width=0.49\textwidth]{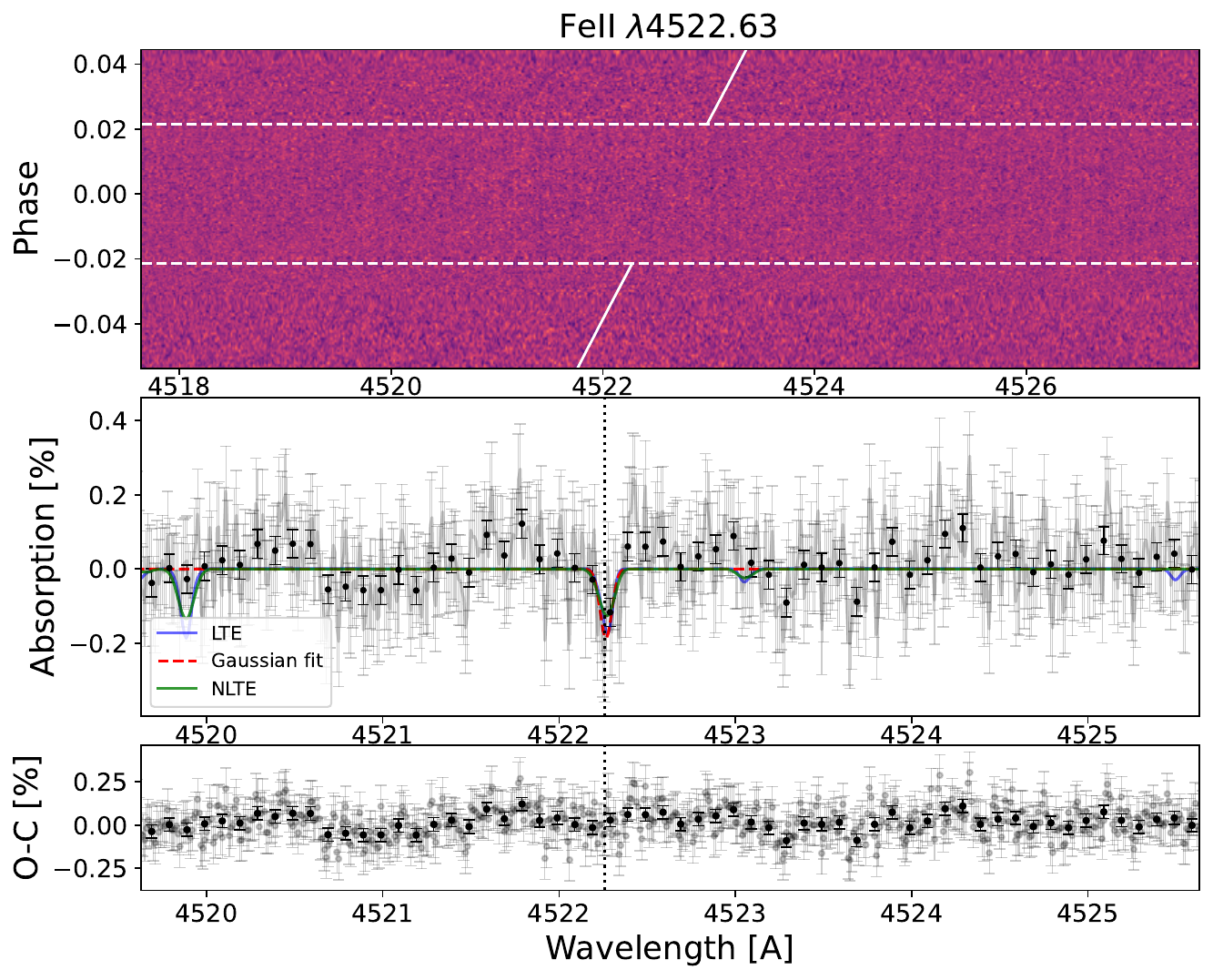}
\includegraphics[width=0.49\textwidth]{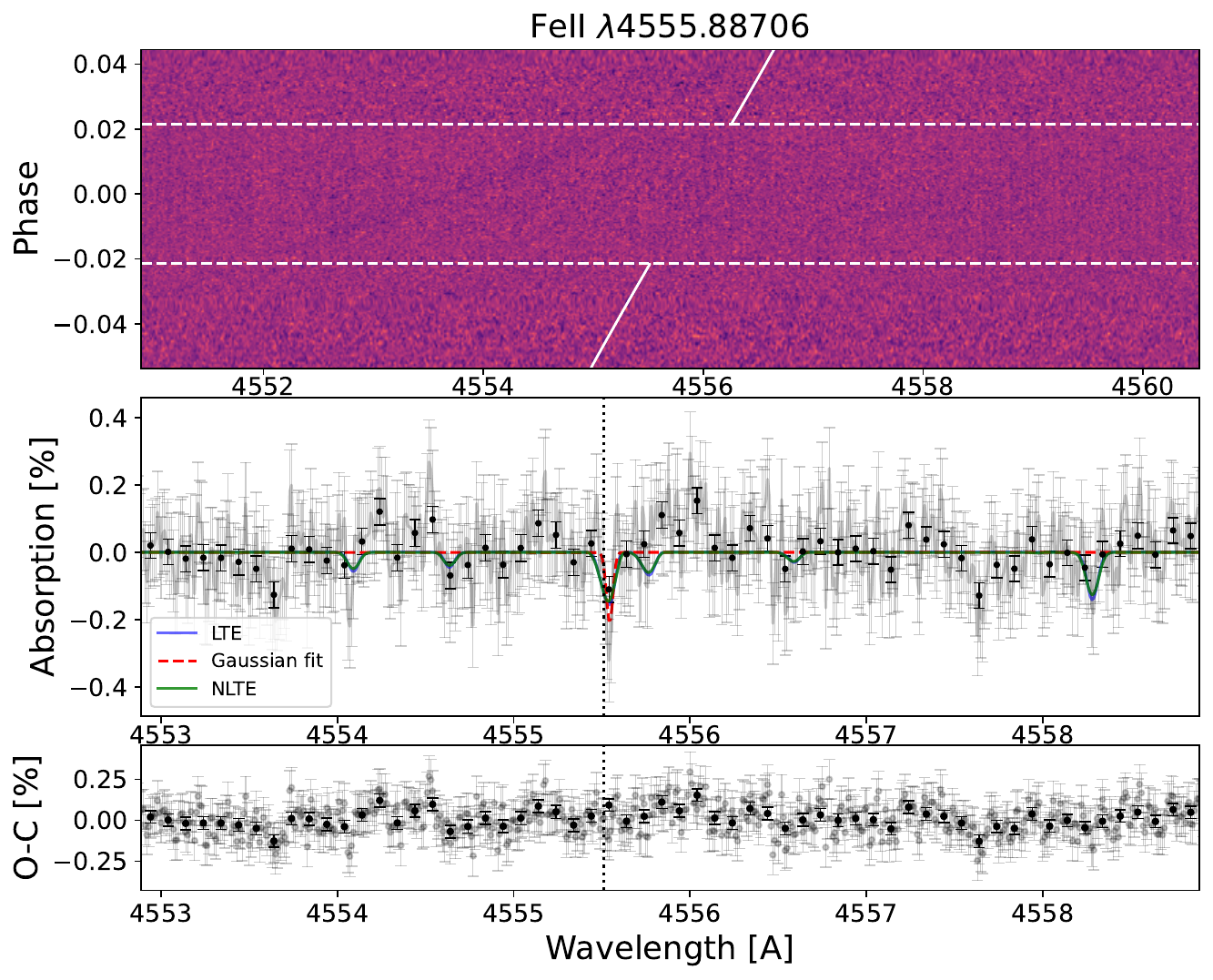}

\includegraphics[width=0.49\textwidth]{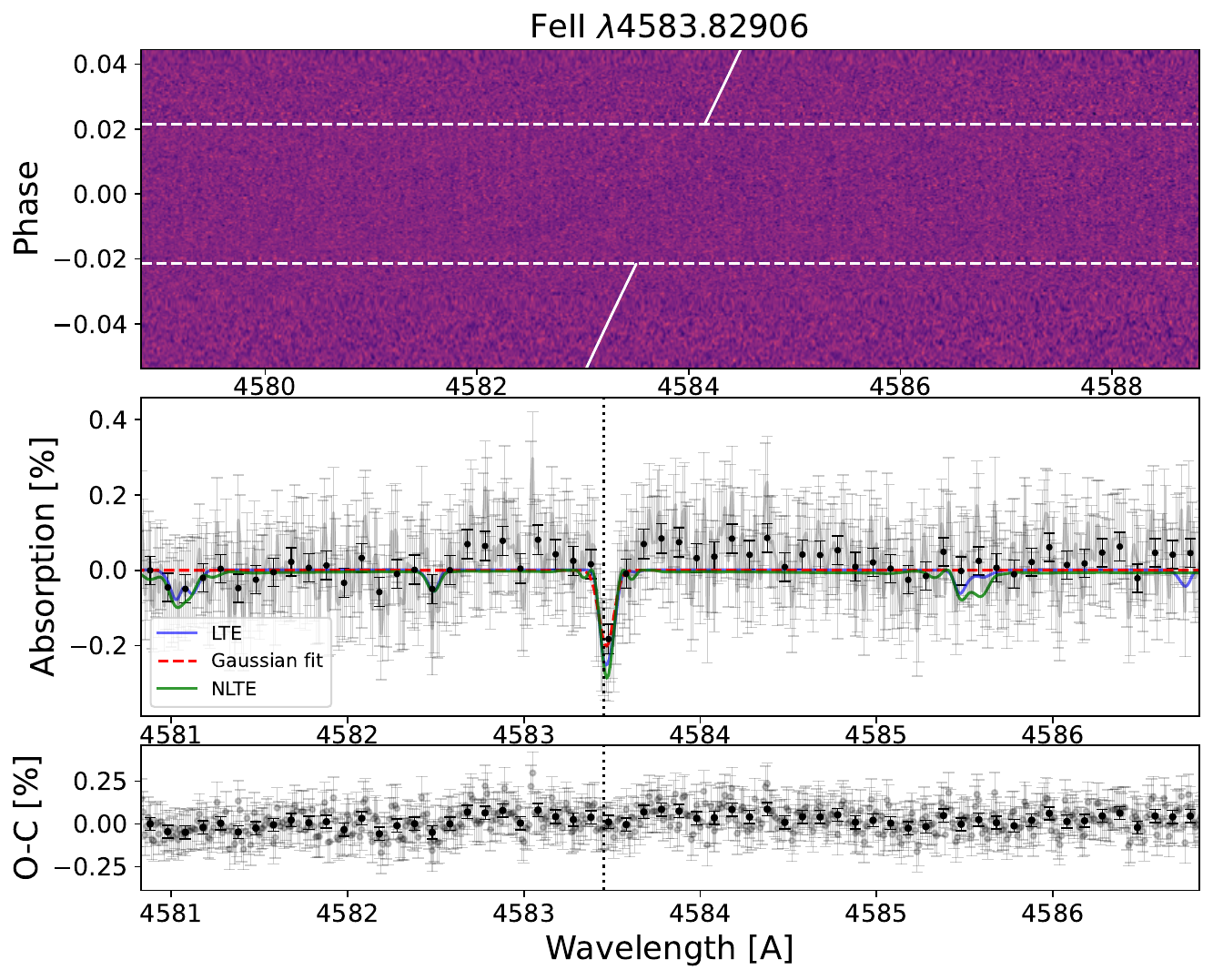}
\includegraphics[width=0.49\textwidth]{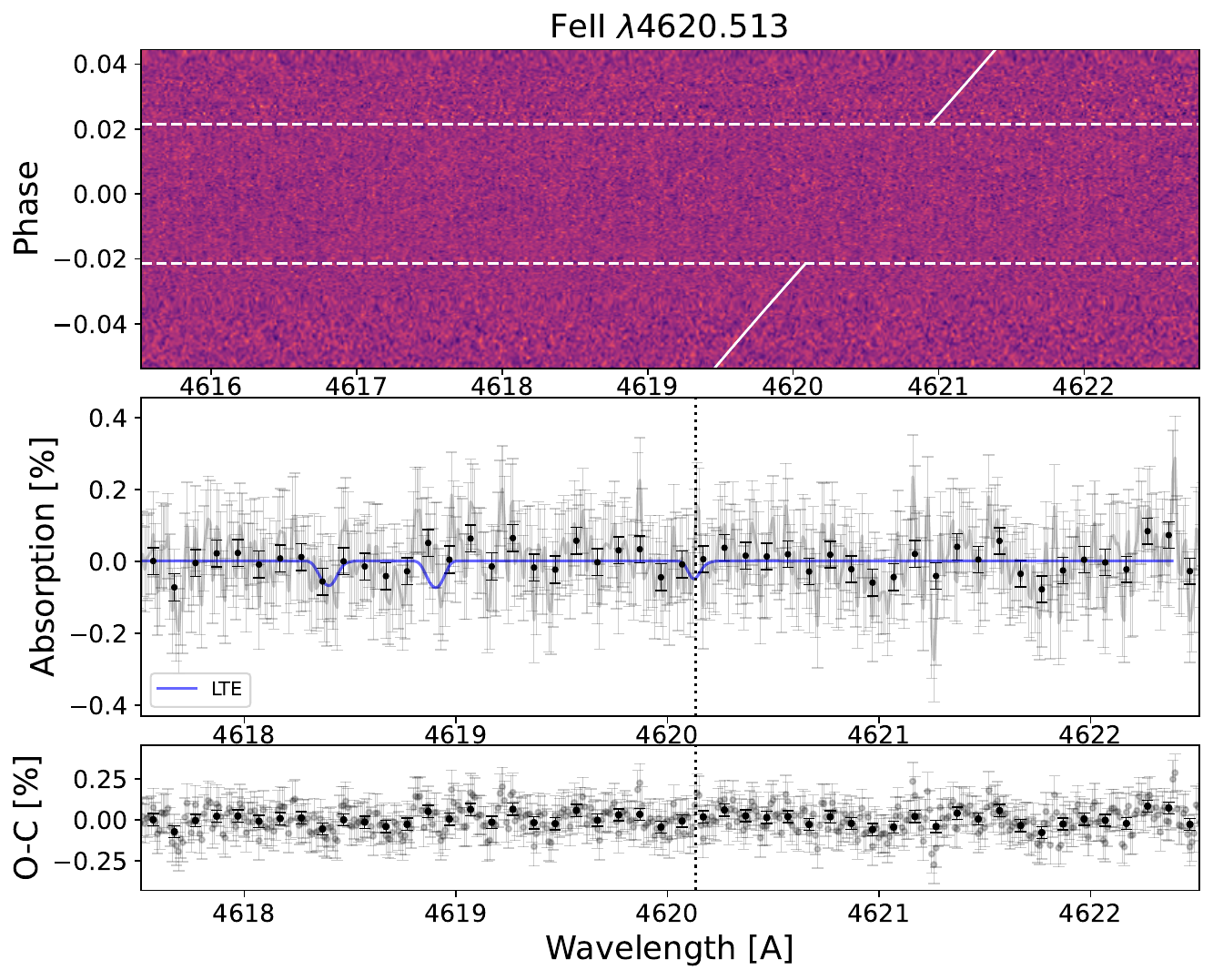}

\caption{Same as Figure \ref{Fig:TS_M2_1}, but for $\lambda$4522, $\lambda$4555, $\lambda$4583,  and $\lambda$4620.}
\label{Fig:TS_M2_3}
\end{figure}

\begin{figure}[h]
\centering
\includegraphics[width=0.49\textwidth]{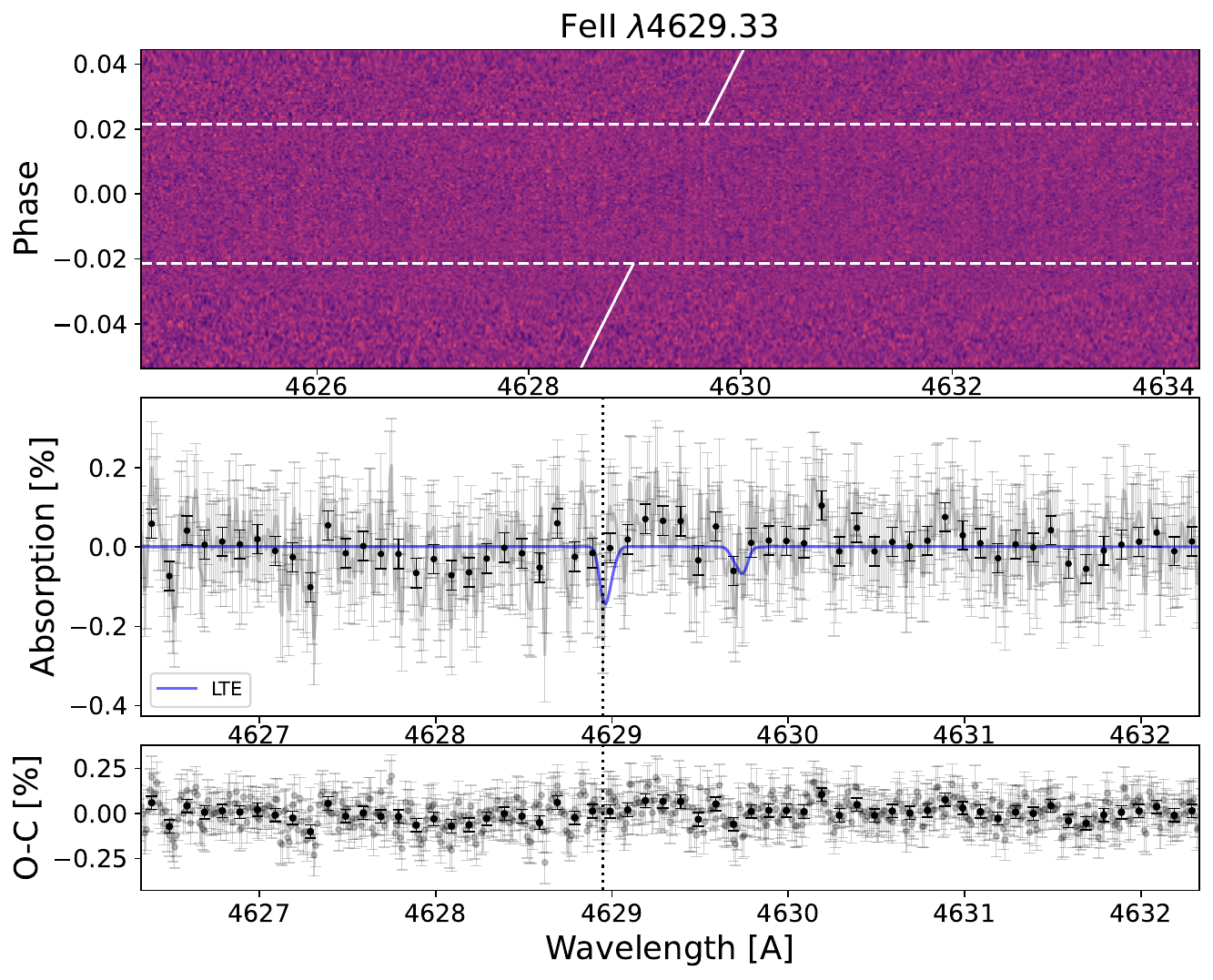}
\includegraphics[width=0.49\textwidth]{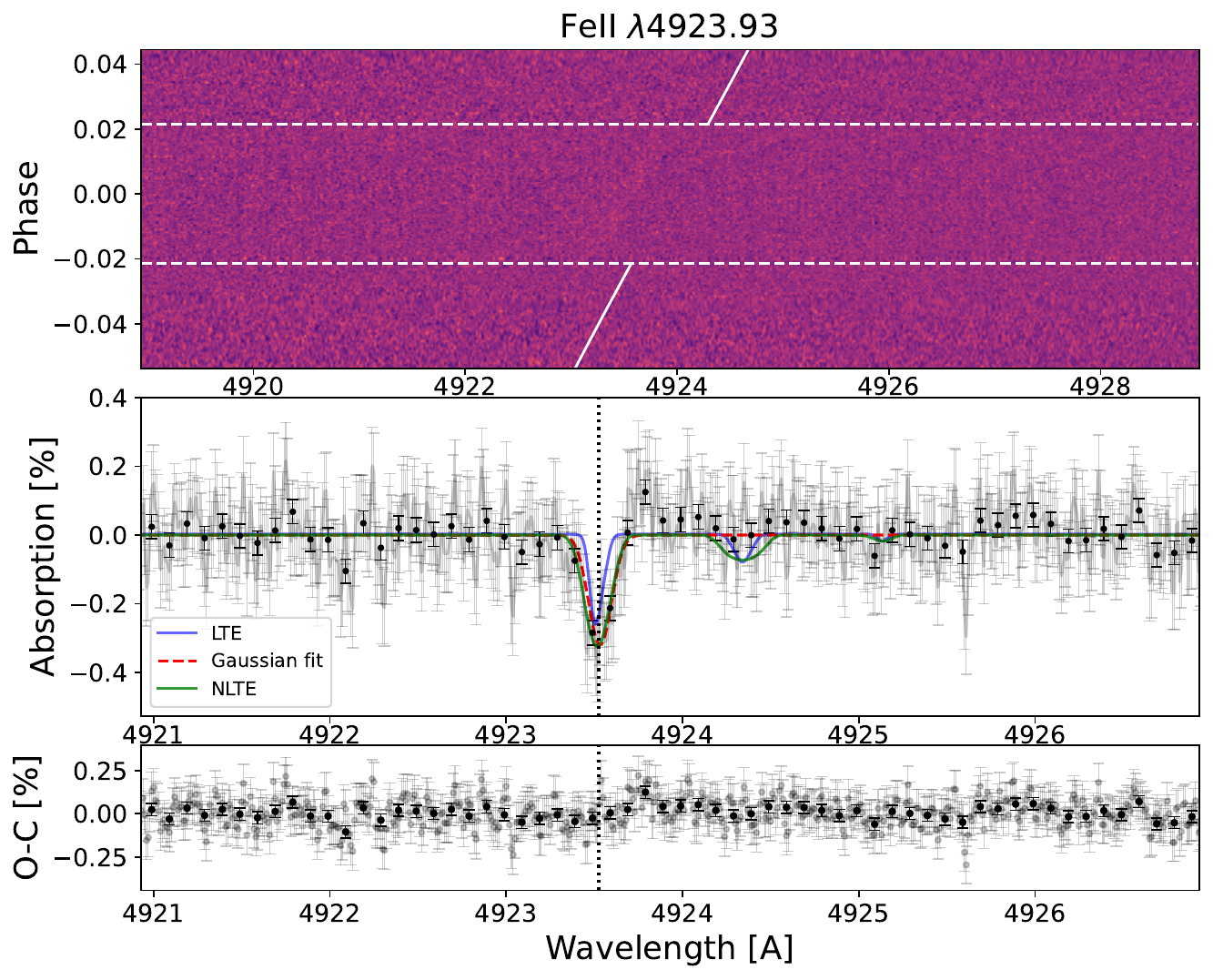}

\includegraphics[width=0.49\textwidth]{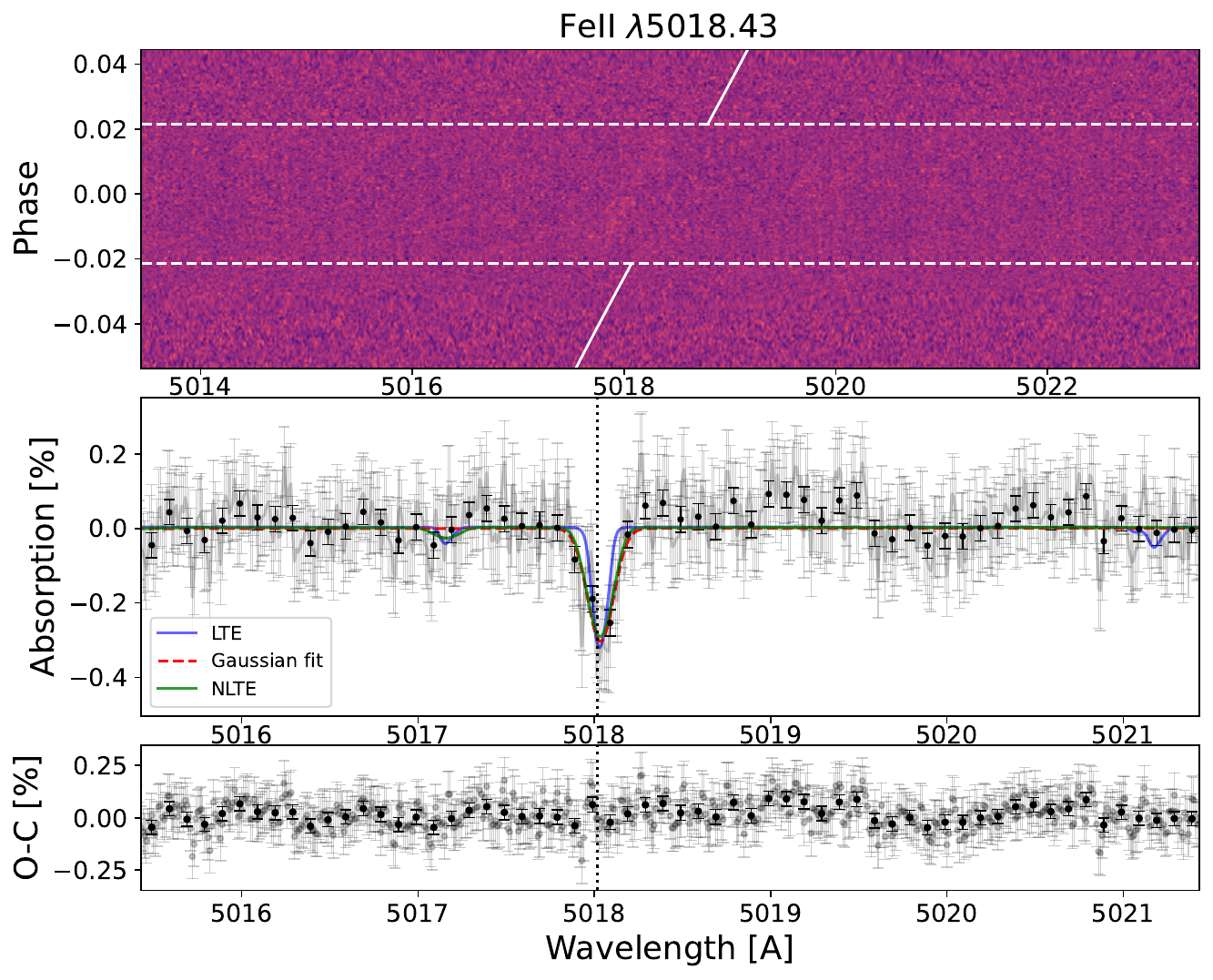}
\includegraphics[width=0.49\textwidth]{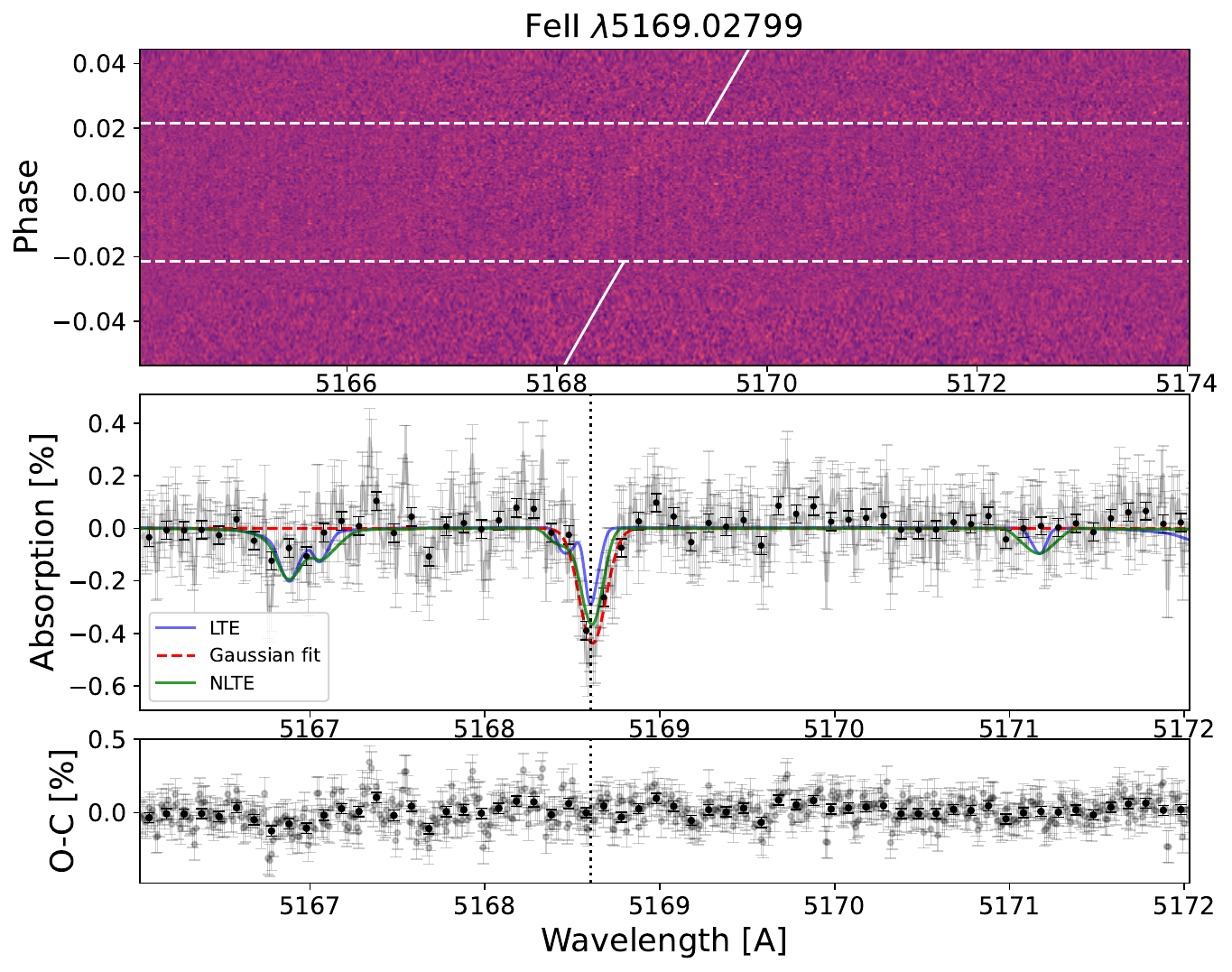}

\caption{Same as Figure \ref{Fig:TS_M2_1}, but for $\lambda$4629, $\lambda$4923, $\lambda$5018, and $\lambda$5169.}
\label{Fig:TS_M2_4}
\end{figure}

\begin{figure}[h]
\centering
\includegraphics[width=0.49\textwidth]{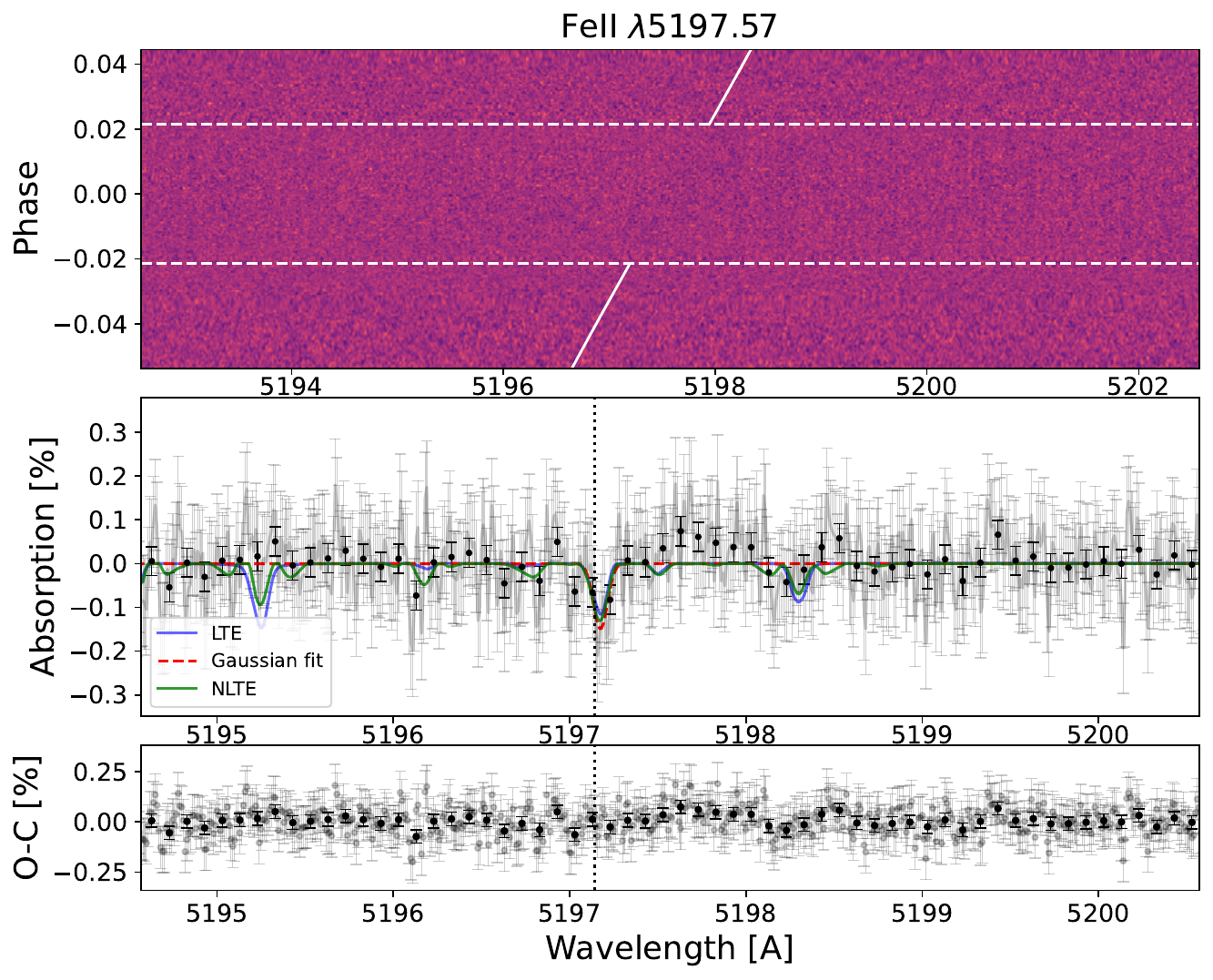}
\includegraphics[width=0.49\textwidth]{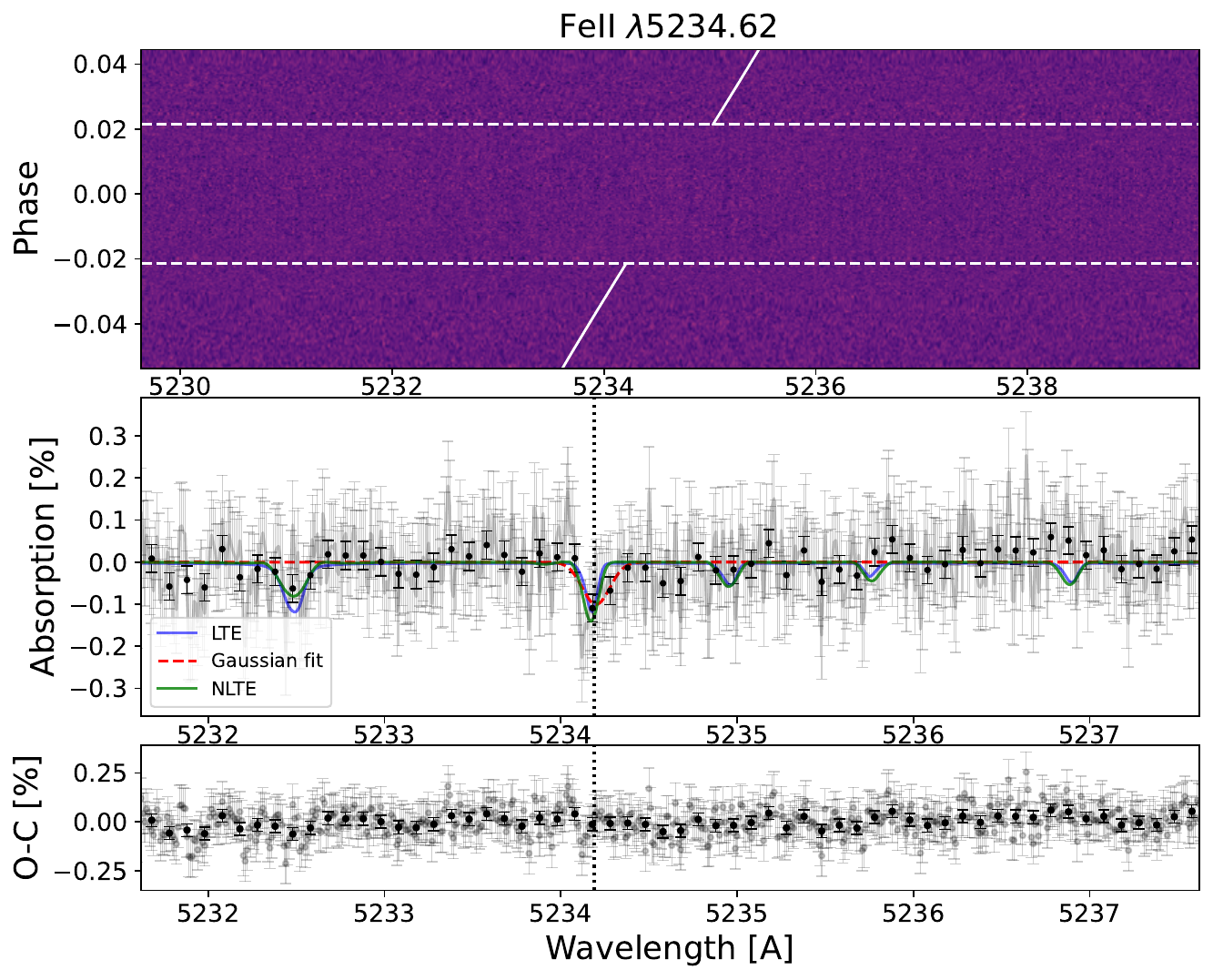}

\includegraphics[width=0.49\textwidth]{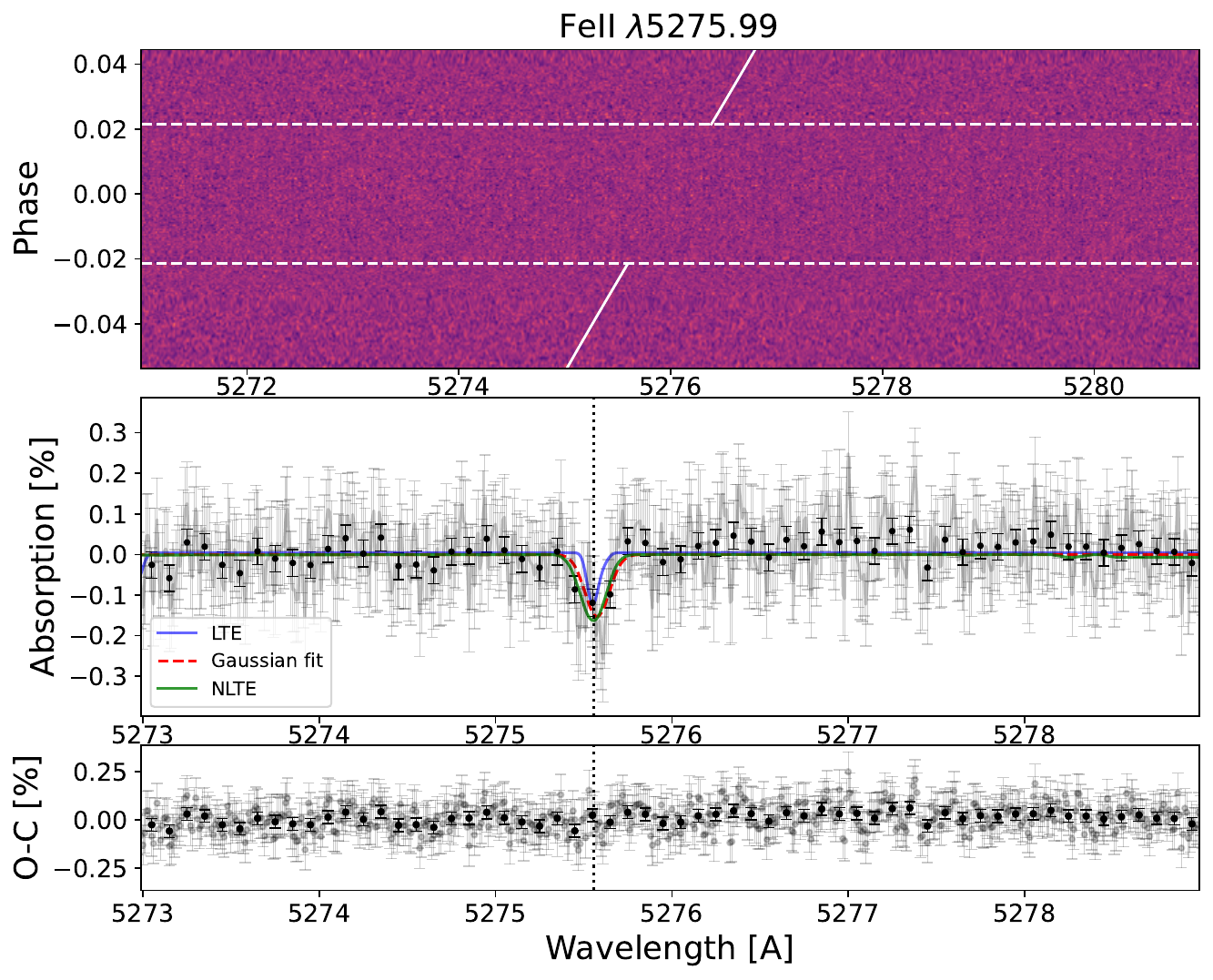}
\includegraphics[width=0.49\textwidth]{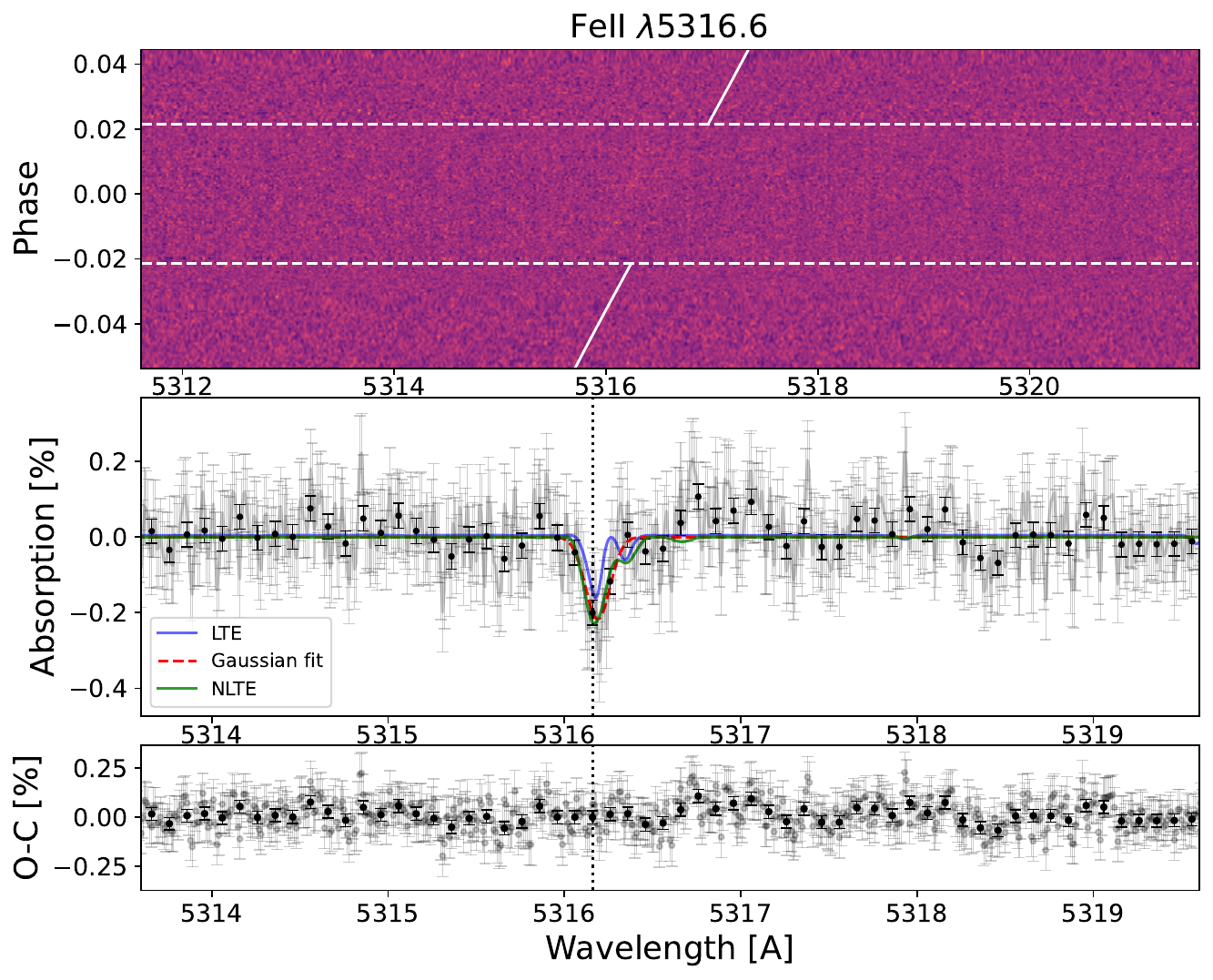}

\caption{Same as Figure \ref{Fig:TS_M2_1}, but for $\lambda$5197, $\lambda$5234, $\lambda$5275.99, and $\lambda$5316.}
\label{Fig:TS_M2_5}
\end{figure}

\begin{figure}[h]
\centering
\includegraphics[width=0.49\textwidth]{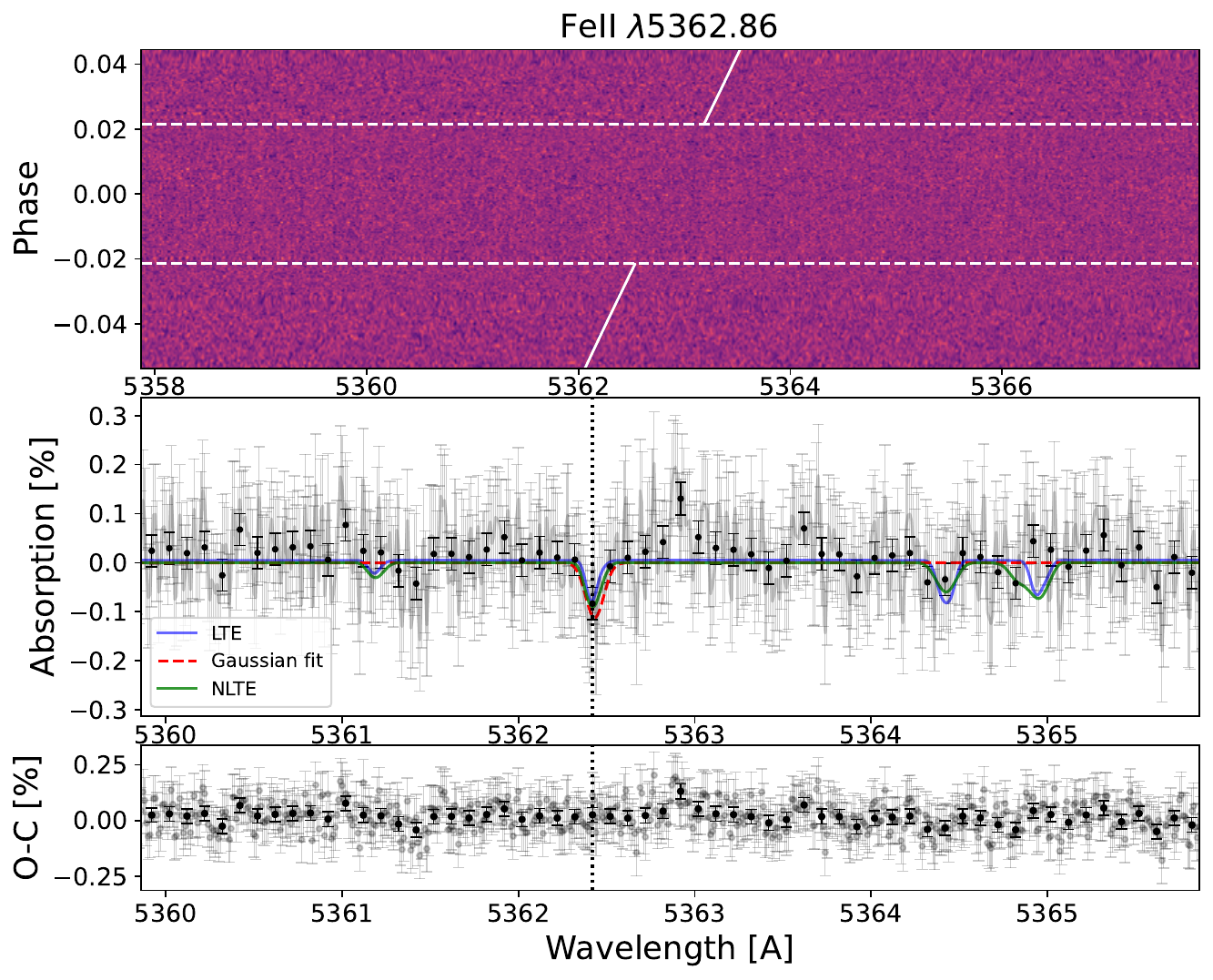}
\caption{Same as Figure \ref{Fig:TS_M2_1}, but for $\lambda$5362.}
\label{Fig:TS_M2_6}
\end{figure}

\begin{figure}[h]
\centering
\includegraphics[width=0.38\textwidth]{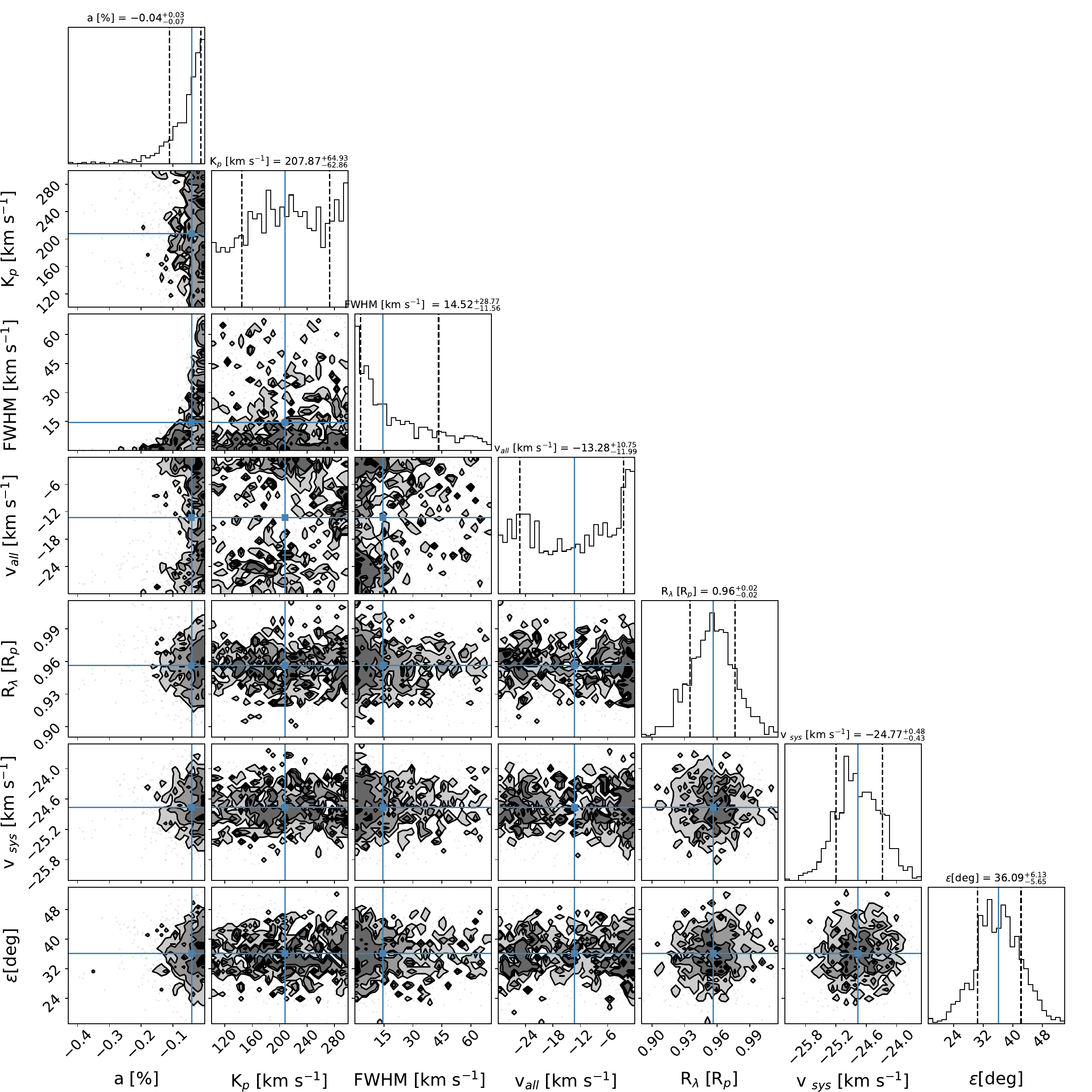}
\includegraphics[width=0.38\textwidth]{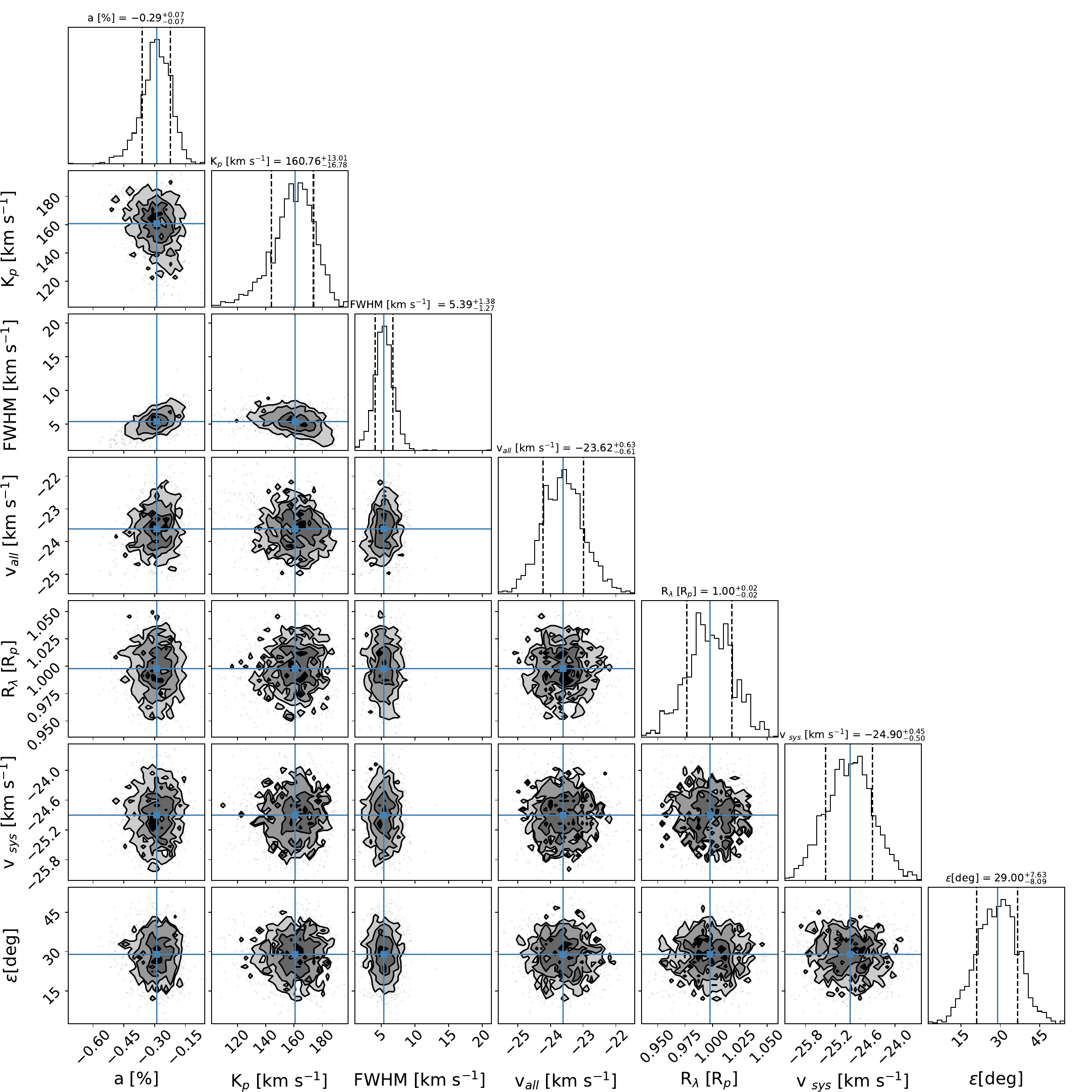}
\caption{The corner plots of the MCMC analysis for KELT-20b of the line \ion{Fe}{ii} $\lambda$4173 and $\lambda$4233. }
\label{fig:corner_k20_1}
\end{figure}

\begin{figure}[h]
\centering
\includegraphics[width=0.38\textwidth]{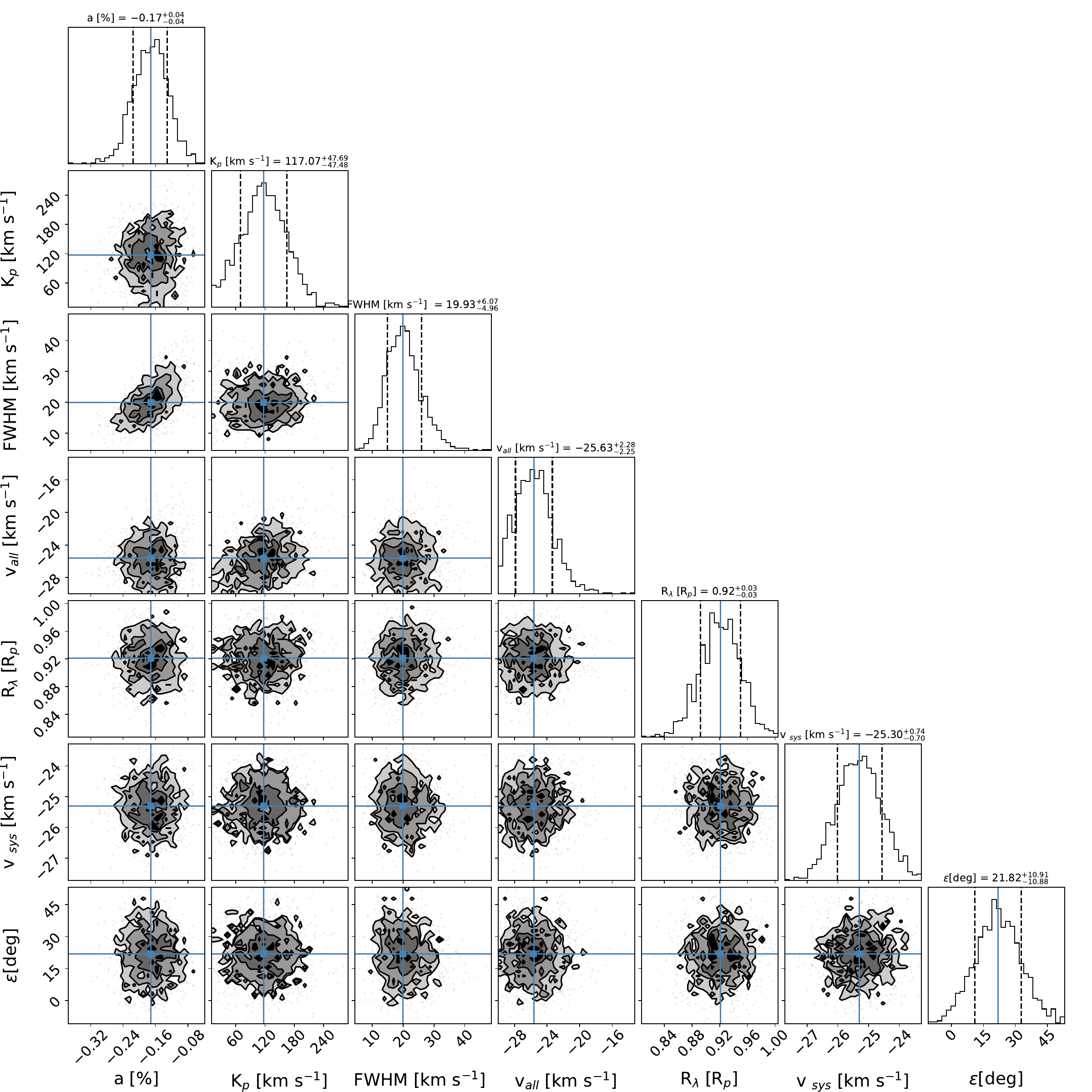}
\includegraphics[width=0.38\textwidth]{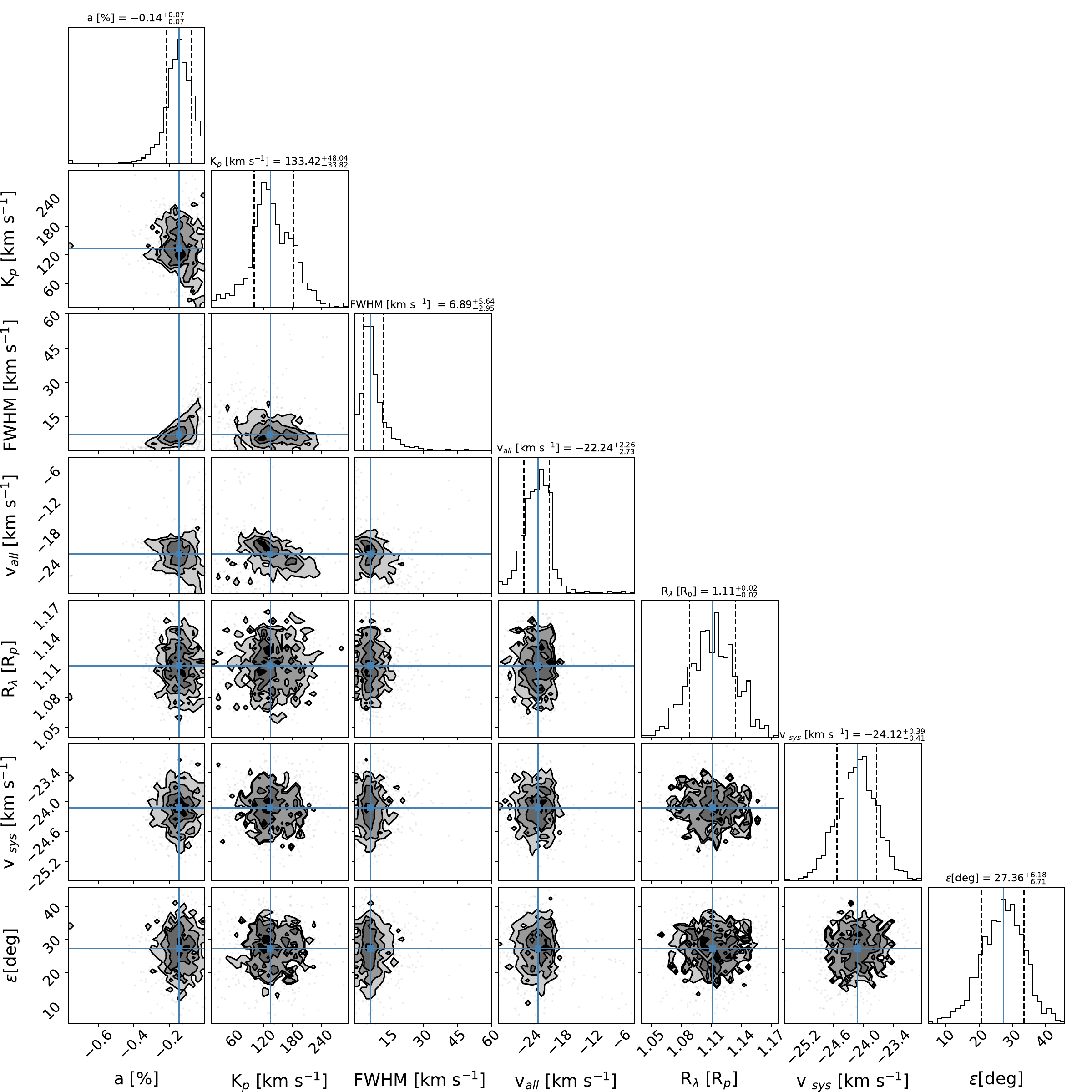}
\caption{Same as Figure \ref{fig:corner_k20_1}, but for \ion{Fe}{ii}$ \lambda$4351 and $\lambda$4385. }
\label{fig:corner_k20_2}
\end{figure}

\begin{figure}[h]
\centering
\includegraphics[width=0.38\textwidth]{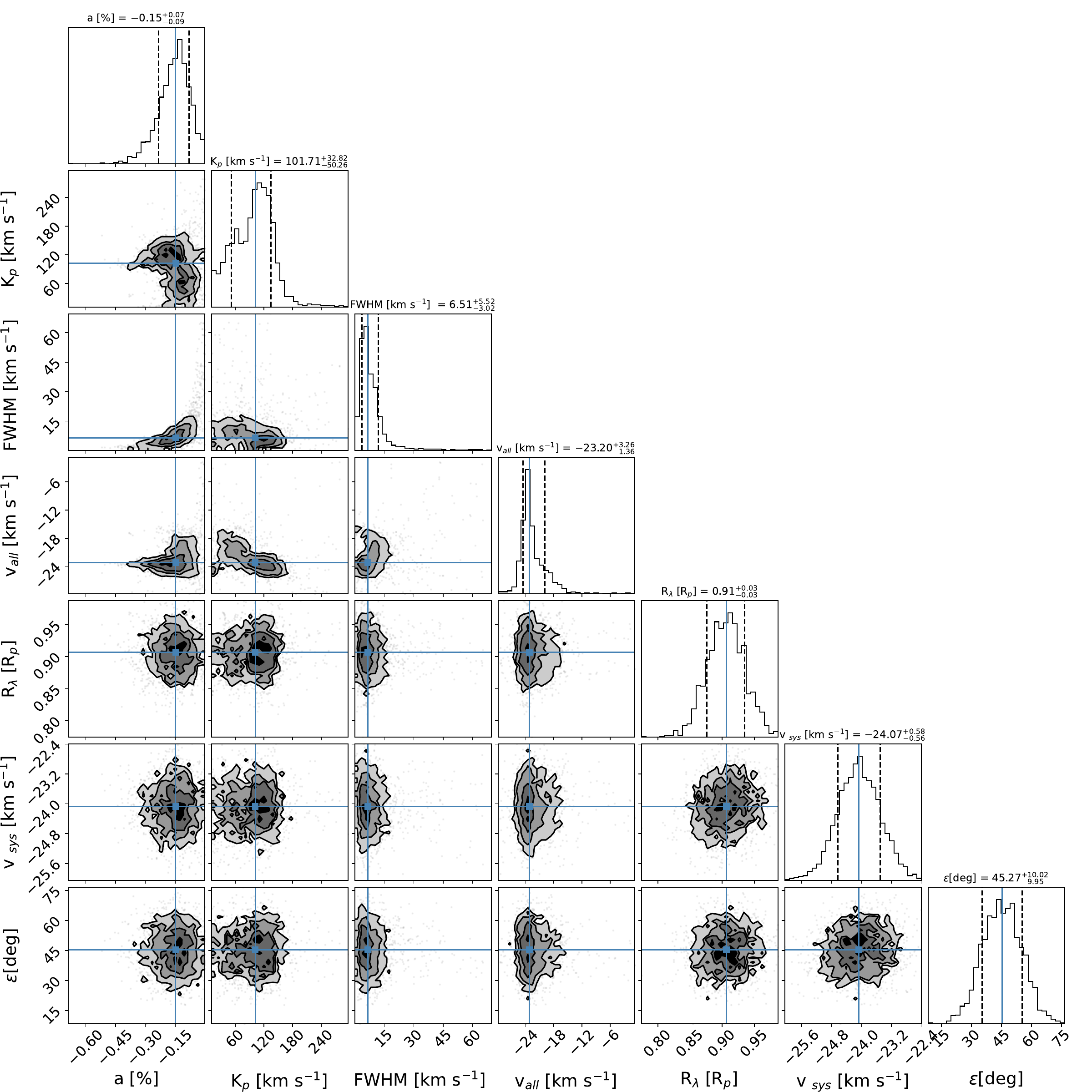}
\includegraphics[width=0.38\textwidth]{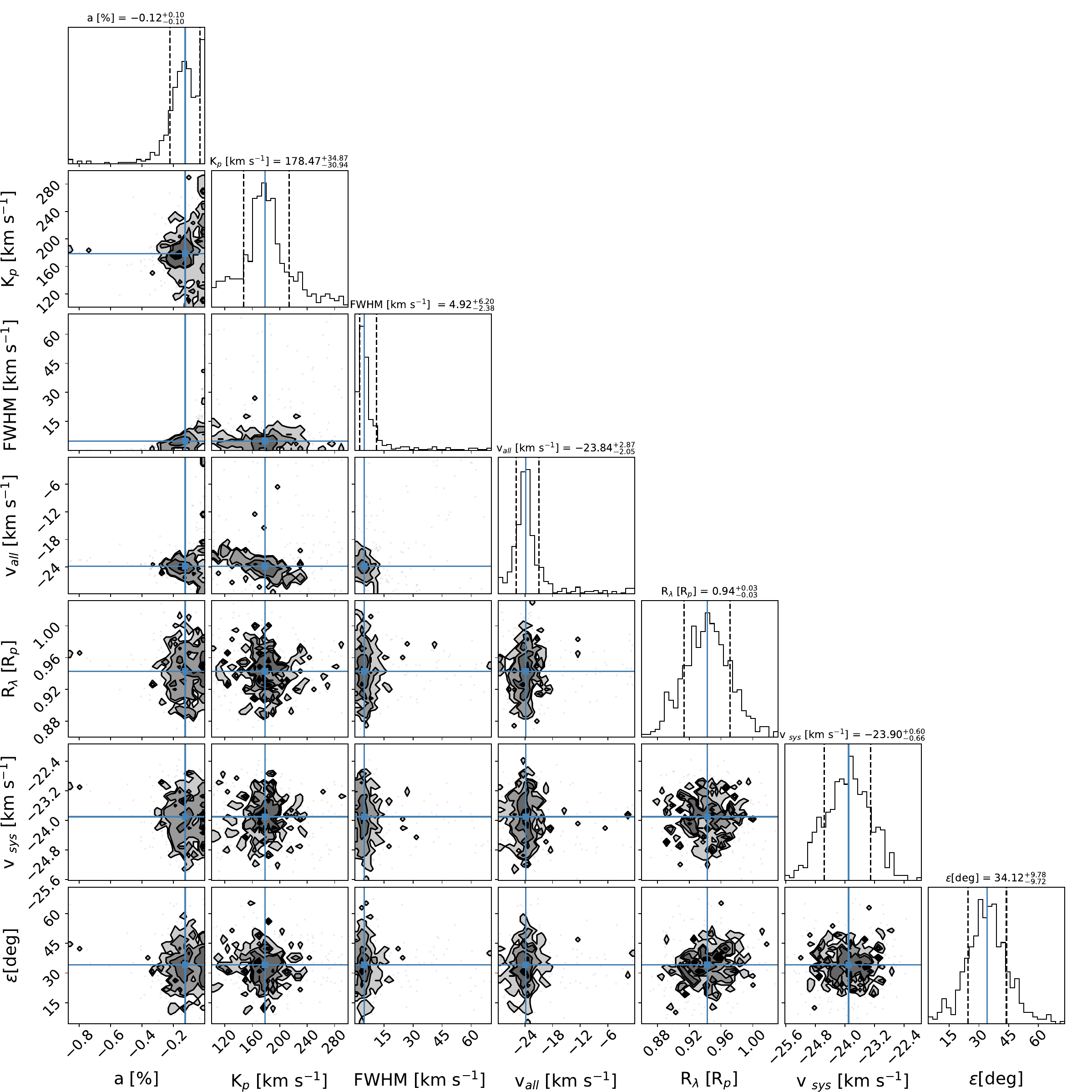}
\caption{Same as Figure \ref{fig:corner_k20_1}, but for  \ion{Fe}{ii} $\lambda$4489 and $\lambda$4508. }
\label{fig:corner_k20_3}
\end{figure}

\begin{figure}[h]
\centering
\includegraphics[width=0.38\textwidth]{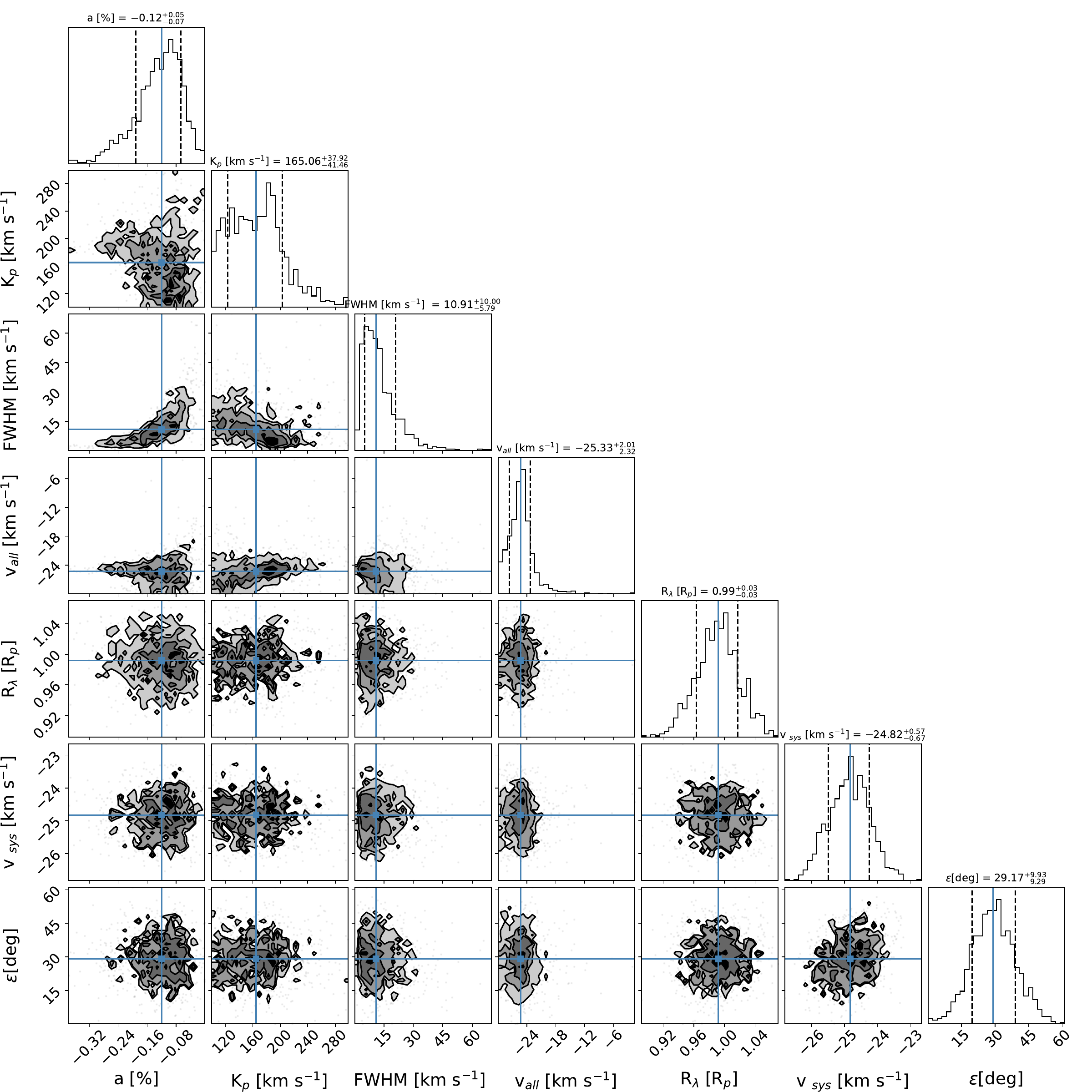}
\includegraphics[width=0.38\textwidth]{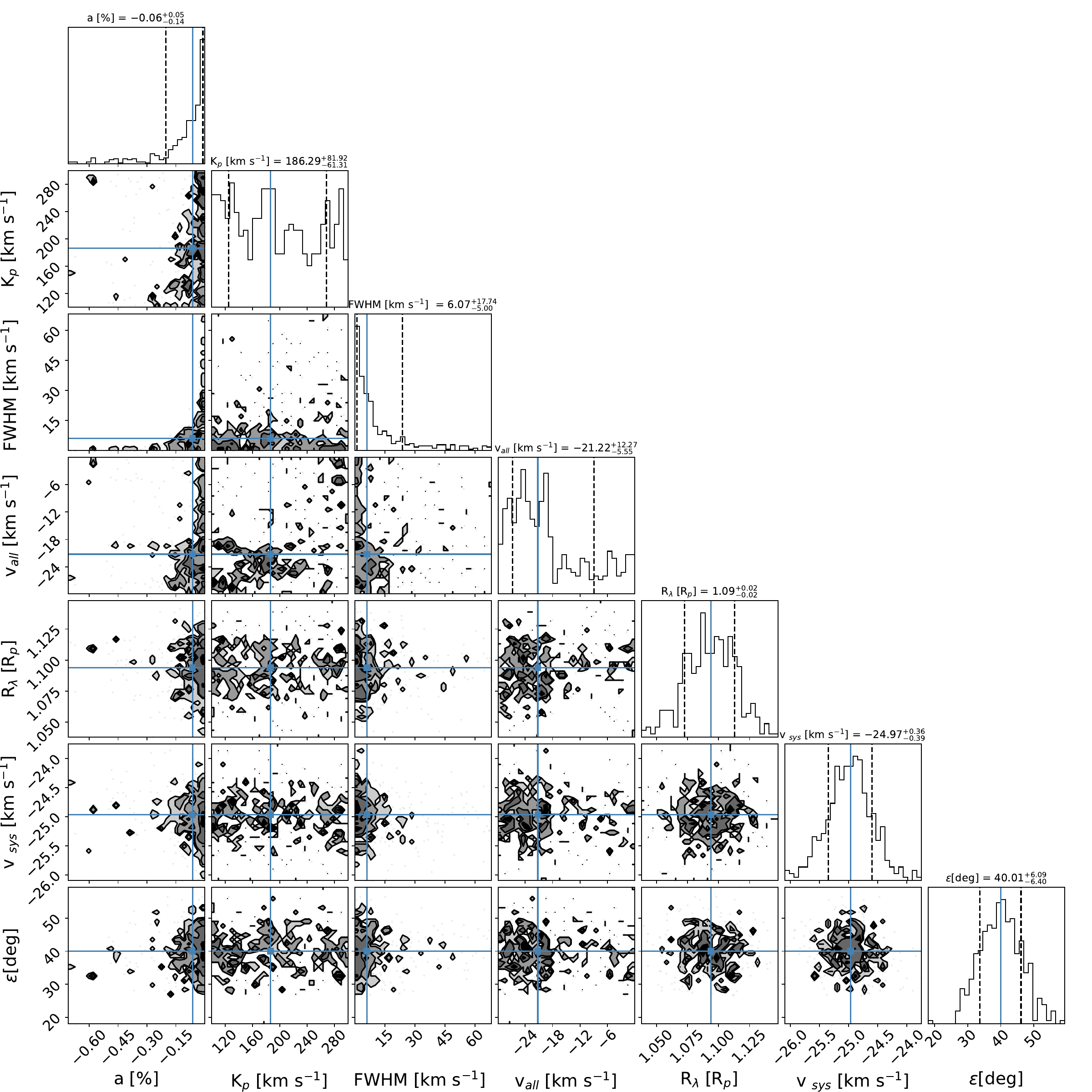}
\caption{Same as Figure \ref{fig:corner_k20_1}, but for \ion{Fe}{ii} $\lambda$4515 and $\lambda$4520.  }
\label{fig:corner_k20_4}
\end{figure}

\begin{figure}[h]
\centering
\includegraphics[width=0.38\textwidth]{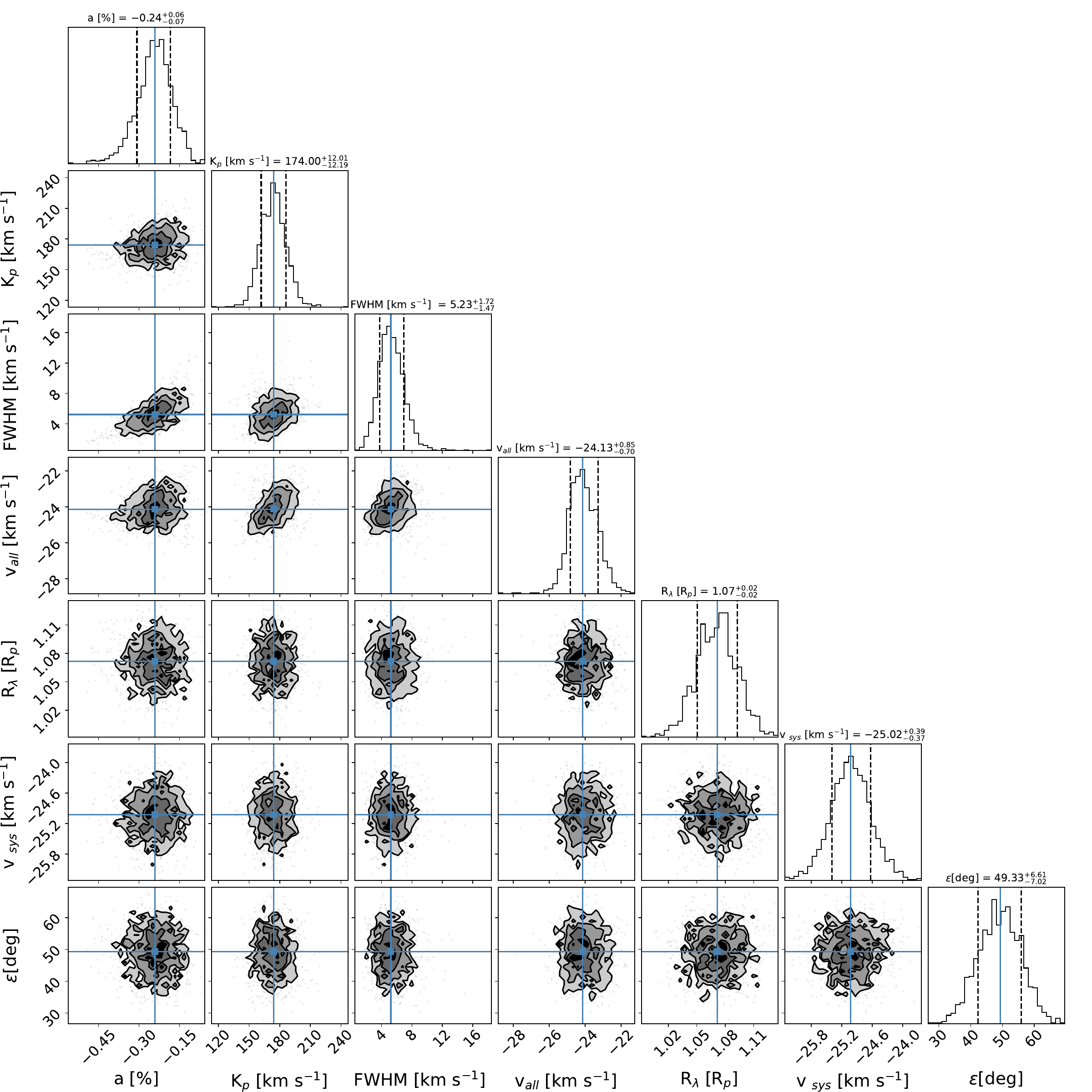}
\includegraphics[width=0.38\textwidth]{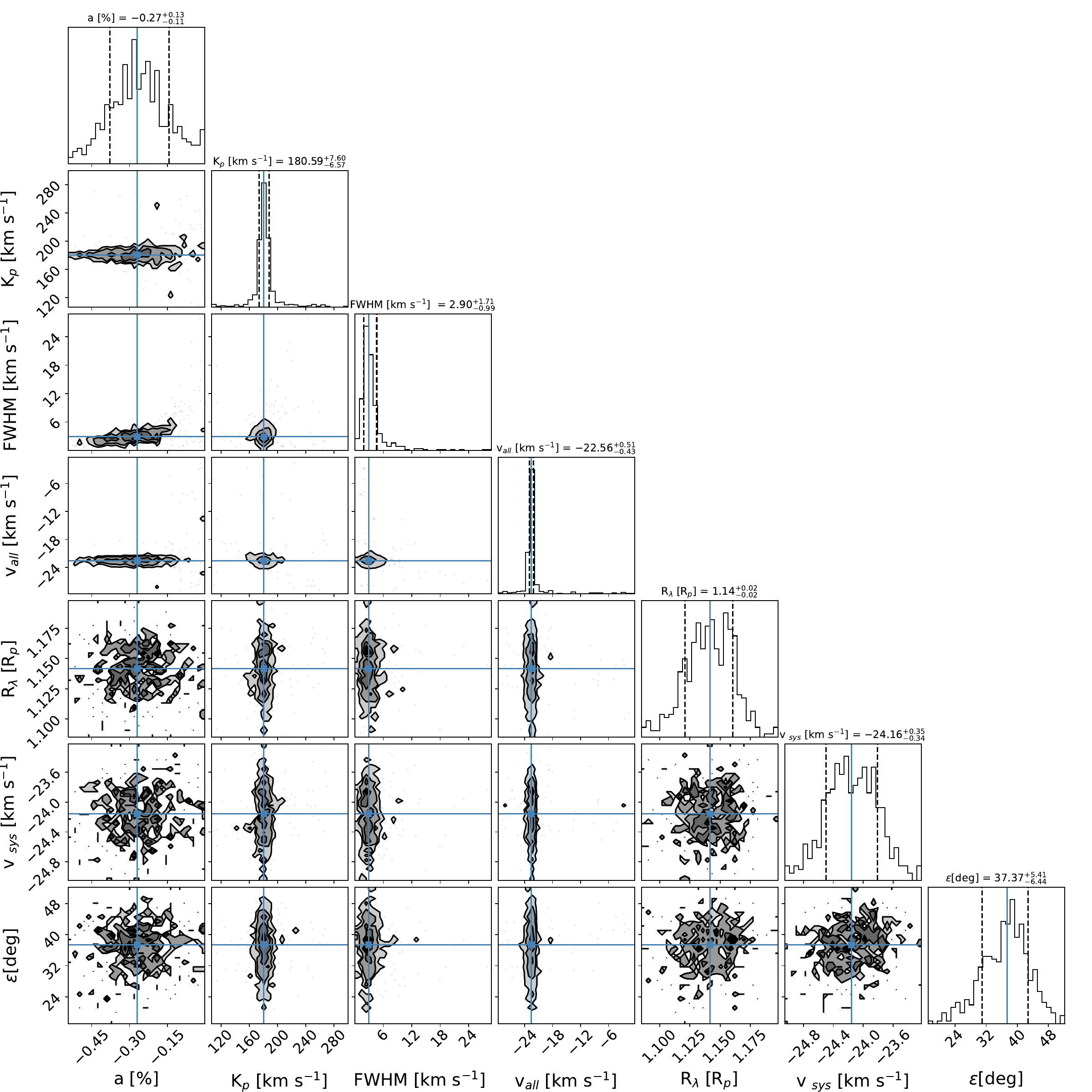}
\caption{Same as Figure \ref{fig:corner_k20_1}, but for  \ion{Fe}{ii} $\lambda$4522 and $\lambda$4555. }
\label{fig:corner_k20_5}
\end{figure}

\begin{figure}[h]
\centering
\includegraphics[width=0.38\textwidth]{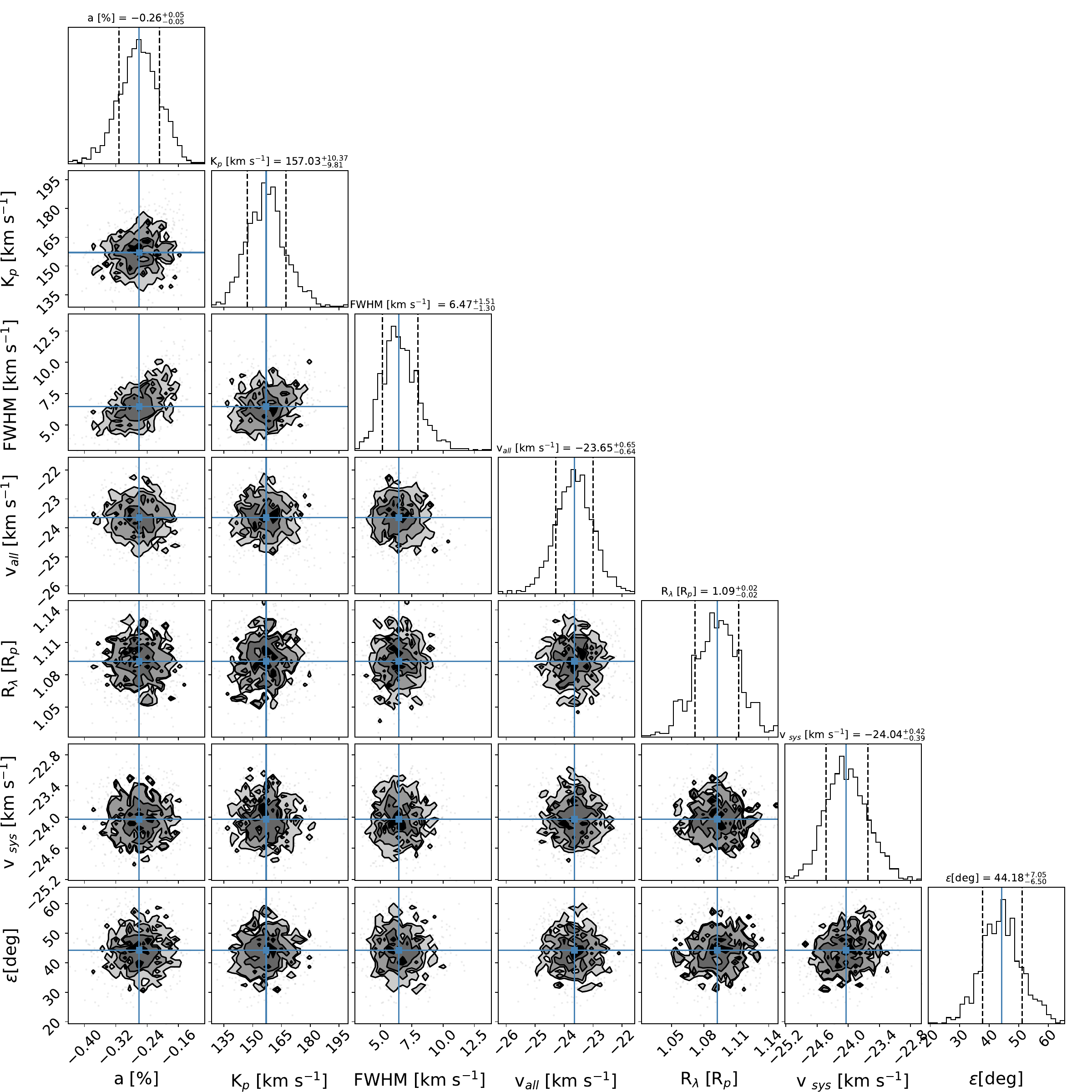}
\includegraphics[width=0.38\textwidth]{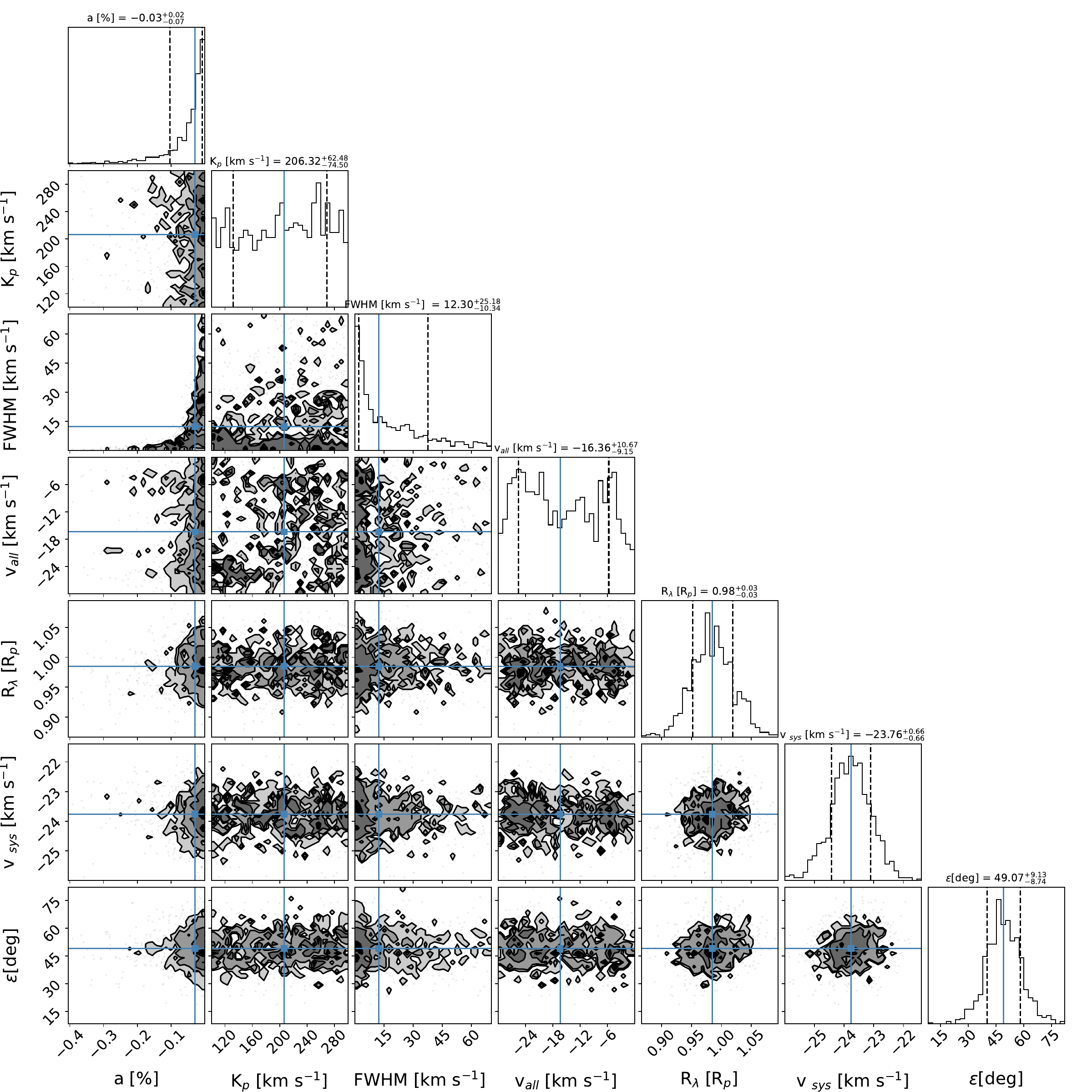}
\caption{Same as Figure \ref{fig:corner_k20_1}, but for \ion{Fe}{ii} $\lambda$4583 and $\lambda$4620. }
\label{fig:corner_k20_6}
\end{figure}

\begin{figure}[h]
\centering
\includegraphics[width=0.38\textwidth]{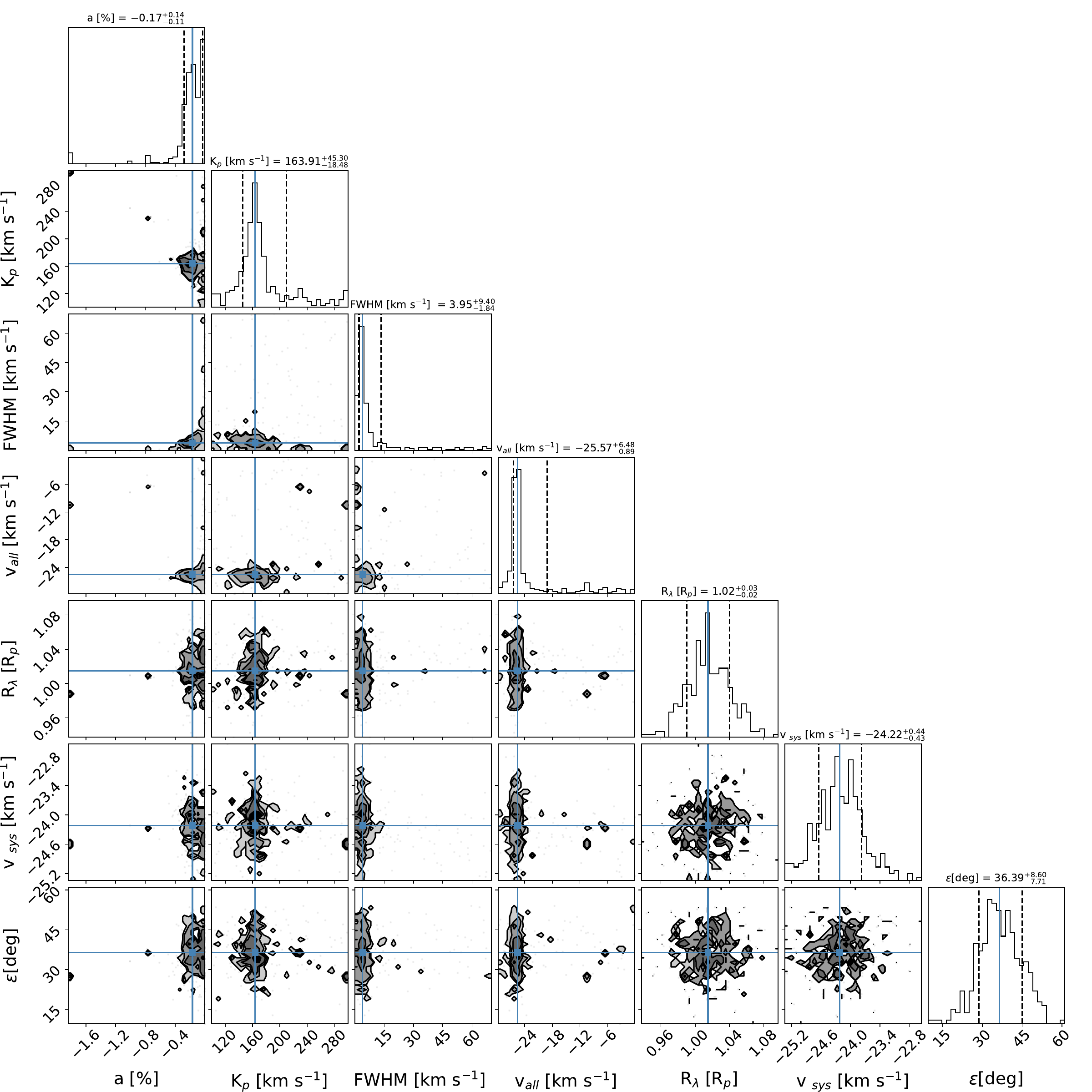}
\includegraphics[width=0.38\textwidth]{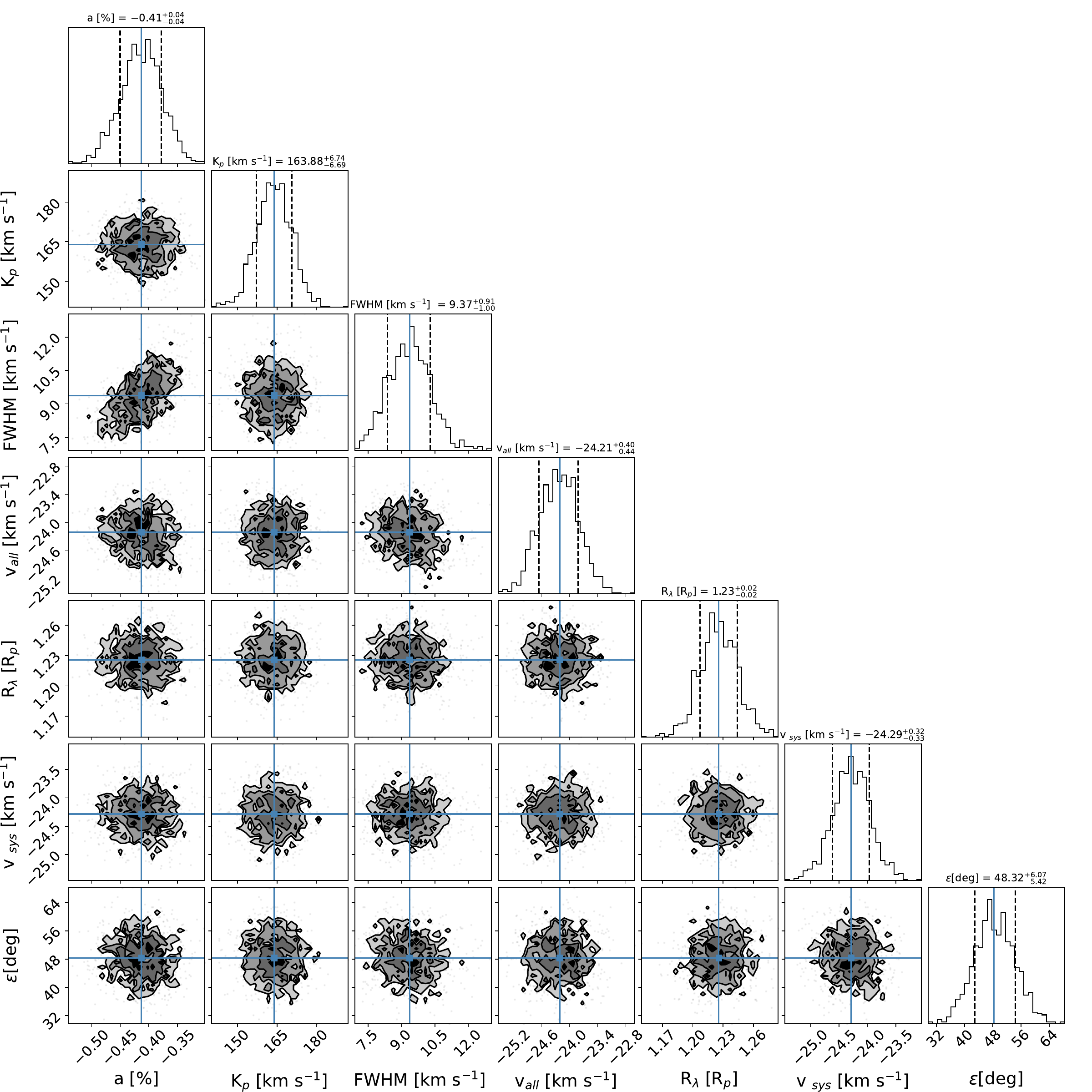}
\caption{Same as Figure \ref{fig:corner_k20_1}, but for \ion{Fe}{ii} $\lambda$4629 and $\lambda$4923. }
\label{fig:corner_k20_7}
\end{figure}

\begin{figure}[h]
\centering
\includegraphics[width=0.38\textwidth]{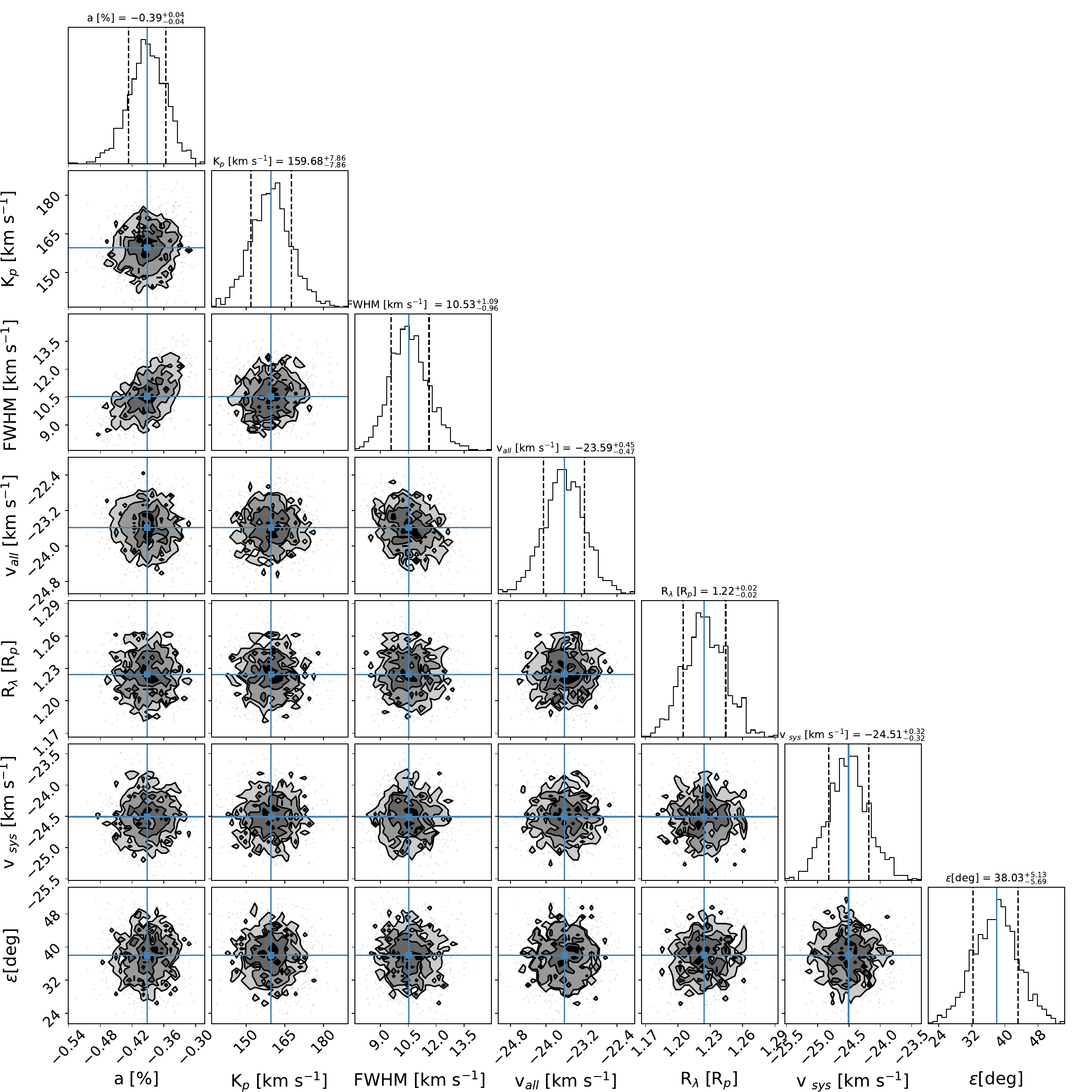}
\includegraphics[width=0.38\textwidth]{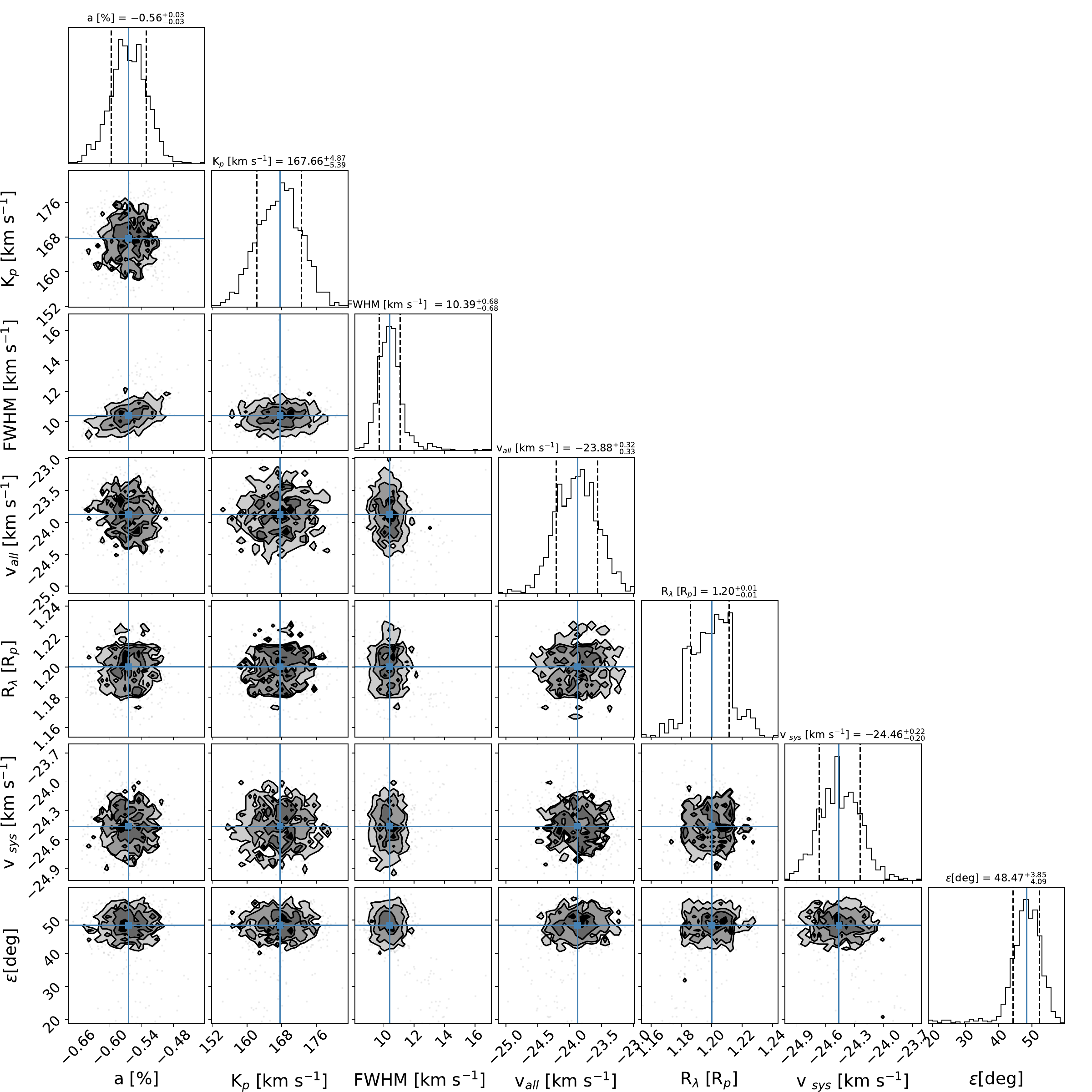}
\caption{Same as Figure \ref{fig:corner_k20_1}, but for \ion{Fe}{ii} $\lambda$5018 and $\lambda$5169. }
\label{fig:corner_k20_8}
\end{figure}

\begin{figure}[h]
\centering
\includegraphics[width=0.38\textwidth]{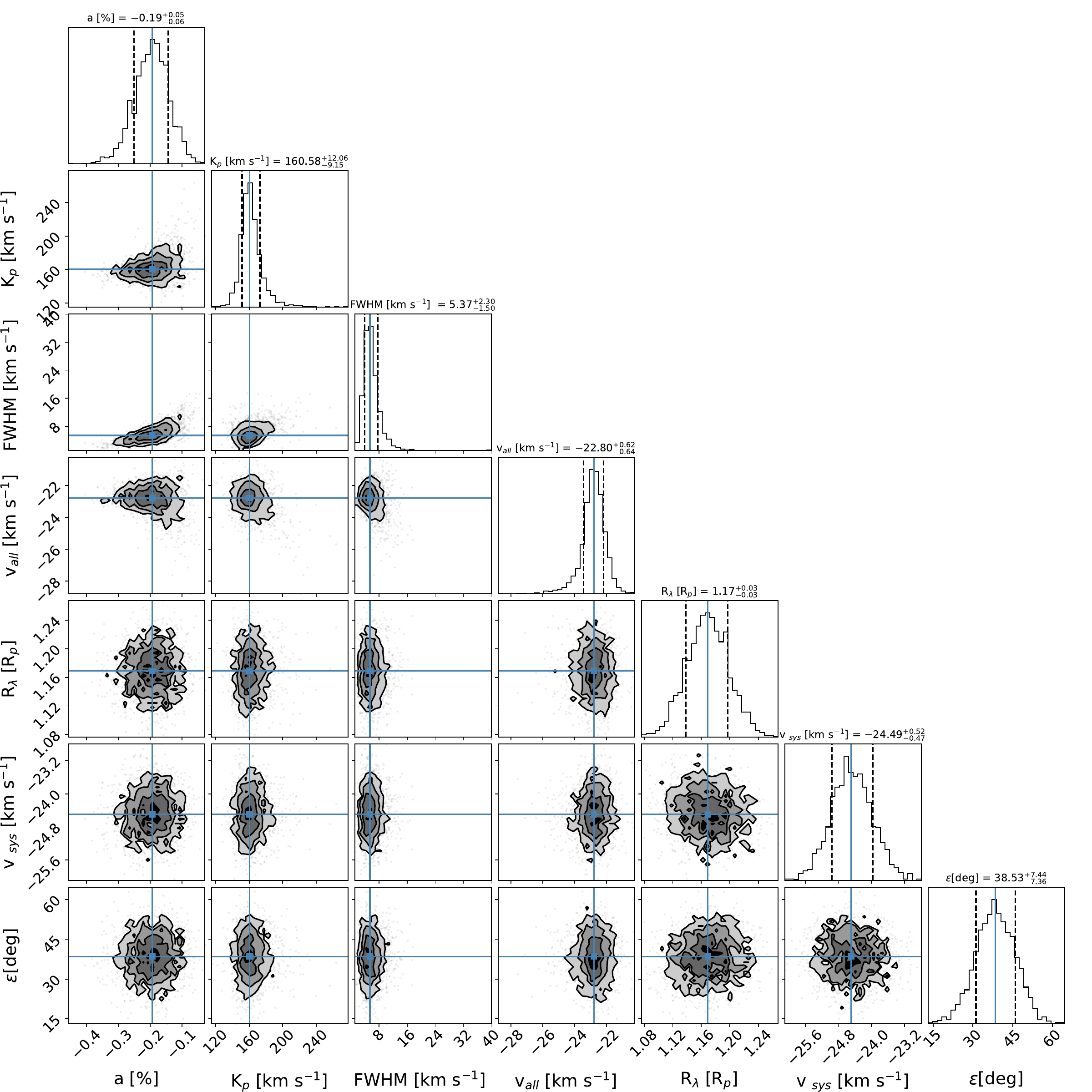}
\includegraphics[width=0.38\textwidth]{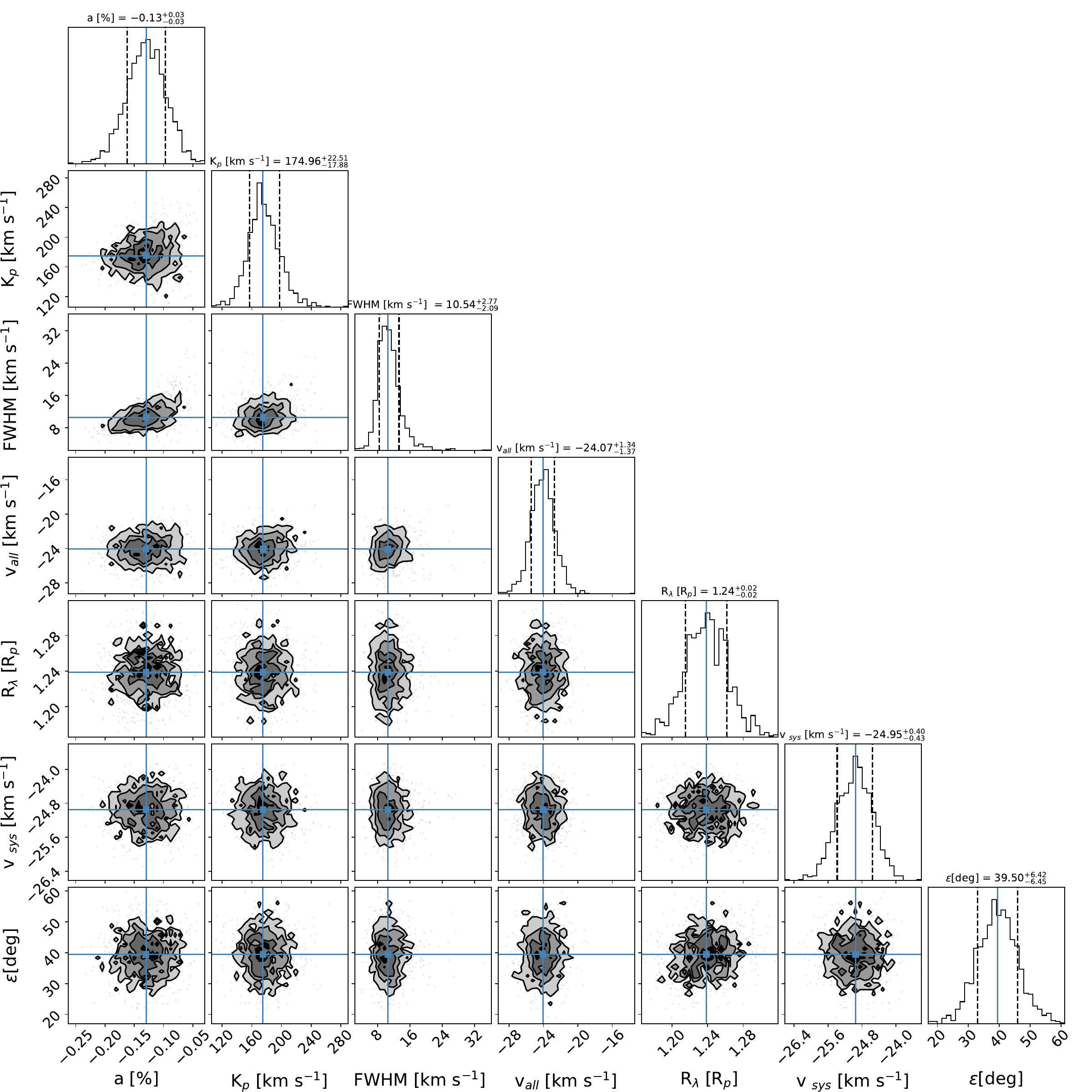}
\caption{Same as Figure \ref{fig:corner_k20_1}, but for \ion{Fe}{ii} $\lambda$5197 and $\lambda$5234. }
\label{fig:corner_k20_9}
\end{figure}

\begin{figure}[h]
\centering
\includegraphics[width=0.38\textwidth]{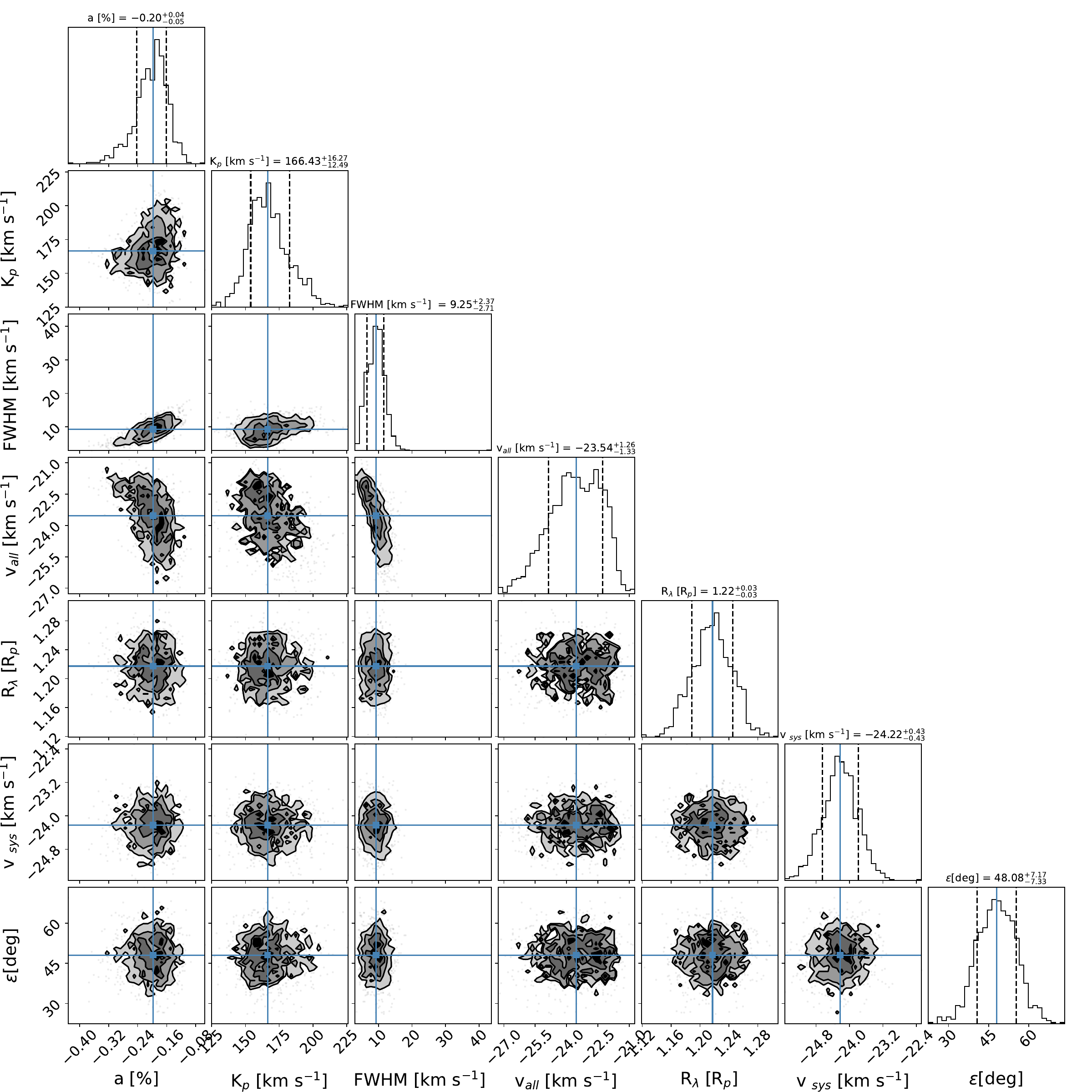}
\includegraphics[width=0.38\textwidth]{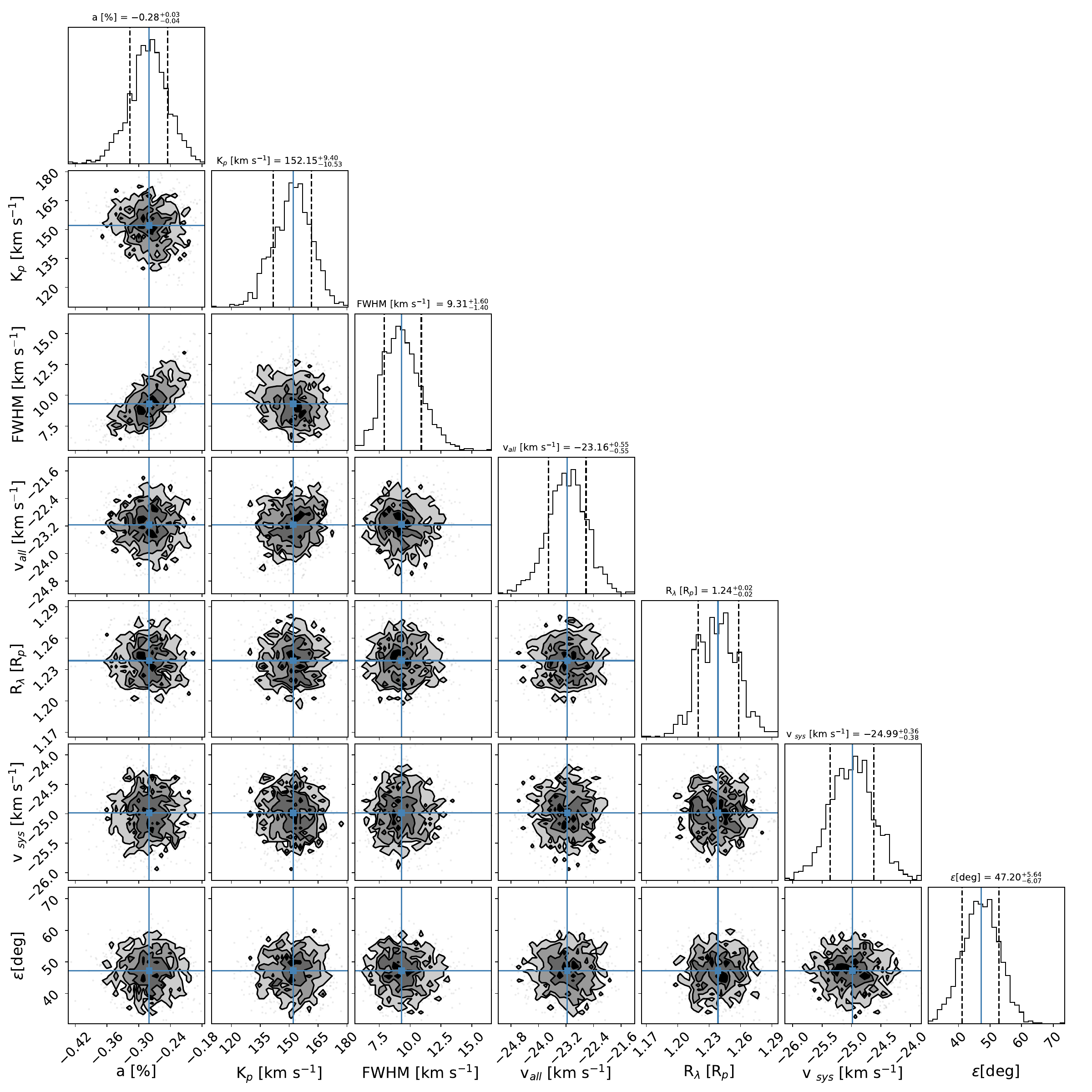}
\caption{Same as Figure \ref{fig:corner_k20_1}, but for \ion{Fe}{ii} $\lambda$5276 and $\lambda$5316. }
\label{fig:corner_k20_10}
\end{figure}

\begin{figure}[h]
\centering
\includegraphics[width=0.38\textwidth]{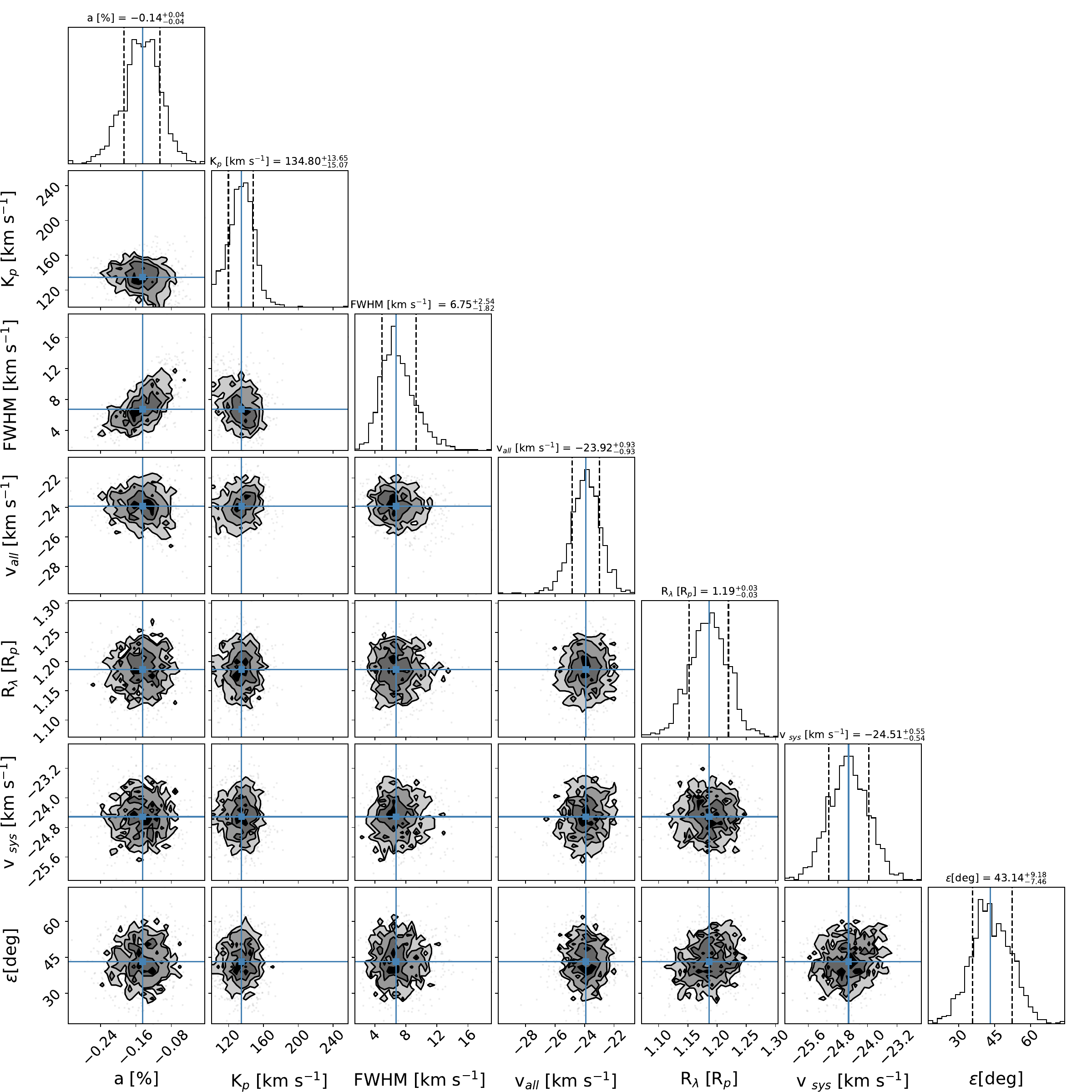}
\caption{Same as Figure \ref{fig:corner_k20_1}, but for \ion{Fe}{ii} $\lambda$5362.}
\label{fig:corner_k20_11}
\end{figure}

\begin{figure}[h]
\centering
\includegraphics[width=0.194\textwidth]{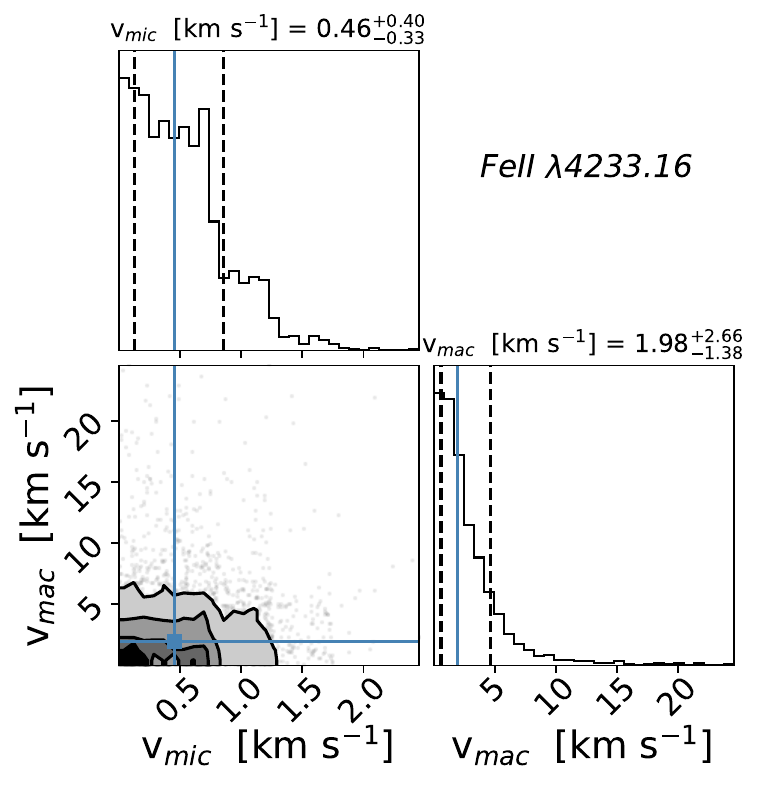}
\includegraphics[width=0.194\textwidth]{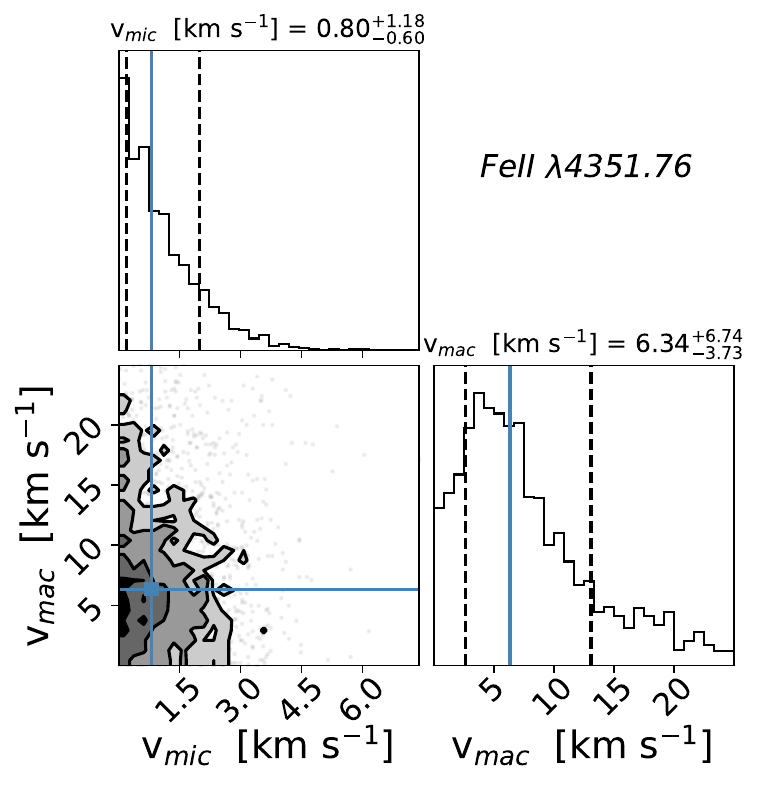}
\includegraphics[width=0.194\textwidth]{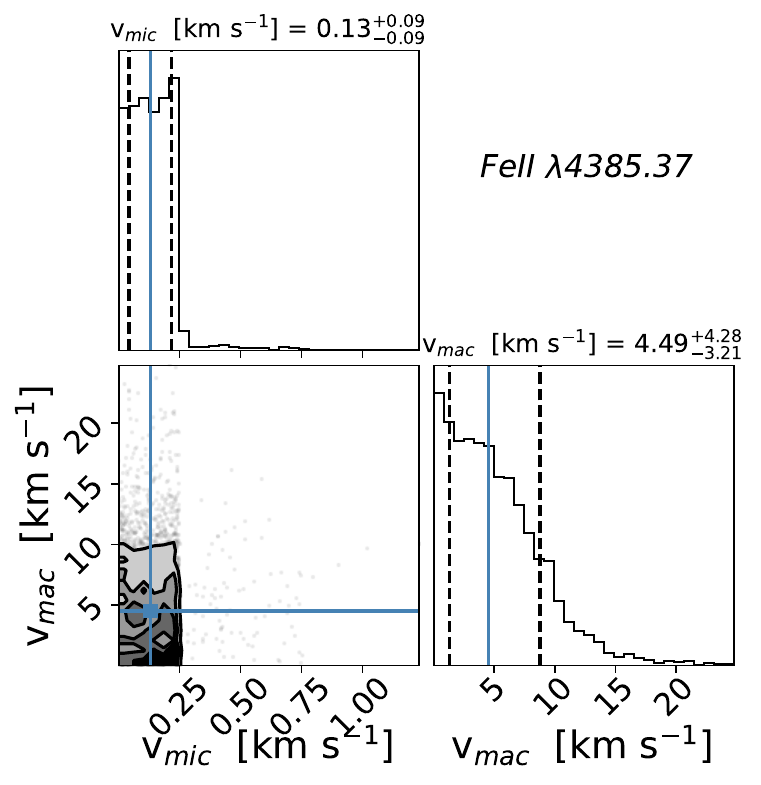}    
\includegraphics[width=0.194\textwidth]{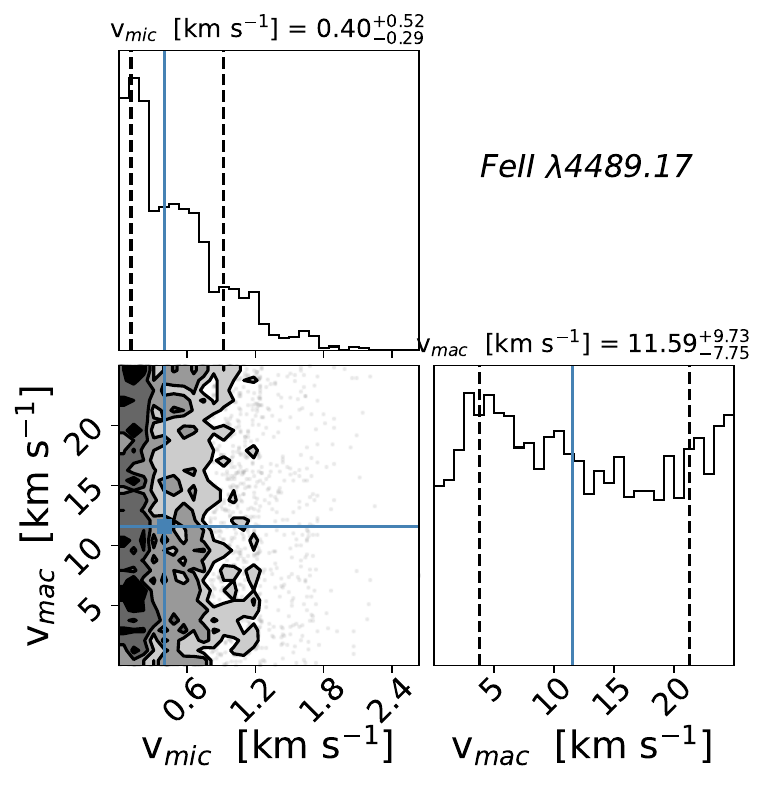}
\includegraphics[width=0.194\textwidth]{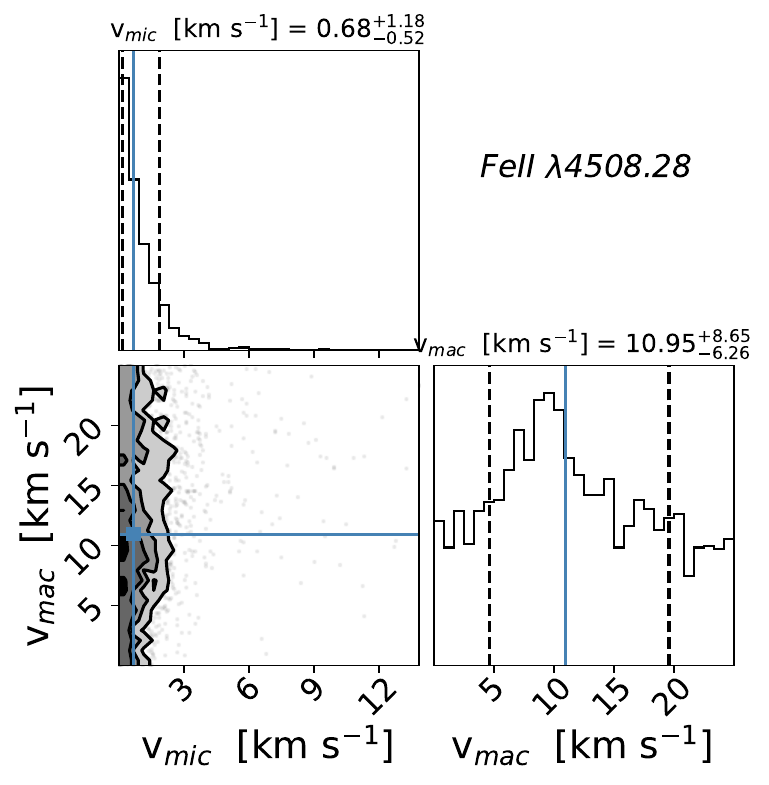}

\includegraphics[width=0.194\textwidth]{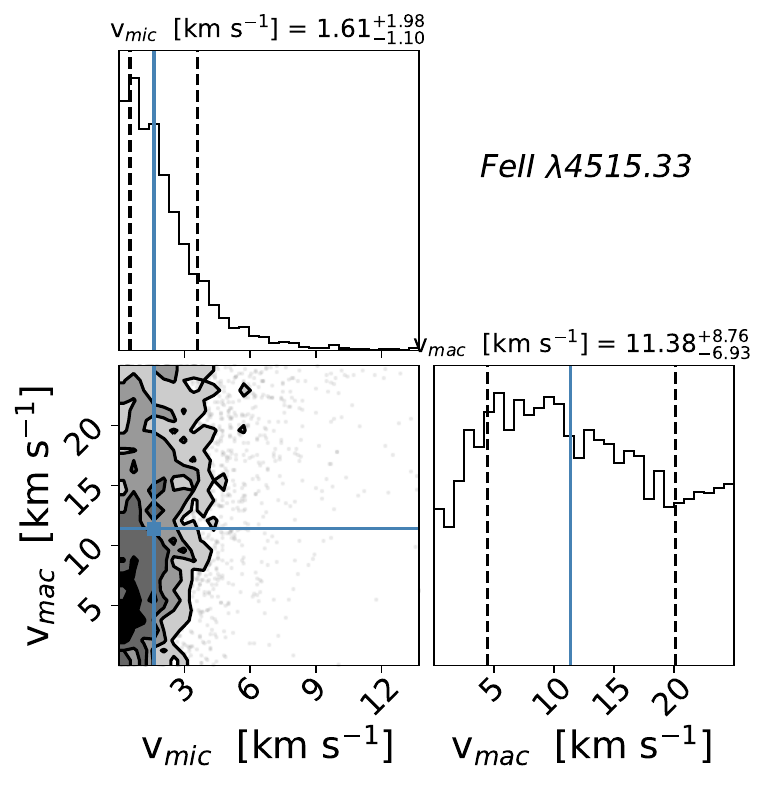}
\includegraphics[width=0.194\textwidth]{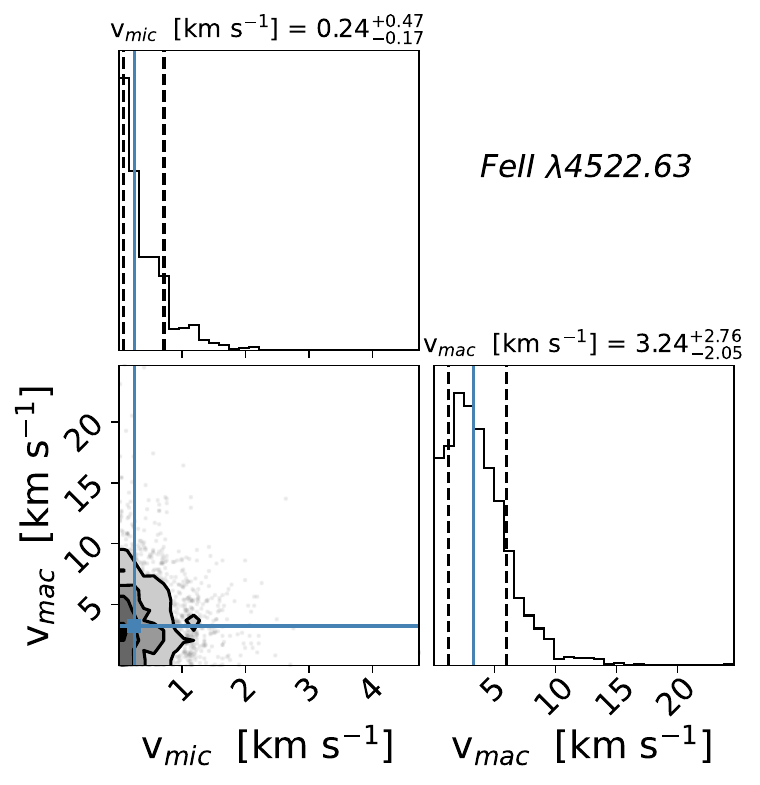}
\includegraphics[width=0.194\textwidth]{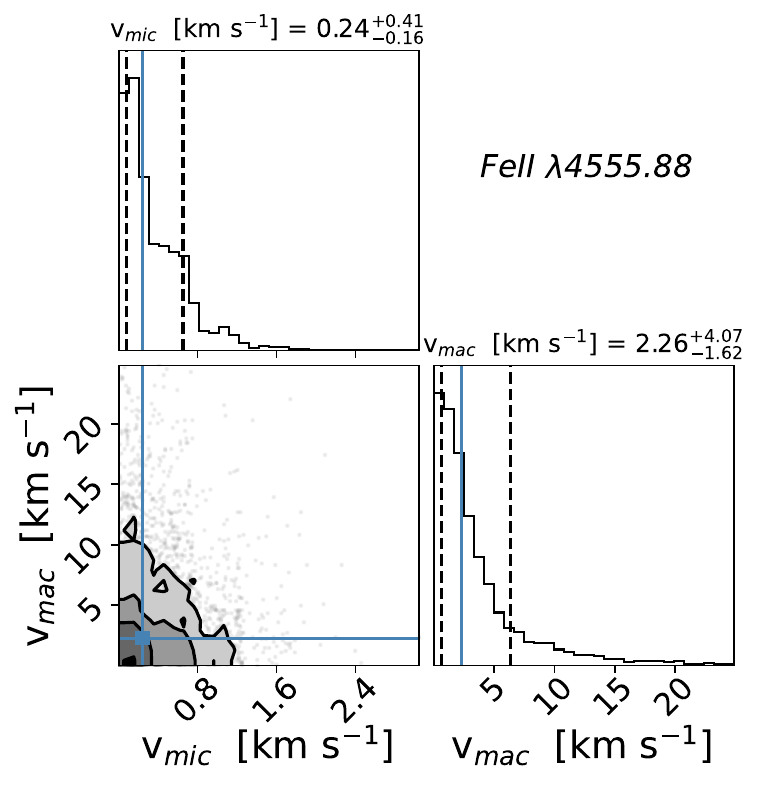}
\includegraphics[width=0.194\textwidth]{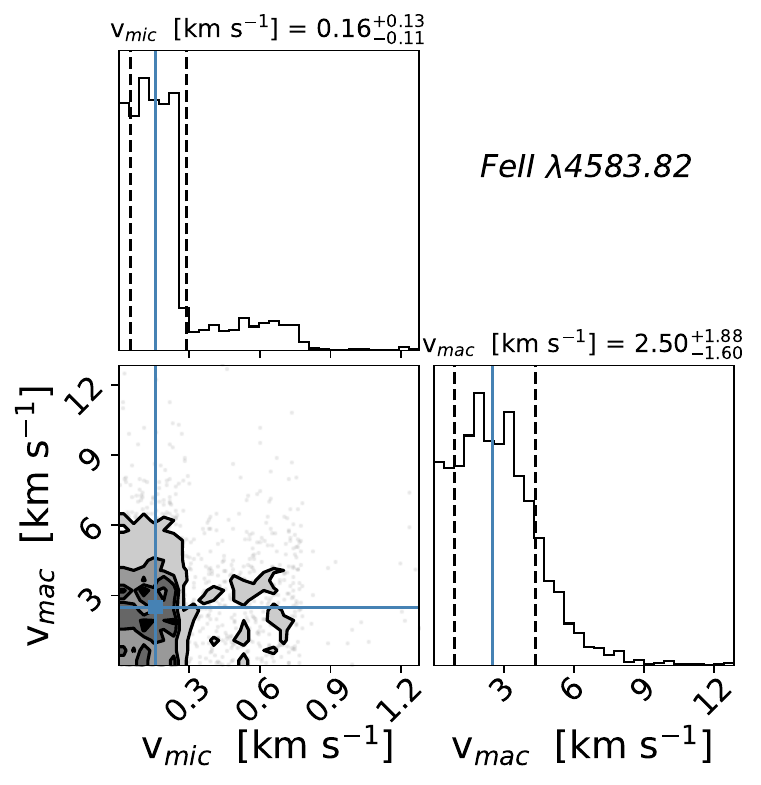}
\includegraphics[width=0.194\textwidth]{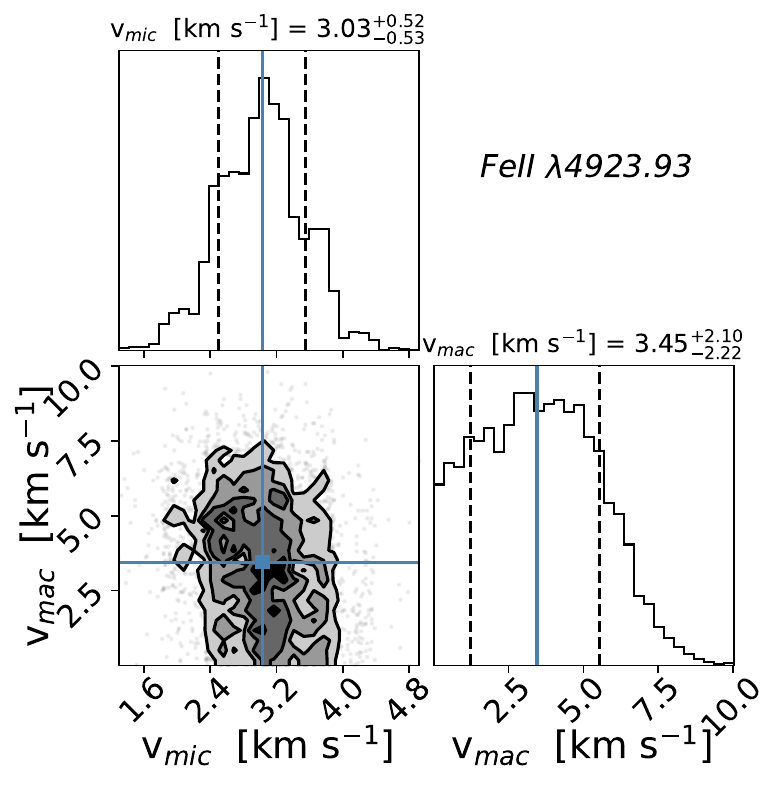}

\includegraphics[width=0.194\textwidth]{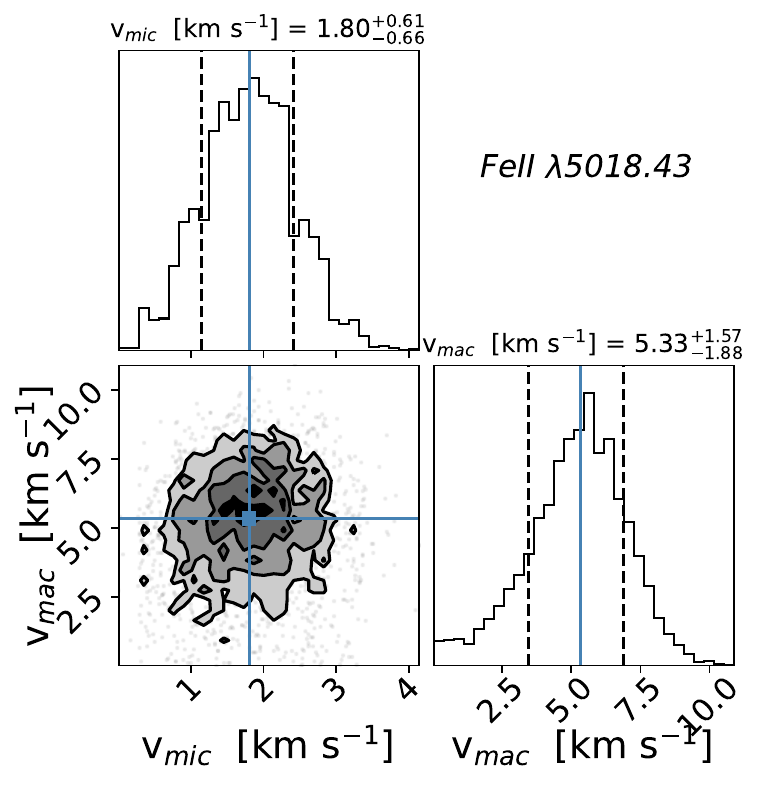}
\includegraphics[width=0.194\textwidth]{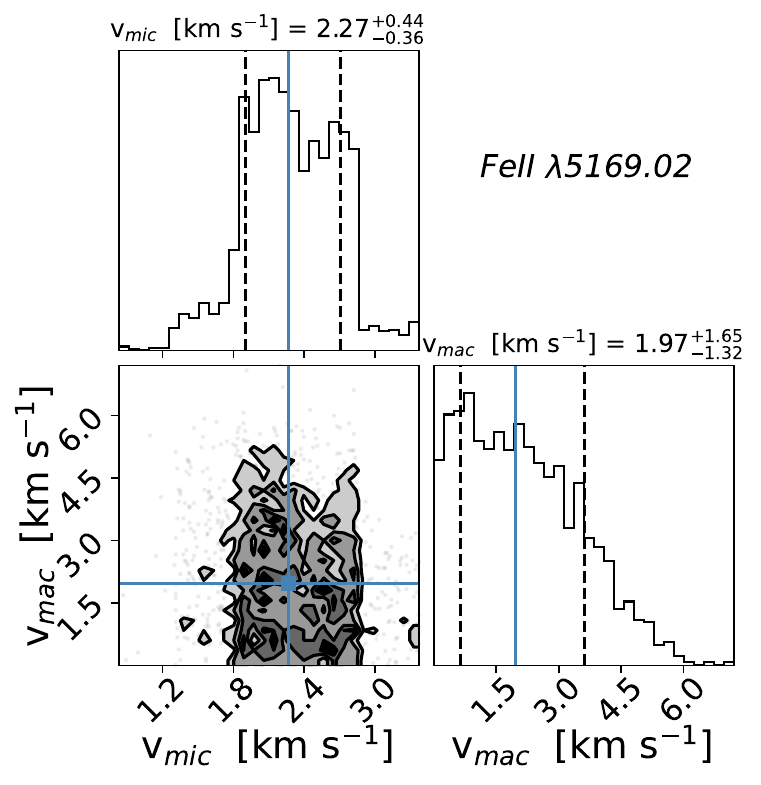}
\includegraphics[width=0.194\textwidth]{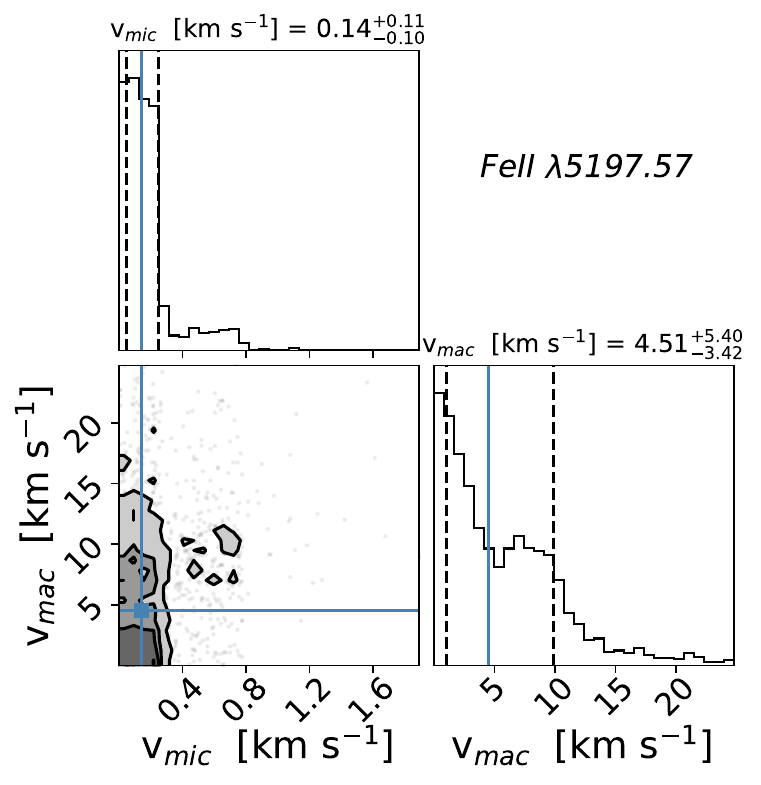}
\includegraphics[width=0.194\textwidth]{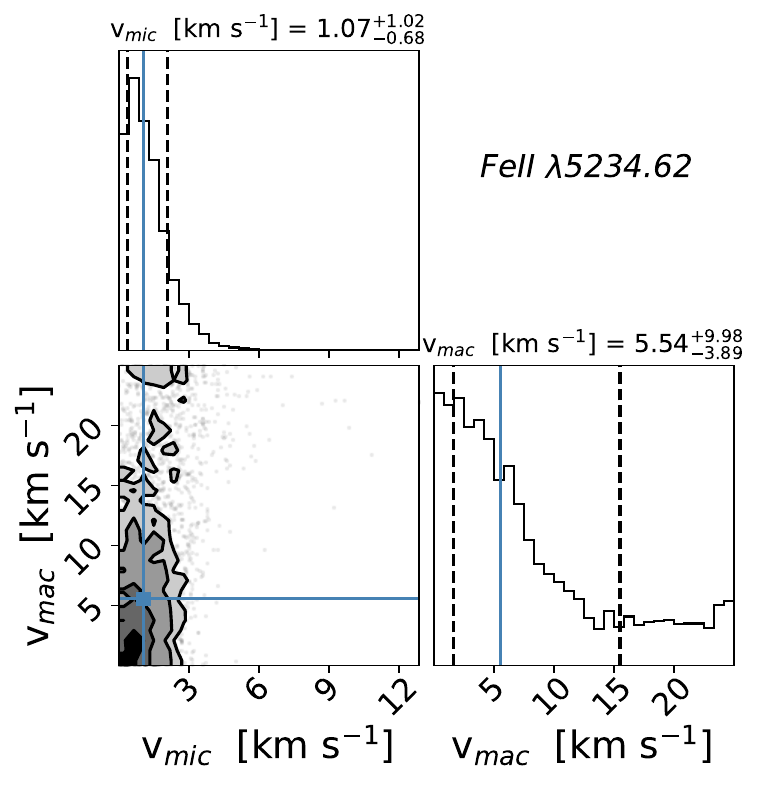}
\includegraphics[width=0.194\textwidth]{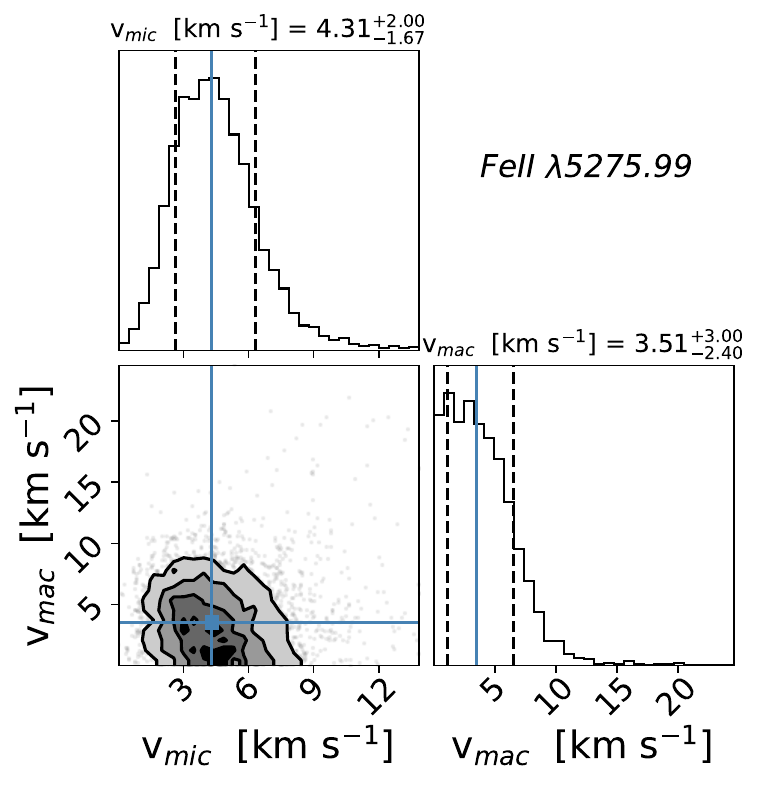}

\includegraphics[width=0.194\textwidth]{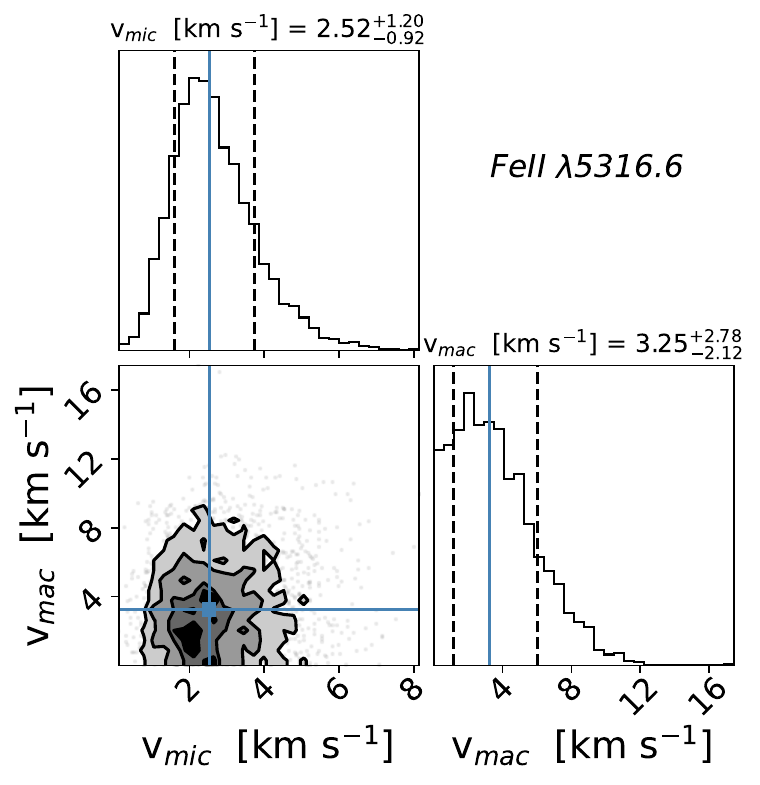}
\includegraphics[width=0.194\textwidth]{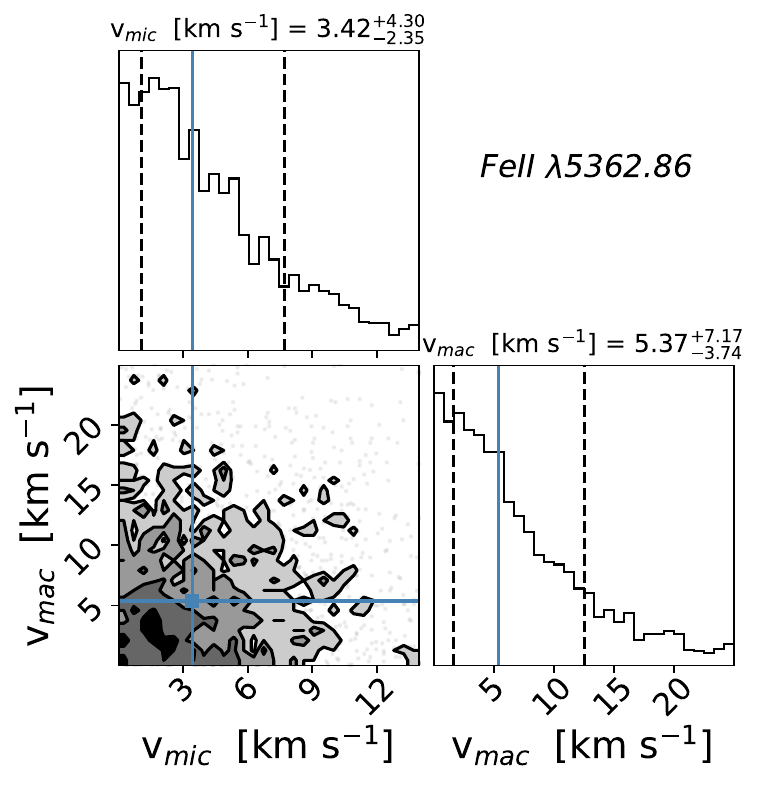}

\caption{The corner plot of the MCMC analysis considering $\nu_{\rm mac}$ and $\nu_{\rm mic}$ for KELT-20b for lines \ion{Fe}{ii} $\lambda$4233.16, \ion{Fe}{ii} $\lambda$4351.7, \ion{Fe}{ii} $\lambda$4385.3,  \ion{Fe}{ii} $\lambda$4489.1  \ion{Fe}{ii} $\lambda$4508.28, \ion{Fe}{ii} $\lambda$4515, \ion{Fe}{ii} $\lambda$4522,  \ion{Fe}{ii} $\lambda$4555, \ion{Fe}{ii} $\lambda$4583, \ion{Fe}{ii} $\lambda$4923, \ion{Fe}{ii} $\lambda$5018, and \ion{Fe}{ii} $\lambda$5169  and \ion{Fe}{ii} $\lambda$5197, \ion{Fe}{ii} $\lambda$5234, \ion{Fe}{ii} $\lambda$5275.66,  \ion{Fe}{ii} $\lambda$5316, and \ion{Fe}{ii} $\lambda$5362.86  }
\label{fig:M2_MCMC_VV_1}
\end{figure}

\begin{figure}[h]
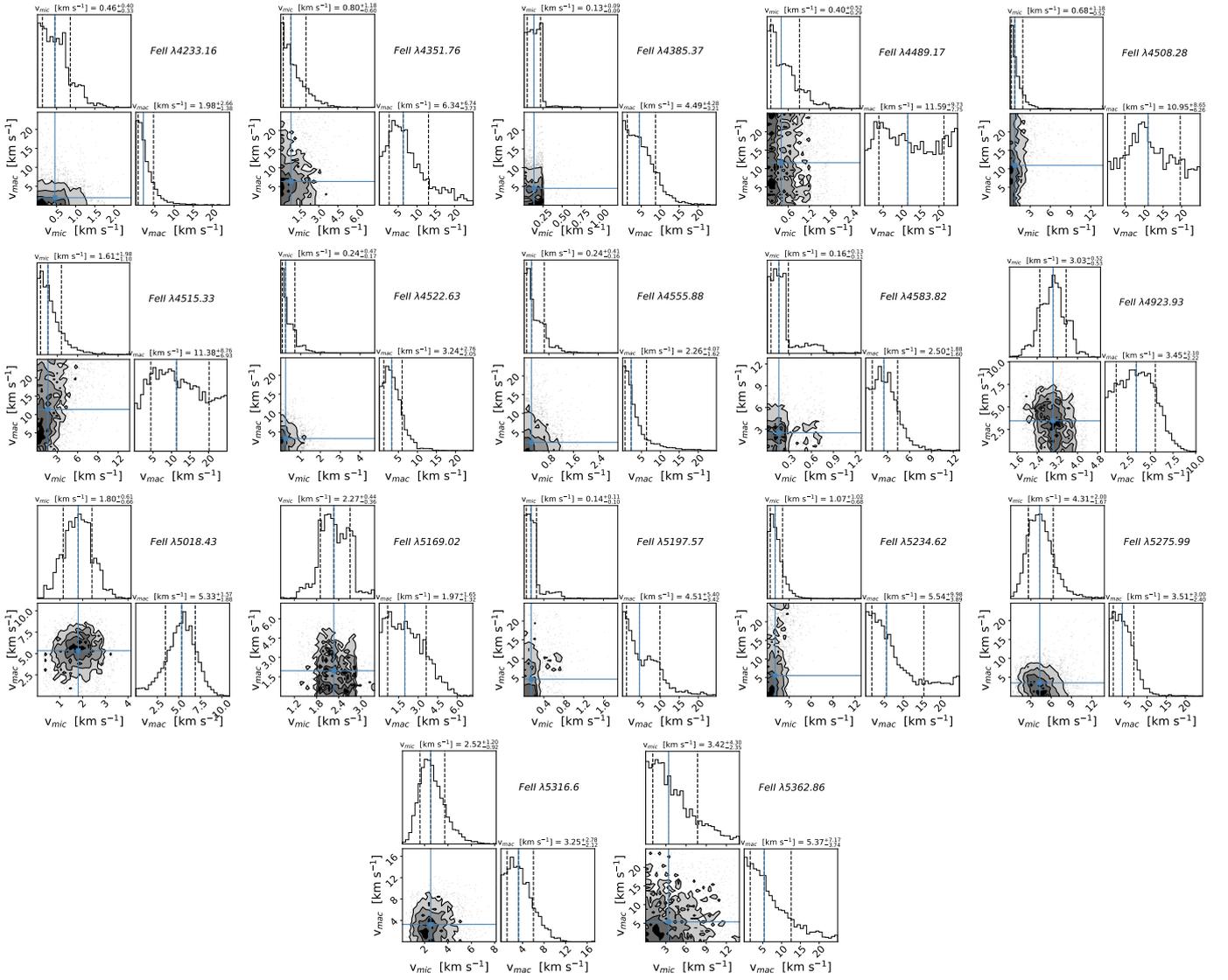

\centering

\caption{Same as Figure \ref{fig:M2_MCMC_VV_1}, but for lines  }
\label{fig:M2_MCMC_VV_2}
\end{figure}

\end{appendix}

\end{document}